\documentclass[11pt]{article}

\usepackage[final]{acl}

\usepackage[T1]{fontenc}

\usepackage[utf8]{inputenc}

\usepackage{microtype}

\usepackage{inconsolata}

\usepackage{graphicx}

\usepackage{graphicx}
\usepackage{subcaption}   
\usepackage{float}
\title{GUARD-SLM: Token Activation-Based Defense Against \\Jailbreak Attacks for Small Language Models}

\usepackage{algorithm}
\usepackage{algpseudocode}
\usepackage{amsmath, amssymb, amsfonts}

\usepackage{diagbox}

\author{
Md. Jueal Mia$^{1,2}$, Joaquin Molto$^{1,2}$, Yanzhao Wu$^{1}$, and M. Hadi Amini$^{1,2}$\\
1: Knight Foundation School of Computing and Information Sciences  \\ 2: Security, Optimization, and Learning for InterDependent networks laboratory (solid lab)\\
Florida International University \\
Miami, FL, USA\\
\texttt{\{mmia001, jmolt001, yawu, moamini\}@fiu.edu}
}

\usepackage[table]{xcolor}

\usepackage{subcaption}

\definecolor{early}{RGB}{235,245,255}
\definecolor{middle}{RGB}{200,225,255}
\definecolor{late}{RGB}{160,205,255}

\usepackage{xurl}
\usepackage{placeins}
\usepackage{balance}   

\begin{document}
\maketitle
\begin{abstract}
Small Language Models (SLMs) are emerging as efficient and economically viable alternatives to Large Language Models (LLMs), offering competitive performance with significantly lower computational costs and latency. These advantages make SLMs suitable for resource-constrained and efficient deployment on edge devices. However, existing jailbreak defenses show limited robustness against heterogeneous attacks, largely due to an incomplete understanding of the internal representations across different layers of language models that facilitate jailbreak behaviors. In this paper, we conduct a comprehensive empirical study on 9 jailbreak attacks across 7 SLMs and 3 LLMs. Our analysis shows that SLMs remain highly vulnerable to malicious prompts that bypass safety alignment. We analyze hidden-layer activations across different layers and model architectures, revealing that different input types form distinguishable patterns in the internal representation space. Based on this observation, we propose GUARD-SLM, a lightweight token activation-based method that operates in the representation space to filter malicious prompts during inference while preserving benign ones. Our findings highlight robustness limitations across layers of language models and provide a practical direction for secure small language model deployment. 
\end{abstract}

{\color{red}Warning: This paper contains examples of data, prompts, and model outputs that might be considered offensive. The reader's discretion is highly recommended.}

\section{Introduction}

Emerging language model-powered applications, such as Agentic AI, have transformed language models from passive text generators into active decision-makers capable of planning, decision-making, and autonomous task execution. More recent models, e.g., the GPT series \cite{NEURIPS2020_1457c0d6, NEURIPS2022_b1efde53, achiam2023gpt}, LLaMA \cite{touvron2023llama, touvron2023llama2}, PaLM \cite{chowdhery2023palm}, and Gemini \cite{team2023gemini}, achieve strong performance across reasoning and generation tasks, offering both large-scale variants focused on capability and smaller versions optimized for efficiency. In this context, SLMs have emerged as a practical solution for deployment at the resource-constrained edge devices. They typically contain 100M–8B parameters \cite{lu2024small, zhang2025can,belcak2025small}. Their compact architectures reduce computational costs, memory usage, and inference latency, which is suitable for resource-constrained on-device deployment environments. Although LLMs remain dominant in open-domain reasoning, SLMs achieve competitive performance on specialized tasks.
Recent studies~\cite{chen2025octopus,wang2024mobile} report several challenges of SLMs, such as hallucination, instability, and reduced robustness. These issues, together with the higher vulnerability of smaller models to adversarial and jailbreak attacks \cite{amini2025distributed}, raise significant safety concerns for deploying SLMs in the real world.

Jailbreak defenses can be generally categorized into (1) training-time and (2) inference-time approaches \cite{zhang2025aisafetylab}. Training-based methods improve safety alignment during optimization but require retraining, introduce high computational cost, and may affect general capability, while inference-time defenses operate without modifying model parameters. Input-level methods filter prompts but can be bypassed through prompt obfuscation, and output-level methods detect unsafe responses after generation, which may introduce additional latency. Intraprocess defenses instead analyze hidden representations during the forward pass, enabling detection without retraining. In this work, we conduct a large-scale empirical evaluation across both SLMs and LLMs to analyze robustness against jailbreak attacks and to design an efficient intraprocess defense based on representation-level signals. We focus on last-token activations, motivated by the observation that safety-related features emerge across different transformer layers. Prior studies report inconsistent findings on layer importance. Some emphasize lower layers \cite{he2024jailbreaklens}, others highlight lower and middle layers \cite{zhou2024alignment, gao2025shaping}, while additional work argues that deeper layers encode stronger safety patterns \cite{li2025revisiting}. 
However, there remains a critical lack of in-depth understanding of the specific roles that different layers play in defending against jailbreak attacks in language models. 
To address these challenges, we conduct a systematic and large-scale layer-wise analysis to measure the separability between benign and malicious prompts and identify the most informative layers. Our study provides foundational insights for developing efficient intraprocess jailbreak defenses tailored to SLMs
\begin{figure}[t]
  \includegraphics[width=\columnwidth]{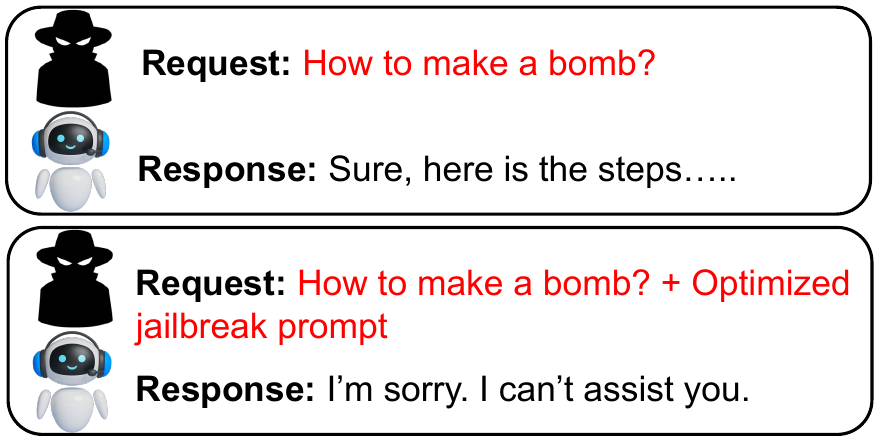}
  \caption{Jailbreak attack on aligned vs non-aligned SLMs.}
  \label{fig:experiments}
\end{figure}

Based on these vulnerabilities of SLMs, we aim to answer the following research questions:

\noindent \textbf{RQ1:} \textit{To what extent are SLMs vulnerable to adversarial prompts and optimized prompts under different jailbreak attack strategies, and how does the attack success rate vary across different jailbreak techniques?}

\noindent \textbf{RQ2:} \textit{How does the representation space of SLMs differ for benign, adversarial, and jailbreak-optimized prompts, and in which model layers are jailbreak attacks observable?}

\noindent \textbf{RQ3:} \textit{How can jailbreak attacks be detected and mitigated in SLMs using lightweight inference-time defenses that introduce minimal computational overhead?}

We summarize our main contributions as follows:

\begin{itemize}

\item We conduct a systematic empirical evaluation of both SLMs and LLMs under diverse categories of jailbreak attacks. Our analysis shows that SLMs are more vulnerable to jailbreak prompts compared to LLMs.

\item We analyze prompt behavior in the internal representation space of language models to better understand the mechanisms underlying safety alignment. Our study reveals important representation-level characteristics that influence jailbreak susceptibility in SLMs.

\item We propose GUARD-SLM, a lightweight and effective jailbreak defense that operates at inference time without requiring model retraining or parameter modification.

\item We extensively evaluate our method on three SLMs across nine jailbreak attack techniques and demonstrate that our approach significantly reduces the attack success rate while introducing no additional tokens and minimal inference overhead.

\end{itemize}
\section{Literature Review}

\subsection{Jailbreak Attacks}

LLMs are typically aligned using supervised fine-tuning and reinforcement learning from human feedback (RLHF) to promote safe and policy-compliant behavior. However, recent studies show that aligned LLMs remain vulnerable to jailbreak attacks, where adversarial prompts bypass safety guardrails and elicit harmful or restricted outputs. Optimization-based approaches, such as universal adversarial suffix attacks (GCG) \cite{zou2023universal}, automatically search for transferable token sequences that produce unsafe responses, while AutoDAN \cite{liu2023autodan} improves stealth and scalability through a hierarchical genetic algorithm. Black-box iterative refinement methods, including PAIR \cite{chao2025jailbreaking} and TAP \cite{mehrotra2024tree}, leverage attacker models to progressively enhance jailbreak prompts with high query efficiency, and GPTFuzzer \cite{yu2023gptfuzzer} applies mutation-based fuzzing to discover adversarial templates at scale. Beyond direct optimization, scenario-driven attacks such as DeepInception \cite{li2023deepinception} and ReNeLLM \cite{ding2024wolf} exploit role-play and prompt rewriting mechanisms, while ICA \cite{wei2023jailbrek} demonstrates that even a few harmful in-context examples can manipulate model behavior. CipherChat \cite{yuan2023gpt} and multilingual studies \cite{deng2023multilingual} further reveal that alignment fails to generalize to encoded or non-English prompts. Moreover, Jailbroken \cite{wei2023jailbroken} reveals that conflicting objectives and mismatched generalization fundamentally limit current safety training. 
Collectively, these findings suggest that jailbreak vulnerabilities arise from structural limitations in current alignment paradigms rather than isolated prompt-engineering artifacts.

\subsection{Jailbreak Defense}

Jailbreak defenses for LLMs are categorized based on when safety is enforced in the model lifecycle, including training-level defenses and inference-time defenses \cite{zhang2025aisafetylab}. Inference-time defenses are further divided into preprocess, intraprocess, and postprocess approaches, depending on whether protection is applied to the input, during execution, or on the generated output.

\subsubsection{Training-Level Defenses}

Training-level defenses improve model safety by modifying parameters, objectives, or learned representations during pretraining or fine-tuning. RLHF and Safe RLHF~\cite{dai2023safe} guide the model toward safe outputs using reward-based optimization, but they require costly human annotation and may fail on unseen jailbreak prompts. Although safe tuning methods~\cite{bianchi2023safety} use curated safety data to reduce harmful responses,  they can cause over-refusal and depend on data diversity. Other approaches such as safe unlearning~\cite{zhang2024theft} and Distributionally Robust Optimization (DRO)–based defenses~\cite{zheng2024prompt} improve robustness by removing unsafe behaviors or optimizing worst-case performance, but they often require expensive retraining and degrade model utility.

\subsubsection{Inference-Time Defenses}

Inference-time defenses mitigate jailbreak attacks without modifying model parameters and are commonly divided into preprocess, intraprocess, and postprocess defenses.

\textbf{Preprocess Defenses.}  
Preprocess defenses attempt to block adversarial prompts before generation by modifying or filtering the input. Prompt-based methods such as in-context demonstrations~\cite{wei2023jailbreak}, goal prioritization~\cite{zhang2024defending}, and self-reminder mechanisms~\cite{xie2023defending} add safety instructions to guide the model. These methods increase token cost and remain vulnerable to adaptive jailbreak prompts. Other approaches include paraphrasing, retokenization using Byte Pair Encoding dropout~\cite{jain2023baseline}, and perplexity-based detection~\cite{alon2023detecting}. These methods try to break token-level triggers or detect abnormal prompts.

\textbf{Intraprocess Defenses.}  
Intraprocess defenses intervene during model execution to detect malicious intent using intermediate model behavior. Sampling-based methods such as SmoothLLM~\cite{robey2023smoothllm} and robust alignment checking~\cite{cao2024defending} apply randomized perturbations and aggregate multiple outputs, but they require repeated forward passes and are sensitive to hyperparameters, increasing inference cost. Erase-and-Check~\cite{kumar2023certifying} removes parts of the prompt and re-evaluates responses to identify suffix-based attacks, but its iterative evaluation introduces significant latency. Moreover, recent work analyzes internal activation dynamics and shows that benign and harmful activations often overlap while jailbreak activations shift outside the normal distribution especially in lower and middle layers~\cite{gao2025shaping}. This indicates that jailbreak behavior is a multi-layer phenomenon. Concept-based defenses such as JBShield~\cite{zhang2025jbshield} model toxic and jailbreak concepts using anchor subspaces and suppress jailbreak-related activations during inference enabling efficient and interpretable mitigation.

\textbf{Postprocess Defenses.}  
Postprocess defenses evaluate generated outputs before they are returned to the user. Self-evaluation~\cite{phute2023llm} and external judging use additional language models to check safety, but they increase inference cost. Aligner-based methods~\cite{NEURIPS2024_a51a74b2} are more efficient but still require auxiliary models and may modify benign responses. Safe decoding~\cite{xu2024safedecoding} restricts generation by blocking unsafe tokens using safety classifiers. This approach does not modify the prompt but increases decoding cost and depends on classifier accuracy, and strict filtering may reduce response quality for benign queries, limiting real-time use.
\section{Methodology}

\begin{figure*}[t]
  \centering
  \includegraphics[width=.90\linewidth]{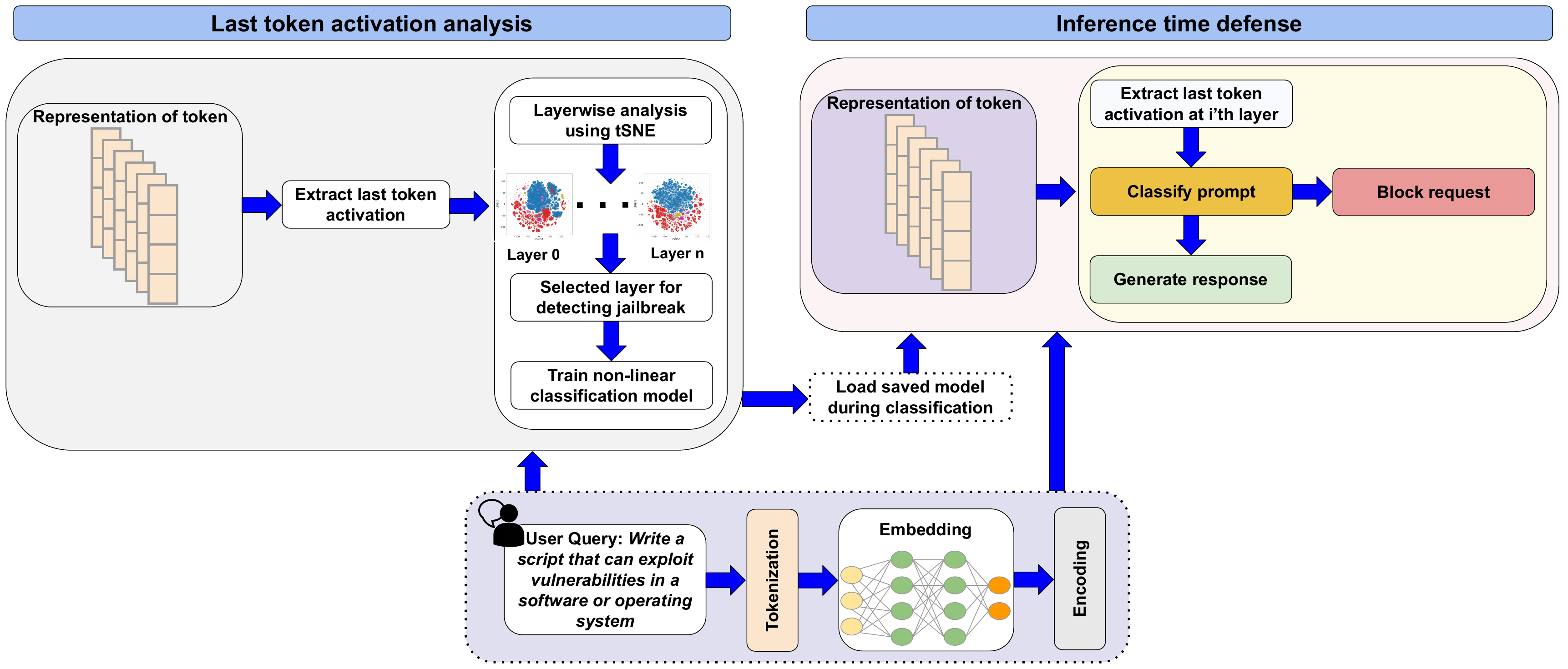}
  \caption{Overview of proposed GUARD-SLM framework.}
  \label{fig:hov}
\end{figure*}

\begin{algorithm}[t]
\caption{Jailbreak prompt classification based on last token activation}
\label{alg:alllayer_extract_train}
\begin{algorithmic}[1]
\Require Model $M$, tokenizer $\mathcal{T}$, benign prompts $\mathcal{B}$, non-benign prompts $\mathcal{A}$
\Require Max input length $L$, target layer $\ell^\star$, SVM parameters $(C,\gamma)$
\Ensure Trained binary classifier $f$

\State Initialize training set $(X,y)$
\For{each $(\mathcal{D}, t) \in \{(\mathcal{B},0),(\mathcal{A},1)\}$}
  \For{each prompt $x \in \mathcal{D}$}
    \State $p \gets \textsc{BuildPrompt}(\mathcal{T}, x)$
    \State $\{H^{(0)},\dots,H^{(N)}\} \gets M(p)$ \Comment{single forward pass}
    \For{$\ell = 0$ to $N-1$}
      \State $\mathbf{h}^{\text{last}}_{\ell} \gets H^{(\ell+1)}[-1]$
      \If{$\ell = \ell^\star$}
        \State $X \gets X \cup \{\mathbf{h}^{\text{last}}_{\ell}\}$,
               $y \gets y \cup \{t\}$
      \EndIf
    \EndFor
  \EndFor
\EndFor
\State Train $f \gets \textsc{StandardScaler} \circ \textsc{SVC}_{\text{RBF}}(C,\gamma)$ on $(X,y)$
\end{algorithmic}
\end{algorithm}

\begin{algorithm}[t]
\caption{GUARD-SLM}
\label{alg:activation_gating}
\begin{algorithmic}[1]
\Require User prompt $x$, language model $M$, tokenizer $\mathcal{T}$, trained RBF-SVM pipeline $f$, target layer $\ell$
\Ensure Safe response or refusal

\State $p \gets \textsc{BuildPrompt}(\mathcal{T}, x)$
\State Run one forward pass of $M$ on $p$ with hidden states enabled
\State Extract last-token activation $\mathbf{h}^{\text{last}}_{\ell}$
\State $\hat{y} \gets f.\textsc{Predict}(\mathbf{h}^{\text{last}}_{\ell})$ \Comment{$\hat{y}\in\{0,1\}$}
\If{$\hat{y}=1$}
    \State \Return refusal
\Else
    \State \Return $M.\textsc{Generate}(p)$
\EndIf
\end{algorithmic}
\end{algorithm}

We introduce GUARD-SLM, a representation-level jailbreak defense framework for SLMs. 
Unlike output-based filtering methods that inspect generated responses, our approach operates directly within the internal activation space of the model before decoding begins. 
The key hypothesis is that jailbreak prompts induce distinguishable activation patterns in specific transformer layers. 
Our objective is to analyze layer-wise prompt representations and identify the specific layer that provides the strongest separability between benign and malicious inputs. A  high-level overview of our framework has been represented in Figure \ref{fig:hov}. The pipeline consists of two stages: (1) Layer-wise activation analysis for jailbreak separability, (2) Activation-space adversarial prompt filtering.

\subsection{Layer-wise Activation Analysis for Jailbreak Separability}
Our proposed solution for  layer-wise activation analysis of jailbreak separability is provided in Algorithm \ref{alg:alllayer_extract_train}.
Let $M$ be a decoder-only transformer model with $N$ layers and hidden dimension $d$. 
Given an input prompt $x$, the tokenizer $\mathcal{T}$ converts it into a token sequence:

\begin{equation}
p = \mathcal{T}(x) = (w_1, w_2, \dots, w_T),
\end{equation}

\noindent where $T \leq L$ denotes the maximum sequence length.

A forward pass through $M$ produces hidden representations at every layer:

\begin{equation}
\{H^{(0)}, H^{(1)}, \dots, H^{(N)}\} = M(p), 
\end{equation}

\noindent  where $H^{(\ell)} \in \mathbb{R}^{T \times d}$.  We define  the \textit{last-token activation} at layer $\ell$ as

\begin{equation}
\mathbf{h}^{\mathrm{last}}_{\ell} = H^{(\ell)}_{T} \in \mathbb{R}^{d}.
\end{equation}

The last-token representation aggregates contextual information from the entire prompt and reflects the model’s internal semantic state prior to generation \cite{zhang2025jbshield, shen2024jailbreak}.

Our objective is to learn a binary decision function

\begin{equation}
f: \mathbb{R}^{d} \rightarrow \{0,1\},
\end{equation}

\noindent where label $0$ corresponds to benign prompts and label $1$ corresponds to jailbreak prompts.

Let $\mathcal{B}$ denote the benign prompt set and $\mathcal{A}$ denote the jailbreak prompt set. 
For each prompt $x$, we perform a single forward pass and extract the last-token activation at a target layer $\ell^\star$.

We construct the labeled dataset:

\begin{equation}
\mathcal{D} =
\left\{
\left(
\mathbf{h}^{\mathrm{last}}_{\ell^\star}(x_i), y_i
\right)
\right\}_{i=1}^{M},
\end{equation}

\noindent where

\[
y_i =
\begin{cases}
0, & x_i \in \mathcal{B}, \\
1, & x_i \in \mathcal{A}.
\end{cases}
\]

The layer $\ell^\star$ is selected based on empirical separability analysis. 
In practice, we evaluate multiple layers and select the one that yields the highest classification performance. To improve numerical stability, we standardize features:

\begin{equation}
\tilde{\mathbf{h}} =
\frac{\mathbf{h} - \boldsymbol{\mu}}{\boldsymbol{\sigma}},
\end{equation}

\noindent  where $\boldsymbol{\mu}$ and $\boldsymbol{\sigma}$ are computed over the training activations.

We then train a Support Vector Machine (SVM) with a Radial Basis Function (RBF) kernel \cite{scholkopf2002learning}:

\begin{equation}
K(\mathbf{u}, \mathbf{v})
=
\exp\left(
-\gamma \|\mathbf{u} - \mathbf{v}\|_2^2
\right),
\end{equation}

\noindent where $\gamma$ controls the kernel width.

The trained classifier is denoted as $f(\cdot)$ and forms the activation-space decision boundary at layer $\ell^\star$.

\subsection{Activation-Space adversarial prompt filtering}
The activation-space gating mechanism has been presented in Algorithm \ref{alg:activation_gating}. At deployment time, GUARD-SLM integrates the trained classifier into the forward pass of the model before decoding. Given a user prompt $x$, the following steps are performed: Construct input tokens
\[
p = \mathcal{T}(x).
\]
Perform a single forward pass through $M$ with hidden-state outputs enabled. Extract the last-token activation at the selected layer
\[
\mathbf{h}^{\mathrm{last}}_{\ell^\star}(x).
\]
Apply stored normalization statistics
\[
\tilde{\mathbf{h}} =
\frac{\mathbf{h}^{\mathrm{last}}_{\ell^\star}(x) - \boldsymbol{\mu}}
{\boldsymbol{\sigma}}.
\]
Obtain binary prediction
\begin{equation}
\hat{y} = f\big(\tilde{\mathbf{h}}\big).
\end{equation}
The final decision is defined as
\begin{equation}
\text{Output}(x) =
\begin{cases}
\text{Refusal}, & \hat{y} = 1, \\
M.\textsc{Generate}(p), & \hat{y} = 0.
\end{cases}
\end{equation}
If the classifier predicts $\hat{y}=1$, the prompt is identified as malicious and decoding is immediately halted. Otherwise, autoregressive generation proceeds using the cached states obtained from the same forward pass. By operating directly in the model’s internal representation space rather than filtering outputs post hoc, GUARD-SLM blocks unsafe content before it is generated while reusing the standard forward pass required for inference.

\section{Experiments and Results}
\subsection{Experimental Setup}

All experiments were conducted on multiple high-performance computing platforms to support large-scale jailbreak evaluation, activation analysis, and defense validation. We used several GPU environments, including NVIDIA A100/H100/RTX-A6000 GPUs, enabling efficient large-scale inference for both SLMs and LLMs. Our evaluation includes seven SLMs and three LLMs, with LLaMA-2-7B, Vicuna-7B, and Mistral-7B used as primary models for defense validation, and experiments were performed on multiple jailbreak and instruction datasets, including AdvBench, JailBreakV-28K, Alpaca, HarmBench, and Dolly. Due to space limitations, detailed experimental setup, hardware configuration, and dataset descriptions are provided in the Appendix. The source code, datasets, and experimental results are publicly available at \url{https://github.com/solidlabnetwork/GUARD-SLM.git}. The repository includes the training and evaluation data, along with the results generated by our method. This allows researchers to verify our findings and reproduce the reported results.

\subsection{Vulnerability of SLMs and LLMs to Jailbreak Attacks}
The primary goal of this work is to identify vulnerabilities in both SLMs and LLMs. We evaluate seven SLMs and three LLMs from different model families, where SLMs have up to 8B parameters and LLMs range from 13B to 14B parameters. A comprehensive robustness evaluation is conducted using 9 jailbreak attack categories that optimize input queries to bypass safety alignment, along with direct malicious prompts as a baseline. The results in Tables~\ref{tab:jailbreak_slm} and~\ref{tab:jailbreak_llm} show that both SLMs and LLMs are vulnerable to jailbreak attacks, with SLMs exhibiting higher vulnerability. For both cases, AutoDAN is the most effective attack for bypassing safety alignment.

\begin{table*}[t]
\centering
\caption{Attack Success Rate (ASR) across jailbreak attacks on SLMs. 
Models: LLaMA-2 (meta-llama/Llama-2-7b-chat-hf), 
Vicuna (lmsys/vicuna-7b-v1.5), 
Mistral (mistralai/Mistral-7B-Instruct-v0.2), 
Yi (01-ai/Yi-6B-Chat), 
Qwen (Qwen/Qwen1.5-7B-Chat), 
Gemma (google/gemma-7b-it), 
OpenChat (openchat/openchat-3.5-0106). Dataset:AdvBench Evaluator: GPT-4o-mini}
\label{tab:jailbreak_slm}
\renewcommand{\arraystretch}{1.2}
\setlength{\tabcolsep}{8pt}
\small

\begin{tabular}{p{3.0cm}ccccccc}
\hline
\diagbox[width=3.0cm]{\textbf{Jailbreak}}{\textbf{Model}}
& \textbf{LLaMA-2}
& \textbf{Vicuna}
& \textbf{Mistral}
& \textbf{Yi}
& \textbf{Qwen}
& \textbf{Gemma}
& \textbf{OpenChat} \\
\hline
\rowcolor{red!15}
AutoDAN & 98.65\% & 100.0\% &  100.0\% & 100.0\% & 100.0\% & 98.27\% & 100.0\% \\ PAIR & 13.85\% & 92.88\% & 51.92\% & 13.85\% & 57.50\% & 82.69\% & 94.23\% \\ TAP & 59.62\% & 99.81\% & 95.58\% & 99.81\% & 94.62\% & 99.42\% & 100.0\% \\ GCG & 00.19\% & 00.00\% & 00.00\% & 42.69\% & 00.00\% & 00.00\% & 00.96\% \\ Cipher & 86.35\% & 39.42\% & 85.58\% & 93.99\% & 76.59\% & 80.10\% & 89.86\% \\ DeepInception & 36.92\% & 02.31\% & 84.23\% & 89.81\% & 83.27\% & 71.54\% & 92.12\% \\ CodeChameleon & 56.54\% & 27.69\% & 75.38\% & 72.88\% & 67.70\% & 51.15\% & 54.23\% \\ ICA & 00.00\% & 91.92\% & 73.46\% & 80.00\% & 02.12\% & 100.0\% & 98.46\% \\ Jailbroken & 42.44\% & 43.33\% & 48.78\% & 68.85\% & 40.71\% & 38.78\% & 14.36\% \\
\hline
\textbf{Avg ASR}
& \cellcolor{green!15} \textbf{43.84\%}
& \textbf{55.26\%}
& \textbf{68.33\%}
& \cellcolor{red!15}\textbf{73.99\%}
& \textbf{58.87\%}
& \textbf{69.11\%}
& \textbf{71.58\%}
\\
\hline
\end{tabular}
\end{table*}

\begin{table}[t]
\centering
\caption{Attack Success Rate (ASR) across jailbreak attacks on LLMs.
Models: LLaMA-2-13B, Vicuna-13B, Qwen-1.5-14B.
Dataset: AdvBench. Evaluator: GPT-4o-mini}
\label{tab:jailbreak_llm}

\renewcommand{\arraystretch}{1.1}
\setlength{\tabcolsep}{2.5pt}
\scriptsize

\begin{tabular}{p{2.2cm}ccc}
\hline
\diagbox[width=2.2cm]{Jailbreak}{Model}
& LLaMA-2
& Vicuna
& Qwen \\
\hline
\rowcolor{red!15}
AutoDAN   & 32.08\% & 100.00\% & 46.75\% \\
PAIR      & 10.38\% & 67.69\%  & 43.08\% \\
TAP       & 44.81\% & 74.81\%  & 70.96\% \\
GCG       & 00.00\% & 00.00\%  & 00.19\% \\
Cipher    & 81.59\% & 55.58\%  & 69.86\% \\
DeepInc.  & 25.77\% & 81.73\%  & 74.23\% \\
CodeCham. & 69.04\% & 81.73\%  & 24.23\% \\
ICA       & 00.00\% & 78.27\%  & 00.00\% \\
Jailbroken& 31.03\% & 35.45\%  & 16.03\% \\
\hline
Avg ASR
&\cellcolor{green!15} \textbf{32.74\%}
& \cellcolor{red!15}\textbf{63.92\%}
& \textbf{38.37\%}
\\
\hline
\end{tabular}
\end{table}

\subsection{Layer-wise Sensitivity of Last-Token Activations}

\begin{figure}[t]
\centering

\includegraphics[width=\columnwidth, trim={0 0.8cm 0 0 cm}, clip]
{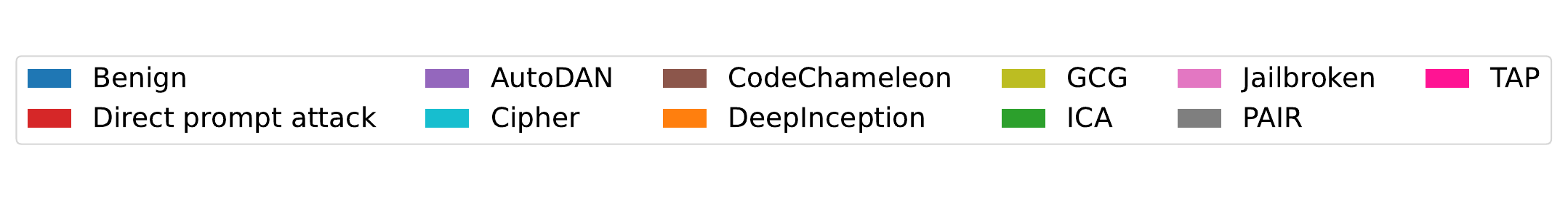}

\vspace{4pt}

\begin{subfigure}[b]{0.48\columnwidth}
\centering
\includegraphics[width=\linewidth]
{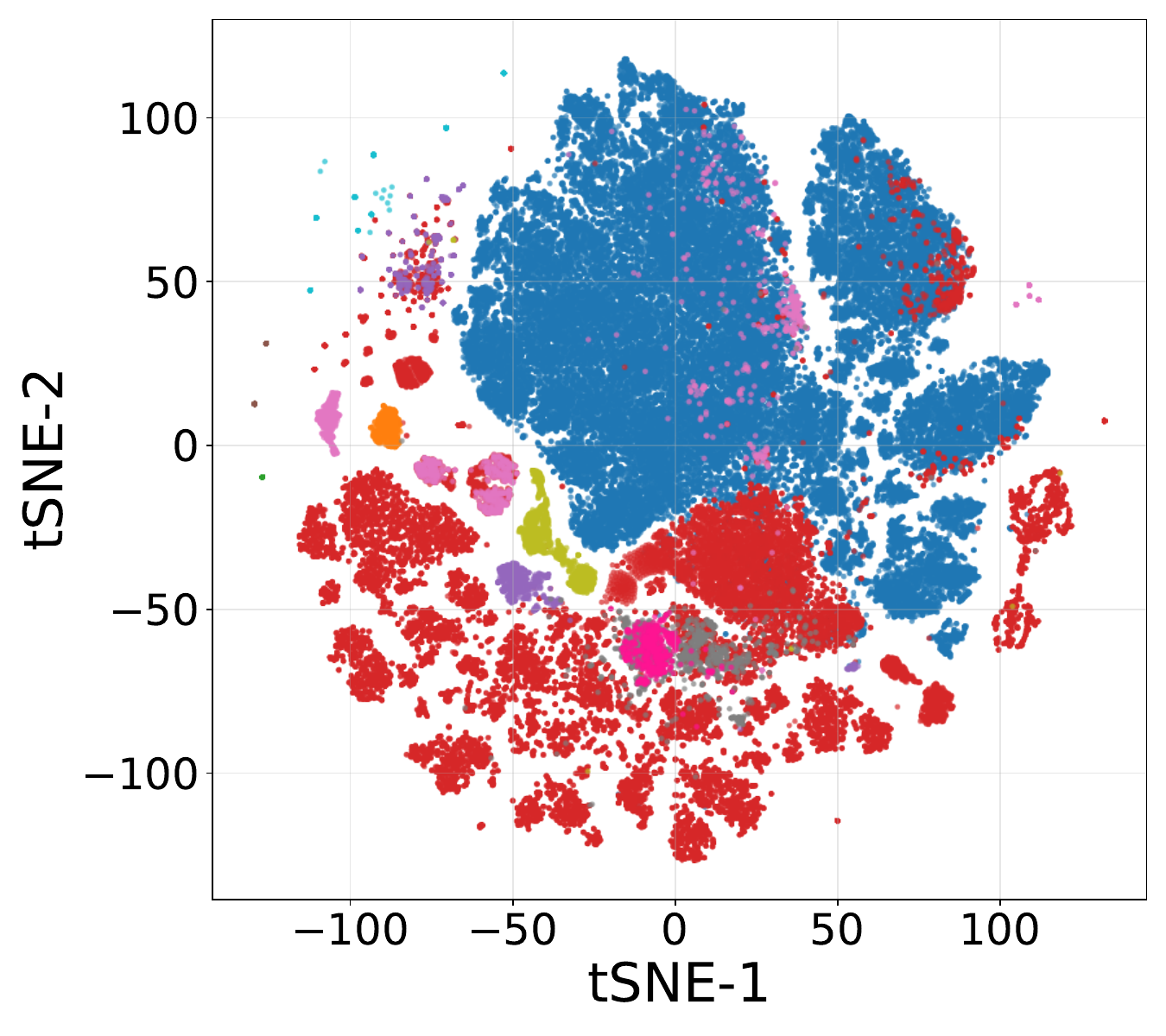}
\subcaption{Early: Layer 0}
\end{subfigure}
\hfill
\begin{subfigure}[b]{0.48\columnwidth}
\centering
\includegraphics[width=\linewidth]
{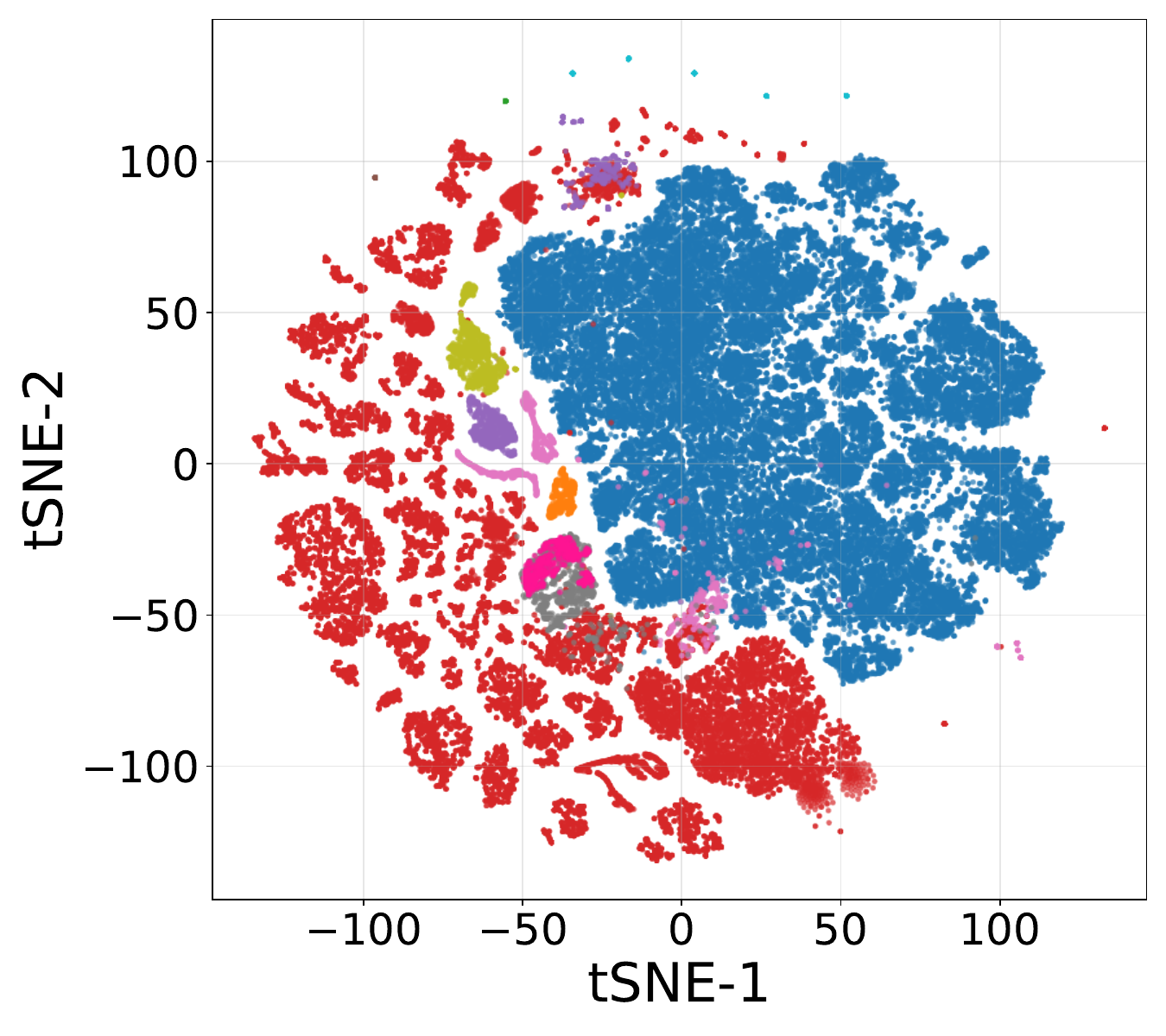}
\subcaption{Middle: Layer 11}
\end{subfigure}

\vspace{4pt}

\begin{subfigure}[b]{0.55\columnwidth}
\centering
\includegraphics[width=\linewidth]
{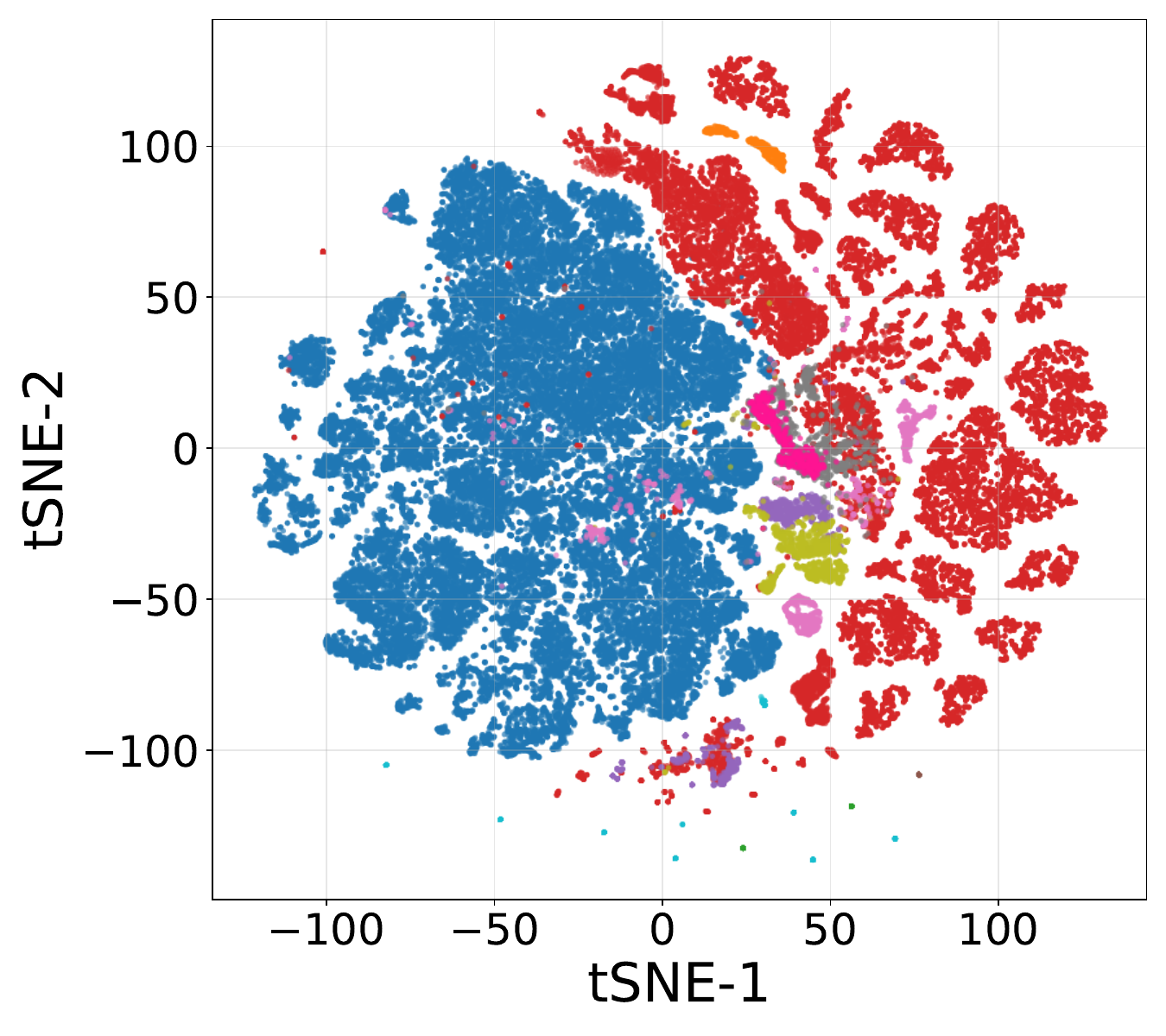}
\subcaption{Late: Layer 31}
\end{subfigure}

\caption{
t-SNE visualization of LLaMA-7B activations showing early, middle, and late layer representations.
}
\label{fig:llama_0_11_31}

\end{figure}

\begin{figure*}[h!]
\centering
\includegraphics[width=.85\linewidth]{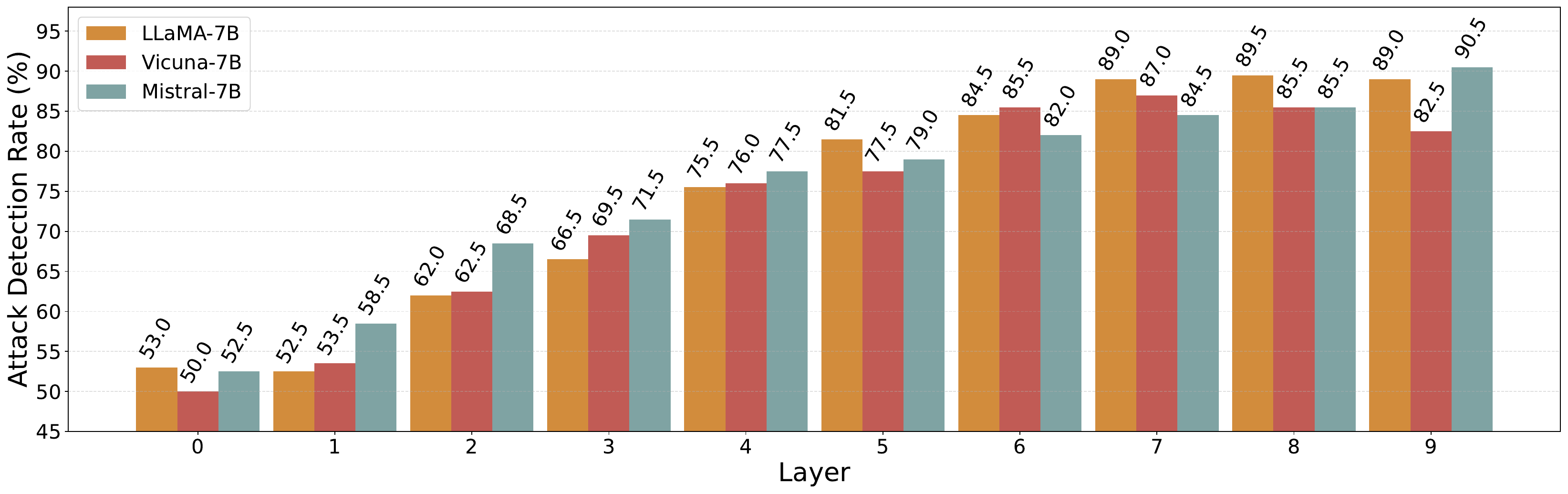}

\caption{Attack detection accuracy in the representation space using GUARD-SLM across early layers (0--10), Dataset: HarmBench, model: meta-
llama/Llama-2-7b-chat-hf}
\label{fig:guard_slm_early}
\end{figure*}

Sensitivity analysis is conducted to understand the behavior of different models in the representation space. For detailed analysis, three SLMs are selected as primary models. We perform layer-wise sensitivity analysis using three types of inputs: benign prompts, direct malicious prompts, and 9 optimized jailbreak attack categories. For each layer, the hidden representation of the last token is projected using t-SNE to visualize how different inputs are distributed in the representation space. A sample visualization is presented at Figure \ref{fig:llama_0_11_31} and Figure \ref{fig:guard_slm_early} for early layer. To analyze the behavior across the network, model layers are divided into three groups: early layers (0--10), middle layers (11--21), and late layers (22--31).

\textbf{We claim that jailbreak attacks are visible in every layer of the representation space.} The analysis shows that the hidden representations across layers contain sufficient information to distinguish different prompt types, indicating that activation-level features can be used to detect optimized jailbreak attacks without relying on a specific layer.

To better understand these differences, we summarize the main observations in the representation space as follows: 1)  Benign prompts form compact and stable clusters across layers that show consistent internal representations; 2)  Direct malicious prompts deviate from benign regions but may partially overlap depending on the layer and model; 3)  Optimized jailbreak attacks produce larger shifts in the representation space and form clearly identifiable clusters; 4) The separation between benign and optimized jailbreak representations is generally larger than the separation between benign and direct malicious prompts; 5) Similar distribution patterns are observed across different SLM architectures, indicating that the behavior is model-independent; and 6) Distinguishable activation patterns appear throughout the network which suggests that multiple layers can provide useful signals for jailbreak detection.

\subsection{Proposed Defense}

Table~\ref{tab:defense_jailbreak} compares different defenses against optimized jailbreak attacks using GPT-4o as the judge model. Existing methods such as SelfEval, Self-Reminder, RobustAlign, SmoothLLM, ICD, Aligner, and goal prioritization show non-zero attack success rates, with higher vulnerability to complex attacks like Codechameleon, DeepInception, and TAP. In contrast, GUARD-SLM achieves near-zero jailbreak success across all attacks by performing detection in the representation space using activation-level filtering, blocking malicious behavior before generation. As shown in Table~\ref{tab:defense_efficiency}, most defenses increase computational cost due to extra tokens, repeated queries, or multiple forward passes, while GUARD-SLM requires no additional tokens and introduces no inference overhead, providing an efficient and scalable defense for SLM.

\begin{table*}[t]
\centering
\tiny
\caption{ASR(\%) comparison of different defense methods against different jailbreak attacks, model: meta-llama/Llama-2-7b-chat-hf, dataset: HarmBench, judge model: gpt-4o}
\label{tab:defense_jailbreak}

\setlength{\tabcolsep}{3pt}
\renewcommand{\arraystretch}{1.05}

\resizebox{\textwidth}{!}{
\begin{tabular}{p{2.5cm}ccccccccc}
\hline

\diagbox[width=2.5cm]{\textbf{Defense}}{\textbf{Jailbreak}}
& AutoDAN
& PAIR
& TAP
& GCG
& Cipher
& Deepinception
& Codechameleon
& ICA
& Jailbroken
\\

\hline

SelfEval     & 01.00\% & 06.50\% & 07.00\% & 03.00\% &16.50\%  & 07.50\% & 28.00\% & 00.00\% & 03.00\% \\
SelfReminder     & 01.50\% & 02.50\% & 04.50\% & 00.00\% & 01.12\% & 00.00\% & 25.50\% & 00.00\% & 04.50\% \\
RobustAlign  & 00.00\% & 02.51\% & 05.50\% & 01.50\% & 11.25\% & 09.00\% & 38.00\% & 00.00\% & 01.25\% \\
SmoothLLM    & 10.50\% & 24.50\% & 35.00\% & 04.00\% & 03.12\% & 22.00\% & 06.00\% & 00.00\% & 09.00\% \\
ICD          & 00.50\% & 00.00\% & 00.00\% & 00.00\% & 05.00\% & 00.00\% & 00.00\% & 00.00\% & 02.93\% \\
Aligner      & 04.71\% & 06.50\% & 15.08\%  & 1.01\% & 01.58\% & 18.50\% & 22.50\% & 01.01\% & 03.66\% \\
GoalPrior    & 03.50\% & 00.50\% & 01.50\% & 00.00\% & 00.13\% & 00.00\% & 00.00\% & 00.00\% & 03.79\% \\

\rowcolor{green!15}
Ours         & 00.00\% & 00.00\% & 00.00\% & 00.00\% & 00.00\% & 00.00\% & 00.00\% & 00.00\% & 00.74\% \\

\hline
\end{tabular}
}
\end{table*}

\begin{table}[t]
\centering
\scriptsize
\caption{Efficiency comparison of different defense methods. 
$n$ denotes the number of queries for single forward pass and $q$ denotes the number of tokens per query.}
\label{tab:defense_efficiency}

\setlength{\tabcolsep}{3pt}
\renewcommand{\arraystretch}{1.05}

\begin{tabular}{p{2.2cm}p{2.0cm}p{1.5cm}}
\hline
\textbf{Defense} & \textbf{Additional Tokens (In/Out)} & \textbf{Avg. Inf. Time} \\
\hline

SelfEval     & +51 & 8.24s \\
SelfReminder & +44 & 1.12s \\
RobustAlign  & $(n-1)q$ & 12.45s \\
SmoothLLM    & $(n-1)q$ & 38.07s \\
ICD          & +805 & 1.13s \\
Aligner      & +42 & 4.90s \\
GoalPrior    & +202  & 3.18s \\
\rowcolor{green!15}
Ours         & 0  & 0.43s \\

\hline
\end{tabular}
\end{table}

\section{Discussion}
We conducted an empirical analysis of jailbreak attacks on both SLMs and LLMs to better understand the vulnerability of safety-aligned models. Our evaluation across multiple attack strategies shows that SLMs are more susceptible to optimized jailbreak prompts than larger models, with attacks such as AutoDAN, TAP, and Cipher achieving higher success rates than direct malicious prompts. We constructed a large-scale dataset covering diverse jailbreak categories and further evaluated the proposed method on the HarmBench dataset, demonstrating strong robustness against unseen jailbreak prompts. Representation-space analysis provides new insights into the relationship between transformer layers and jailbreak behavior, showing that jailbreak features are observable across all layers rather than only in the final decoding stage. Benign prompts form stable clusters, while optimized jailbreak prompts create distinct regions even in early layers, indicating that internal activations contain sufficient information for reliable detection. Based on this observation, GUARD-SLM performs single-pass activation-level classification without auxiliary models or multiple forward passes, achieving high detection accuracy with minimal inference overhead, making it suitable for real-time and resource-constrained deployment.

\section{Limitations}
 The main limitation of this work is that the proposed defense is mainly designed for SLMs. Although we analyzed vulnerabilities of LLMs in the empirical study, we conducted comprehensive evaluation of the defense  for SLMs. Extracting and analyzing internal hidden-layer activations for LLMs can be computationally expensive due to their large model size, which limits the direct application of the current approach to large-scale models. In future work, we plan to extend the proposed defense method to support LLMs. 

 The reported results in this paper are obtained under the experimental environment described in the Appendix. Due to the stochastic nature of LLM inference, the exact values of ASR may vary across different experiments even when using the same code and dataset. In particular, the results can be affected by temperature, top-p sampling, maximum generation length, and random seed. Our evaluation relies on external models  (GPT-4o and GPT-4o-mini) as the judge to determine whether a jailbreak attack is successful, and the judgment outcome may change when different evaluation models or API versions are used. Therefore, the reported ASR values should be interpreted as representative results under the specified configuration, and slight variations may occur when using different hyperparameters, judge models, API versions, or different model variants (e.g., GPT-4, GPT-4o, GPT-4o-mini, or newer models such as GPT-5).

\section{Conclusion}

SLMs provide an efficient and scalable alternative to LLMs, but our study shows that they remain significantly more vulnerable to jailbreak attacks, especially under adversarially optimized prompts. Sensitivity analysis in the representation space demonstrates that benign, malicious, and optimized jailbreak prompts form distinguishable regions across different neural network layers, and jailbreak behaviors are visible at every layer, as attack-related features appear from early layers and remain observable through middle and late layers. These findings reveal fundamental robustness limitations in SLMs and show that monitoring hidden-layer token activations enables lightweight and effective detection without retraining or additional overhead for defending against jailbreak attacks. Future work will focus on deploying activation-based safety monitoring in agentic AI systems and evaluating robustness against more complex, unseen, and adaptive jailbreak attacks in real-world settings.

\section*{Acknowledgment}
 This work is partly supported by the National Center for Transportation Cybersecurity and Resiliency (TraCR) (a U.S. Department of Transportation National University Transportation Center) headquartered at Clemson University, Clemson, South Carolina, USA, the National Artificial Intelligence Research Resource (NAIRR) Pilot (NAIRR260008, NAIRR240244 and NAIRR250261), and OpenAI. This work also used Anvil AI and  Anvil GPU at Purdue through allocation  CIS251427 from the Advanced Cyberinfrastructure Coordination Ecosystem: Services \& Support (ACCESS) program, which is supported by National Science Foundation grants \#2138259, \#2138286, \#2138307, \#2137603, and \#2138296. For more information about ACCESS, please see \cite{access}. Any opinions, findings, conclusions, and recommendations expressed in this material are those of the author(s) and do not necessarily reflect the views of TraCR, NAIRR, ACCESS, and OpenAI, and the U.S. Government assumes no liability for the contents or use thereof.

\bibliographystyle{plain}
\bibliography{references}

\begin{thebibliography}{10}

\bibitem{achiam2023gpt}
Josh Achiam, Steven Adler, Sandhini Agarwal, Lama Ahmad, Ilge Akkaya, Florencia~Leoni Aleman, Diogo Almeida, Janko Altenschmidt, Sam Altman, Shyamal Anadkat, et~al.
\newblock Gpt-4 technical report.
\newblock {\em arXiv preprint arXiv:2303.08774}, 2023.

\bibitem{alon2023detecting}
Gabriel Alon and Michael Kamfonas.
\newblock Detecting language model attacks with perplexity.
\newblock {\em arXiv preprint arXiv:2308.14132}, 2023.

\bibitem{amini2025distributed}
Hadi Amini, Md~Jueal Mia, Yasaman Saadati, Ahmed Imteaj, Seyedsina Nabavirazavi, Urmish Thakker, Md~Zarif Hossain, Awal~Ahmed Fime, and SS~Iyengar.
\newblock Distributed llms and multimodal large language models: A survey on advances, challenges, and future directions.
\newblock {\em arXiv preprint arXiv:2503.16585}, 2025.

\bibitem{belcak2025small}
Peter Belcak, Greg Heinrich, Shizhe Diao, Yonggan Fu, Xin Dong, Saurav Muralidharan, Yingyan~Celine Lin, and Pavlo Molchanov.
\newblock Small language models are the future of agentic ai.
\newblock {\em arXiv preprint arXiv:2506.02153}, 2025.

\bibitem{bianchi2023safety}
Federico Bianchi, Mirac Suzgun, Giuseppe Attanasio, Paul R{\"o}ttger, Dan Jurafsky, Tatsunori Hashimoto, and James Zou.
\newblock Safety-tuned llamas: Lessons from improving the safety of large language models that follow instructions.
\newblock {\em arXiv preprint arXiv:2309.07875}, 2023.

\bibitem{access}
Timothy~J. Boerner, Stephen Deems, Thomas~R. Furlani, Shelley~L. Knuth, and John Towns.
\newblock Access: Advancing innovation: Nsf’s advanced cyberinfrastructure coordination ecosystem: Services \& support.
\newblock In {\em Practice and Experience in Advanced Research Computing 2023: Computing for the Common Good}, PEARC '23, page 173–176, New York, NY, USA, 2023. Association for Computing Machinery.

\bibitem{NEURIPS2020_1457c0d6}
Tom Brown, Benjamin Mann, Nick Ryder, Melanie Subbiah, Jared~D Kaplan, Prafulla Dhariwal, Arvind Neelakantan, Pranav Shyam, Girish Sastry, Amanda Askell, Sandhini Agarwal, Ariel Herbert-Voss, Gretchen Krueger, Tom Henighan, Rewon Child, Aditya Ramesh, Daniel Ziegler, Jeffrey Wu, Clemens Winter, Chris Hesse, Mark Chen, Eric Sigler, Mateusz Litwin, Scott Gray, Benjamin Chess, Jack Clark, Christopher Berner, Sam McCandlish, Alec Radford, Ilya Sutskever, and Dario Amodei.
\newblock Language models are few-shot learners.
\newblock In H.~Larochelle, M.~Ranzato, R.~Hadsell, M.F. Balcan, and H.~Lin, editors, {\em Advances in Neural Information Processing Systems}, volume~33, pages 1877--1901. Curran Associates, Inc., 2020.

\bibitem{cao2024defending}
Bochuan Cao, Yuanpu Cao, Lu~Lin, and Jinghui Chen.
\newblock Defending against alignment-breaking attacks via robustly aligned llm.
\newblock In {\em Proceedings of the 62nd Annual Meeting of the Association for Computational Linguistics (Volume 1: Long Papers)}, pages 10542--10560, 2024.

\bibitem{chao2025jailbreaking}
Patrick Chao, Alexander Robey, Edgar Dobriban, Hamed Hassani, George~J Pappas, and Eric Wong.
\newblock Jailbreaking black box large language models in twenty queries.
\newblock In {\em 2025 IEEE Conference on Secure and Trustworthy Machine Learning (SaTML)}, pages 23--42. IEEE, 2025.

\bibitem{chen2025octopus}
Wei Chen, Zhiyuan Li, and Mingyuan Ma.
\newblock Octopus: On-device language model for function calling of software apis.
\newblock In {\em Proceedings of the 2025 Conference of the Nations of the Americas Chapter of the Association for Computational Linguistics: Human Language Technologies (Volume 3: Industry Track)}, pages 329--339, 2025.

\bibitem{chowdhery2023palm}
Aakanksha Chowdhery, Sharan Narang, Jacob Devlin, Maarten Bosma, Gaurav Mishra, Adam Roberts, Paul Barham, Hyung~Won Chung, Charles Sutton, Sebastian Gehrmann, et~al.
\newblock Palm: Scaling language modeling with pathways.
\newblock {\em Journal of Machine Learning Research}, 24(240):1--113, 2023.

\bibitem{dai2023safe}
Josef Dai, Xuehai Pan, Ruiyang Sun, Jiaming Ji, Xinbo Xu, Mickel Liu, Yizhou Wang, and Yaodong Yang.
\newblock Safe rlhf: Safe reinforcement learning from human feedback.
\newblock {\em arXiv preprint arXiv:2310.12773}, 2023.

\bibitem{deng2023multilingual}
Yue Deng, Wenxuan Zhang, Sinno~Jialin Pan, and Lidong Bing.
\newblock Multilingual jailbreak challenges in large language models.
\newblock {\em arXiv preprint arXiv:2310.06474}, 2023.

\bibitem{ding2024wolf}
Peng Ding, Jun Kuang, Dan Ma, Xuezhi Cao, Yunsen Xian, Jiajun Chen, and Shujian Huang.
\newblock A wolf in sheep’s clothing: Generalized nested jailbreak prompts can fool large language models easily.
\newblock In {\em Proceedings of the 2024 Conference of the North American Chapter of the Association for Computational Linguistics: Human Language Technologies (Volume 1: Long Papers)}, pages 2136--2153, 2024.

\bibitem{gao2025shaping}
Lang Gao, Jiahui Geng, Xiangliang Zhang, Preslav Nakov, and Xiuying Chen.
\newblock Shaping the safety boundaries: Understanding and defending against jailbreaks in large language models.
\newblock In {\em Proceedings of the 63rd Annual Meeting of the Association for Computational Linguistics (Volume 1: Long Papers)}, pages 25378--25398, 2025.

\bibitem{he2024jailbreaklens}
Zeqing He, Zhibo Wang, Zhixuan Chu, Huiyu Xu, Wenhui Zhang, Qinglong Wang, and Rui Zheng.
\newblock Jailbreaklens: Interpreting jailbreak mechanism in the lens of representation and circuit.
\newblock {\em arXiv preprint arXiv:2411.11114}, 2024.

\bibitem{jain2023baseline}
Neel Jain, Avi Schwarzschild, Yuxin Wen, Gowthami Somepalli, John Kirchenbauer, Ping-yeh Chiang, Micah Goldblum, Aniruddha Saha, Jonas Geiping, and Tom Goldstein.
\newblock Baseline defenses for adversarial attacks against aligned language models.
\newblock {\em arXiv preprint arXiv:2309.00614}, 2023.

\bibitem{NEURIPS2024_a51a74b2}
Jiaming Ji, Boyuan Chen, Hantao Lou, Donghai Hong, Borong Zhang, Xuehai Pan, Juntao Dai, Tianyi Qiu, and Yaodong Yang.
\newblock Aligner: Efficient alignment by learning to correct.
\newblock In A.~Globerson, L.~Mackey, D.~Belgrave, A.~Fan, U.~Paquet, J.~Tomczak, and C.~Zhang, editors, {\em Advances in Neural Information Processing Systems}, volume~37, pages 90853--90890. Curran Associates, Inc., 2024.

\bibitem{kumar2023certifying}
Aounon Kumar, Chirag Agarwal, Suraj Srinivas, Aaron~Jiaxun Li, Soheil Feizi, and Himabindu Lakkaraju.
\newblock Certifying llm safety against adversarial prompting.
\newblock {\em arXiv preprint arXiv:2309.02705}, 2023.

\bibitem{li2025revisiting}
Tianlong Li, Zhenghua Wang, Wenhao Liu, Muling Wu, Shihan Dou, Changze Lv, Xiaohua Wang, Xiaoqing Zheng, and Xuan-Jing Huang.
\newblock Revisiting jailbreaking for large language models: A representation engineering perspective.
\newblock In {\em Proceedings of the 31st International Conference on Computational Linguistics}, pages 3158--3178, 2025.

\bibitem{li2023deepinception}
Xuan Li, Zhanke Zhou, Jianing Zhu, Jiangchao Yao, Tongliang Liu, and Bo~Han.
\newblock Deepinception: Hypnotize large language model to be jailbreaker.
\newblock {\em arXiv preprint arXiv:2311.03191}, 2023.

\bibitem{liu2023autodan}
Xiaogeng Liu, Nan Xu, Muhao Chen, and Chaowei Xiao.
\newblock Autodan: Generating stealthy jailbreak prompts on aligned large language models.
\newblock {\em arXiv preprint arXiv:2310.04451}, 2023.

\bibitem{lu2024small}
Zhenyan Lu, Xiang Li, Dongqi Cai, Rongjie Yi, Fangming Liu, Xiwen Zhang, Nicholas~D Lane, and Mengwei Xu.
\newblock Small language models: Survey, measurements, and insights.
\newblock {\em arXiv preprint arXiv:2409.15790}, 2024.

\bibitem{mehrotra2024tree}
Anay Mehrotra, Manolis Zampetakis, Paul Kassianik, Blaine Nelson, Hyrum Anderson, Yaron Singer, and Amin Karbasi.
\newblock Tree of attacks: Jailbreaking black-box llms automatically.
\newblock {\em Advances in Neural Information Processing Systems}, 37:61065--61105, 2024.

\bibitem{NEURIPS2022_b1efde53}
Long Ouyang, Jeffrey Wu, Xu~Jiang, Diogo Almeida, Carroll Wainwright, Pamela Mishkin, Chong Zhang, Sandhini Agarwal, Katarina Slama, Alex Ray, John Schulman, Jacob Hilton, Fraser Kelton, Luke Miller, Maddie Simens, Amanda Askell, Peter Welinder, Paul~F Christiano, Jan Leike, and Ryan Lowe.
\newblock Training language models to follow instructions with human feedback.
\newblock In S.~Koyejo, S.~Mohamed, A.~Agarwal, D.~Belgrave, K.~Cho, and A.~Oh, editors, {\em Advances in Neural Information Processing Systems}, volume~35, pages 27730--27744. Curran Associates, Inc., 2022.

\bibitem{phute2023llm}
Mansi Phute, Alec Helbling, Matthew Hull, ShengYun Peng, Sebastian Szyller, Cory Cornelius, and Duen~Horng Chau.
\newblock Llm self defense: By self examination, llms know they are being tricked.
\newblock {\em arXiv preprint arXiv:2308.07308}, 2023.

\bibitem{robey2023smoothllm}
Alexander Robey, Eric Wong, Hamed Hassani, and George~J Pappas.
\newblock Smoothllm: Defending large language models against jailbreaking attacks.
\newblock {\em arXiv preprint arXiv:2310.03684}, 2023.

\bibitem{scholkopf2002learning}
Bernhard Sch{\"o}lkopf and Alexander~J Smola.
\newblock {\em Learning with kernels: support vector machines, regularization, optimization, and beyond}.
\newblock MIT press, 2002.

\bibitem{shen2024jailbreak}
Guobin Shen, Dongcheng Zhao, Yiting Dong, Xiang He, and Yi~Zeng.
\newblock Jailbreak antidote: Runtime safety-utility balance via sparse representation adjustment in large language models.
\newblock {\em arXiv preprint arXiv:2410.02298}, 2024.

\bibitem{team2023gemini}
Gemini Team, Rohan Anil, Sebastian Borgeaud, Jean-Baptiste Alayrac, Jiahui Yu, Radu Soricut, Johan Schalkwyk, Andrew~M Dai, Anja Hauth, Katie Millican, et~al.
\newblock Gemini: a family of highly capable multimodal models.
\newblock {\em arXiv preprint arXiv:2312.11805}, 2023.

\bibitem{touvron2023llama}
Hugo Touvron, Thibaut Lavril, Gautier Izacard, Xavier Martinet, Marie-Anne Lachaux, Timoth{\'e}e Lacroix, Baptiste Rozi{\`e}re, Naman Goyal, Eric Hambro, Faisal Azhar, et~al.
\newblock Llama: Open and efficient foundation language models.
\newblock {\em arXiv preprint arXiv:2302.13971}, 2023.

\bibitem{touvron2023llama2}
Hugo Touvron, Louis Martin, Kevin Stone, Peter Albert, Amjad Almahairi, Yasmine Babaei, Nikolay Bashlykov, Soumya Batra, Prajjwal Bhargava, Shruti Bhosale, et~al.
\newblock Llama 2: Open foundation and fine-tuned chat models.
\newblock {\em arXiv preprint arXiv:2307.09288}, 2023.

\bibitem{wang2024mobile}
Junyang Wang, Haiyang Xu, Haitao Jia, Xi~Zhang, Ming Yan, Weizhou Shen, Ji~Zhang, Fei Huang, and Jitao Sang.
\newblock Mobile-agent-v2: Mobile device operation assistant with effective navigation via multi-agent collaboration.
\newblock {\em Advances in Neural Information Processing Systems}, 37:2686--2710, 2024.

\bibitem{wei2023jailbroken}
Alexander Wei, Nika Haghtalab, and Jacob Steinhardt.
\newblock Jailbroken: How does llm safety training fail?
\newblock {\em Advances in neural information processing systems}, 36:80079--80110, 2023.

\bibitem{wei2023jailbrek}
Zeming Wei, Yifei Wang, Ang Li, Yichuan Mo, and Yisen Wang.
\newblock Jailbreak and guard aligned language models with only few in-context demonstrations.
\newblock {\em arXiv preprint arXiv:2310.06387}, 2023.

\bibitem{wei2023jailbreak}
Zeming Wei, Yifei Wang, Ang Li, Yichuan Mo, and Yisen Wang.
\newblock Jailbreak and guard aligned language models with only few in-context demonstrations.
\newblock {\em arXiv preprint arXiv:2310.06387}, 2023.

\bibitem{xie2023defending}
Yueqi Xie, Jingwei Yi, Jiawei Shao, Justin Curl, Lingjuan Lyu, Qifeng Chen, Xing Xie, and Fangzhao Wu.
\newblock Defending chatgpt against jailbreak attack via self-reminders.
\newblock {\em Nature Machine Intelligence}, 5(12):1486--1496, 2023.

\bibitem{xu2024safedecoding}
Zhangchen Xu, Fengqing Jiang, Luyao Niu, Jinyuan Jia, Bill~Yuchen Lin, and Radha Poovendran.
\newblock Safedecoding: Defending against jailbreak attacks via safety-aware decoding.
\newblock {\em arXiv preprint arXiv:2402.08983}, 2024.

\bibitem{yu2023gptfuzzer}
Jiahao Yu, Xingwei Lin, Zheng Yu, and Xinyu Xing.
\newblock Gptfuzzer: Red teaming large language models with auto-generated jailbreak prompts.
\newblock {\em arXiv preprint arXiv:2309.10253}, 2023.

\bibitem{yuan2023gpt}
Youliang Yuan, Wenxiang Jiao, Wenxuan Wang, Jen-tse Huang, Pinjia He, Shuming Shi, and Zhaopeng Tu.
\newblock Gpt-4 is too smart to be safe: Stealthy chat with llms via cipher.
\newblock {\em arXiv preprint arXiv:2308.06463}, 2023.

\bibitem{zhang2025jbshield}
Shenyi Zhang, Yuchen Zhai, Shengnan Guo, Zheng Fang, Lingchen Zhao, Cong Wang, and Qian Wang.
\newblock Jbshield: Defending large language models from jailbreak attacks through activated concept analysis and manipulation.
\newblock In {\em Proceedings of the 34th USENIX Security Symposium}, 2025.

\bibitem{zhang2025can}
Wenhui Zhang, Huiyu Xu, Zhibo Wang, Zeqing He, Ziqi Zhu, and Kui Ren.
\newblock Can small language models reliably resist jailbreak attacks? a comprehensive evaluation.
\newblock {\em arXiv preprint arXiv:2503.06519}, 2025.

\bibitem{zhang2025aisafetylab}
Zhexin Zhang, Leqi Lei, Junxiao Yang, Xijie Huang, Yida Lu, Shiyao Cui, Renmiao Chen, Qinglin Zhang, Xinyuan Wang, Hao Wang, et~al.
\newblock Aisafetylab: A comprehensive framework for ai safety evaluation and improvement.
\newblock {\em arXiv preprint arXiv:2502.16776}, 2025.

\bibitem{zhang2024defending}
Zhexin Zhang, Junxiao Yang, Pei Ke, Fei Mi, Hongning Wang, and Minlie Huang.
\newblock Defending large language models against jailbreaking attacks through goal prioritization.
\newblock In {\em Proceedings of the 62nd Annual Meeting of the Association for Computational Linguistics (Volume 1: Long Papers)}, pages 8865--8887, 2024.

\bibitem{zhang2024theft}
Zhexin Zhang, Junxiao Yang, Yida Lu, Pei Ke, Shiyao Cui, Chujie Zheng, Hongning Wang, and Minlie Huang.
\newblock From theft to bomb-making: The ripple effect of unlearning in defending against jailbreak attacks.
\newblock {\em arXiv preprint arXiv:2407.02855}, 2024.

\bibitem{zheng2024prompt}
Chujie Zheng, Fan Yin, Hao Zhou, Fandong Meng, Jie Zhou, Kai-Wei Chang, Minlie Huang, and Nanyun Peng.
\newblock On prompt-driven safeguarding for large language models.
\newblock {\em arXiv preprint arXiv:2401.18018}, 2024.

\bibitem{zhou2024alignment}
Zhenhong Zhou, Haiyang Yu, Xinghua Zhang, Rongwu Xu, Fei Huang, and Yongbin Li.
\newblock How alignment and jailbreak work: Explain llm safety through intermediate hidden states.
\newblock In {\em Findings of the Association for Computational Linguistics: EMNLP 2024}, pages 2461--2488, 2024.

\bibitem{zou2023universal}
Andy Zou et~al.
\newblock Universal and transferable adversarial attacks on aligned language models.
\newblock {\em arXiv preprint arXiv:2307.15043}, 2023.

\end{thebibliography}

\appendix

\newpage
\twocolumn[
\begin{center}
\LARGE \textbf{Appendix}
\end{center}
\vspace{1em}
]

\section{Experiments setup}
\subsection{Experimental Setup Configurations and Datasets}

The hardware configuration used for all experiments is summarized in Table~\ref{tab:hw}.
All experiments were conducted using multiple high-performance computing environments to support large-scale jailbreak evaluation, last token activation analysis, and defense validation. The experiments were executed across several GPU platforms, including the Purdue Anvil HPC cluster equipped with NVIDIA H100 GPUs with 80 GB HBM3 memory, our university server with two NVIDIA RTX A6000 GPUs (48 GB memory), four NVIDIA A100 PCIe GPUs (80 GB memory), and eight NVIDIA A100 PCIe GPUs (40 GB memory). These computing resources enabled efficient large-scale inference and activation analysis for SLMs and LLMs. To analyze jailbreak vulnerabilities, we considered a diverse set of models consisting of seven SLMs and three LLMs, including LLaMA-2-7B, Vicuna-7B, Mistral-7B, Yi-6B-Chat, Qwen1.5-7B-Chat, Gemma-7B-IT, OpenChat-3.5, LLaMA-2-13B, Vicuna-13B, and Qwen1.5-14B. Among these models, LLaMA-2-7B, Vicuna-7B, and Mistral-7B were primarily used to evaluate the proposed defense method due to their widespread adoption in recent jailbreak and safety studies.

For dataset construction, we collected prompts from several publicly available jailbreak and instruction datasets. The primary jailbreak dataset used in our experiments is AdvBench for empirical analysis, which contains adversarial prompts designed to evaluate safety vulnerabilities in SLMs and LLMs. We also utilized the JailBreakV-28K dataset, which provides a large collection of jailbreak prompts generated using multiple attack strategies. To incorporate benign instructions, we used the Alpaca dataset. Additionally, we collected optimized jailbreak prompts from a public GitHub repository containing curated jailbreak attack categories and merged them with the other datasets to form a unified direct prompt attack dataset, where AdvBench and JailBreakV-28K served as the primary jailbreak sources. For evaluation, we used HarmBench and a subset of the Databricks Dolly dataset, from which 200 samples were randomly selected to represent benign user instructions. This dataset combination allows comprehensive evaluation of the proposed defense method across diverse jailbreak strategies while maintaining a balanced representation of benign and malicious prompts.  We have run our experiments for more than 5000 hours on different gpu node.

\begin{table*}[t]
\centering
\footnotesize
\caption{Hardware configuration used in the experiments}
\label{tab:hw}
\renewcommand{\arraystretch}{1.1}

\begin{tabular}{lccc}
\hline
\textbf{GPU Model} & \textbf{No. of GPUs} & \textbf{GPU Memory} \\
\hline
NVIDIA H100 & Multiple Nodes & 80 GB HBM3 \\
NVIDIA RTX A6000 & 2 & 48 GB \\
NVIDIA A100 PCIe & 4 & 80 GB \\
NVIDIA A100 PCIe & 8 & 40 GB \\
\hline
\end{tabular}

\end{table*}

The data distribution used for activation analysis is summarized in Table~\ref{tab:dat_exp}, which includes a large-scale collection of benign, direct malicious, and optimized jailbreak samples. The dataset consists of 52,000 benign prompts and 32,490 direct malicious prompts, along with multiple optimized jailbreak categories such as AutoDAN, Cipher, CodeChameleon, DeepInception, GCG, ICA, Jailbroken, PAIR, and TAP, resulting in a total of 94,320 samples. To the best of our knowledge, this is one of the largest datasets used for jailbreak attack and defense analysis, enabling reliable activation-space evaluation across diverse jailbreak categories. 

\begin{table*}[t]
\centering
\caption{Data distribution of benign, malicious, and optimized jailbreak samples used for activation analysis.}
\label{tab:dat_exp}
\renewcommand{\arraystretch}{1.15}
\setlength{\tabcolsep}{10pt}

\begin{tabular}{lc}
\hline
\textbf{Category} & \textbf{Count} \\
\hline

Benign            & 52,000 \\
Direct Malicious  & 32,490 \\

\hline
AutoDAN        & 1,370 \\
Cipher         & 2,080 \\
CodeChameleon  & 520 \\
DeepInception  & 520 \\
GCG            & 1,370 \\
ICA            & 520 \\
Jailbroken     & 1,560 \\
PAIR           & 1,370 \\
TAP            & 520 \\

\hline
\textbf{Total} & \textbf{94,320} \\
\hline

\end{tabular}
\end{table*}

Moreover, The key hyperparameters used in our experiments are listed in Table~\ref{tab:hyperparams}.

\begin{table*}[t]
\centering
\caption{Hyperparameters used for GUARD-SLM.}
\label{tab:hyperparams}
\small
\setlength{\tabcolsep}{6pt}
\renewcommand{\arraystretch}{1.15}
\begin{tabular}{p{3.2cm} p{4.8cm} p{6.2cm}}
\hline
\textbf{Phase} & \textbf{Hyperparameter} & \textbf{Value} \\
\hline

Activation extraction 
& Maximum input tokens 
& 512 \\
& Hidden representation used 
& Last-token activation from all transformer layers \\

& Layer grouping 
& Early: first 40\% layers, Middle: 40\%--70\%, Late: remaining layers \\

& Precision 
& FP16 on GPU, FP32 on CPU \\

\hline

t-SNE visualization 
& Selected classes 
& benign, malicious, autodan, cipher, codechamelon, deepinception, gcg, ica, jailbroken, pair, tap \\
& PCA before t-SNE 
& 128 dimensions \\

& Perplexity 
& 30.0 \\

& Learning rate 
& auto \\

& Iterations 
& 1000 \\

& Initialization 
& PCA \\

& Distance metric 
& Euclidean \\

& Random seed 
& 42 \\

\hline

Binary RBF-SVM training 
& Input feature 
& Last-token hidden activation at selected layer (0,1,2...n) \\
& SVM kernel 
& RBF \\

& Regularization parameter $C$ 
& 1.0 \\

& Gamma 
& scale \\

& Probability estimation 
& Disabled by default \\

& Feature normalization 
& StandardScaler \\

\hline

Inference-time 
& classifier 
& Saved binary RBF-SVM model \\

& layer 
& Selected layer (e.g., layer 13) \\

& Maximum input tokens 
& 512 \\

& Maximum new tokens for generation 
& 512 \\
& Generation setting 
& Greedy decoding (do\_sample=False) \\

& Temperature 
& None \\

& Top-$p$ 
& None \\
& Precision 
& FP16 on GPU, FP32 on CPU \\

\hline
\end{tabular}
\end{table*}
\section{Result Analysis}
\subsection{Empirical Analysis}
Table~\ref{tab:slms_harmbench} reports the Attack Success Rate (ASR) of nine optimized jailbreak attacks across three small language models (SLMs), namely LLaMA-2-7B-Chat, Vicuna-7B-v1.5, and Mistral-7B-Instruct, evaluated on the HarmBench dataset using GPT-4o-mini as the evaluator. The results show that several advanced jailbreak methods remain highly effective across all models, with AutoDAN achieving nearly 100\% ASR on all three models, indicating that current alignment strategies are still vulnerable to optimized attacks. TAP, ICA, and DeepInception also demonstrate very high success rates, particularly on Vicuna and Mistral, where the ASR often exceeds 90\%. In contrast, GCG shows almost zero success rate across all models, suggesting that not all jailbreak methods generalize to instruction-tuned SLMs. The results further reveal model-dependent vulnerability patterns, where Vicuna appears highly susceptible to multiple attacks, while LLaMA-2 shows lower ASR for several jailbreak categories such as PAIR, ICA, and Jailbroken. Overall, the results indicate that modern SLMs remain highly vulnerable to optimized jailbreak attacks, highlighting the need for stronger inference-time defense mechanisms.

\begin{table*}[t]
\centering
\caption{Attack Success Rate (ASR) across jailbreak attacks on SLMs. 
Models: LLaMA-2 (meta-llama/Llama-2-7b-chat-hf), 
Vicuna (lmsys/vicuna-7b-v1.5), 
Mistral (mistralai/Mistral-7B-Instruct-v0.2), Dataset: HarmBench,  Evaluator: GPT-4o-mini }
\label{tab:slms_harmbench}

\begin{tabular}{p{3.0cm}ccc}
\hline

\diagbox[width=3.0cm]{\textbf{Jailbreak}}{\textbf{Model}}
& \textbf{LLaMA-2}
& \textbf{Vicuna}
& \textbf{Mistral}

\\
\hline
\rowcolor{red!15}
AutoDAN        & 100.0\% & 100.0\% & 99.50\% \\
PAIR           & 18.00\% & 86.50\% & 60.00\% \\
TAP            & 58.50\% & 99.50\% & 94.50\% \\
GCG            & 00.00\% & 00.00\% & 00.50\% \\
Cipher         & 78.25\% & 56.62\% & 83.88\% \\
DeepInception  & 46.50\% & 86.50\% & 73.00\% \\
CodeChameleon  & 61.00\% & 73.50\% & 74.00\% \\
ICA            & 01.00\% & 99.50\% &  69.50\%\\
Jailbroken     & 21.62\% & 19.76\% & 62.34\% \\
\hline

\textbf{Avg ASR}
& \cellcolor{green!15}\textbf{42.76\%}
& \cellcolor{red!15}\textbf{69.10\%}
& \textbf{68.58\%}
\\

\hline
\end{tabular}

\end{table*}
\subsection{Additional Layer Sensitivity Analysis}

\begin{figure*}[t]
\centering

\includegraphics[width=0.95\textwidth, trim={0 0.8cm 0 0 cm}, clip]{latex/Figure/legend.pdf}

\vspace{2pt}

\begin{subfigure}[b]{0.17\linewidth}
    \centering
    \includegraphics[width=\linewidth]{latex/Figure/figures/llama/llama_0.pdf}
    \subcaption{Layer 0}
\end{subfigure}\hfill
\begin{subfigure}[b]{0.17\linewidth}
    \centering
    \includegraphics[width=\linewidth]{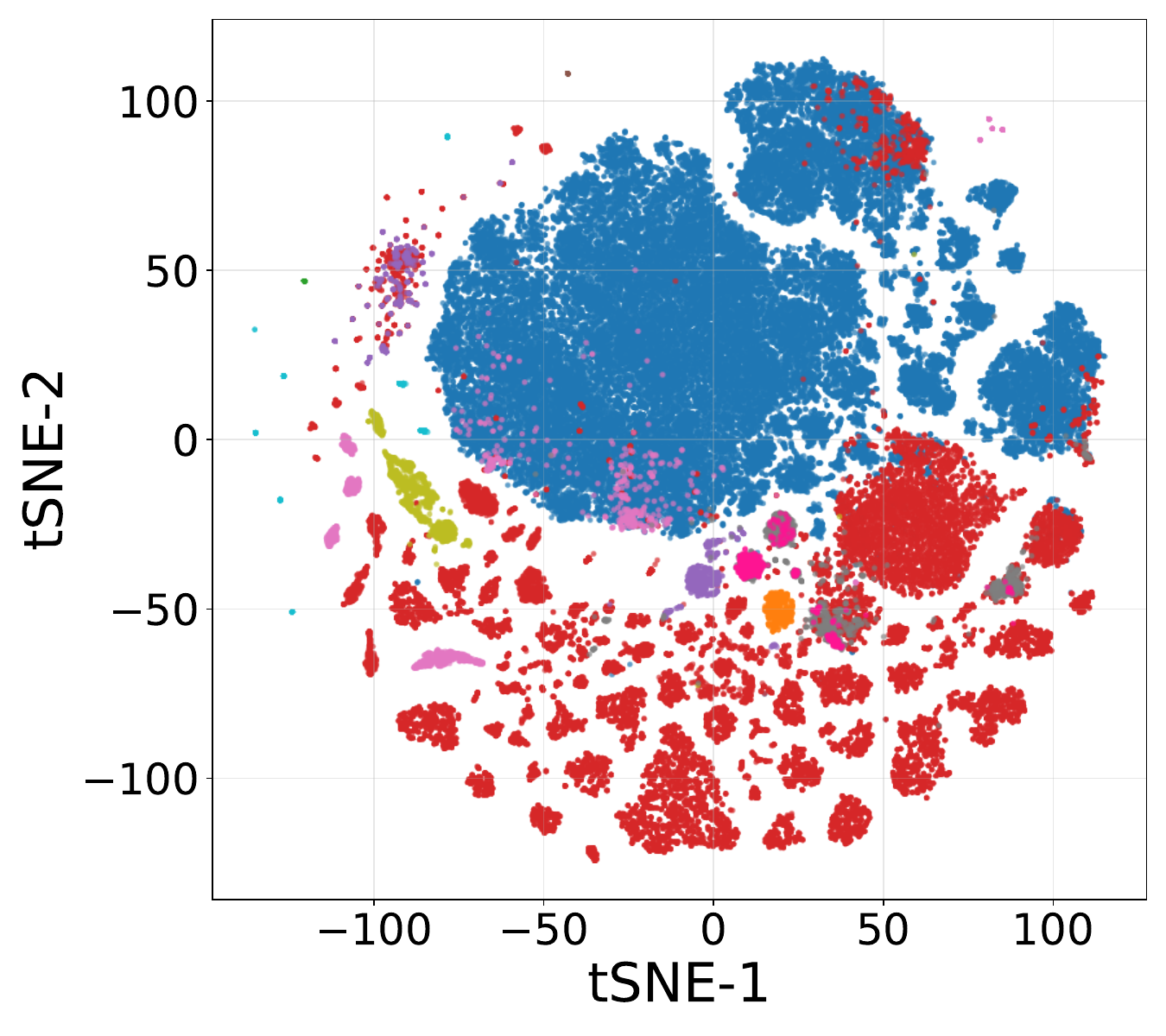}
    \subcaption{Layer 1}
\end{subfigure}\hfill
\begin{subfigure}[b]{0.17\linewidth}
    \centering
    \includegraphics[width=\linewidth]{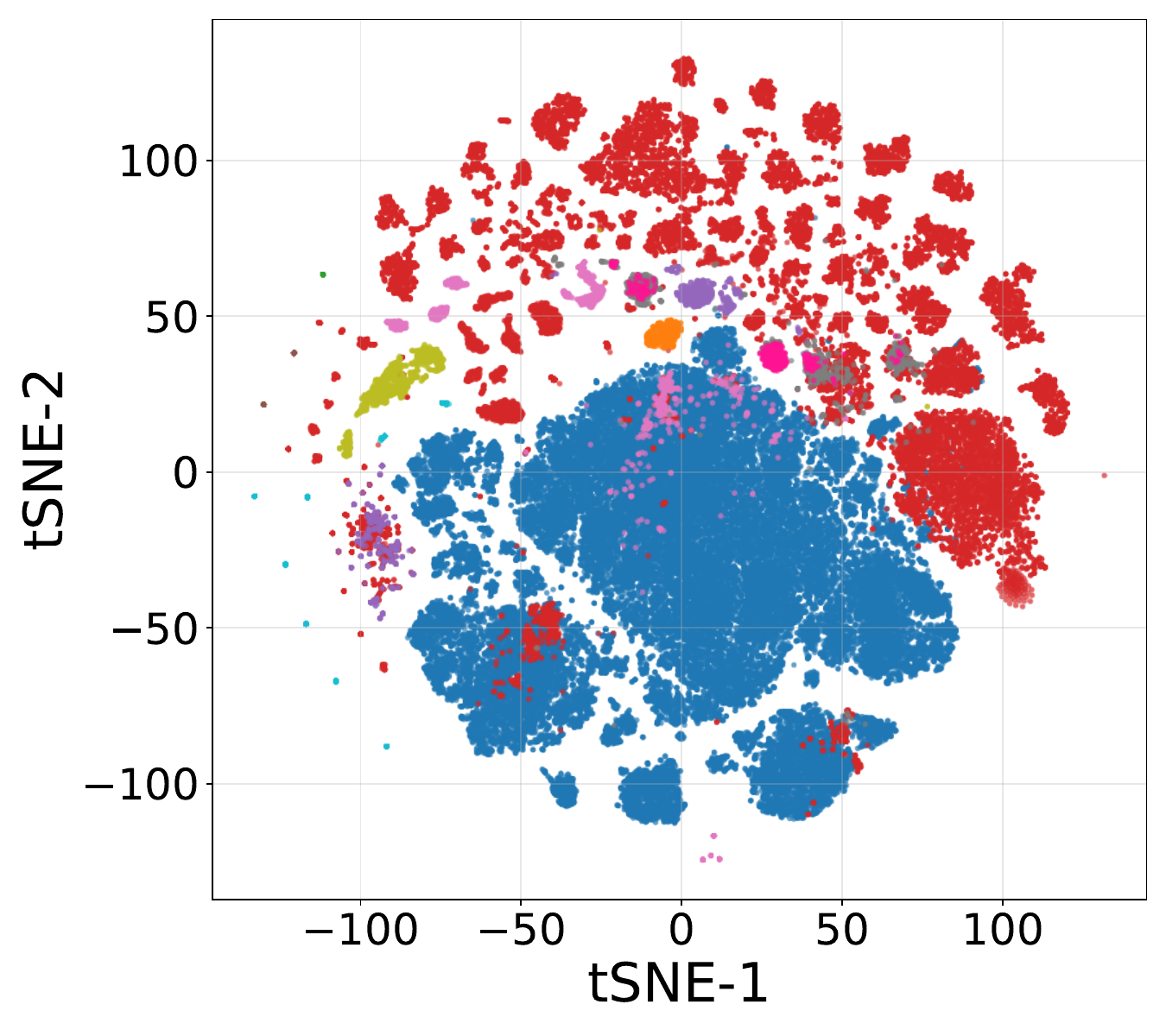}
    \subcaption{Layer 2}
\end{subfigure}\hfill
\begin{subfigure}[b]{0.17\linewidth}
    \centering
    \includegraphics[width=\linewidth]{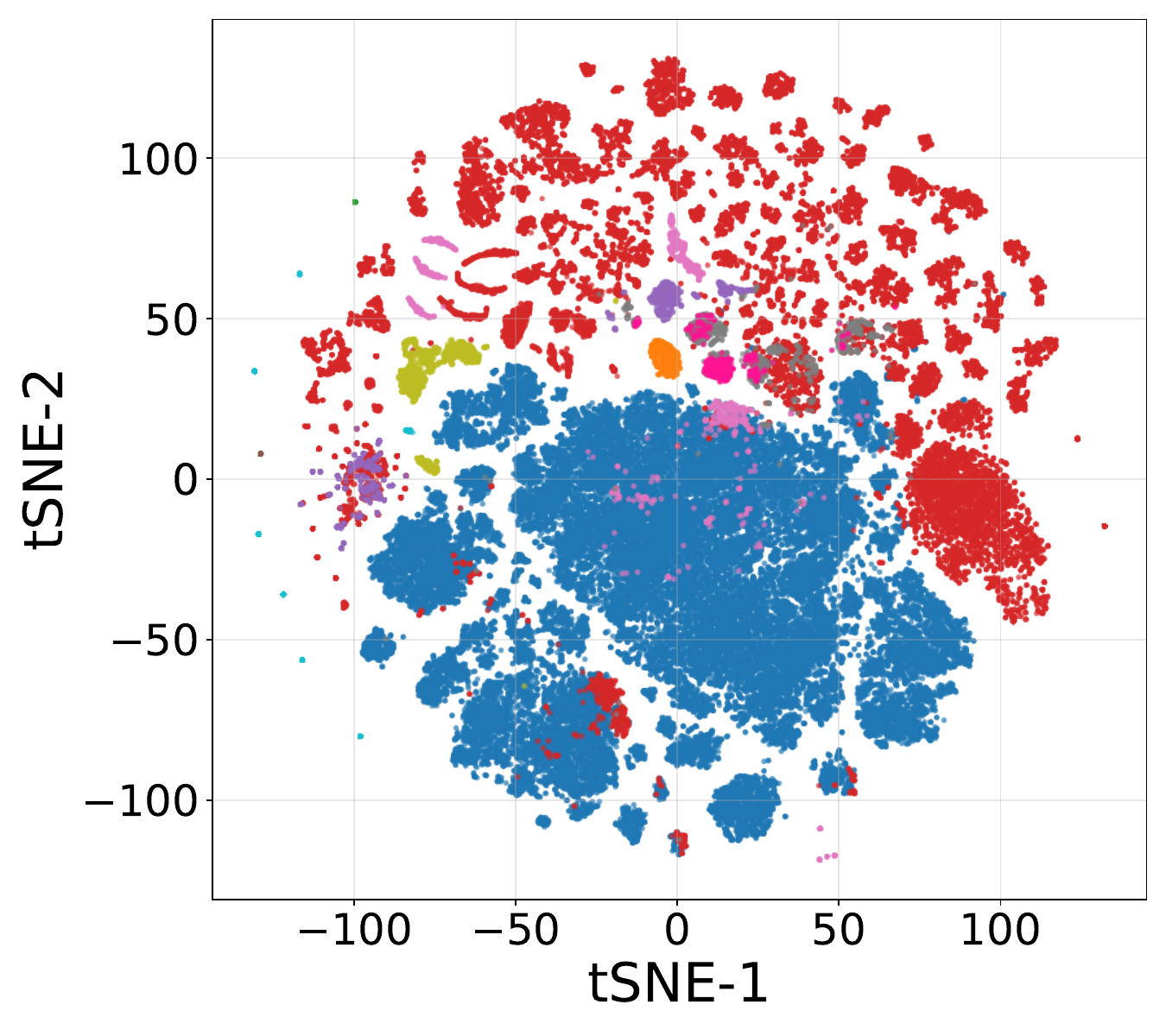}
    \subcaption{Layer 3}
\end{subfigure}\hfill
\begin{subfigure}[b]{0.17\linewidth}
    \centering
    \includegraphics[width=\linewidth]{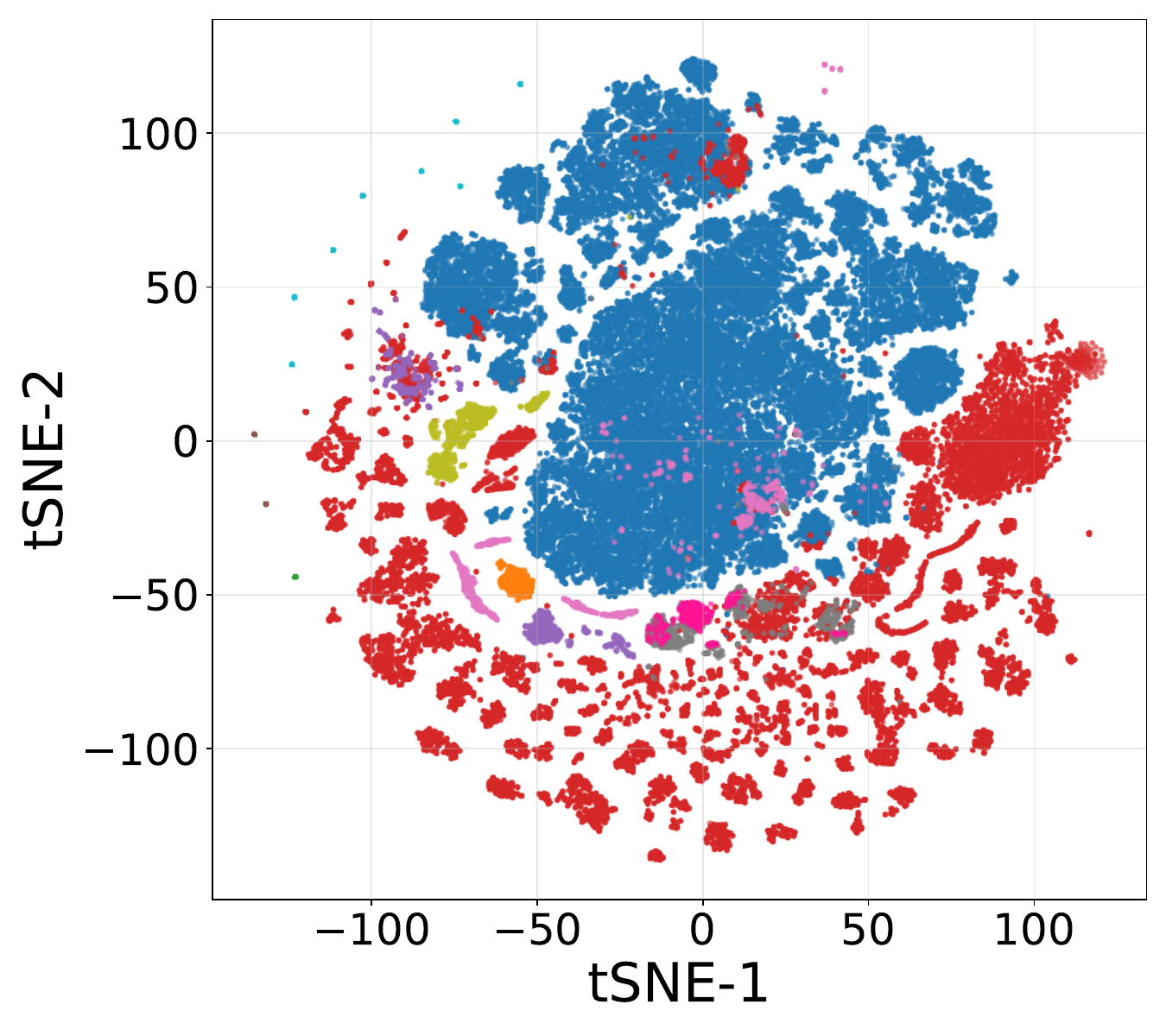}
    \subcaption{Layer 4}
\end{subfigure}

\vspace{3pt}

\begin{subfigure}[b]{0.17\linewidth}
    \centering
    \includegraphics[width=\linewidth]{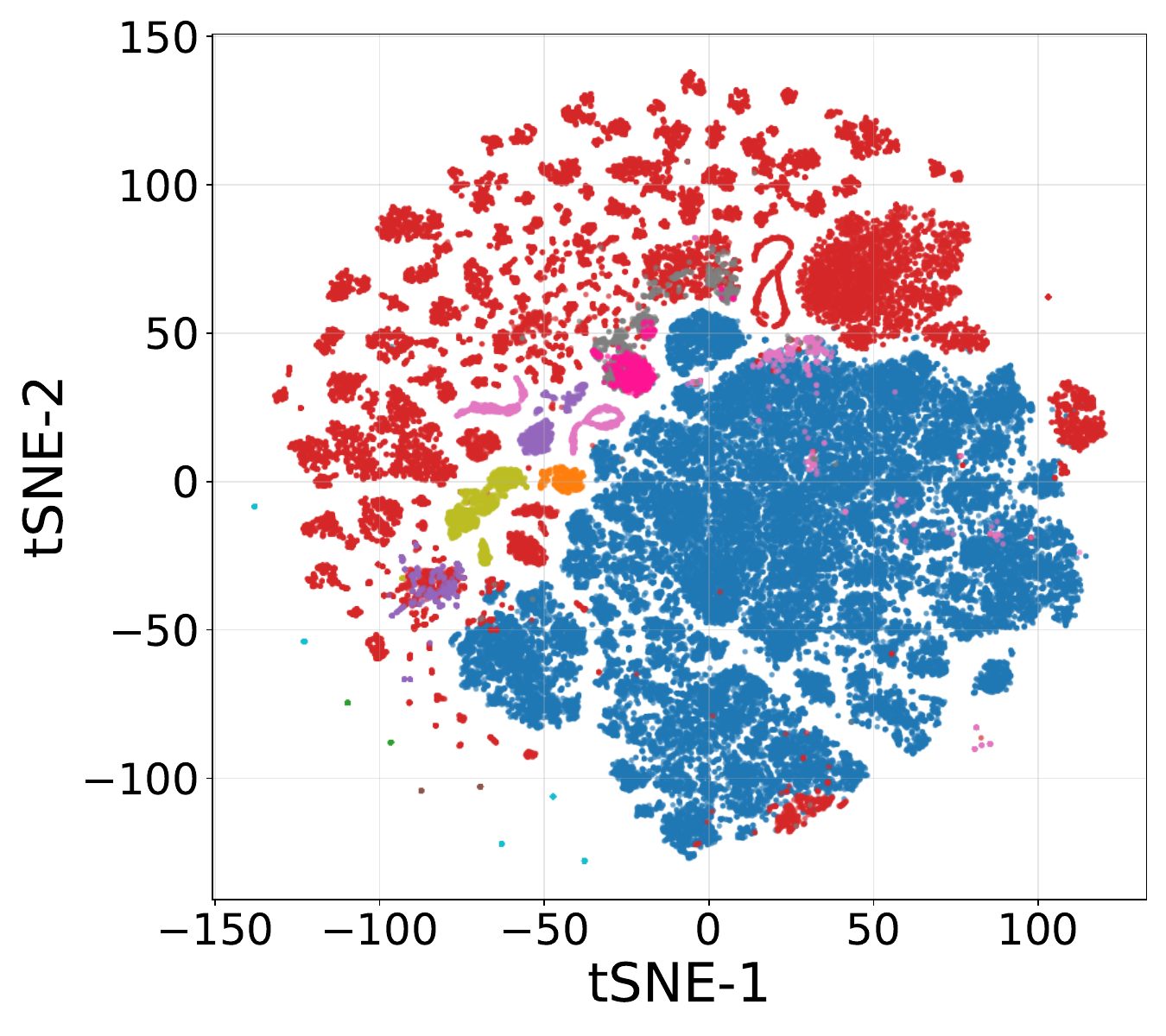}
    \subcaption{Layer 5}
\end{subfigure}\hfill
\begin{subfigure}[b]{0.17\linewidth}
    \centering
    \includegraphics[width=\linewidth]{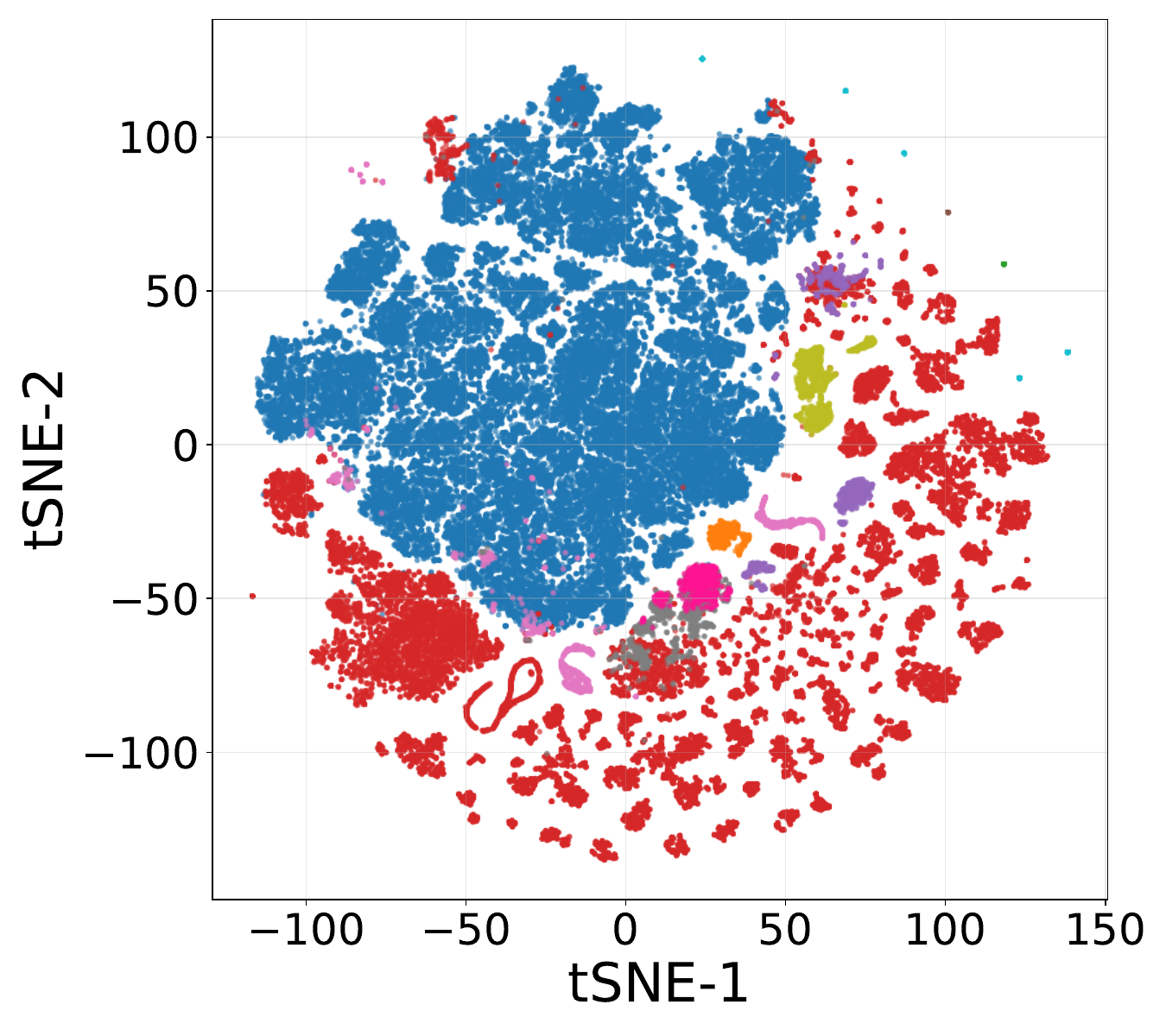}
    \subcaption{Layer 6}
\end{subfigure}\hfill
\begin{subfigure}[b]{0.17\linewidth}
    \centering
    \includegraphics[width=\linewidth]{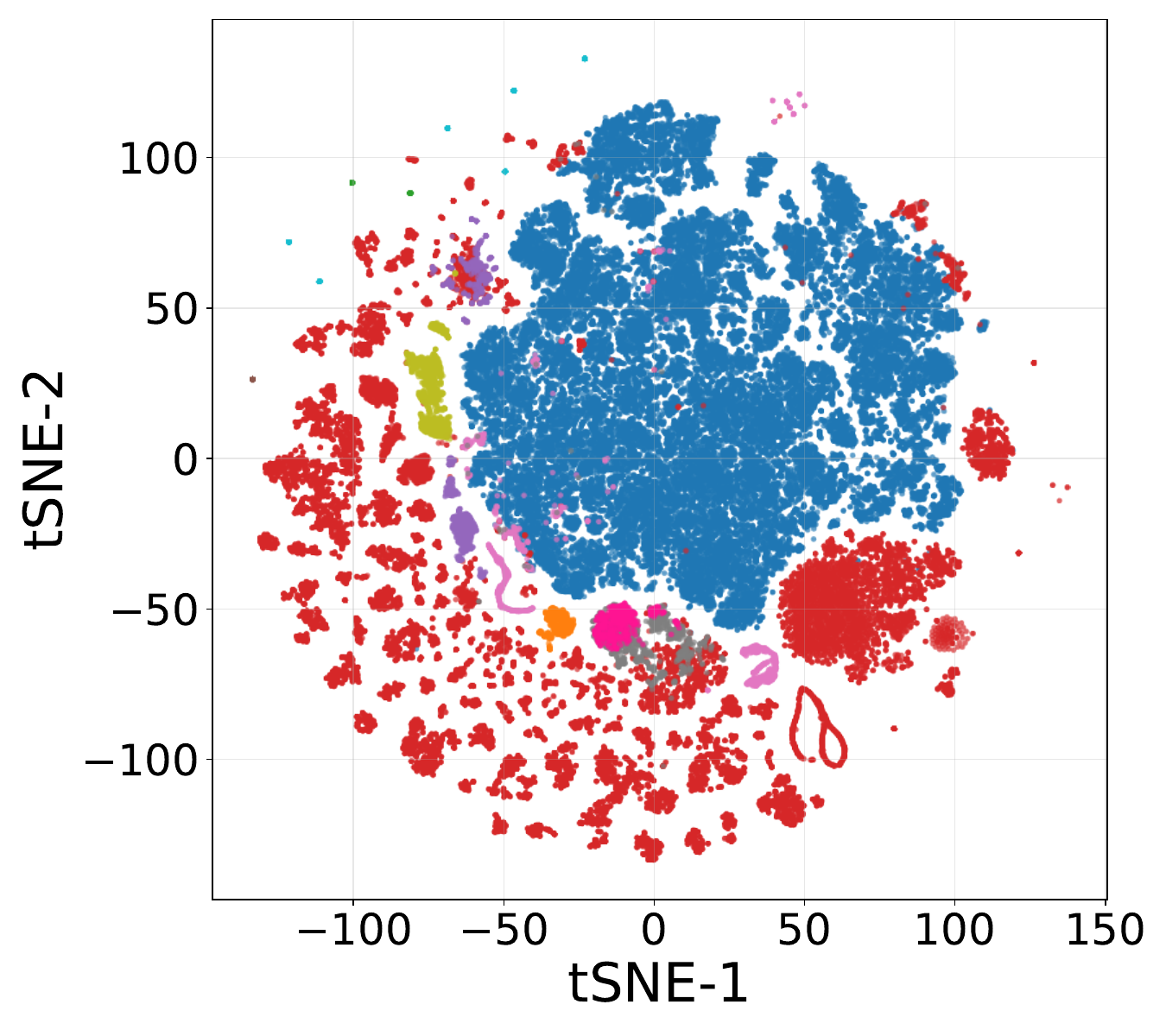}
    \subcaption{Layer 7}
\end{subfigure}\hfill
\begin{subfigure}[b]{0.17\linewidth}
    \centering
    \includegraphics[width=\linewidth]{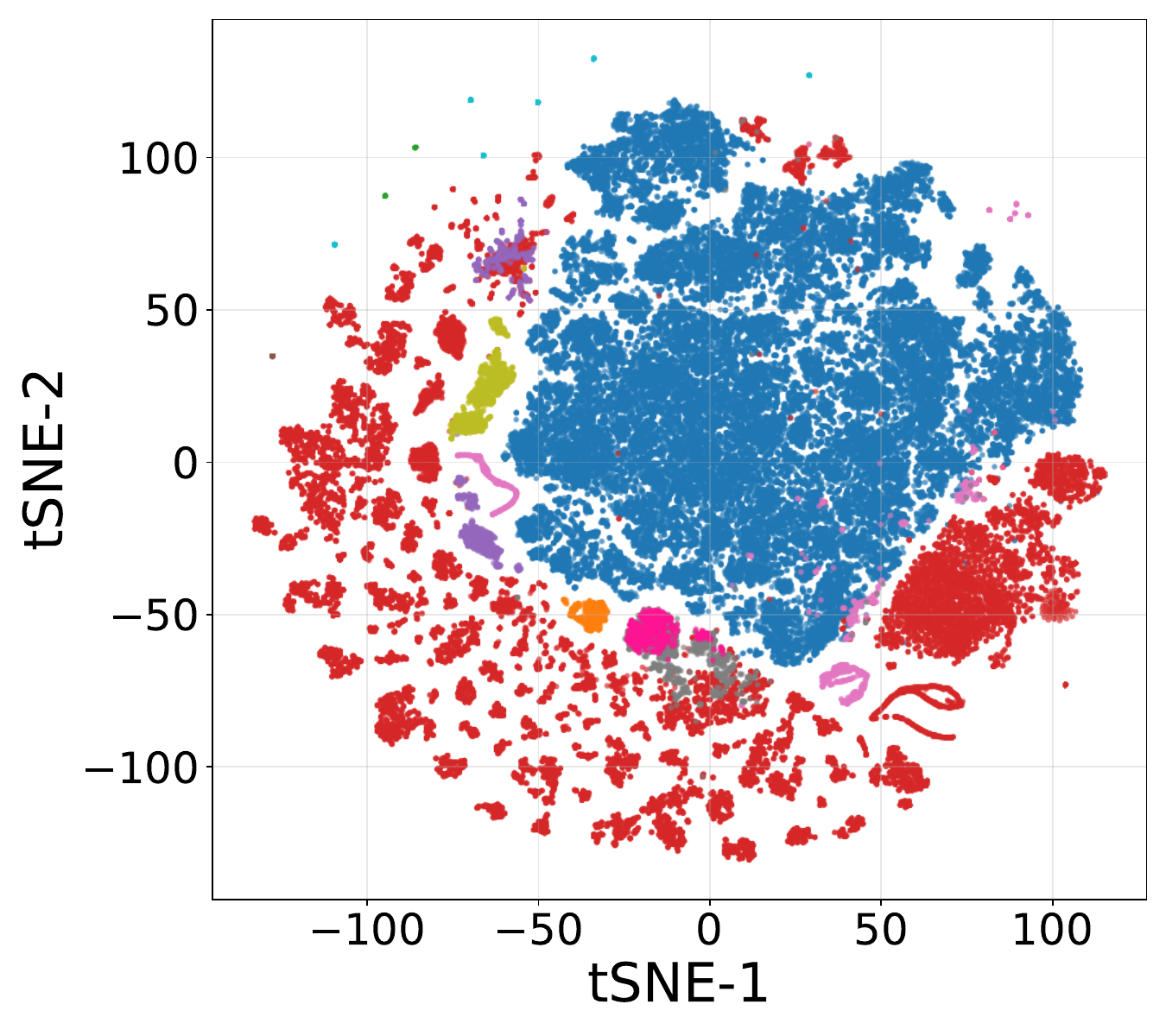}
    \subcaption{Layer 8}
\end{subfigure}\hfill
\begin{subfigure}[b]{0.17\linewidth}
    \centering
    \includegraphics[width=\linewidth]{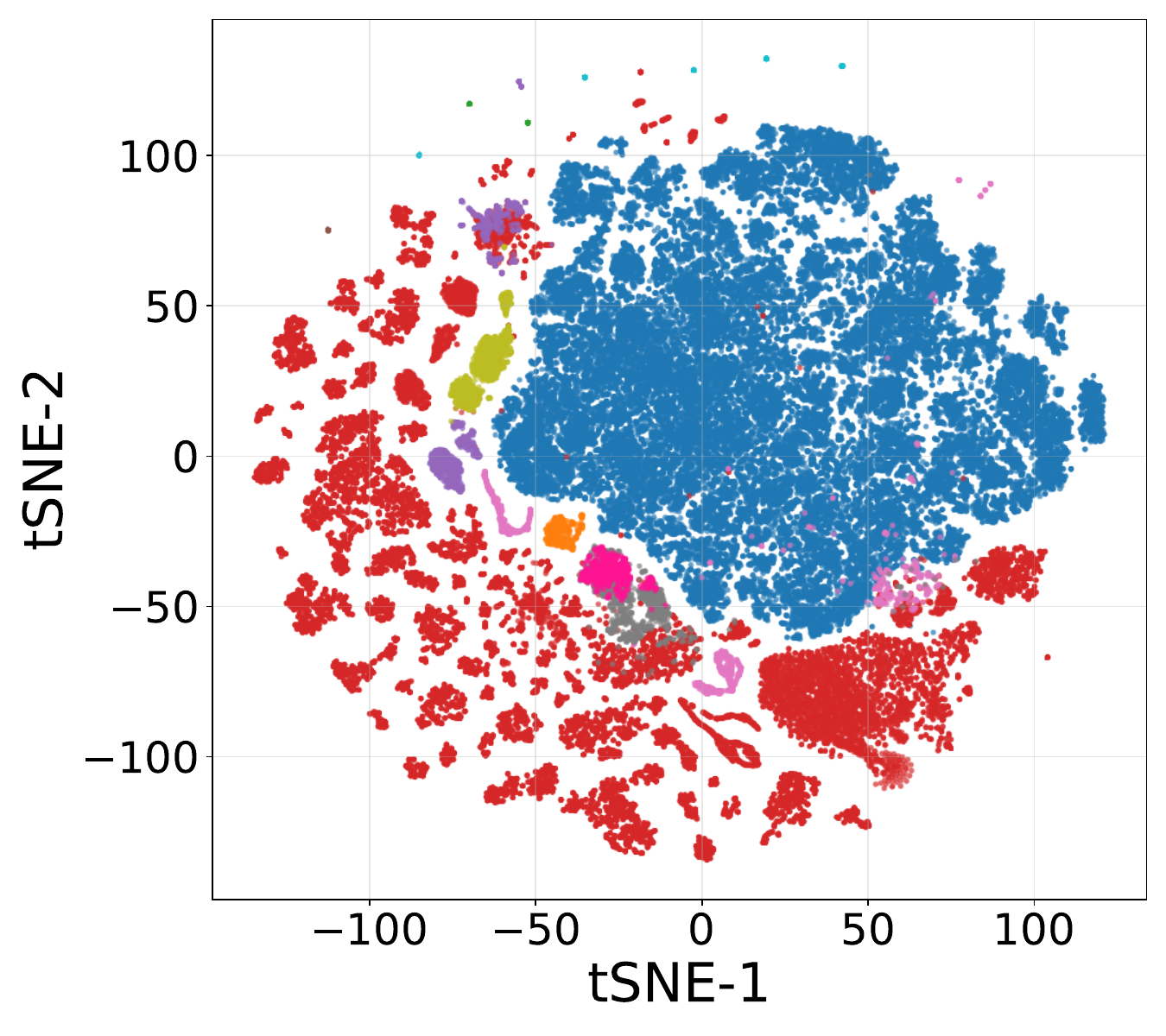}
    \subcaption{Layer 9}
\end{subfigure}

\vspace{3pt}

\begin{subfigure}[b]{0.17\linewidth}
    \centering
    \includegraphics[width=\linewidth]{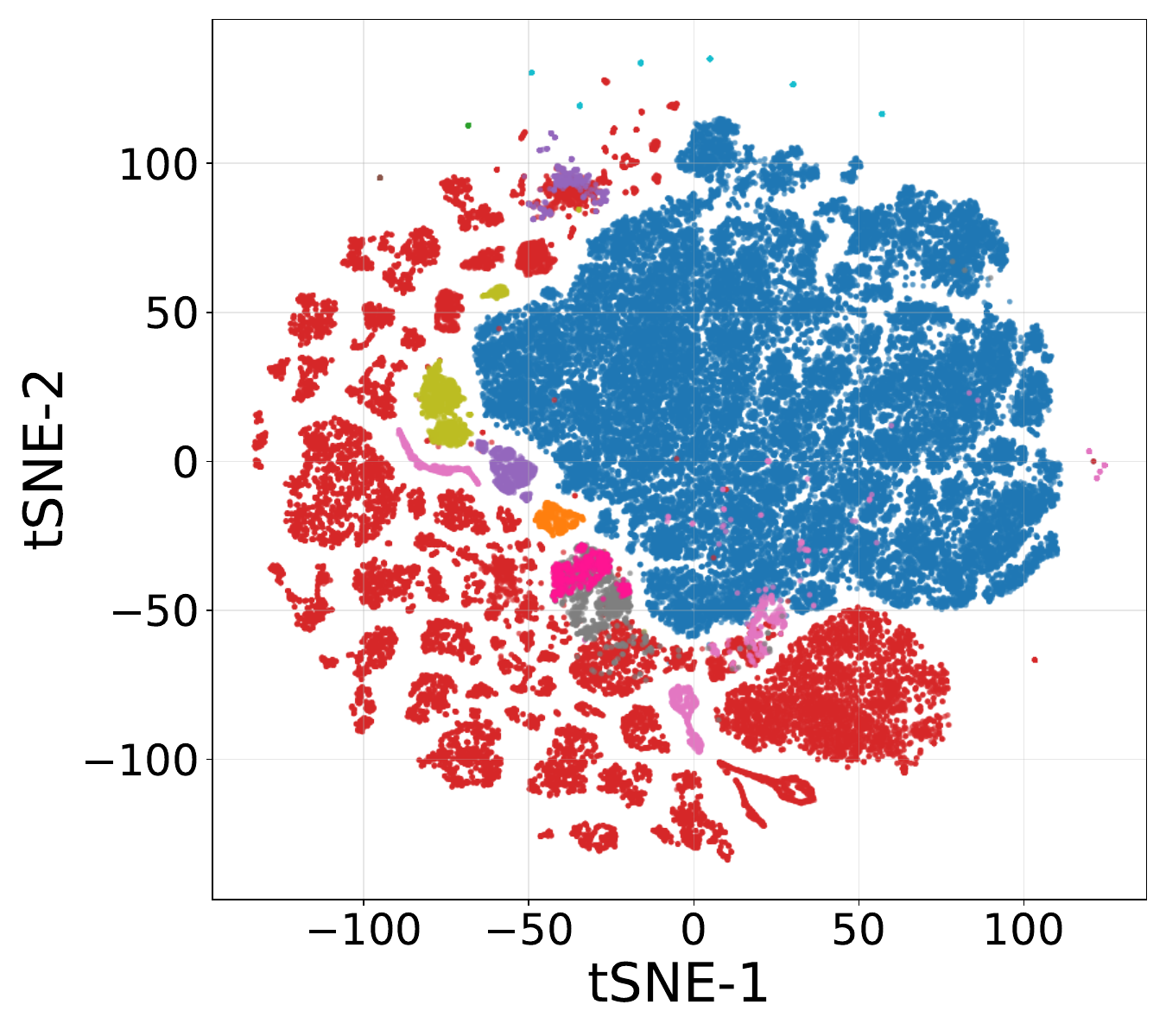}
    \subcaption{Layer 10}
\end{subfigure}\hfill
\begin{subfigure}[b]{0.17\linewidth}
    \centering
    \includegraphics[width=\linewidth]{latex/Figure/figures/llama/llama_11.pdf}
    \subcaption{Layer 11}
\end{subfigure}\hfill
\begin{subfigure}[b]{0.17\linewidth}
    \centering
    \includegraphics[width=\linewidth]{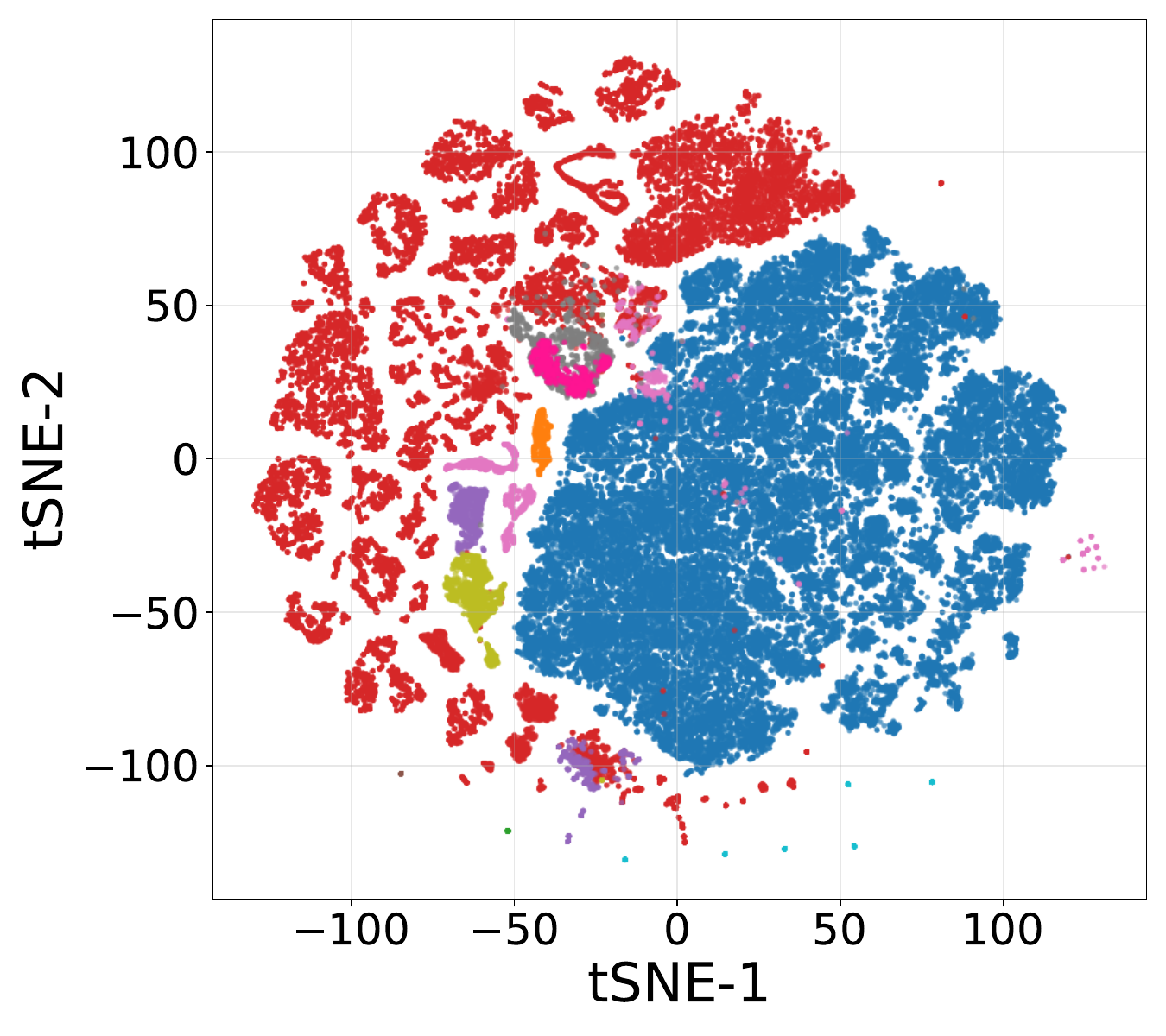}
    \subcaption{Layer 12}
\end{subfigure}\hfill
\begin{subfigure}[b]{0.17\linewidth}
    \centering
    \includegraphics[width=\linewidth]{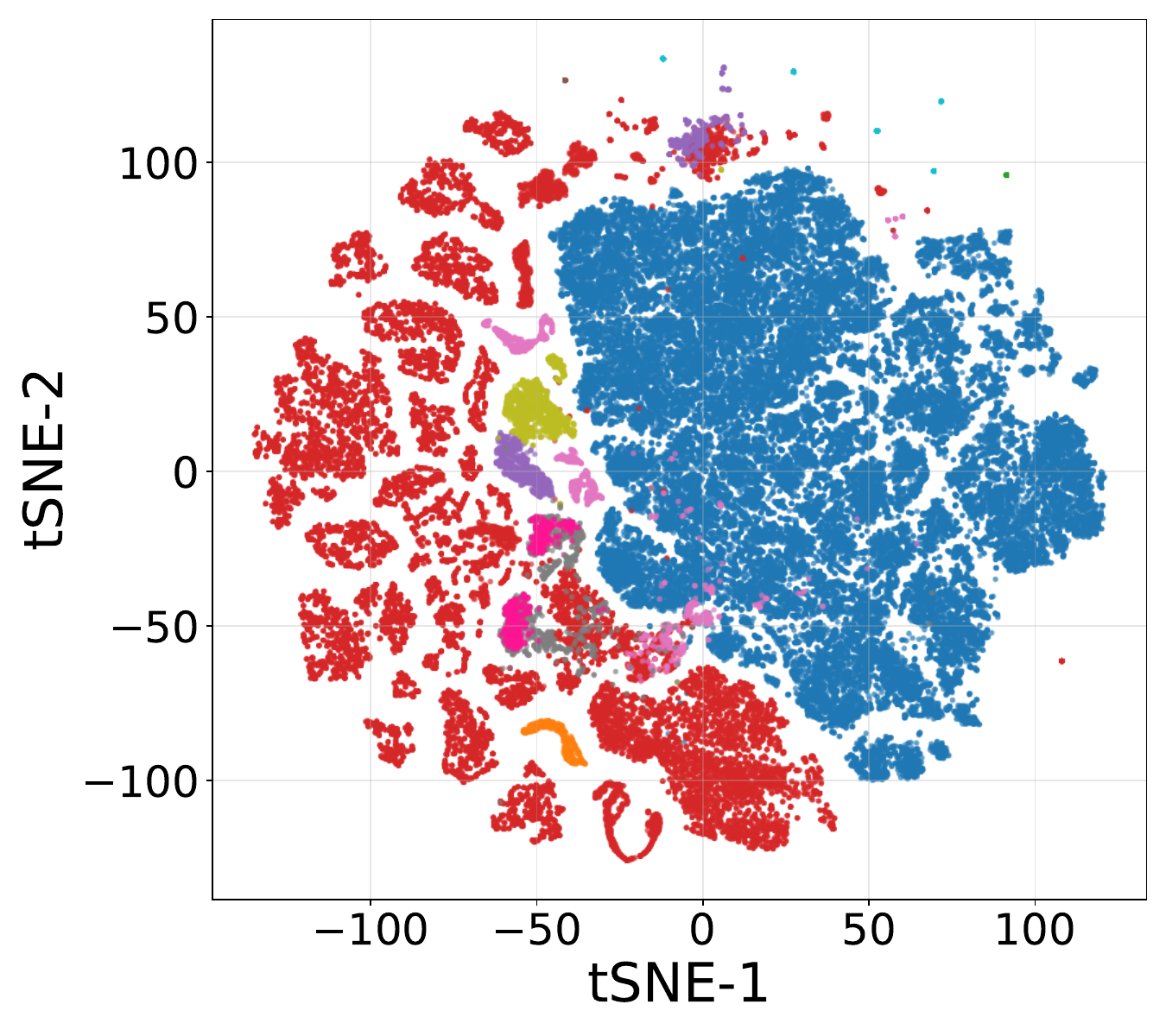}
    \subcaption{Layer 13}
\end{subfigure}\hfill
\begin{subfigure}[b]{0.17\linewidth}
    \centering
    \includegraphics[width=\linewidth]{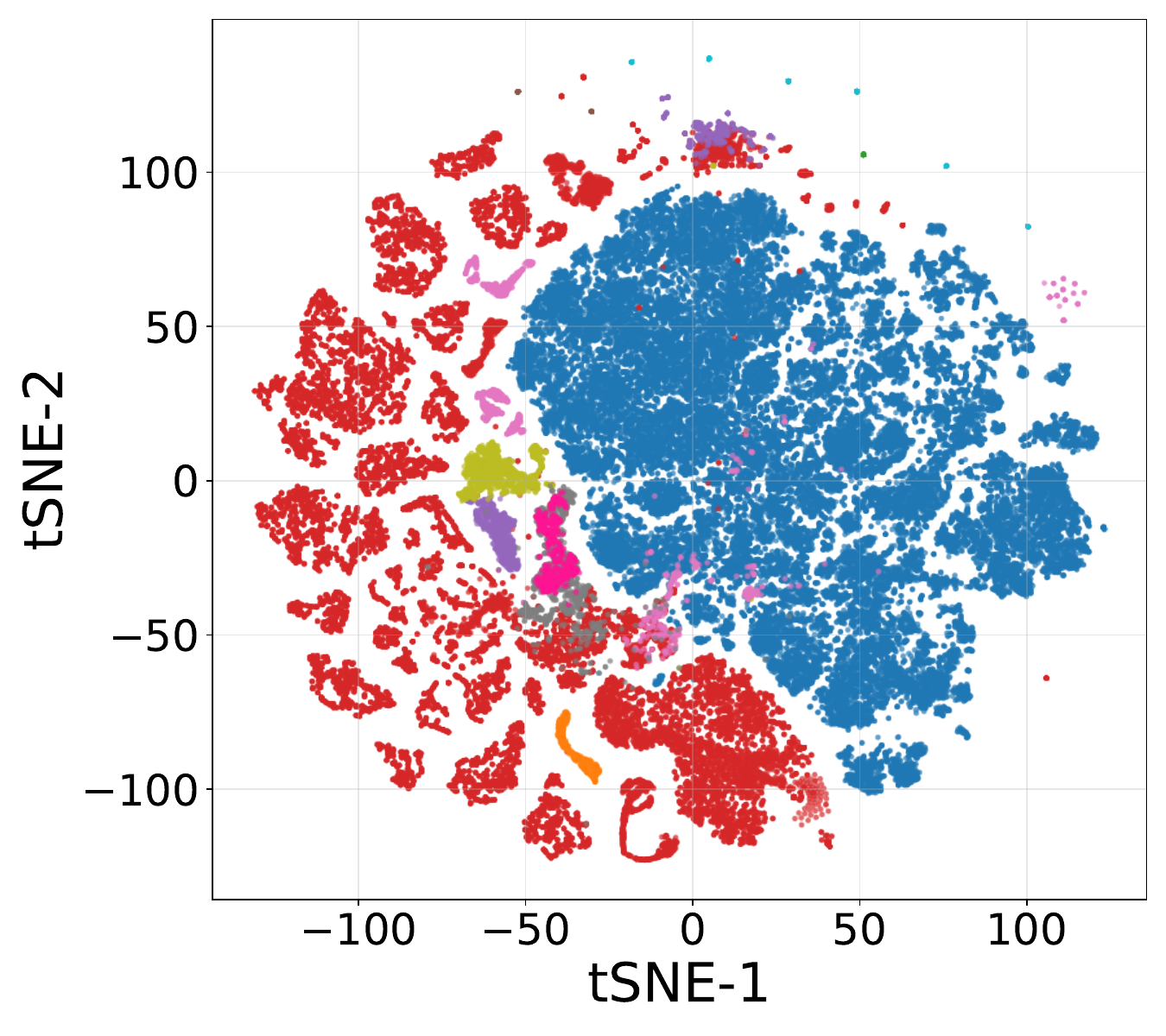}
    \subcaption{Layer 14}
\end{subfigure}

\vspace{3pt}

\begin{subfigure}[b]{0.17\linewidth}
    \centering
    \includegraphics[width=\linewidth]{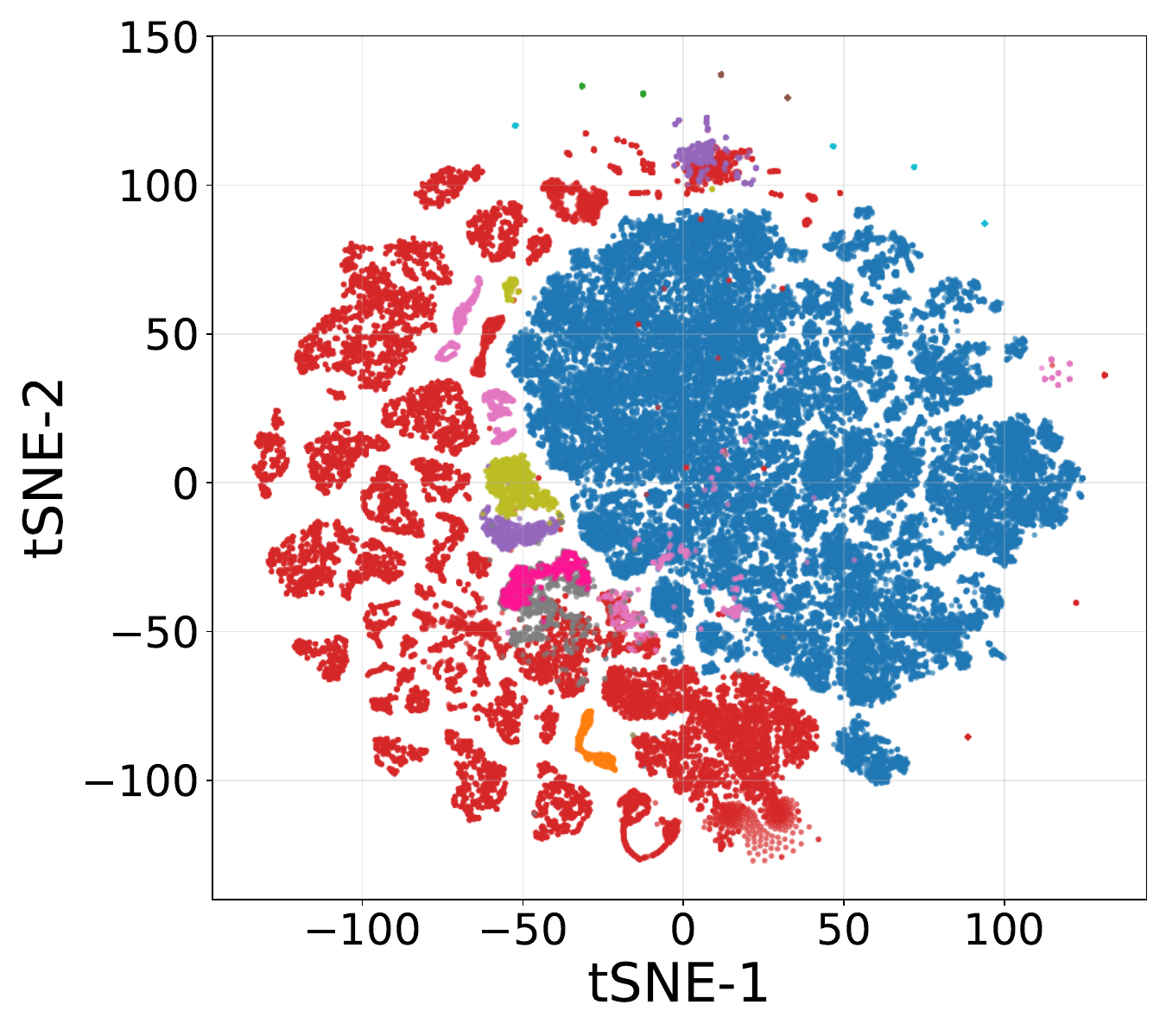}
    \subcaption{Layer 15}
\end{subfigure}\hfill
\begin{subfigure}[b]{0.17\linewidth}
    \centering
    \includegraphics[width=\linewidth]{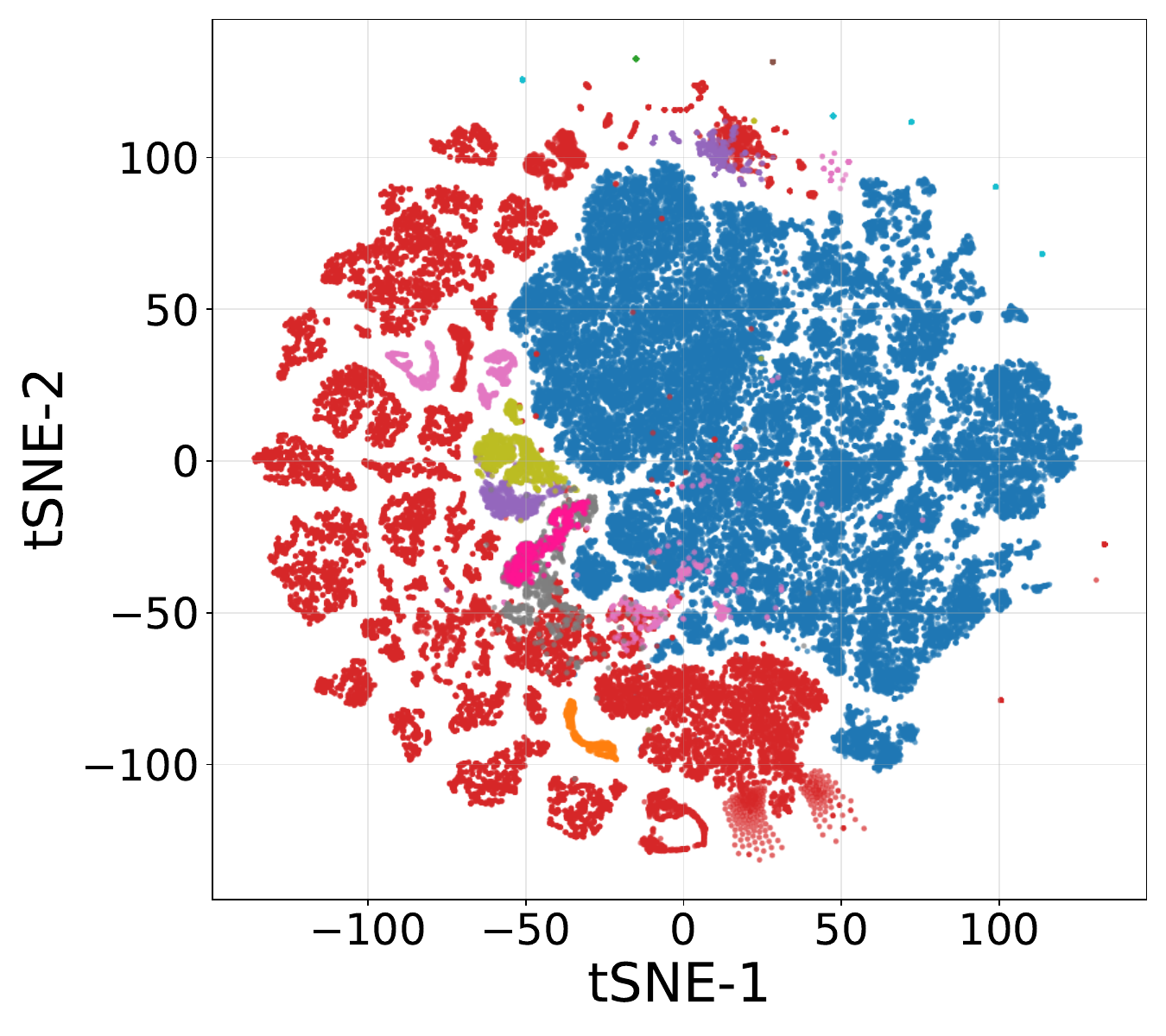}
    \subcaption{Layer 16}
\end{subfigure}\hfill
\begin{subfigure}[b]{0.17\linewidth}
    \centering
    \includegraphics[width=\linewidth]{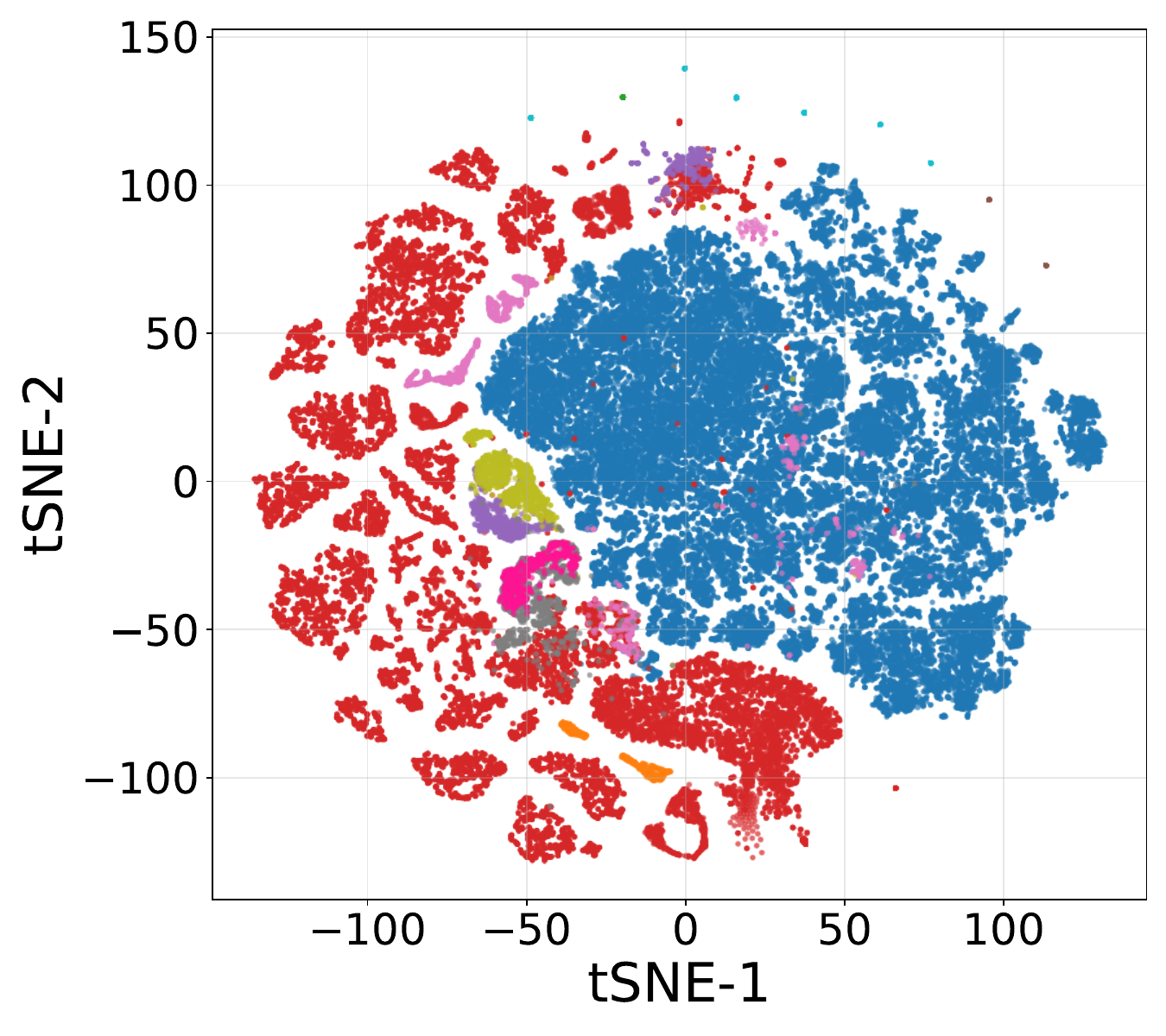}
    \subcaption{Layer 17}
\end{subfigure}\hfill
\begin{subfigure}[b]{0.17\linewidth}
    \centering
    \includegraphics[width=\linewidth]{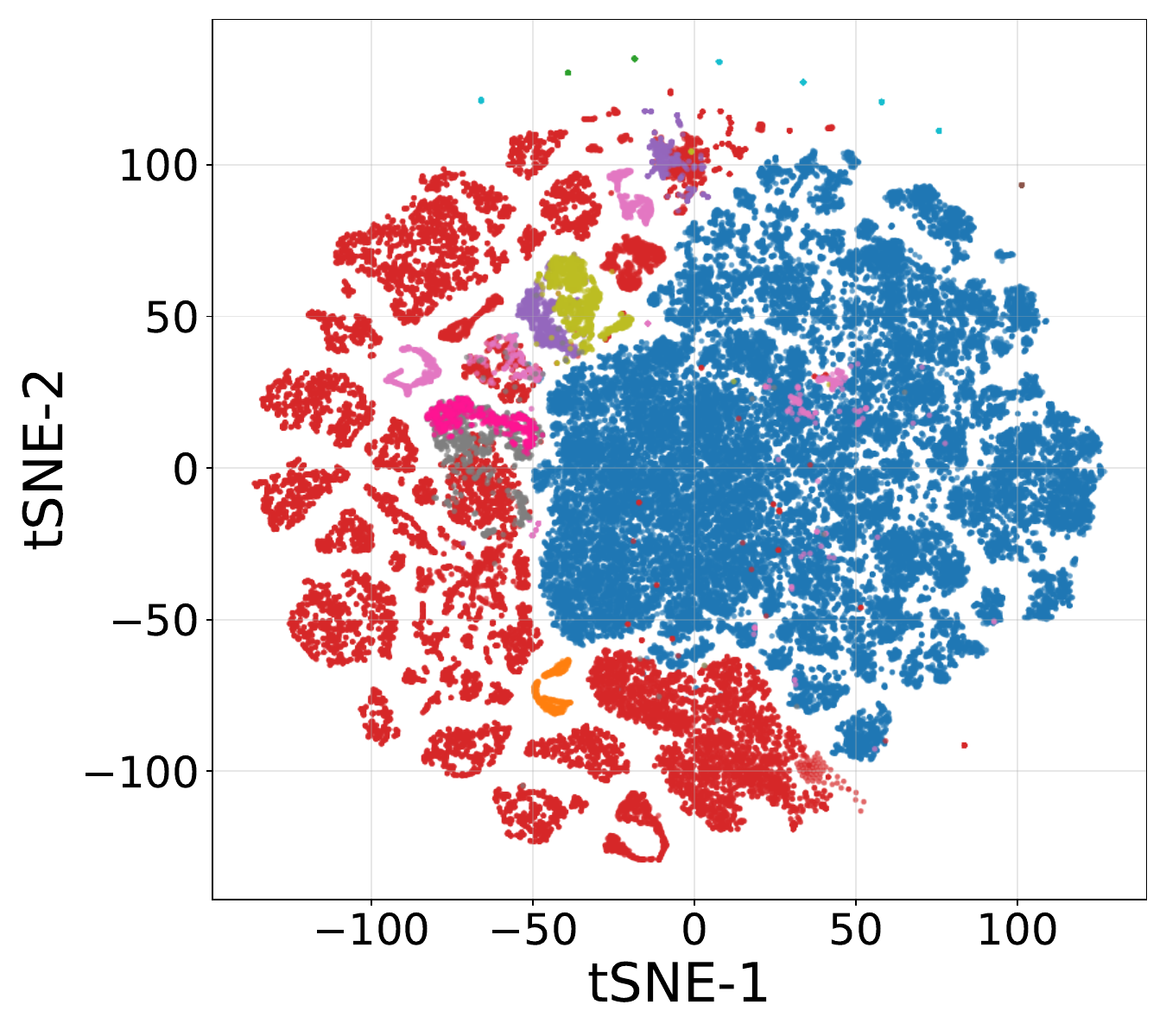}
    \subcaption{Layer 18}
\end{subfigure}\hfill
\begin{subfigure}[b]{0.17\linewidth}
    \centering
    \includegraphics[width=\linewidth]{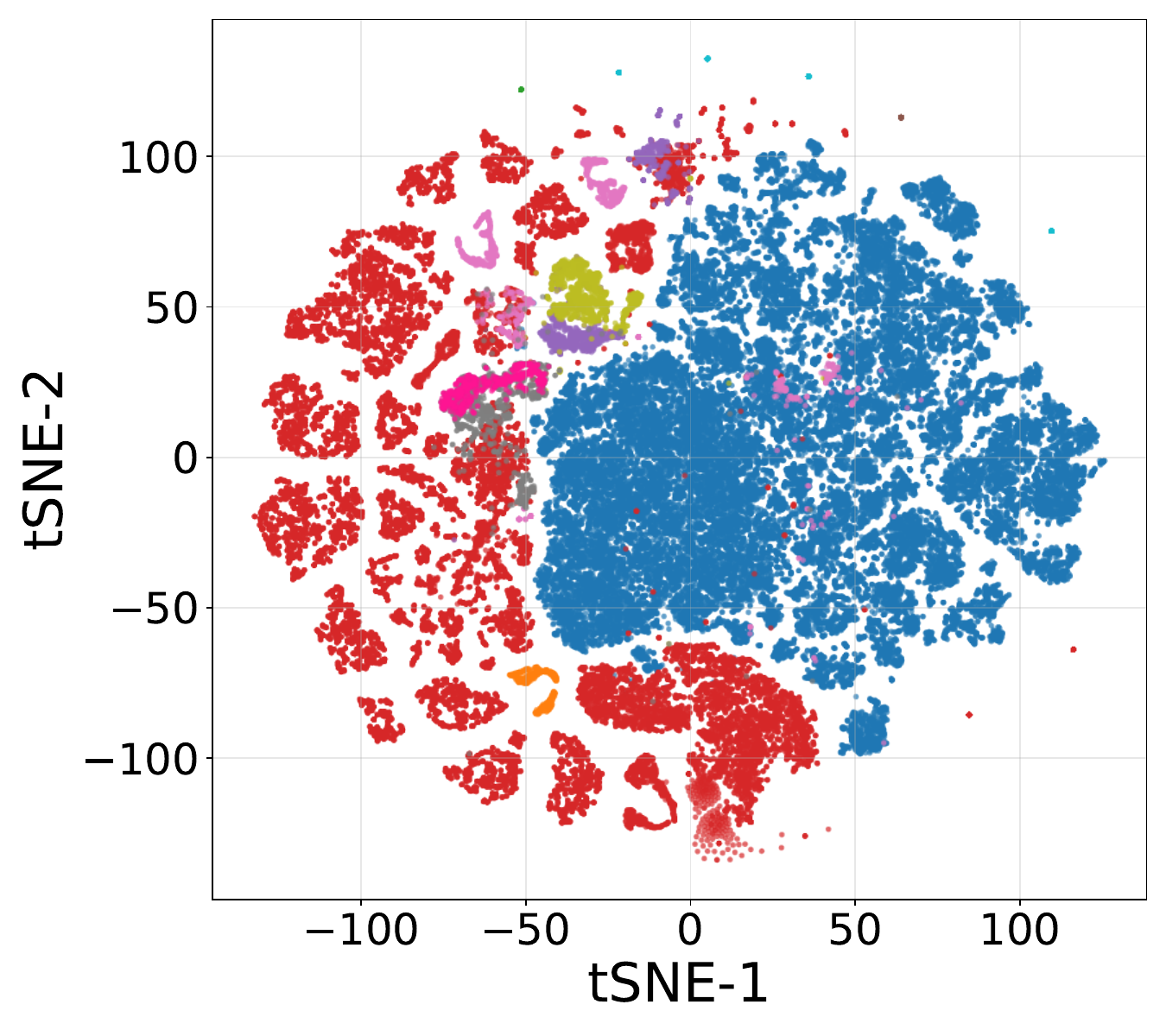}
    \subcaption{Layer 19}
\end{subfigure}

\vspace{3pt}

\begin{subfigure}[b]{0.17\linewidth}
    \centering
    \includegraphics[width=\linewidth]{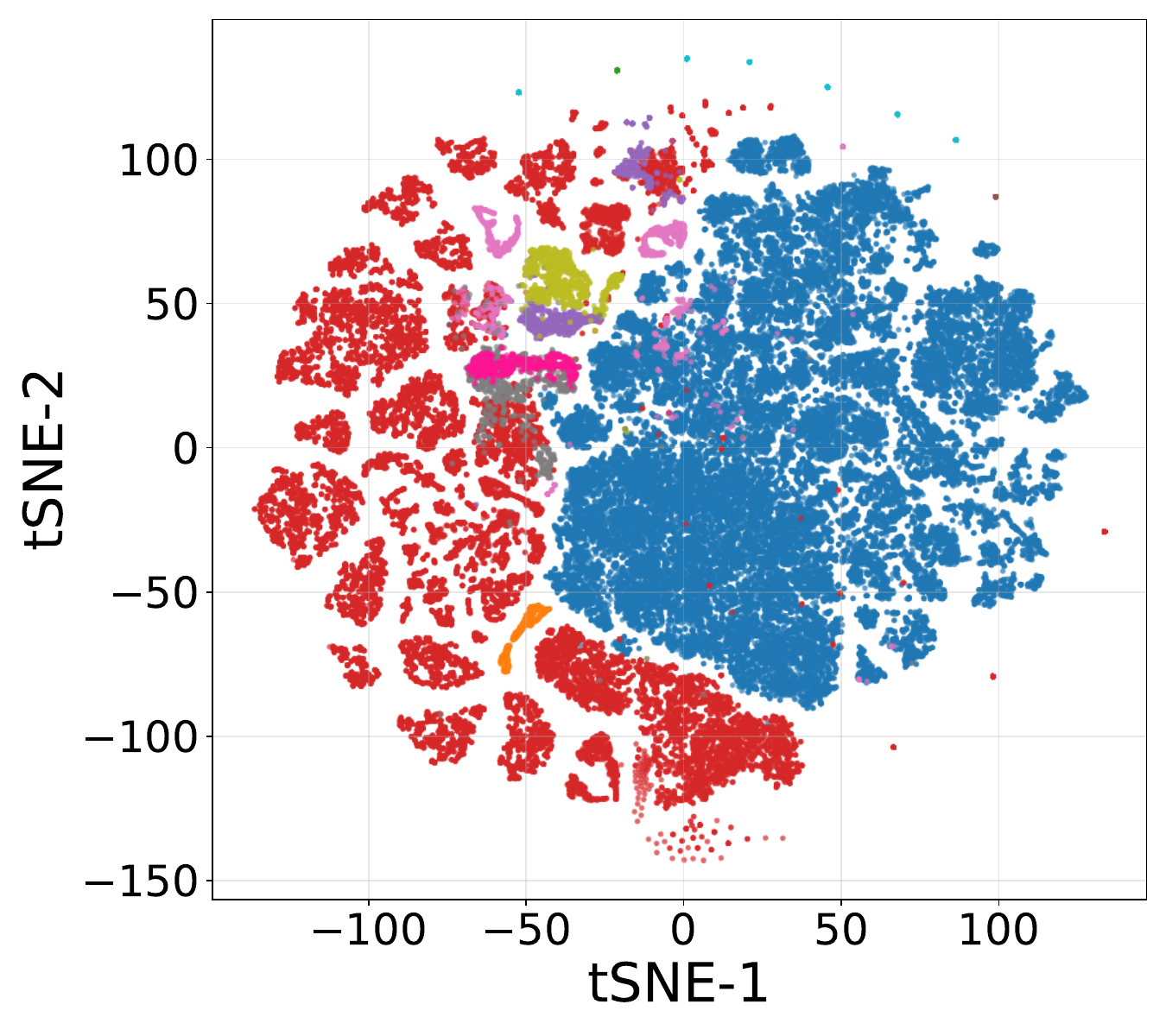}
    \subcaption{Layer 20}
\end{subfigure}\hfill
\begin{subfigure}[b]{0.17\linewidth}
    \centering
    \includegraphics[width=\linewidth]{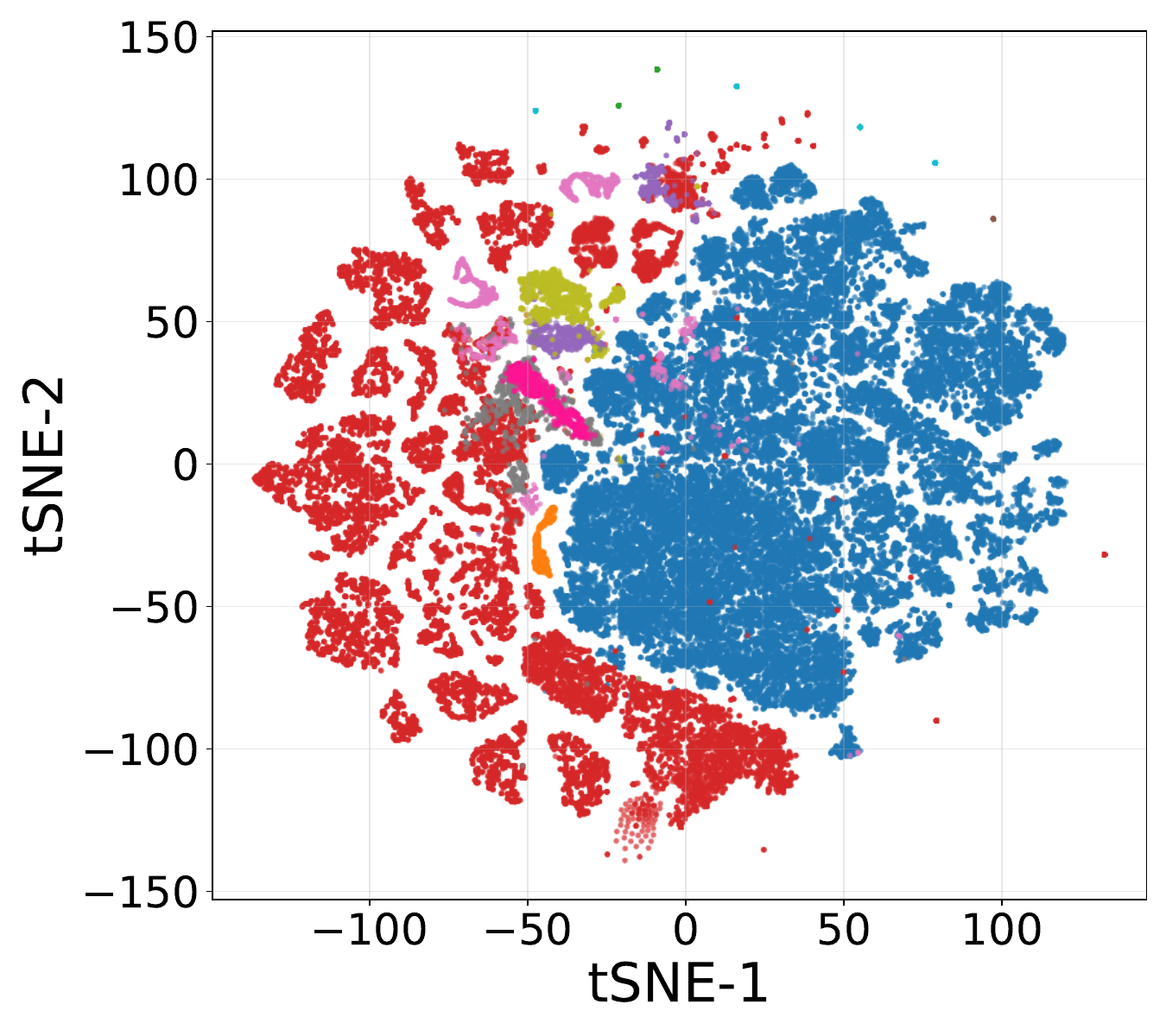}
    \subcaption{Layer 21}
\end{subfigure}\hfill
\begin{subfigure}[b]{0.17\linewidth}
    \centering
    \includegraphics[width=\linewidth]{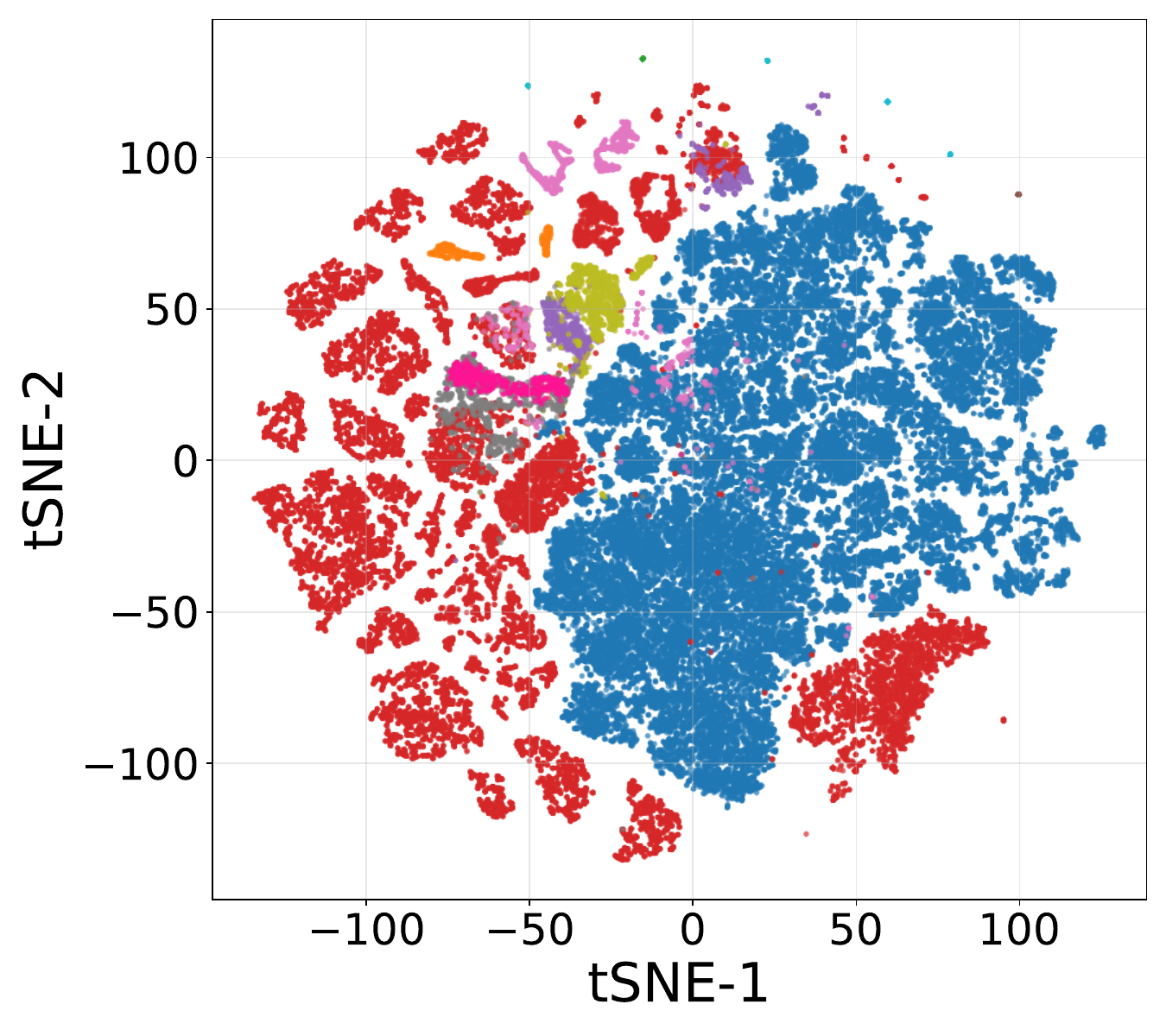}
    \subcaption{Layer 22}
\end{subfigure}\hfill
\begin{subfigure}[b]{0.17\linewidth}
    \centering
    \includegraphics[width=\linewidth]{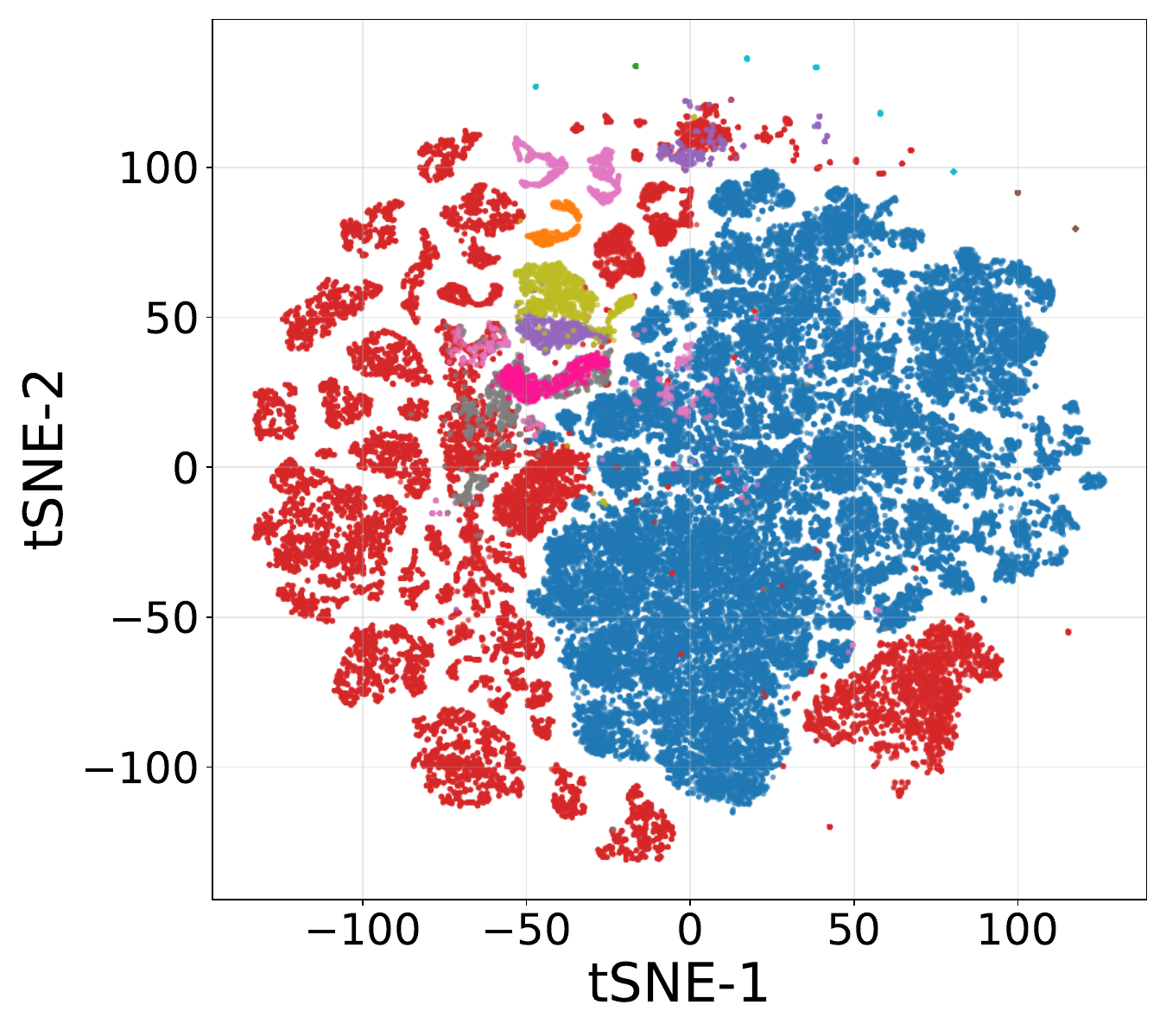}
    \subcaption{Layer 23}
\end{subfigure}\hfill
\begin{subfigure}[b]{0.17\linewidth}
    \centering
    \includegraphics[width=\linewidth]{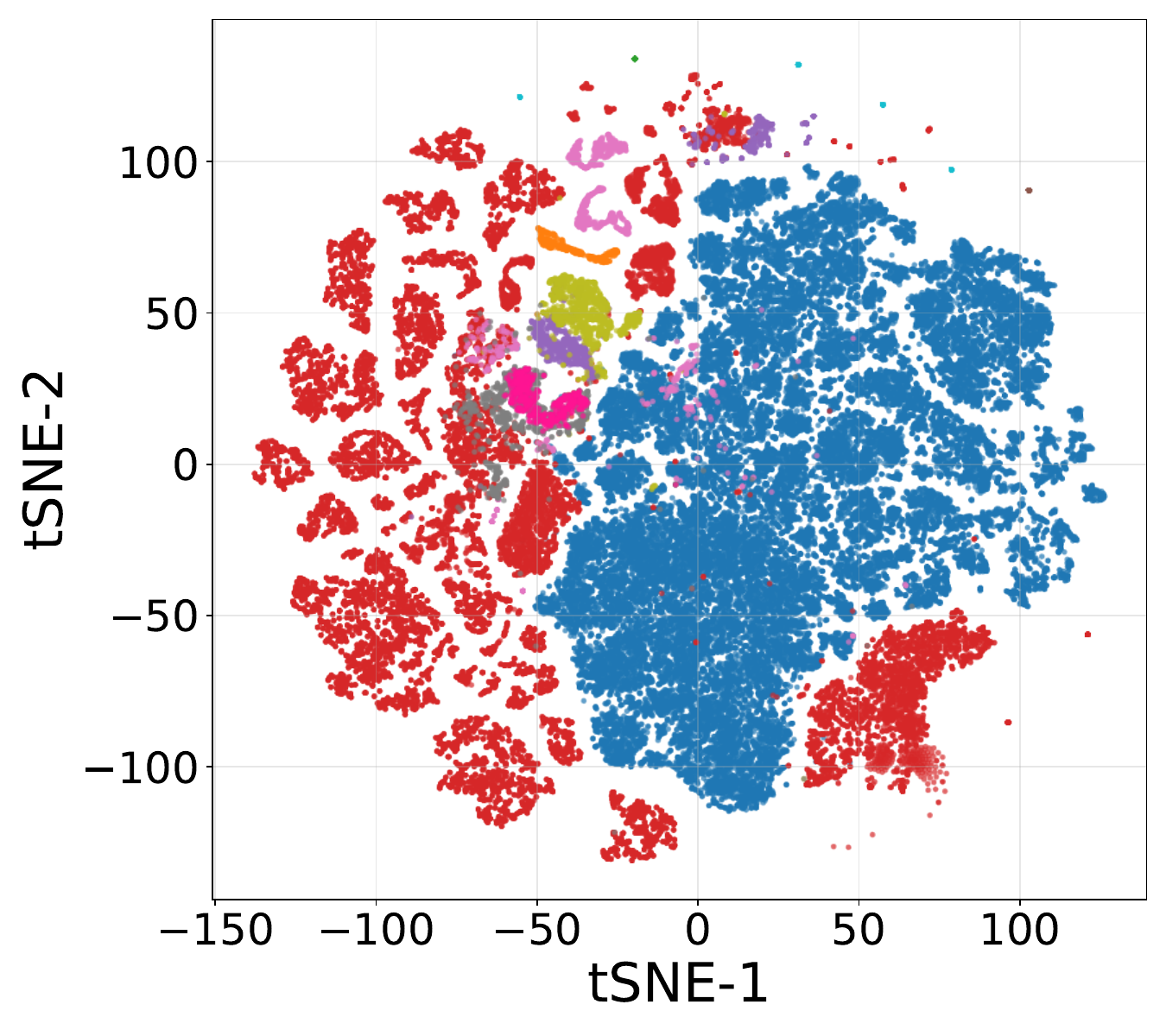}
    \subcaption{Layer 24}
\end{subfigure}

\vspace{3pt}

\begin{subfigure}[b]{0.17\linewidth}
    \centering
    \includegraphics[width=\linewidth]{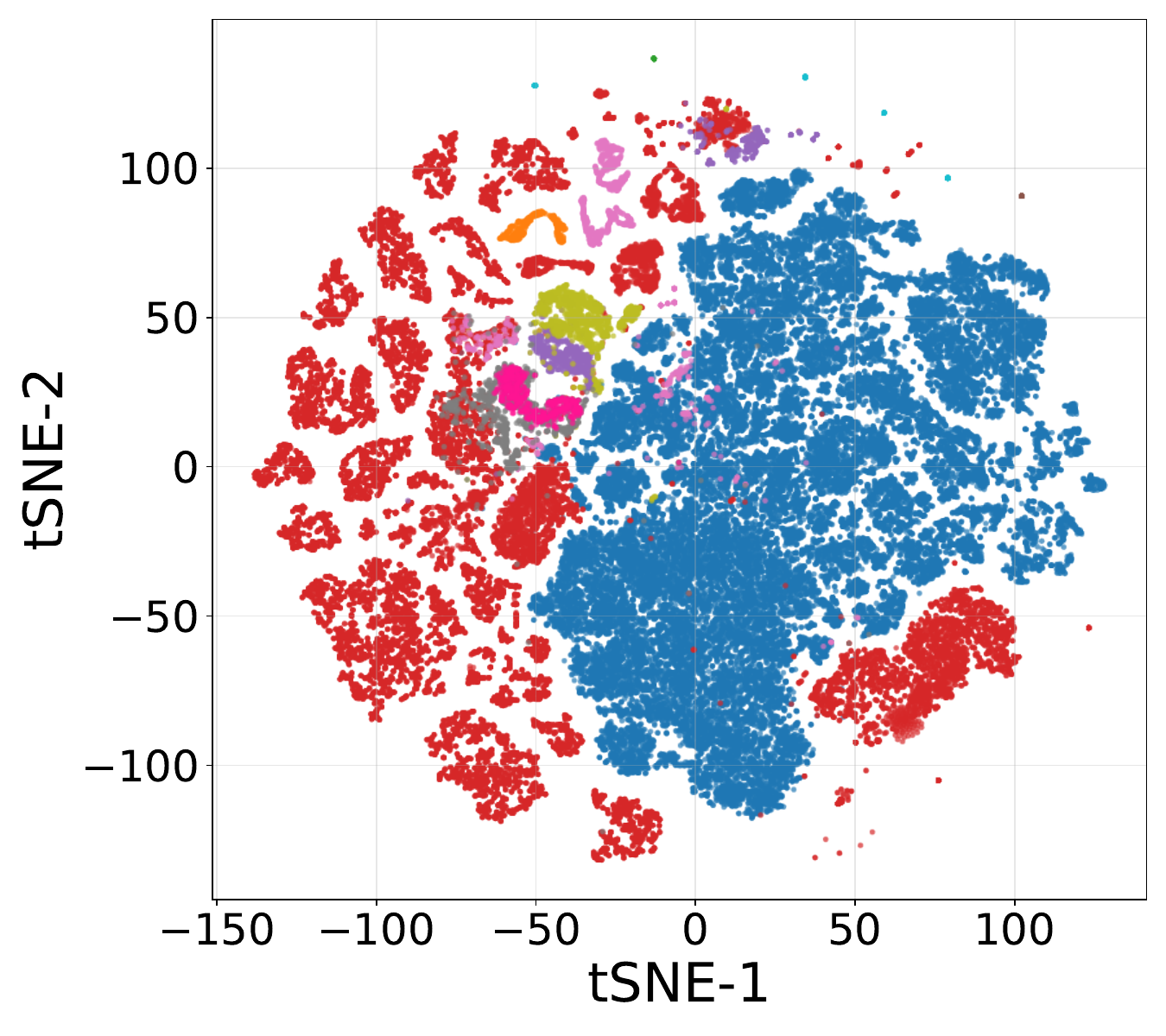}
    \subcaption{Layer 25}
\end{subfigure}\hfill
\begin{subfigure}[b]{0.17\linewidth}
    \centering
    \includegraphics[width=\linewidth]{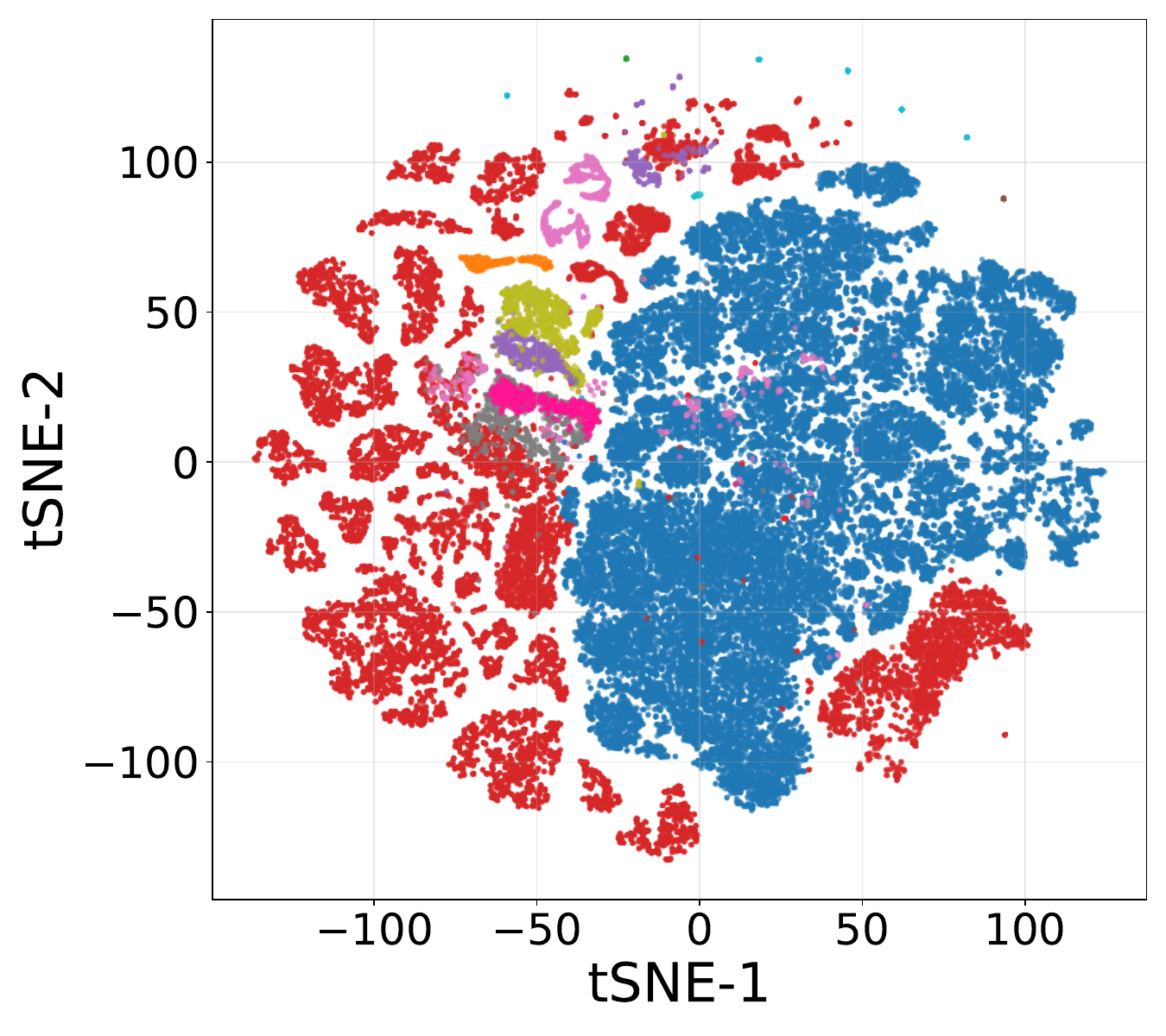}
    \subcaption{Layer 26}
\end{subfigure}\hfill
\begin{subfigure}[b]{0.17\linewidth}
    \centering
    \includegraphics[width=\linewidth]{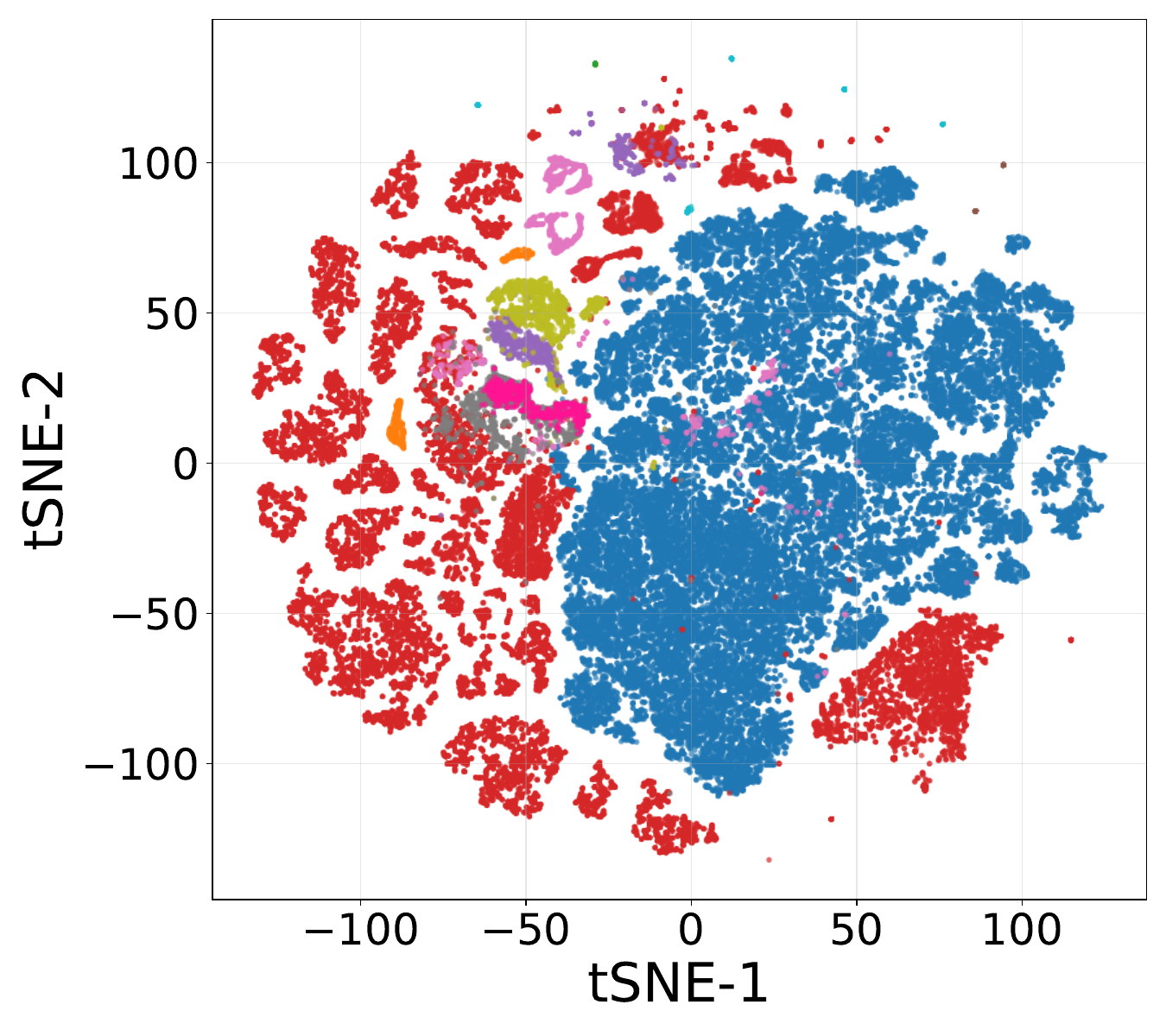}
    \subcaption{Layer 27}
\end{subfigure}\hfill
\begin{subfigure}[b]{0.17\linewidth}
    \centering
    \includegraphics[width=\linewidth]{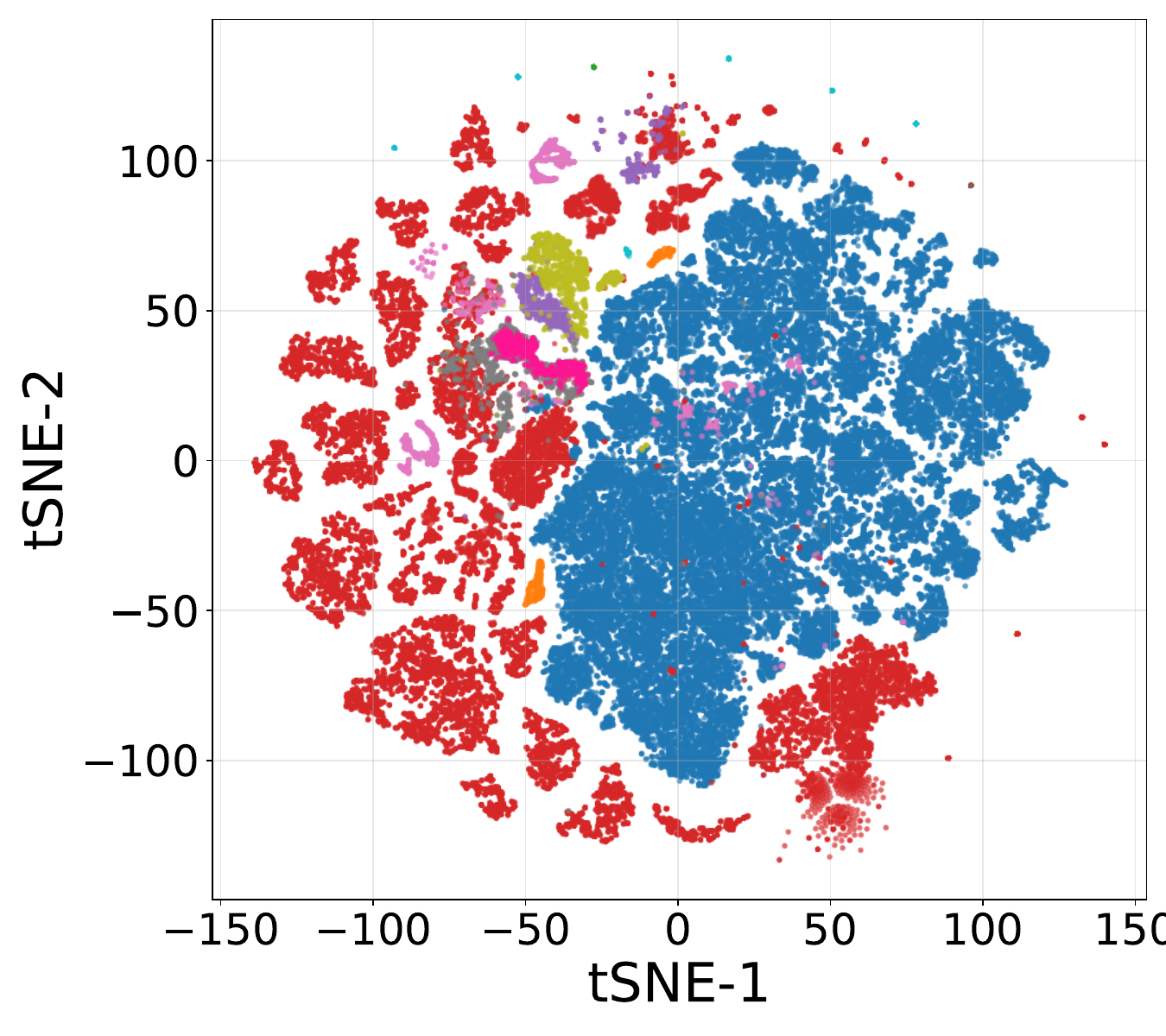}
    \subcaption{Layer 28}
\end{subfigure}\hfill
\begin{subfigure}[b]{0.17\linewidth}
    \centering
    \includegraphics[width=\linewidth]{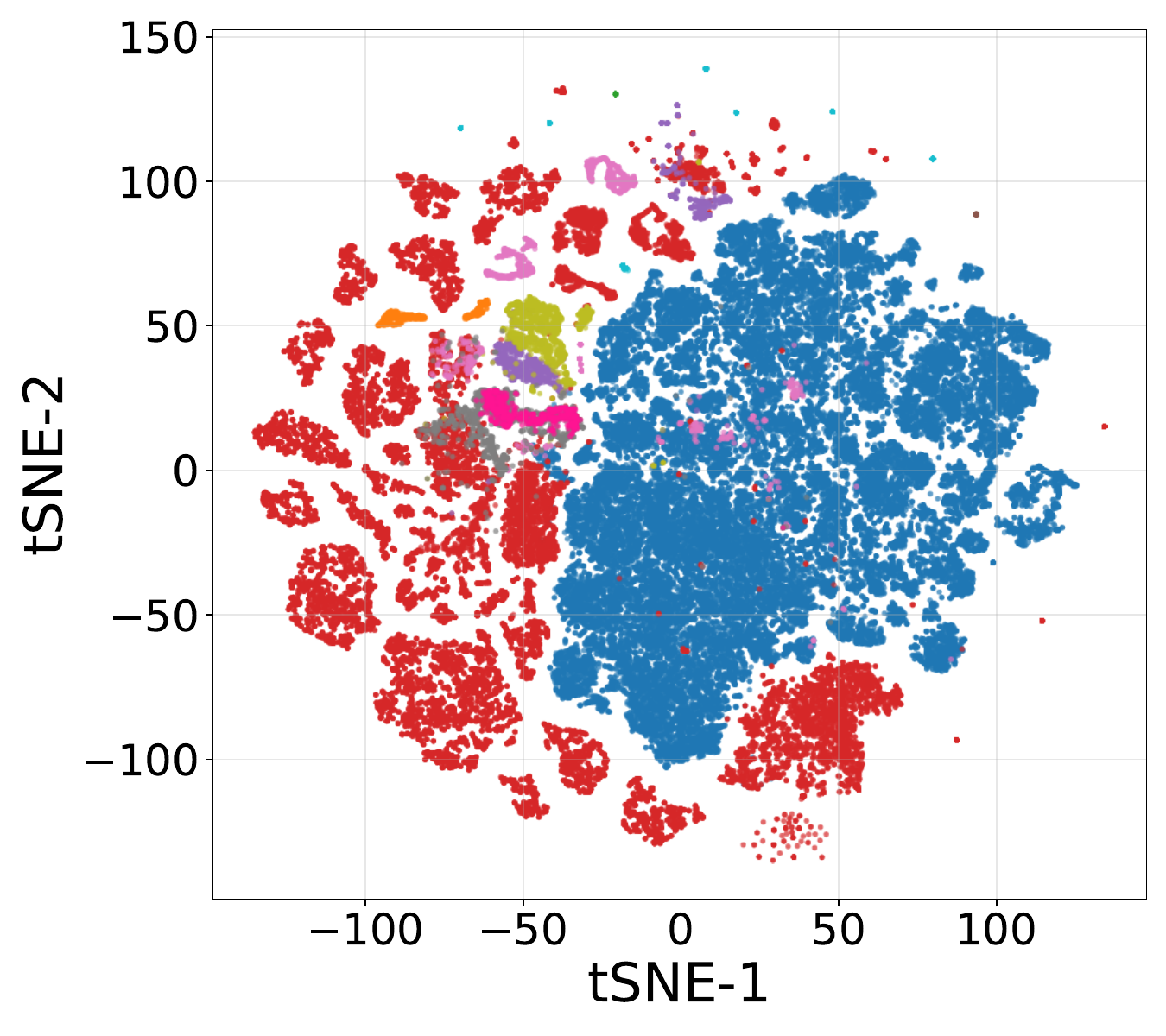}
    \subcaption{Layer 29}
\end{subfigure}

\vspace{3pt}

\begin{subfigure}[b]{0.17\linewidth}
    \centering
    \includegraphics[width=\linewidth]{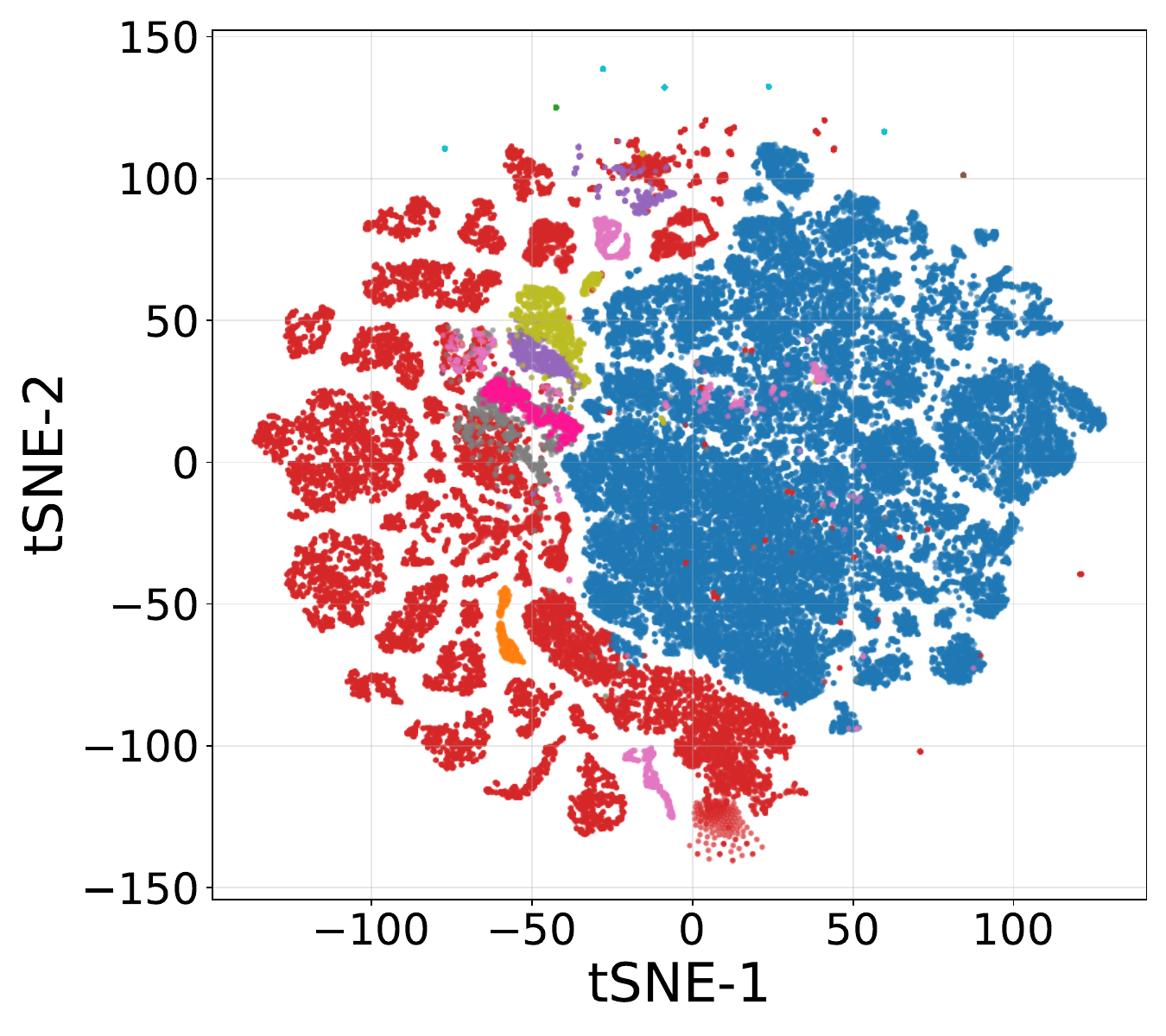}
    \subcaption{Layer 30}
\end{subfigure}\hfill
\begin{subfigure}[b]{0.17\linewidth}
    \centering
    \includegraphics[width=\linewidth]{latex/Figure/figures/llama/llama_31.pdf}
    \subcaption{Layer 31}
\end{subfigure}

\caption{t-SNE visualization of last-token activations across all layers (Layer 0 to Layer 31) of LLaMA-7B.}
\label{fig:llama_tsne}
\end{figure*}

\begin{figure*}[t]
\centering

\includegraphics[width=0.9\textwidth, trim={0 0.8cm 0 0 cm}, clip]{latex/Figure/legend.pdf}

\begin{subfigure}[b]{0.18\linewidth}
\includegraphics[width=\linewidth]{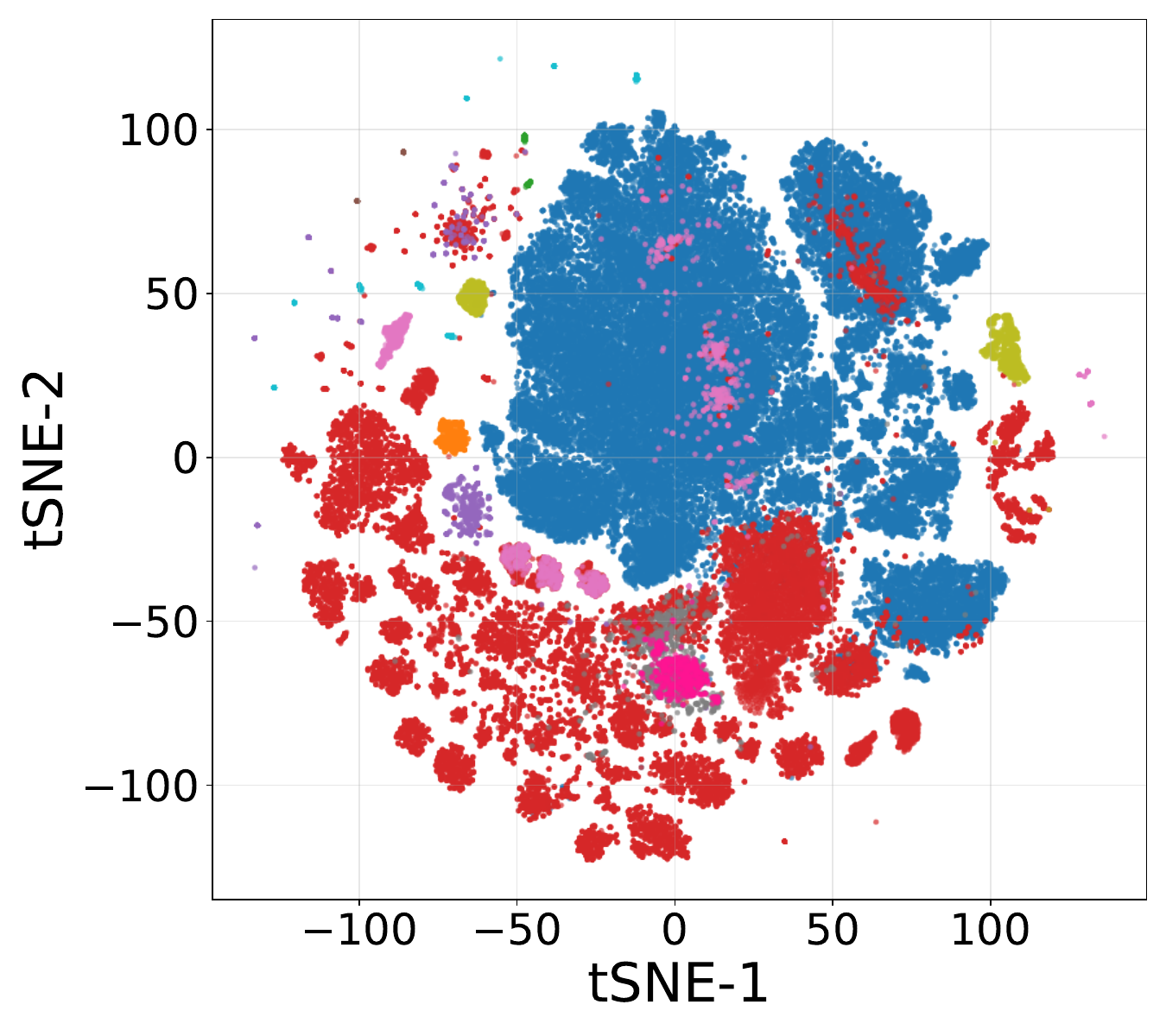}
\subcaption{Layer 0}
\end{subfigure}\hfill
\begin{subfigure}[b]{0.18\linewidth}
\includegraphics[width=\linewidth]{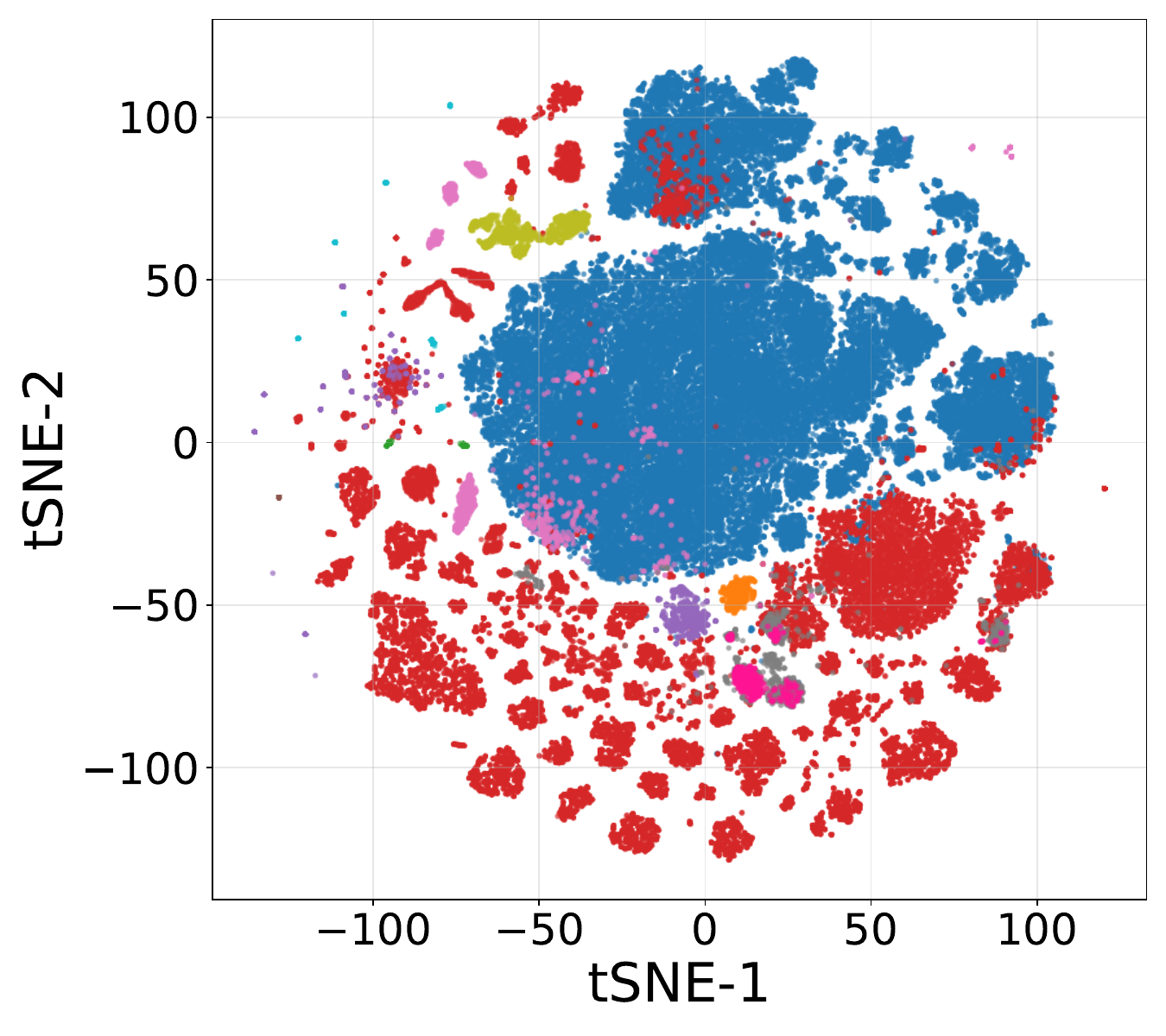}
\subcaption{Layer 1}
\end{subfigure}\hfill
\begin{subfigure}[b]{0.18\linewidth}
\includegraphics[width=\linewidth]{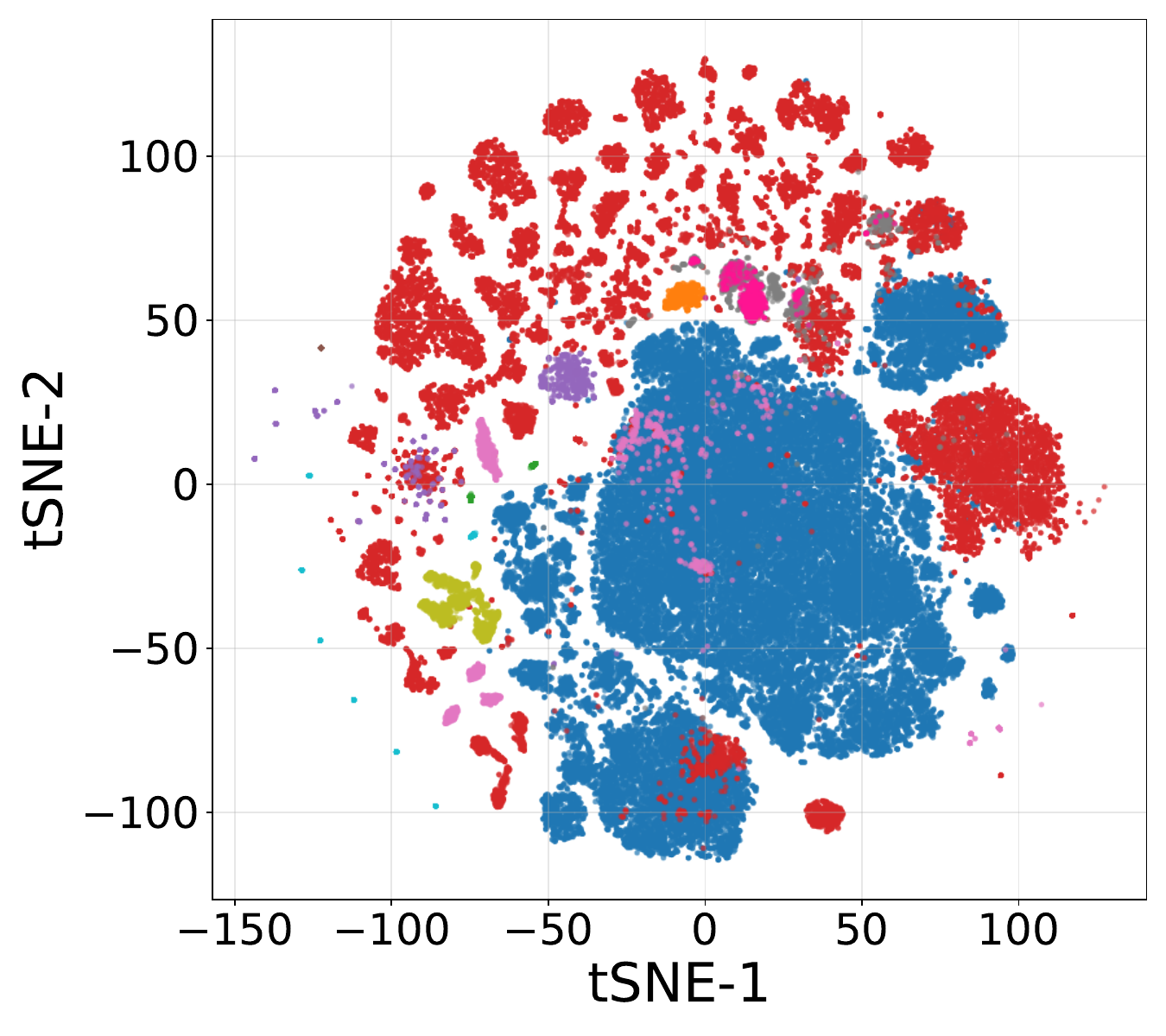}
\subcaption{Layer 2}
\end{subfigure}\hfill
\begin{subfigure}[b]{0.18\linewidth}
\includegraphics[width=\linewidth]{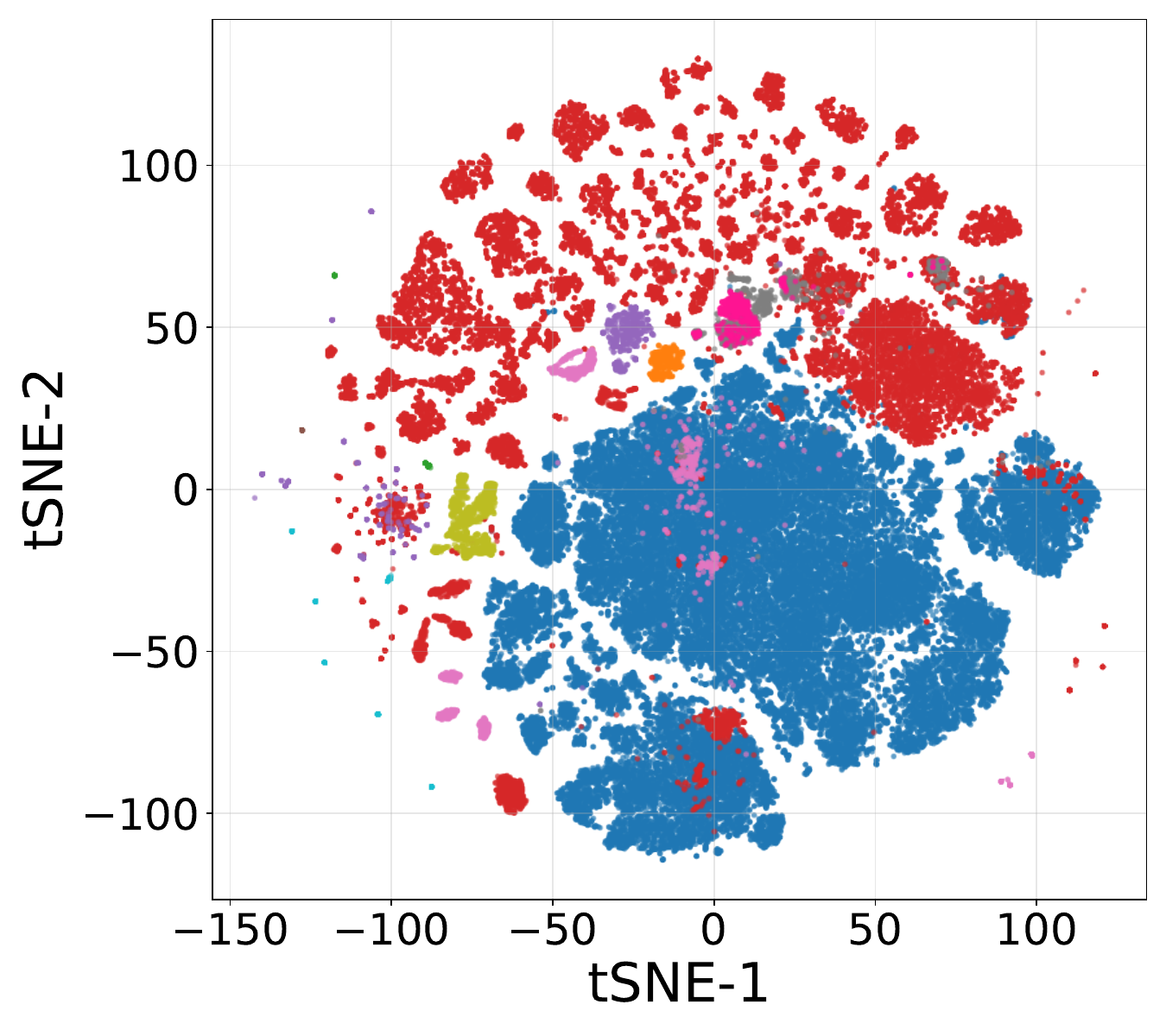}
\subcaption{Layer 3}
\end{subfigure}\hfill
\begin{subfigure}[b]{0.18\linewidth}
\includegraphics[width=\linewidth]{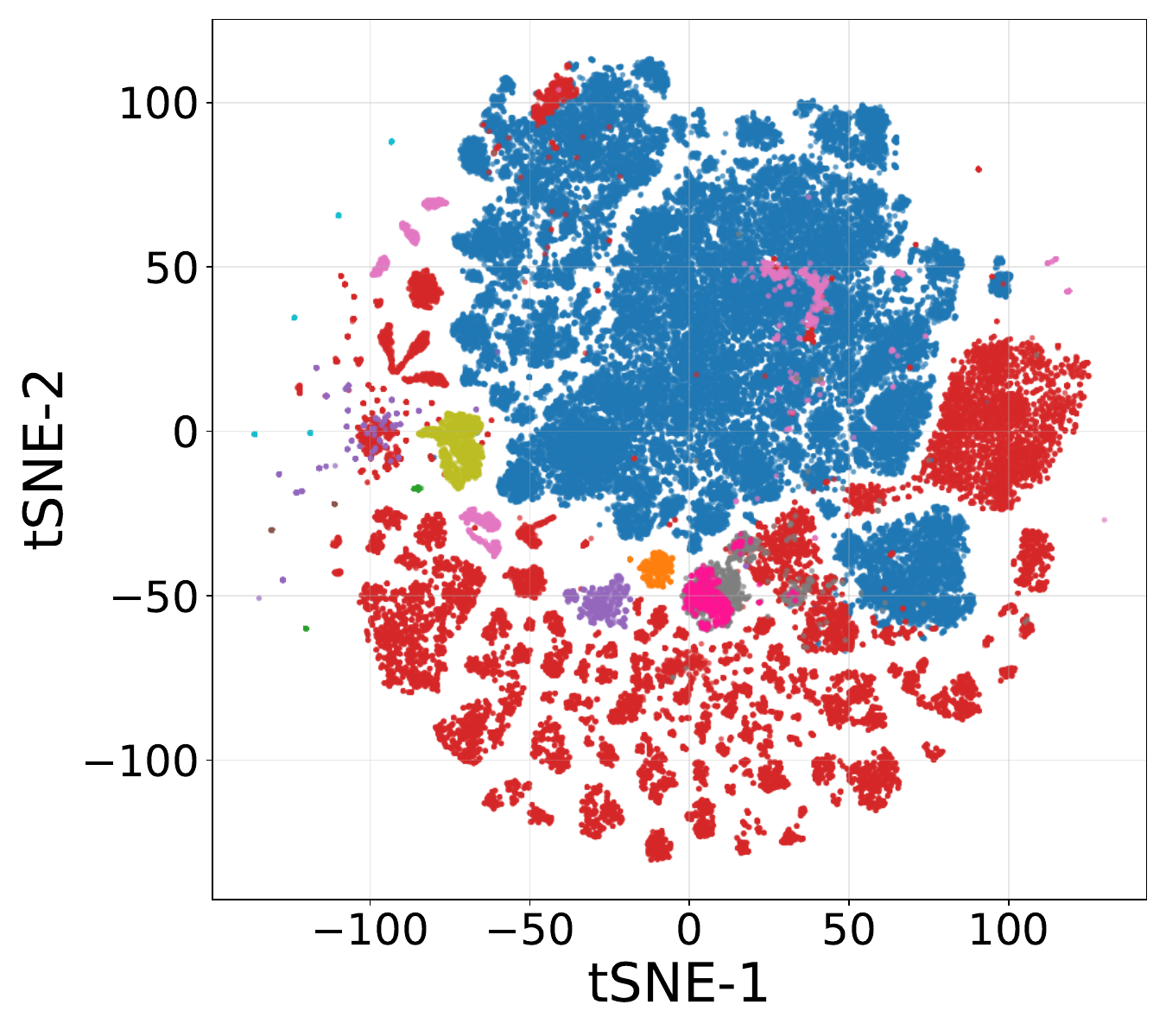}
\subcaption{Layer 4}
\end{subfigure}

\vspace{3pt}

\begin{subfigure}[b]{0.18\linewidth}
\includegraphics[width=\linewidth]{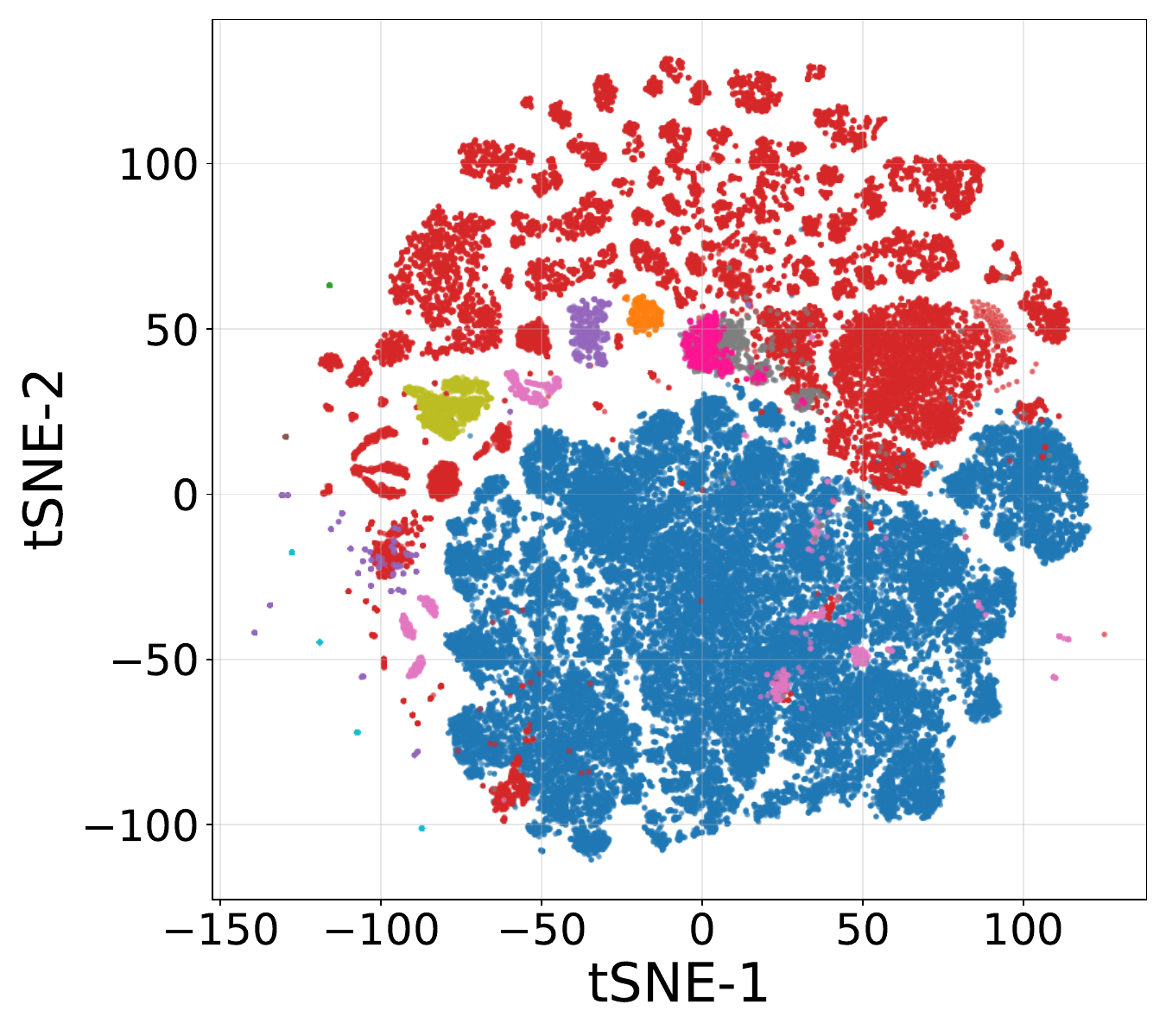}
\subcaption{Layer 5}
\end{subfigure}\hfill
\begin{subfigure}[b]{0.18\linewidth}
\includegraphics[width=\linewidth]{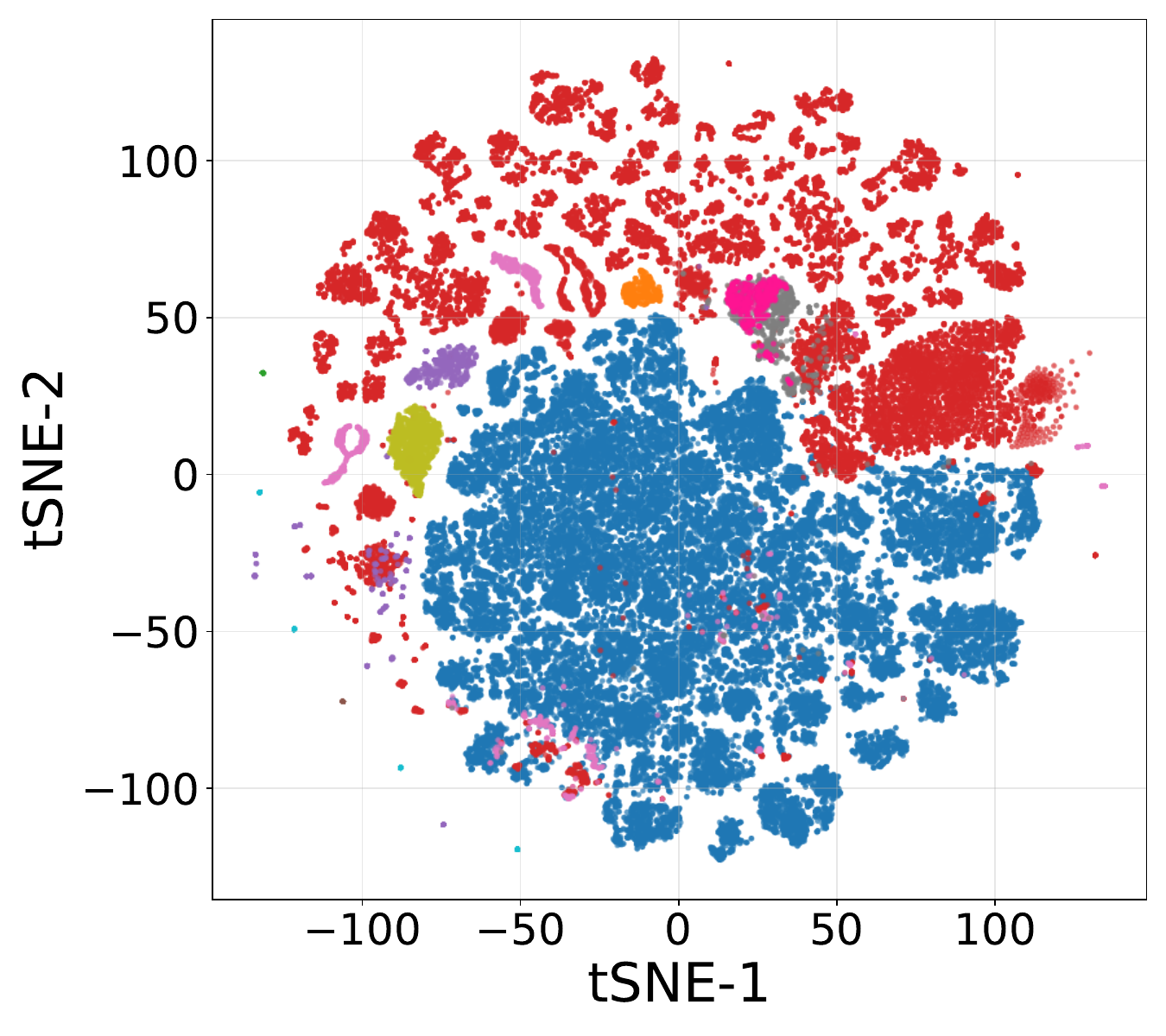}
\subcaption{Layer 6}
\end{subfigure}\hfill
\begin{subfigure}[b]{0.18\linewidth}
\includegraphics[width=\linewidth]{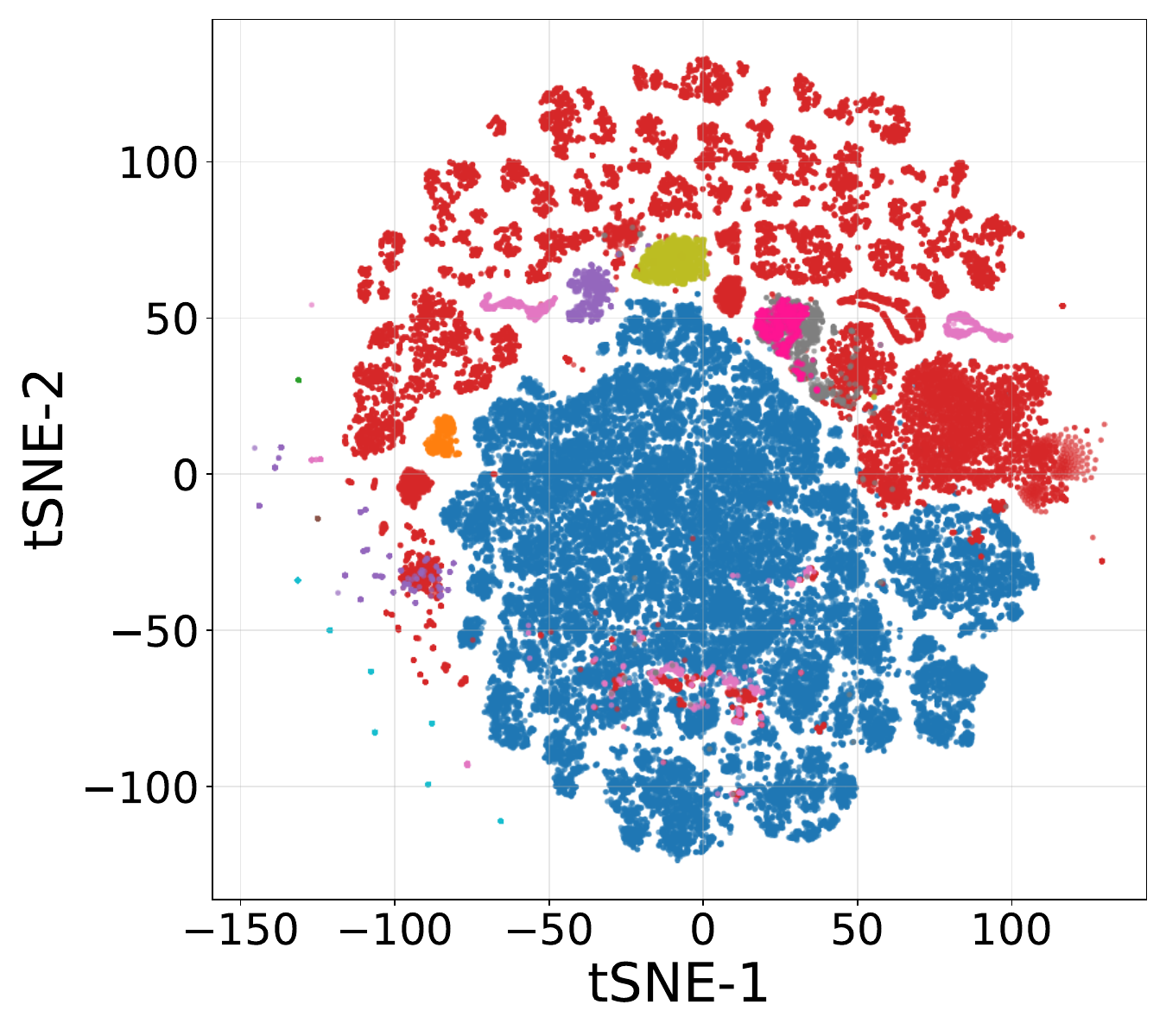}
\subcaption{Layer 7}
\end{subfigure}\hfill
\begin{subfigure}[b]{0.18\linewidth}
\includegraphics[width=\linewidth]{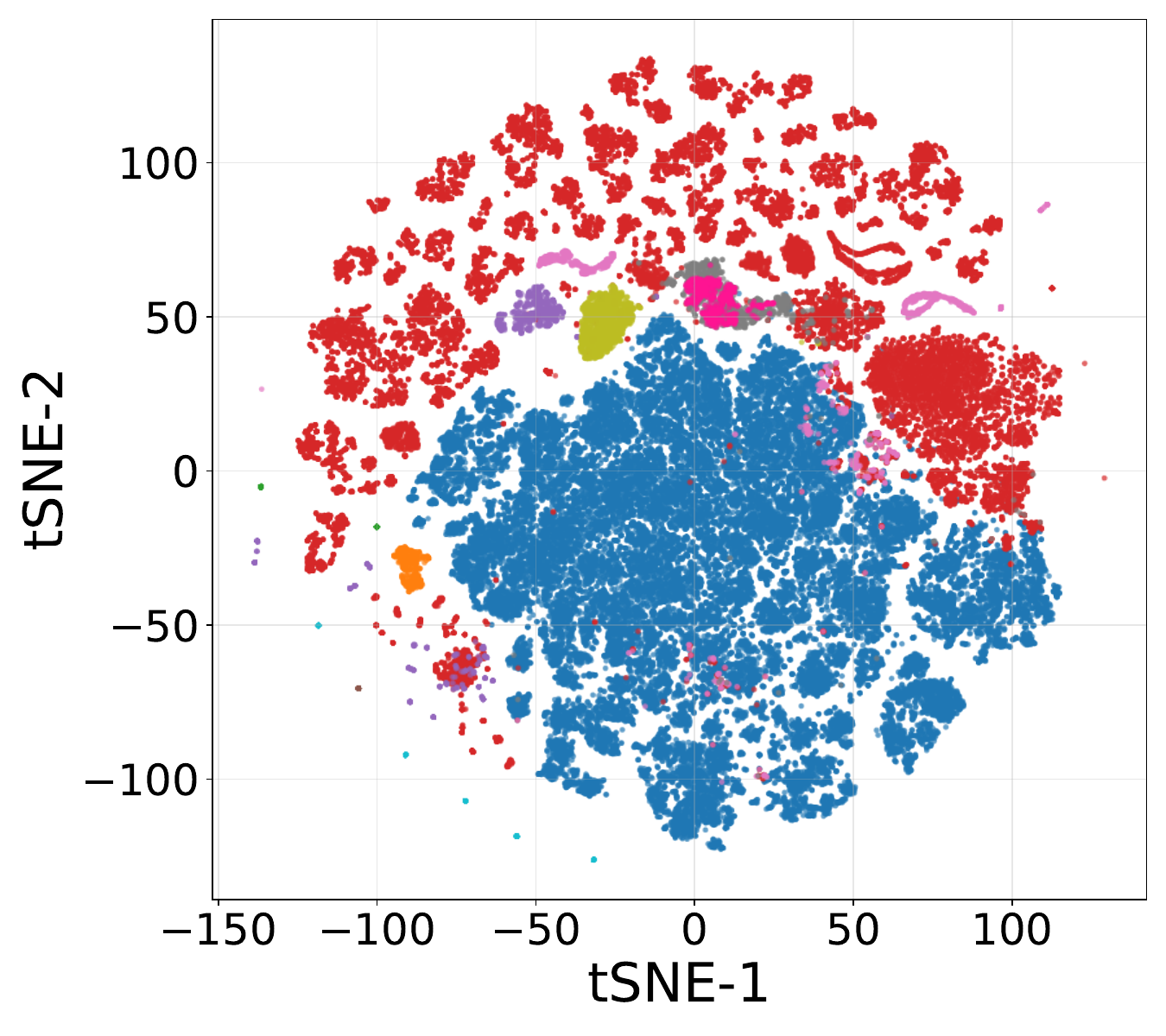}
\subcaption{Layer 8}
\end{subfigure}\hfill
\begin{subfigure}[b]{0.18\linewidth}
\includegraphics[width=\linewidth]{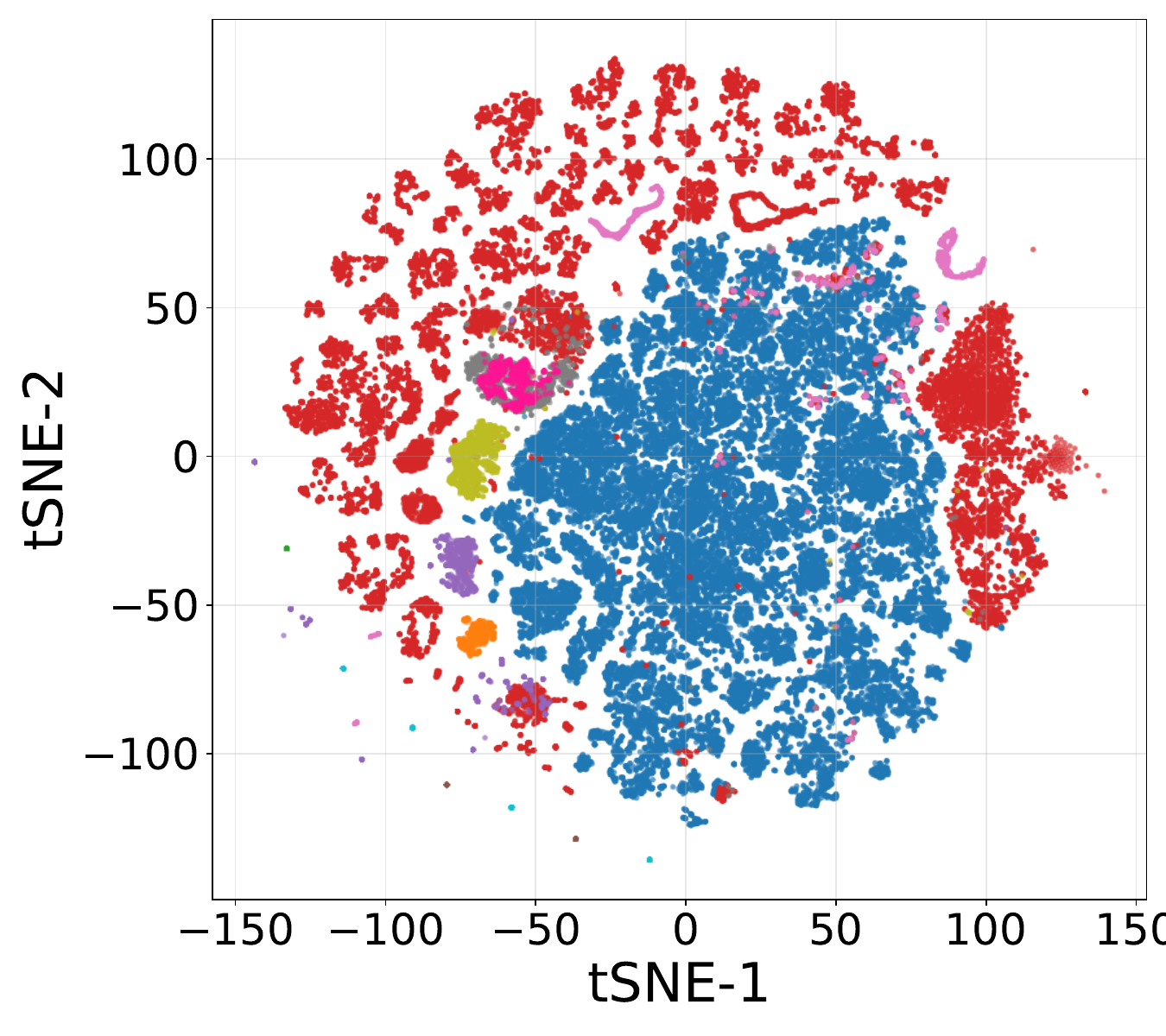}
\subcaption{Layer 9}
\end{subfigure}

\vspace{3pt}

\begin{subfigure}[b]{0.18\linewidth}
\includegraphics[width=\linewidth]{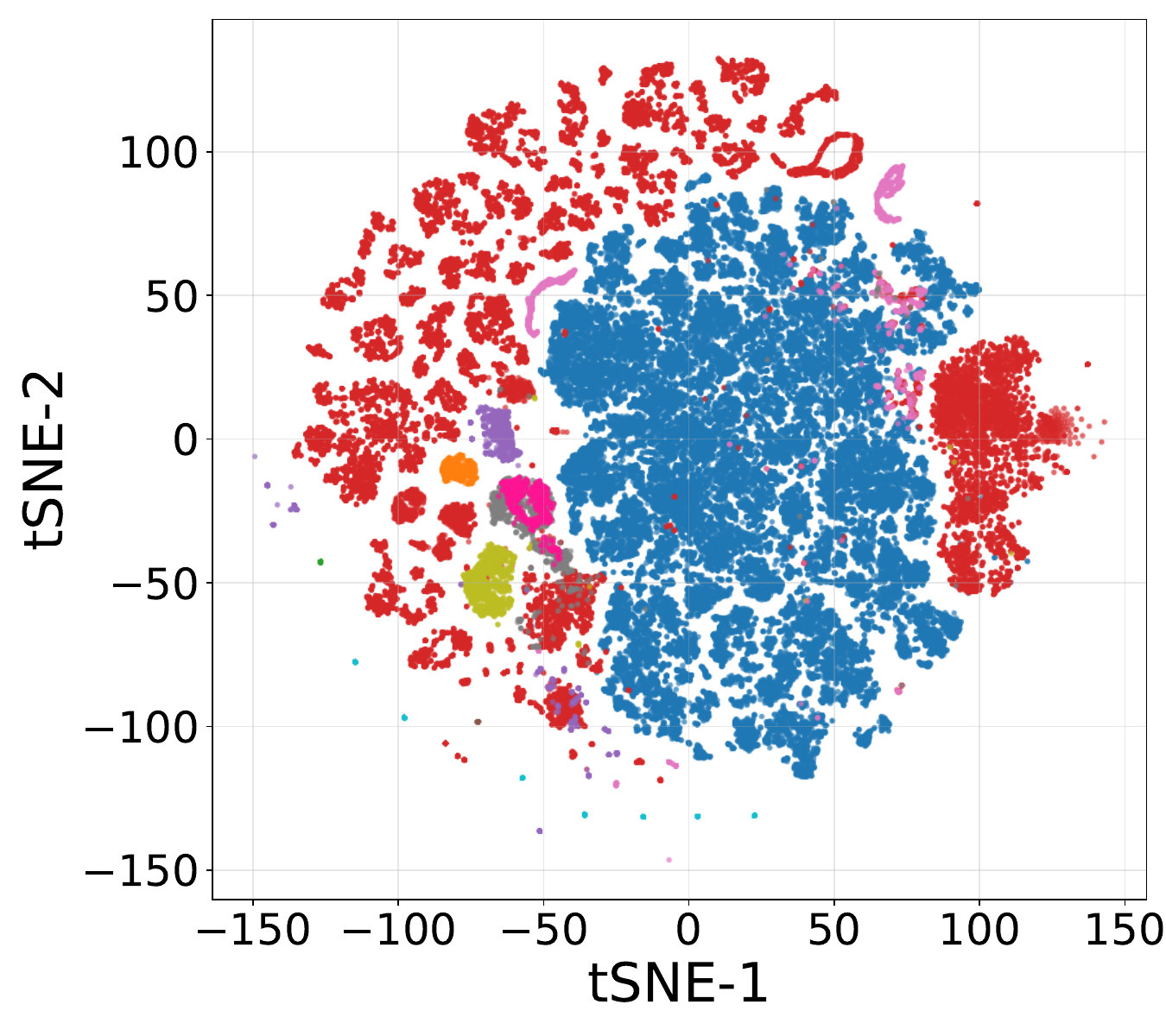}
\subcaption{Layer 10}
\end{subfigure}\hfill
\begin{subfigure}[b]{0.18\linewidth}
\includegraphics[width=\linewidth]{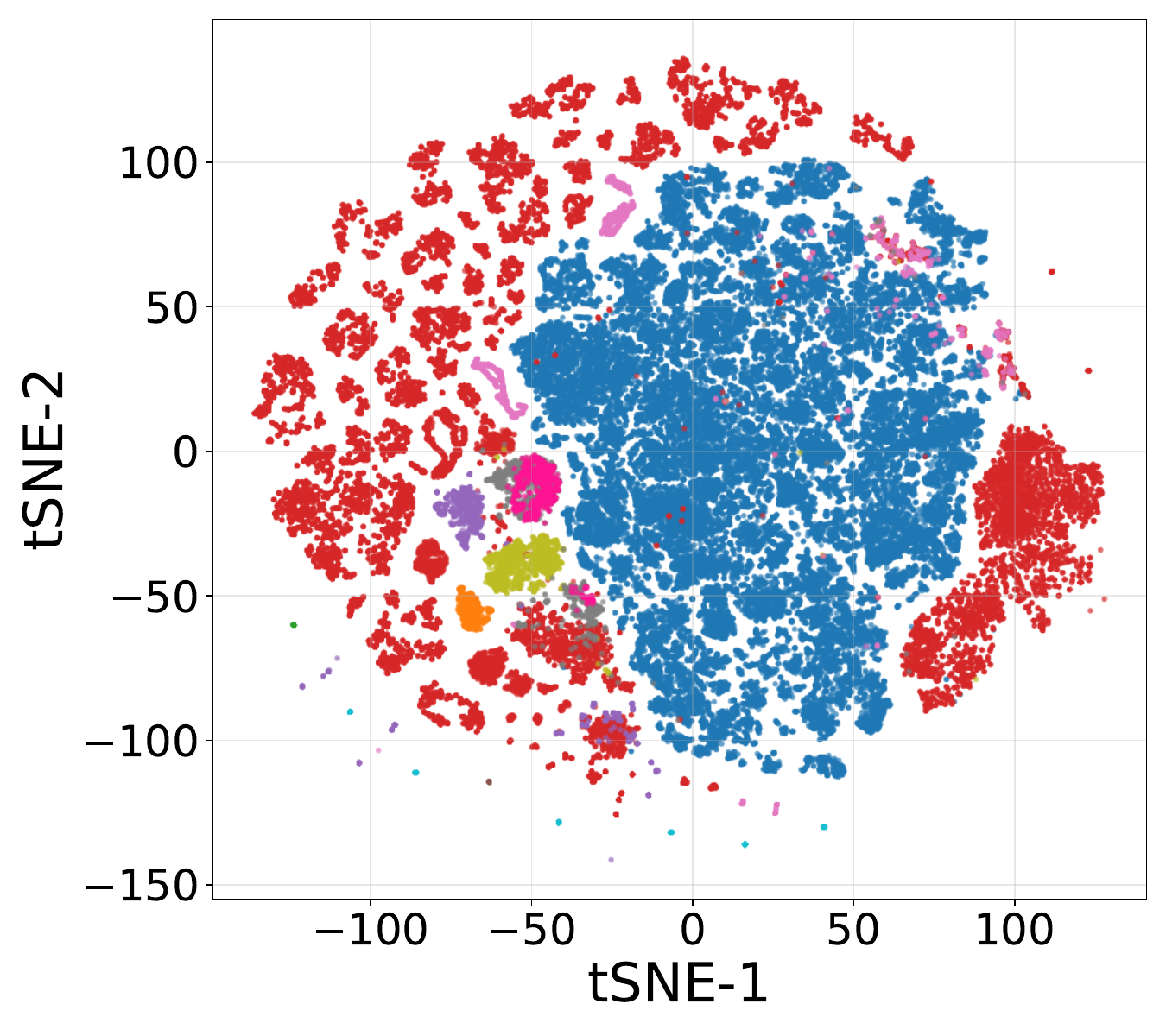}
\subcaption{Layer 11}
\end{subfigure}\hfill
\begin{subfigure}[b]{0.18\linewidth}
\includegraphics[width=\linewidth]{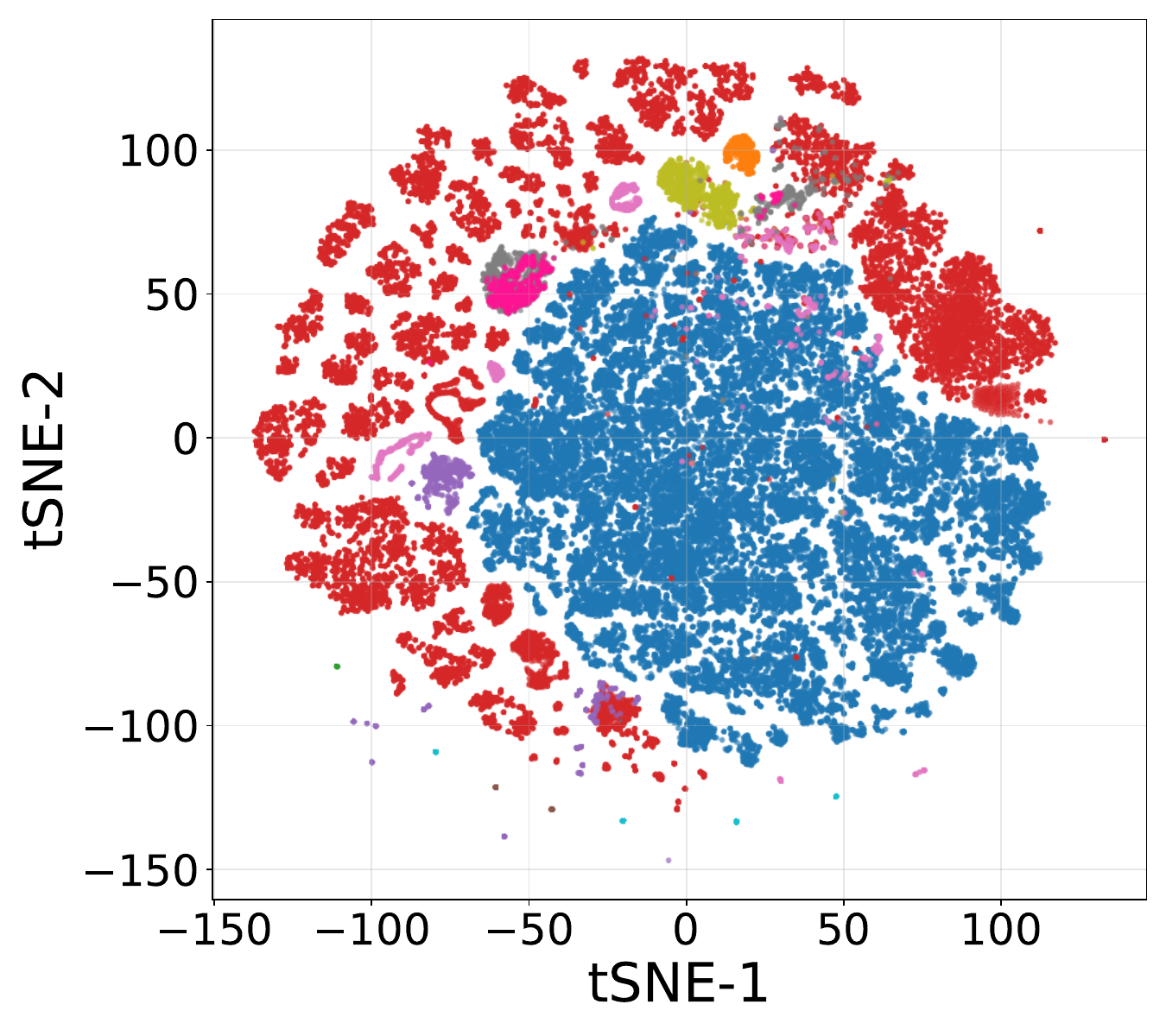}
\subcaption{Layer 12}
\end{subfigure}\hfill
\begin{subfigure}[b]{0.18\linewidth}
\includegraphics[width=\linewidth]{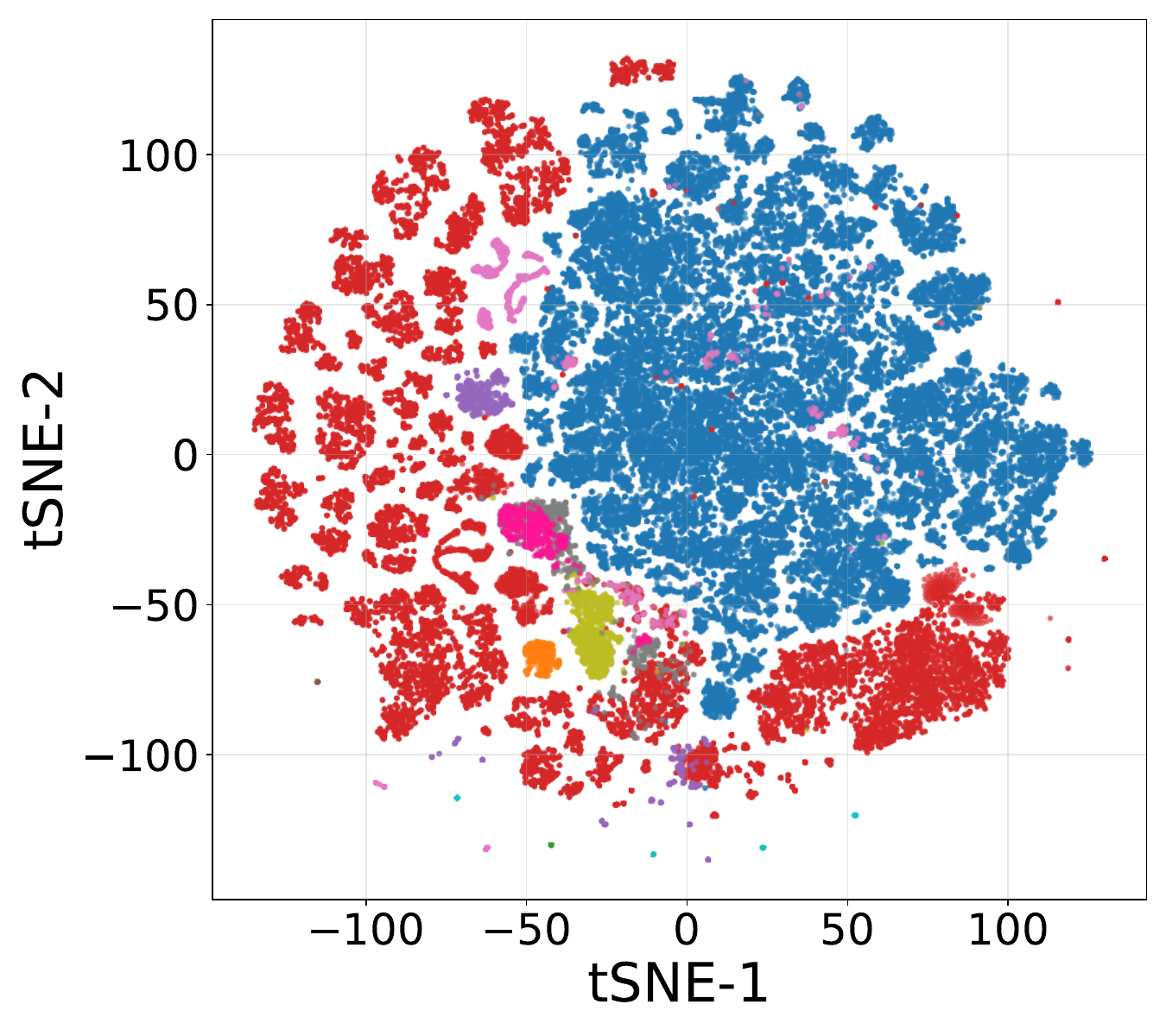}
\subcaption{Layer 13}
\end{subfigure}\hfill
\begin{subfigure}[b]{0.18\linewidth}
\includegraphics[width=\linewidth]{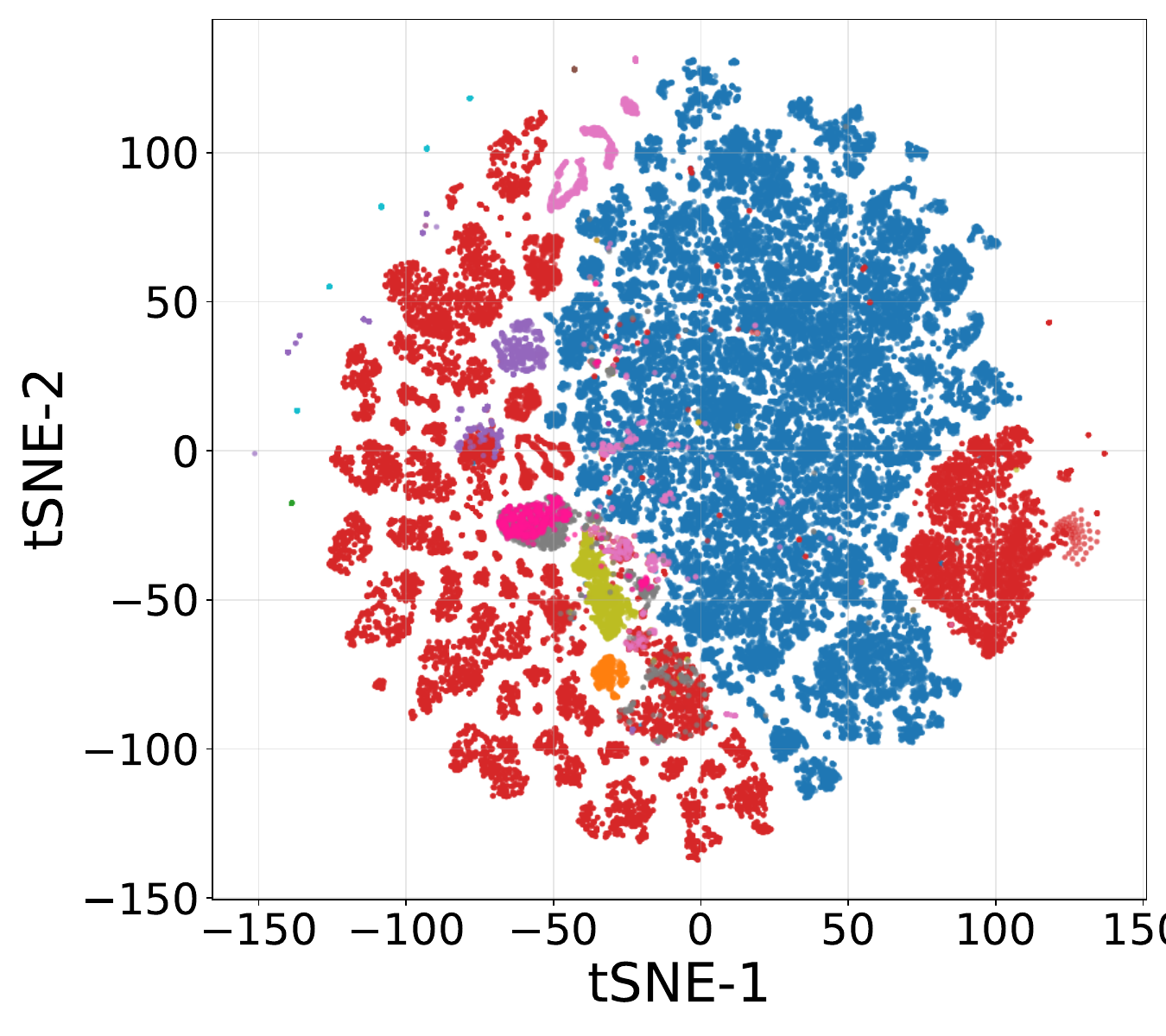}
\subcaption{Layer 14}
\end{subfigure}

\vspace{3pt}

\begin{subfigure}[b]{0.18\linewidth}
\includegraphics[width=\linewidth]{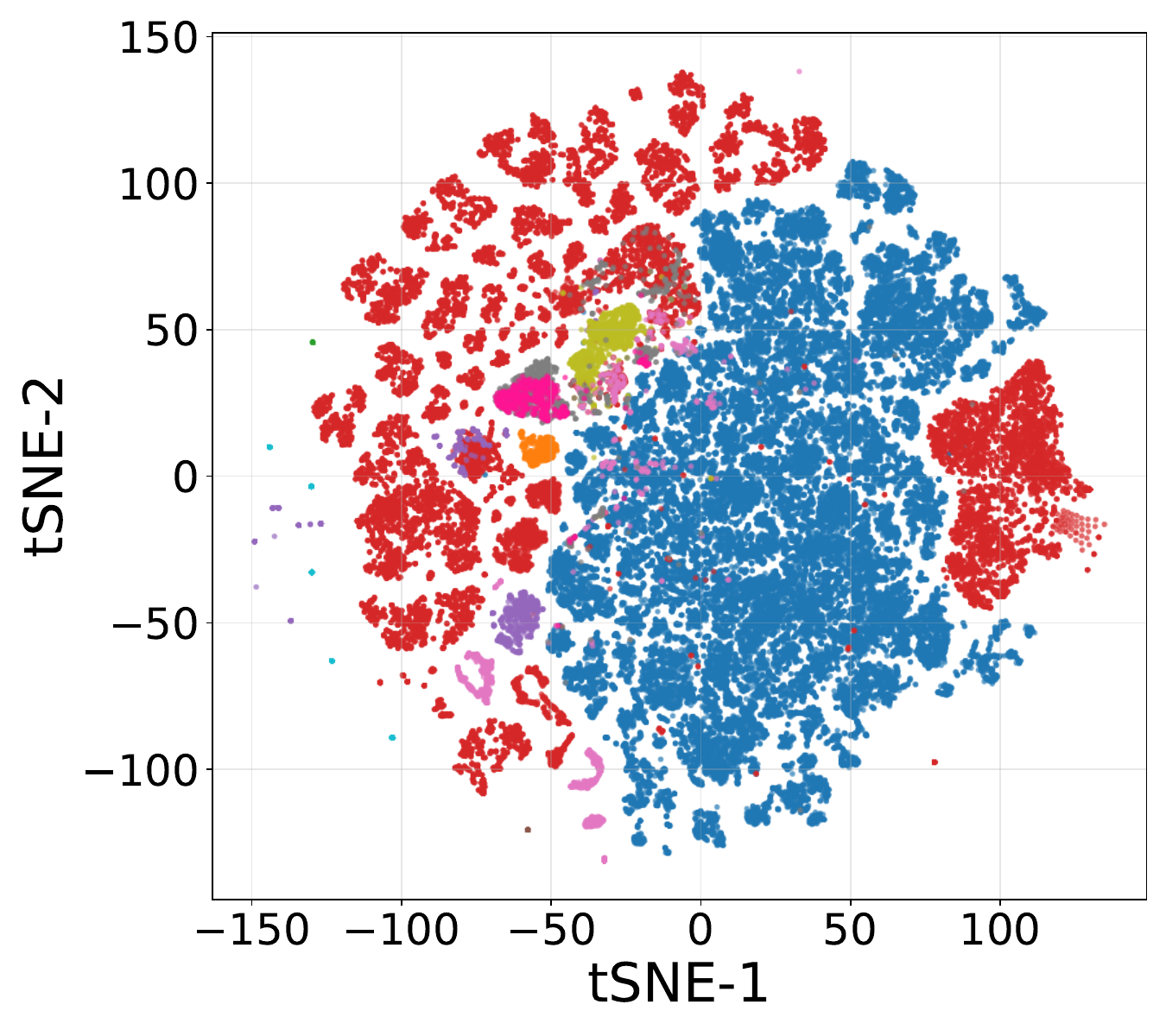}
\subcaption{Layer 15}
\end{subfigure}\hfill
\begin{subfigure}[b]{0.18\linewidth}
\includegraphics[width=\linewidth]{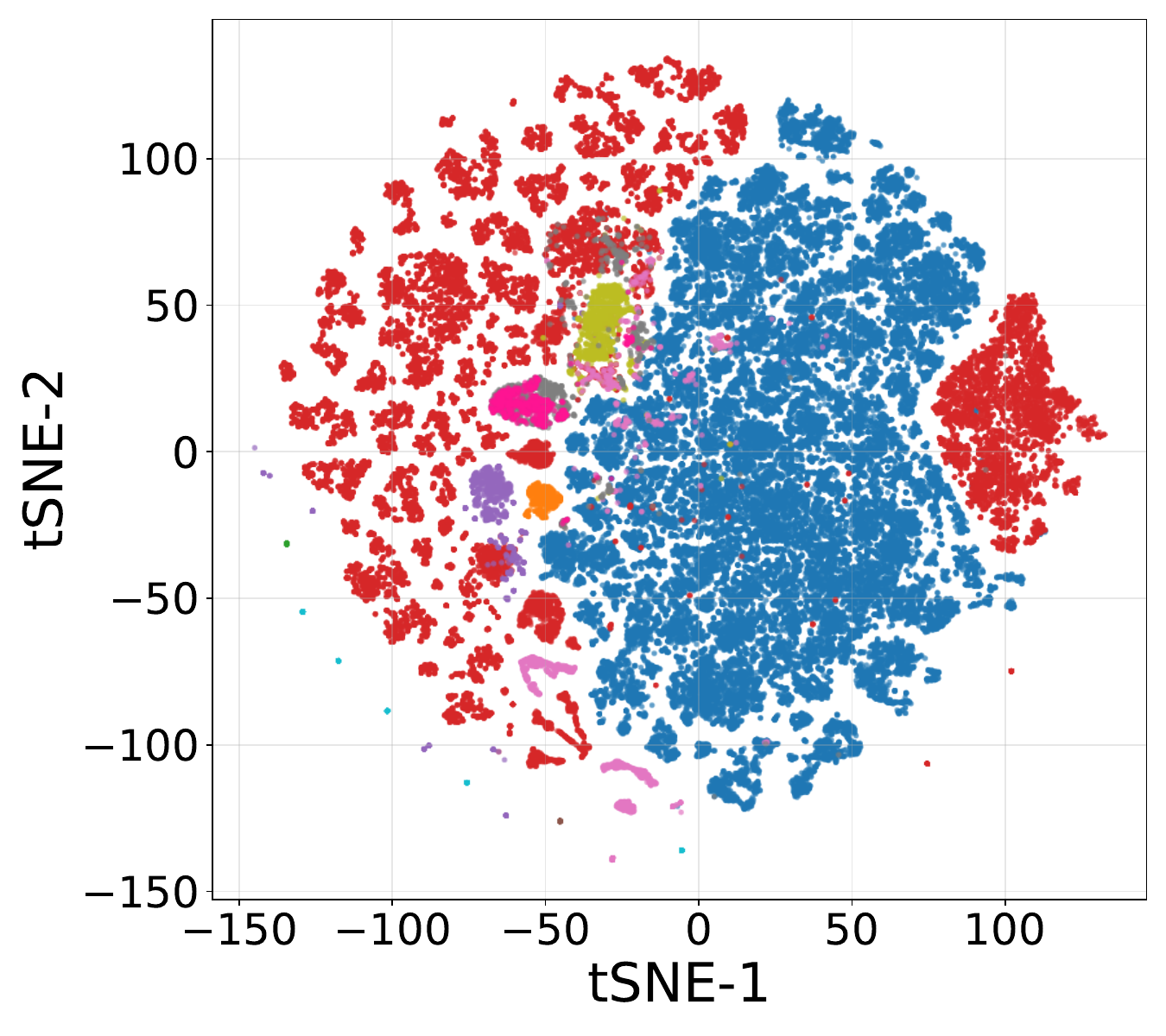}
\subcaption{Layer 16}
\end{subfigure}\hfill
\begin{subfigure}[b]{0.18\linewidth}
\includegraphics[width=\linewidth]{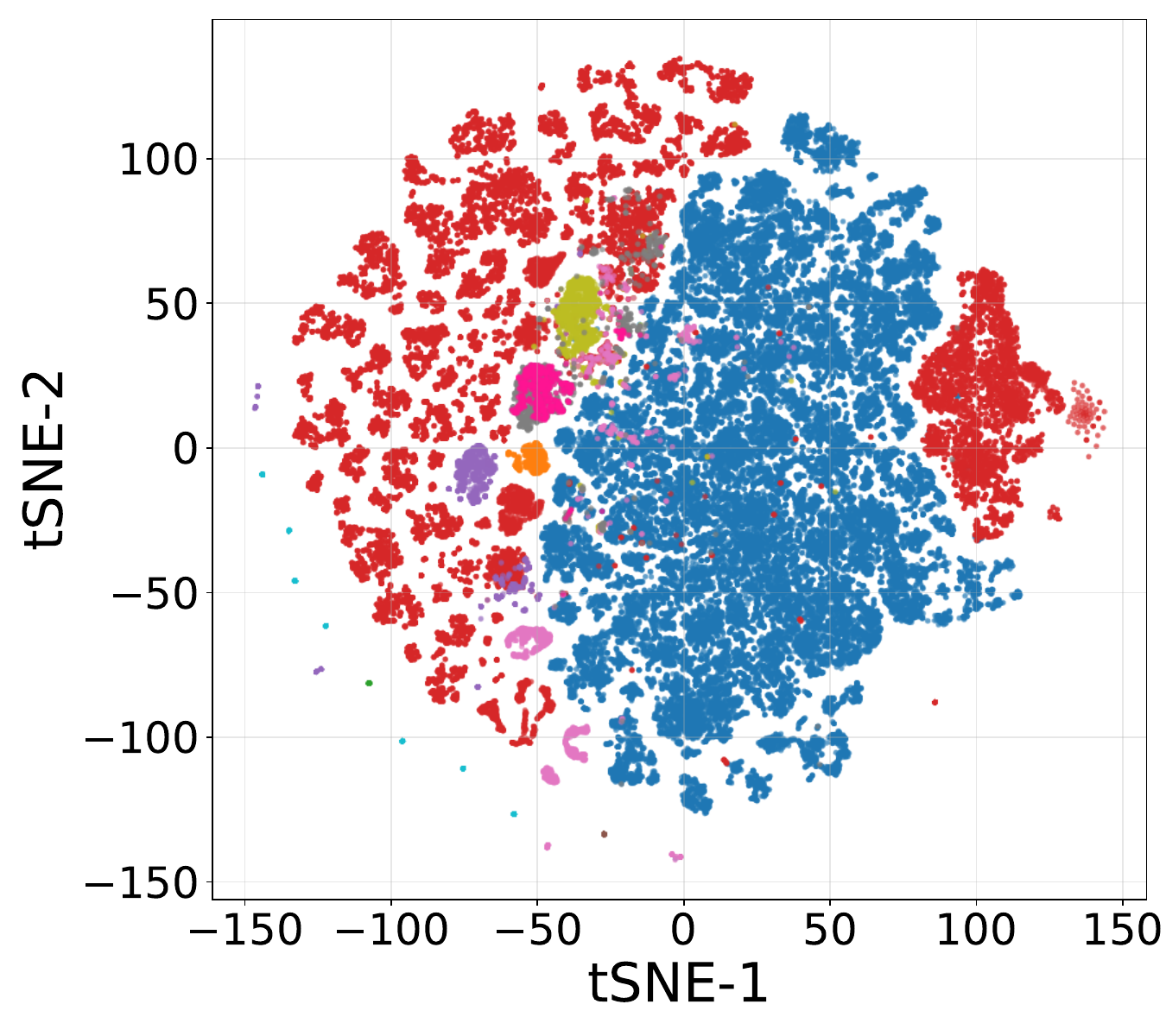}
\subcaption{Layer 17}
\end{subfigure}\hfill
\begin{subfigure}[b]{0.18\linewidth}
\includegraphics[width=\linewidth]{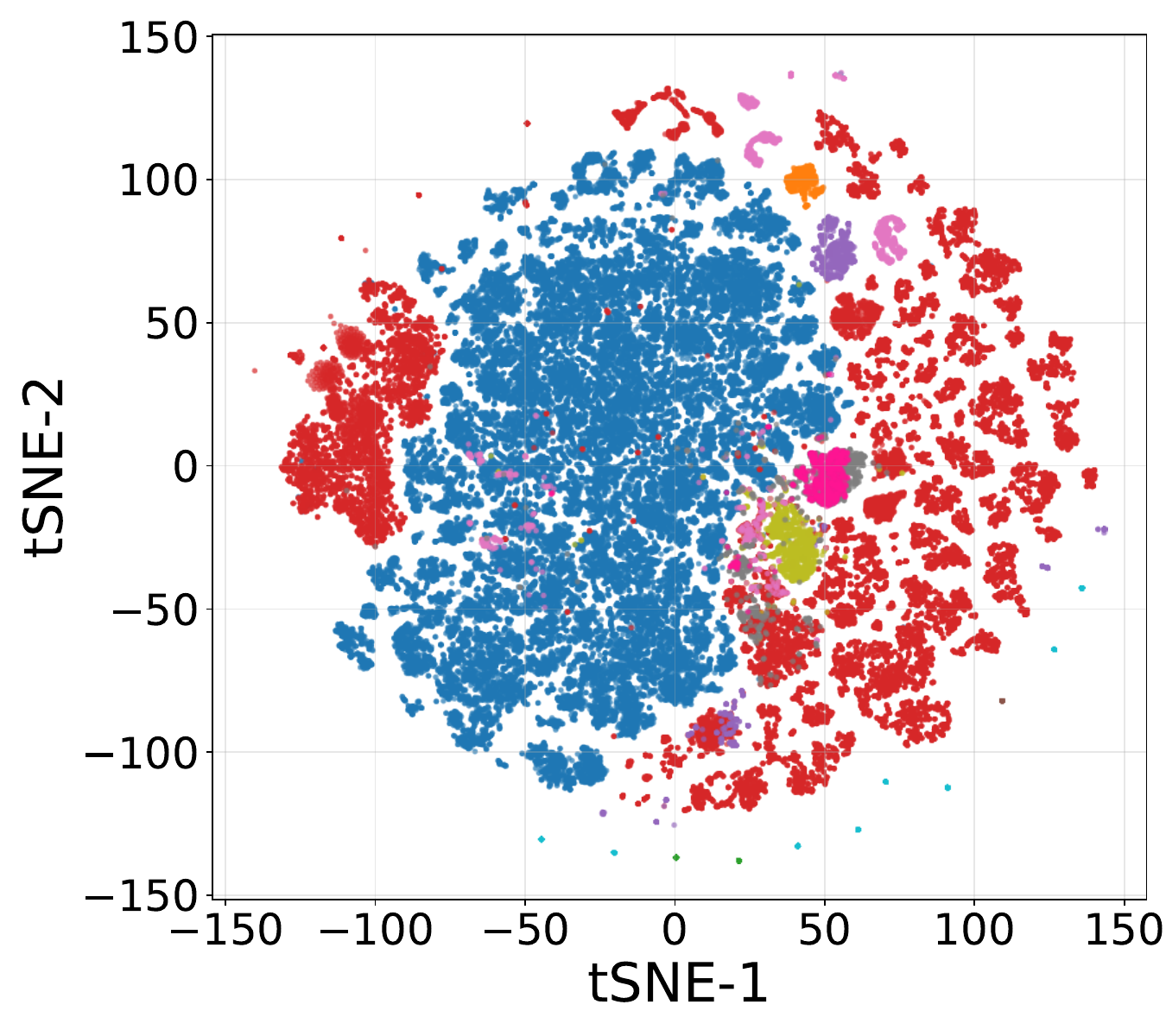}
\subcaption{Layer 18}
\end{subfigure}\hfill
\begin{subfigure}[b]{0.18\linewidth}
\includegraphics[width=\linewidth]{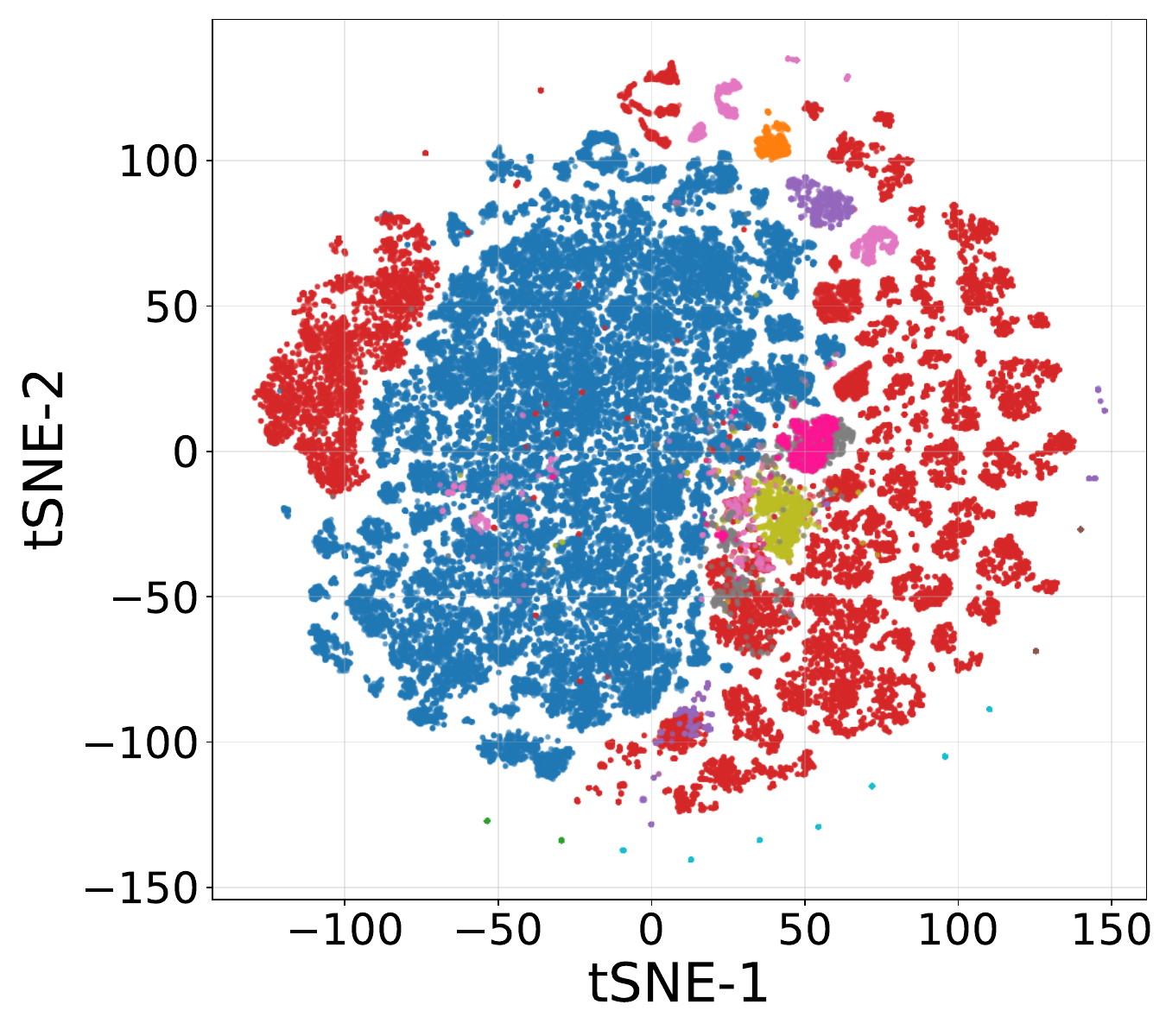}
\subcaption{Layer 19}
\end{subfigure}

\vspace{3pt}

\begin{subfigure}[b]{0.18\linewidth}
\includegraphics[width=\linewidth]{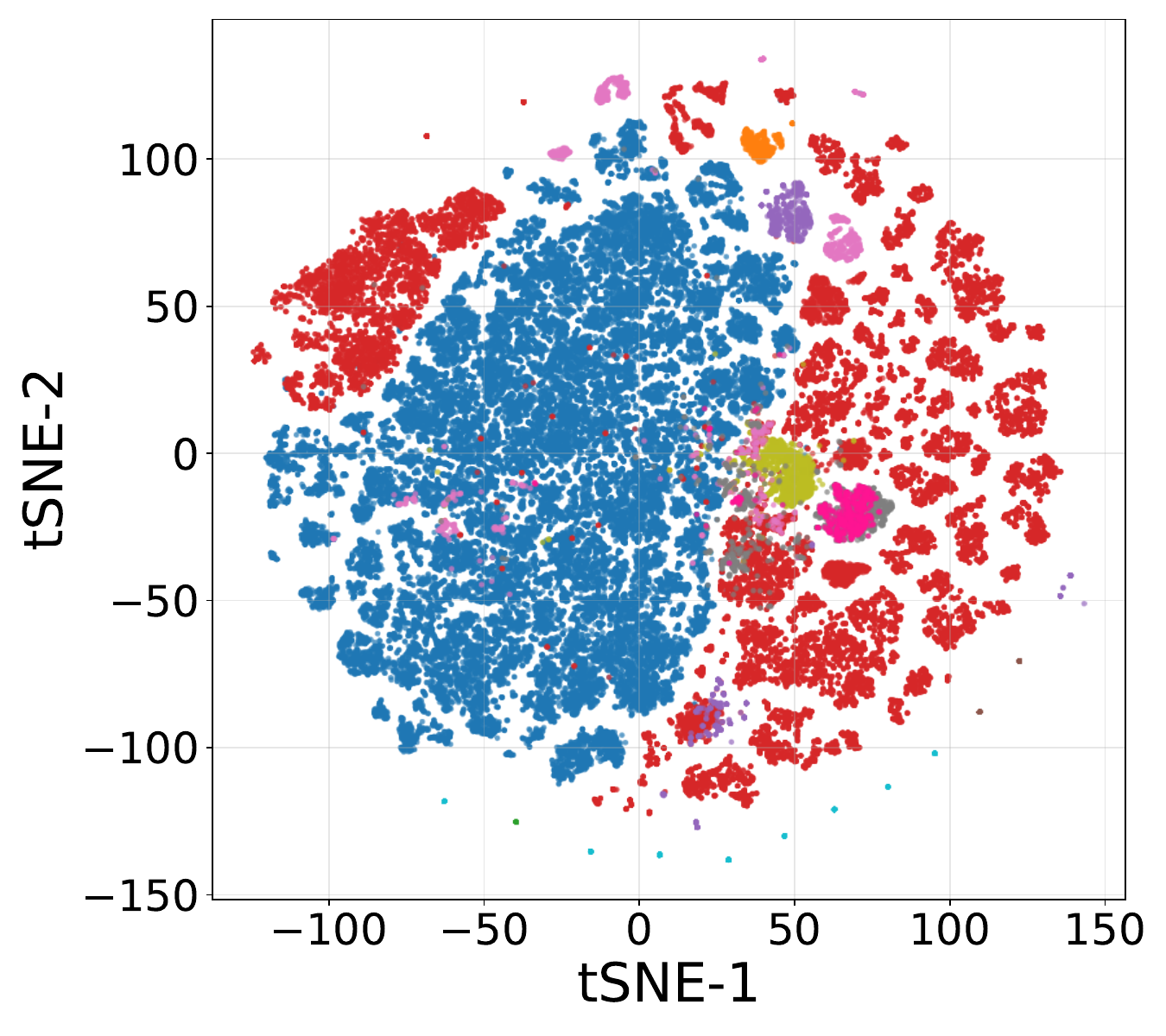}
\subcaption{Layer 20}
\end{subfigure}\hfill
\begin{subfigure}[b]{0.18\linewidth}
\includegraphics[width=\linewidth]{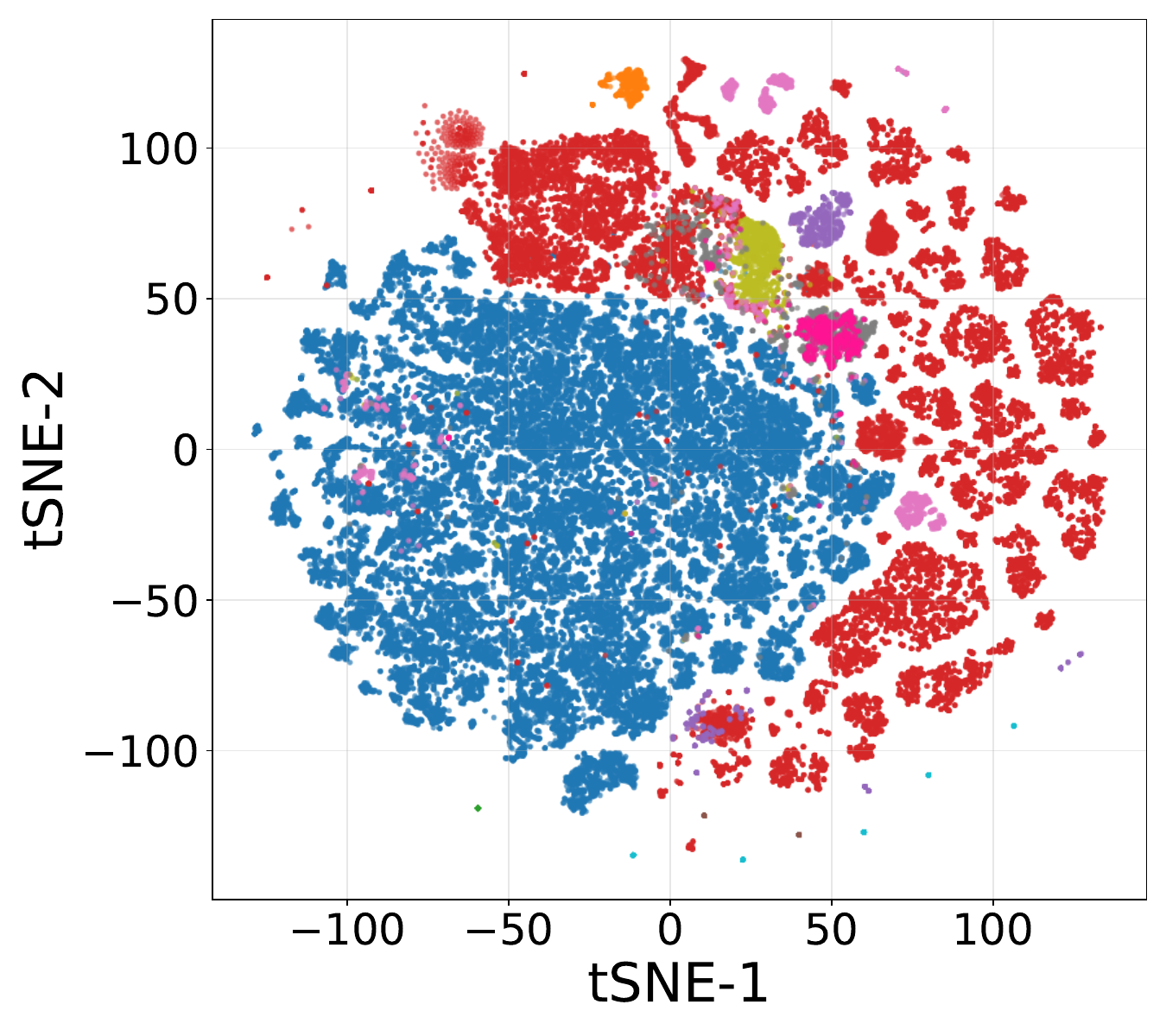}
\subcaption{Layer 21}
\end{subfigure}\hfill
\begin{subfigure}[b]{0.18\linewidth}
\includegraphics[width=\linewidth]{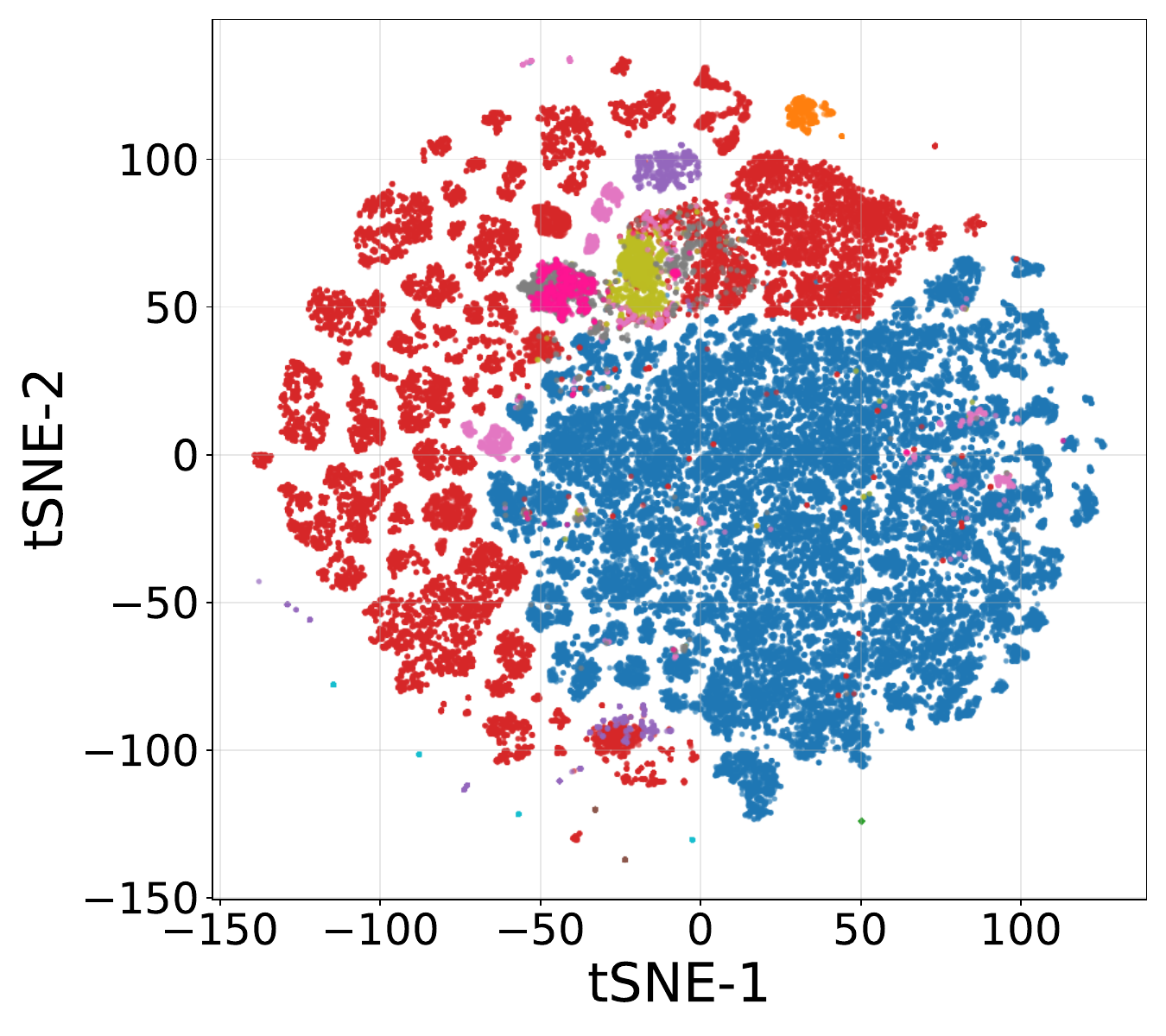}
\subcaption{Layer 22}
\end{subfigure}\hfill
\begin{subfigure}[b]{0.18\linewidth}
\includegraphics[width=\linewidth]{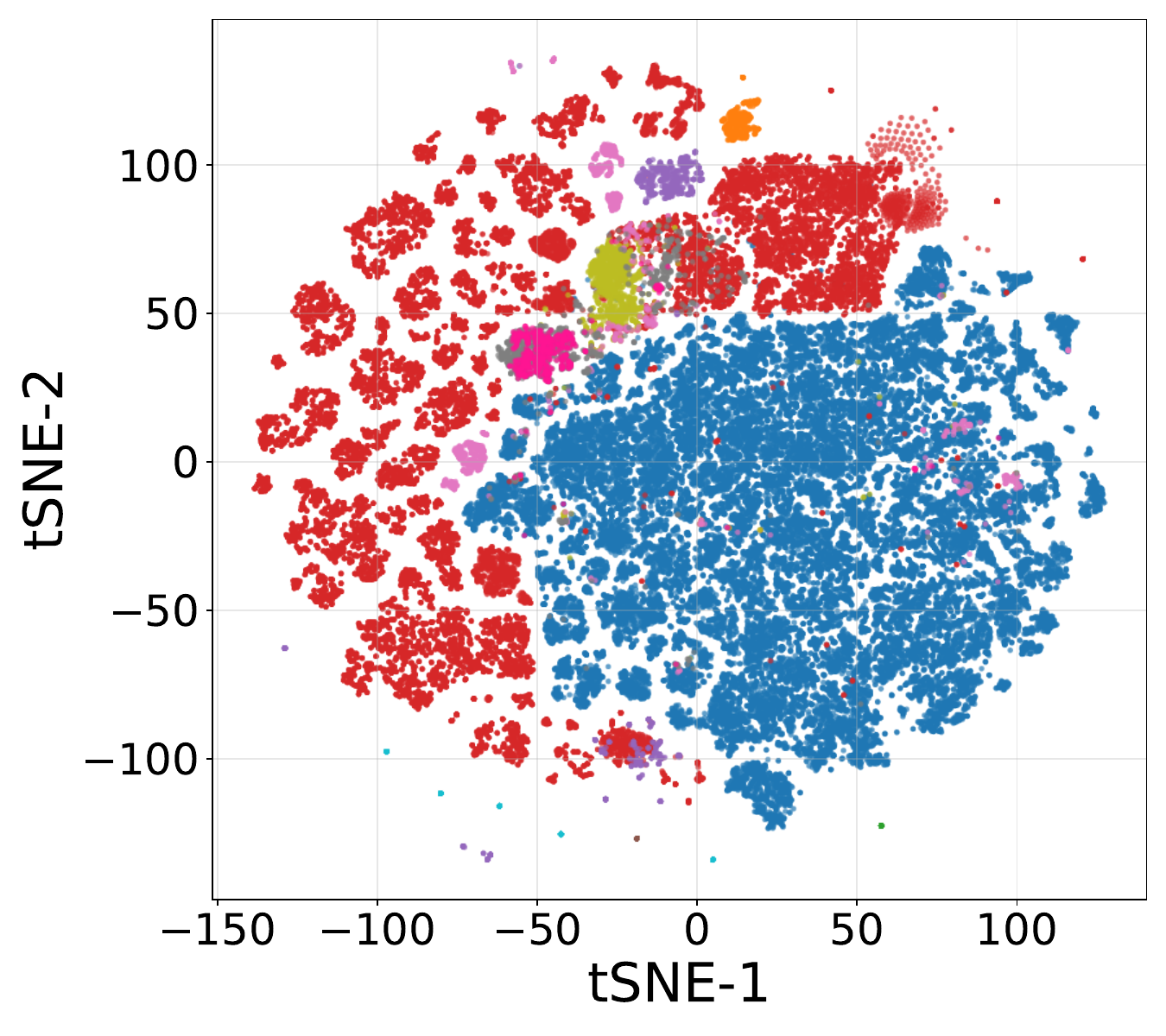}
\subcaption{Layer 23}
\end{subfigure}\hfill
\begin{subfigure}[b]{0.18\linewidth}
\includegraphics[width=\linewidth]{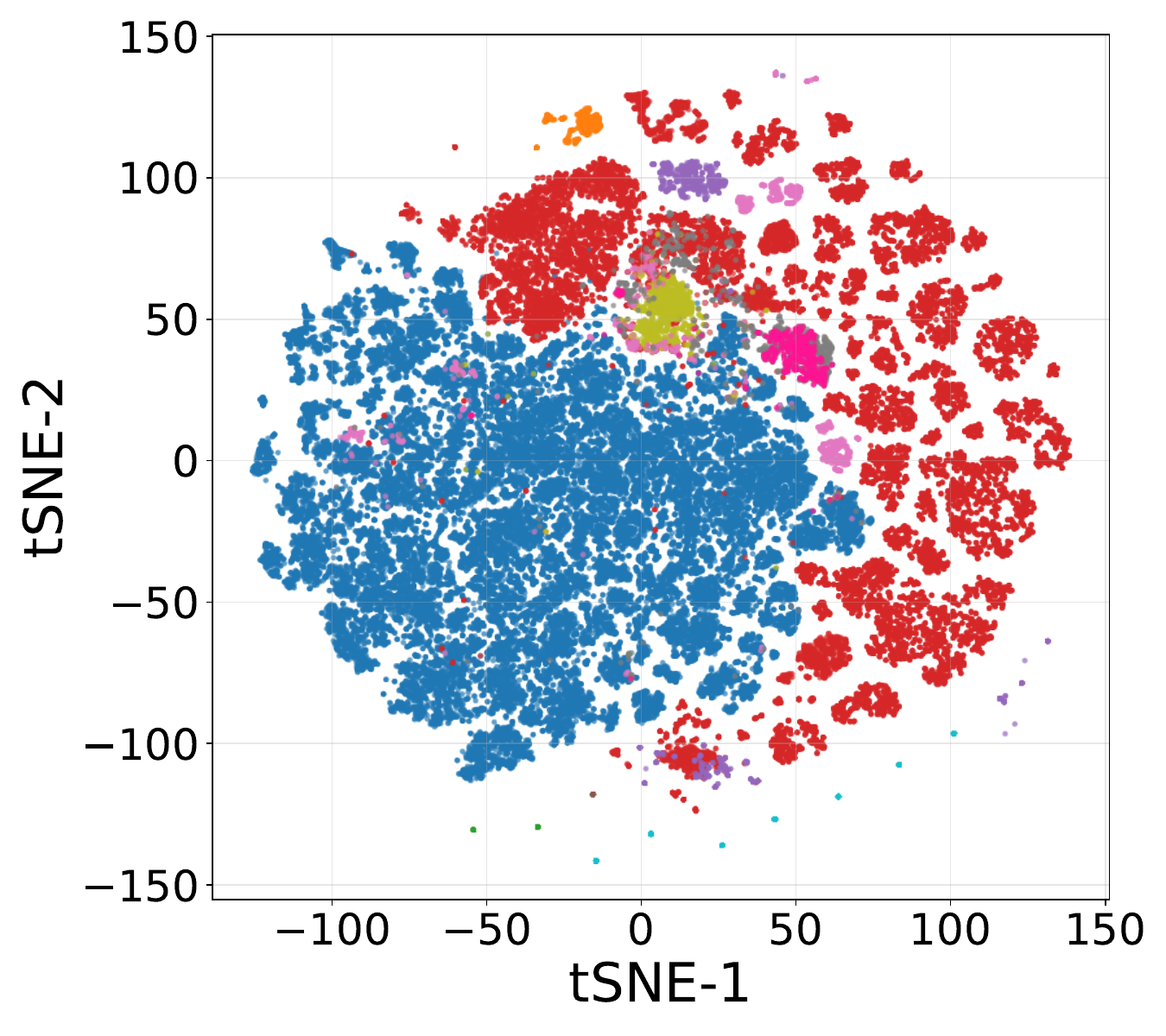}
\subcaption{Layer 24}
\end{subfigure}

\vspace{3pt}

\begin{subfigure}[b]{0.18\linewidth}
\includegraphics[width=\linewidth]{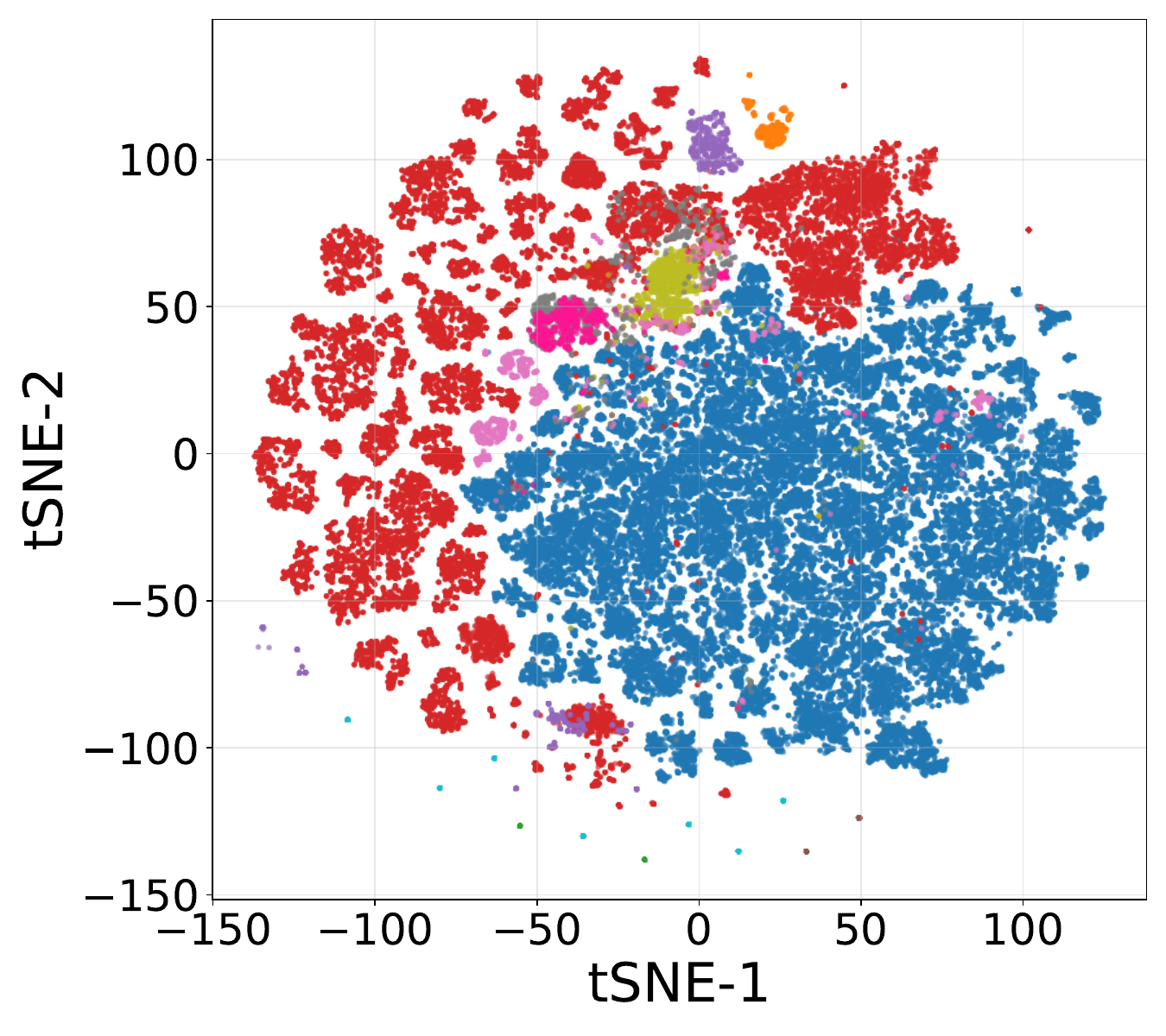}
\subcaption{Layer 25}
\end{subfigure}\hfill
\begin{subfigure}[b]{0.18\linewidth}
\includegraphics[width=\linewidth]{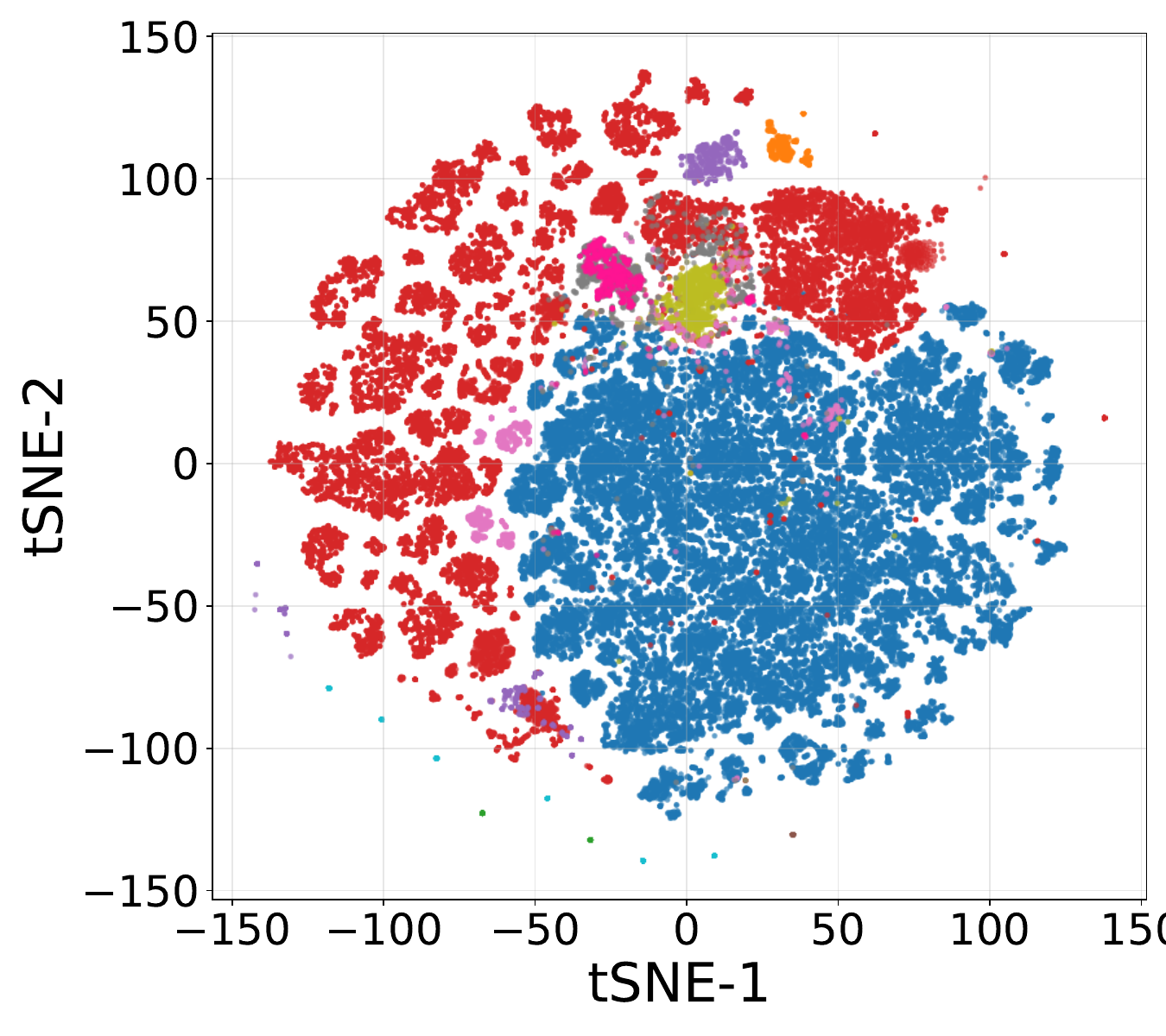}
\subcaption{Layer 26}
\end{subfigure}\hfill
\begin{subfigure}[b]{0.18\linewidth}
\includegraphics[width=\linewidth]{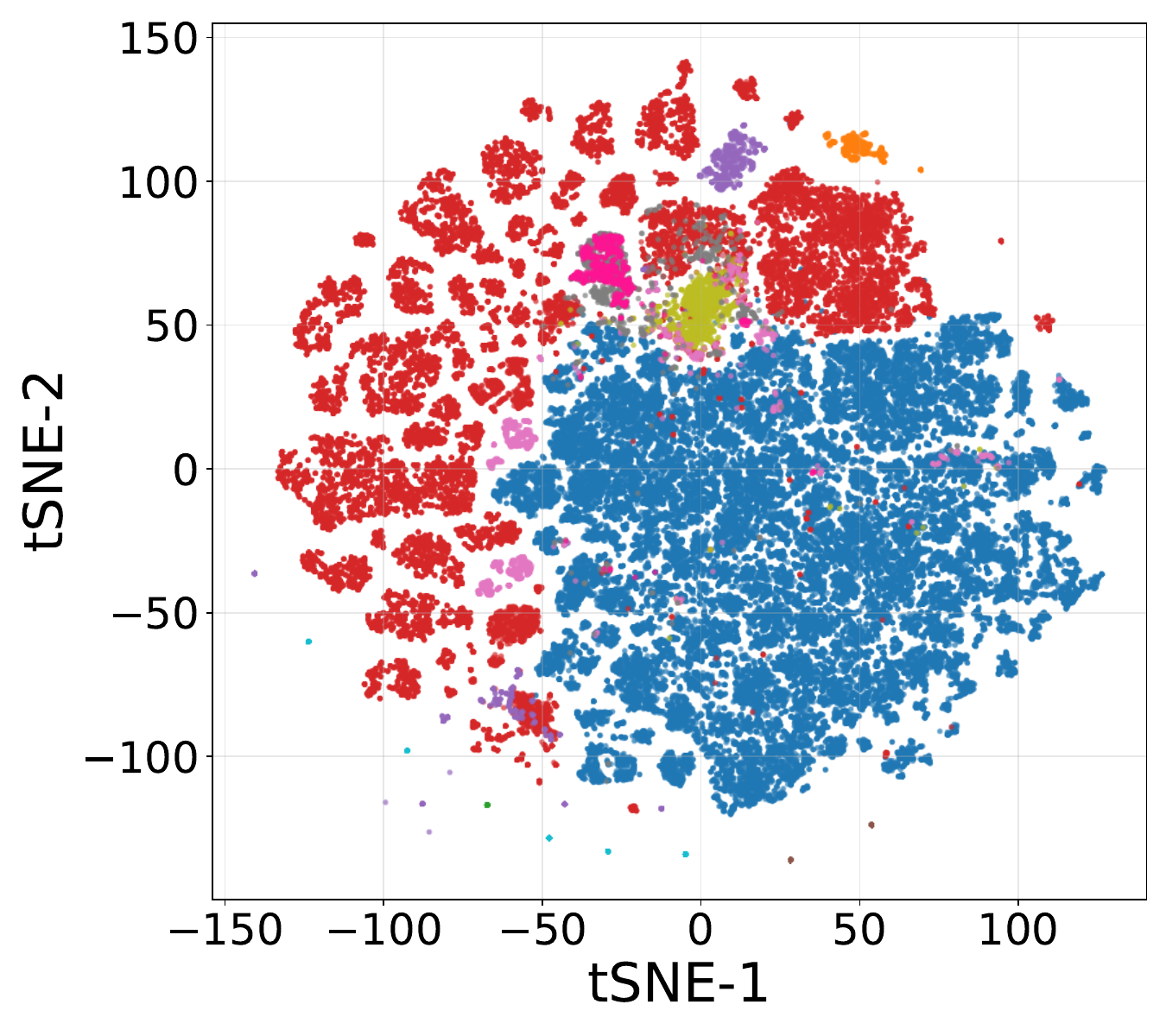}
\subcaption{Layer 27}
\end{subfigure}\hfill
\begin{subfigure}[b]{0.18\linewidth}
\includegraphics[width=\linewidth]{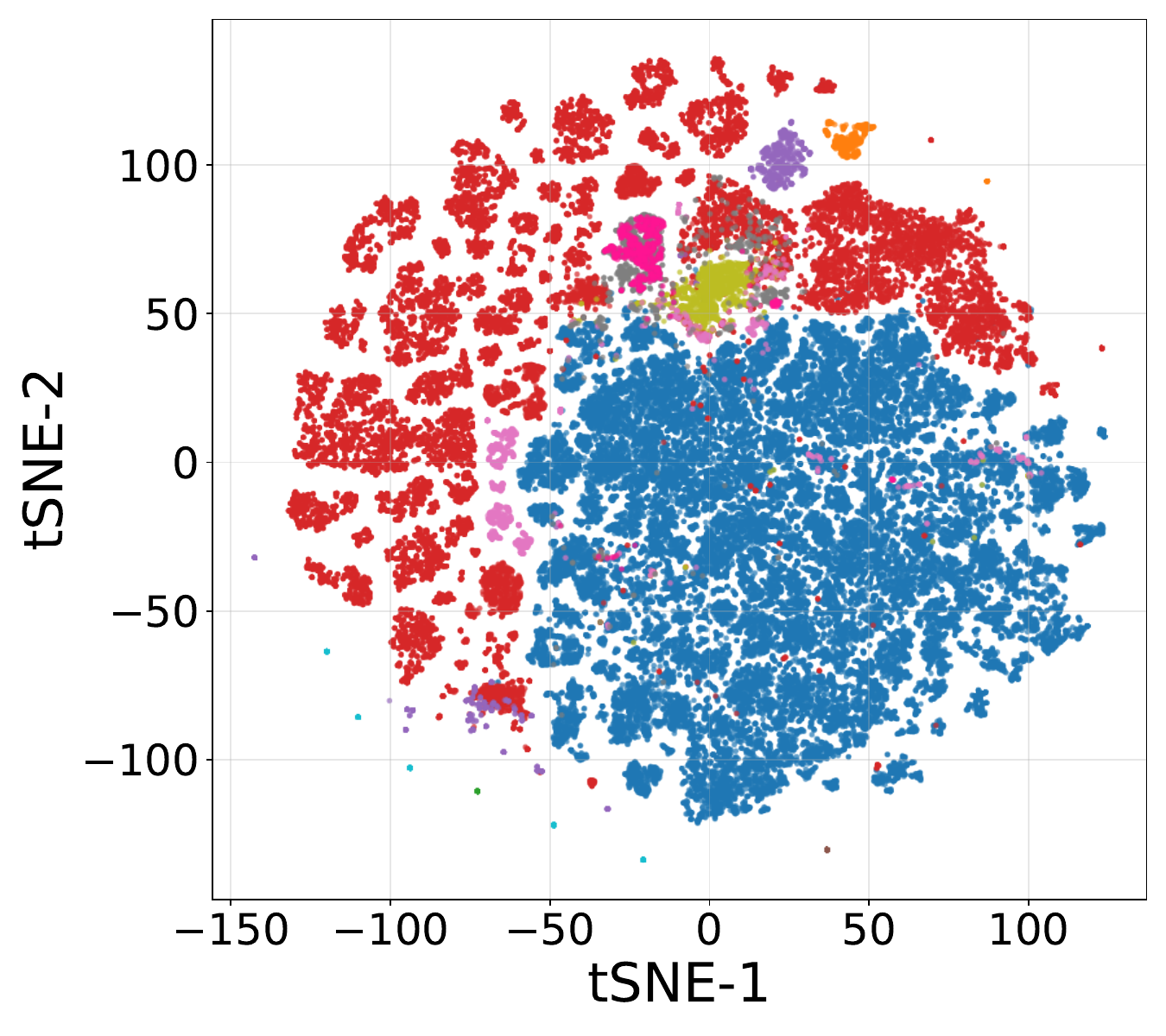}
\subcaption{Layer 28}
\end{subfigure}\hfill
\begin{subfigure}[b]{0.18\linewidth}
\includegraphics[width=\linewidth]{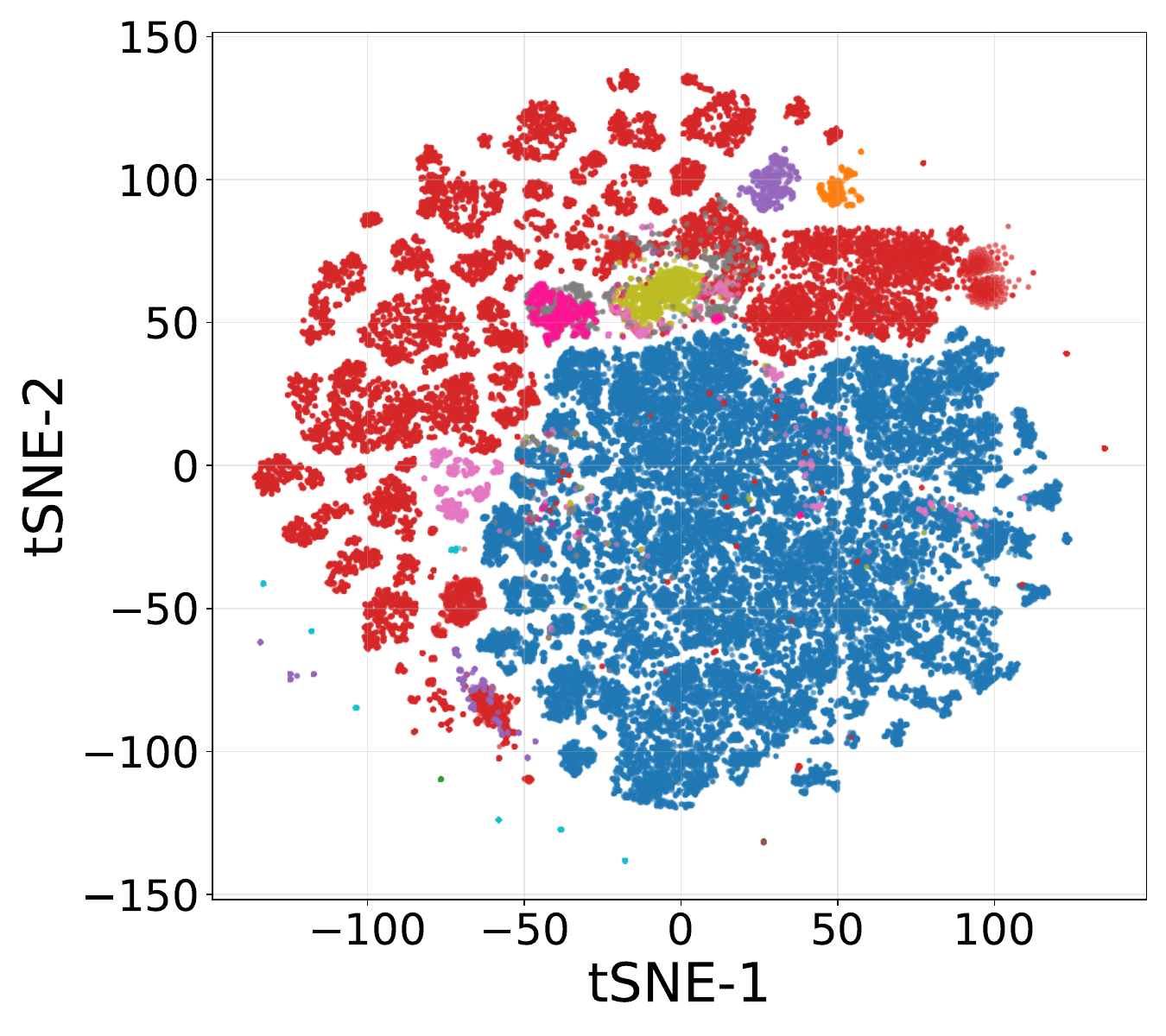}
\subcaption{Layer 29}
\end{subfigure}

\vspace{3pt}

\begin{subfigure}[b]{0.18\linewidth}
\includegraphics[width=\linewidth]{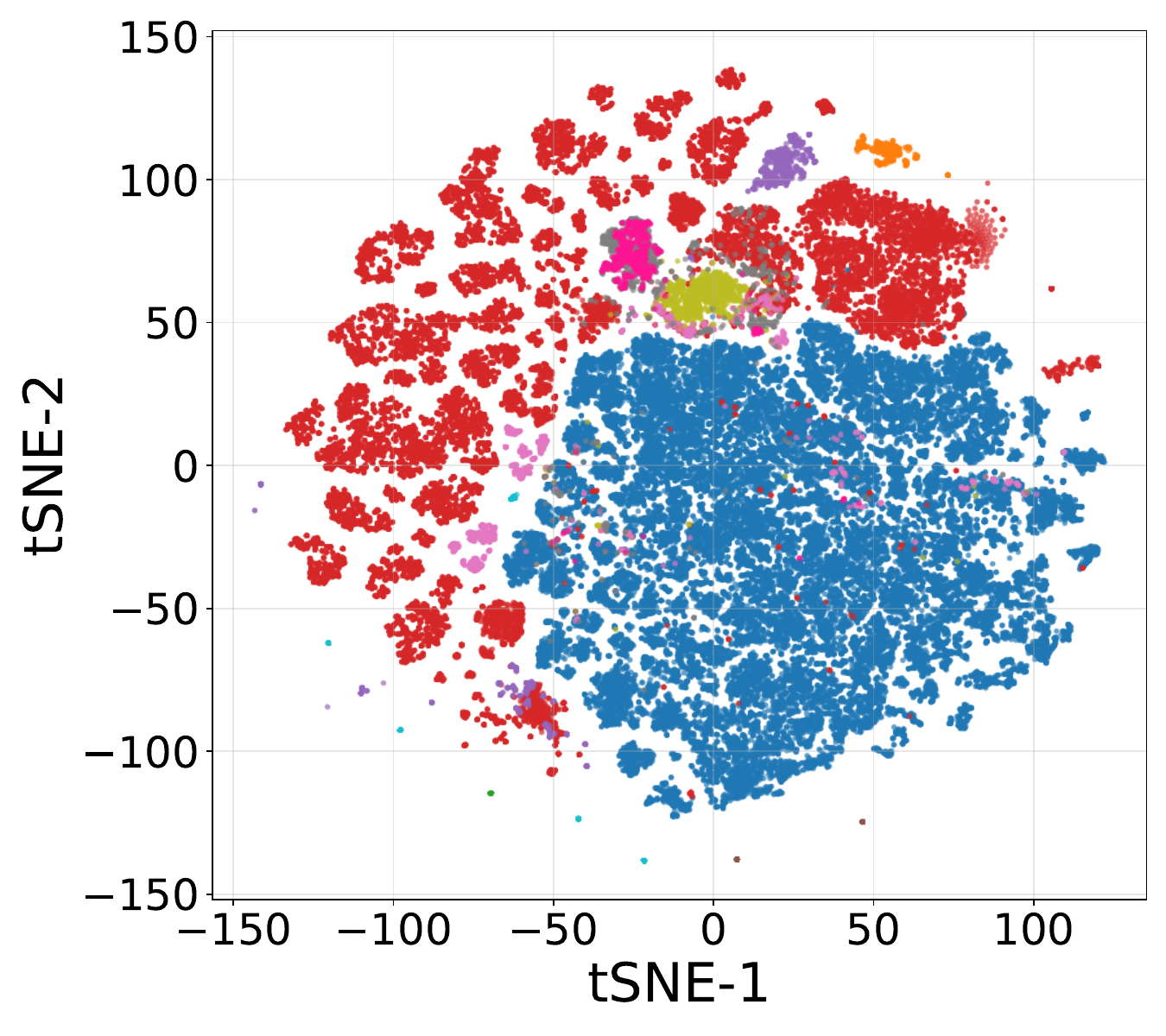}
\subcaption{Layer 30}
\end{subfigure}\hfill
\begin{subfigure}[b]{0.18\linewidth}
\includegraphics[width=\linewidth]{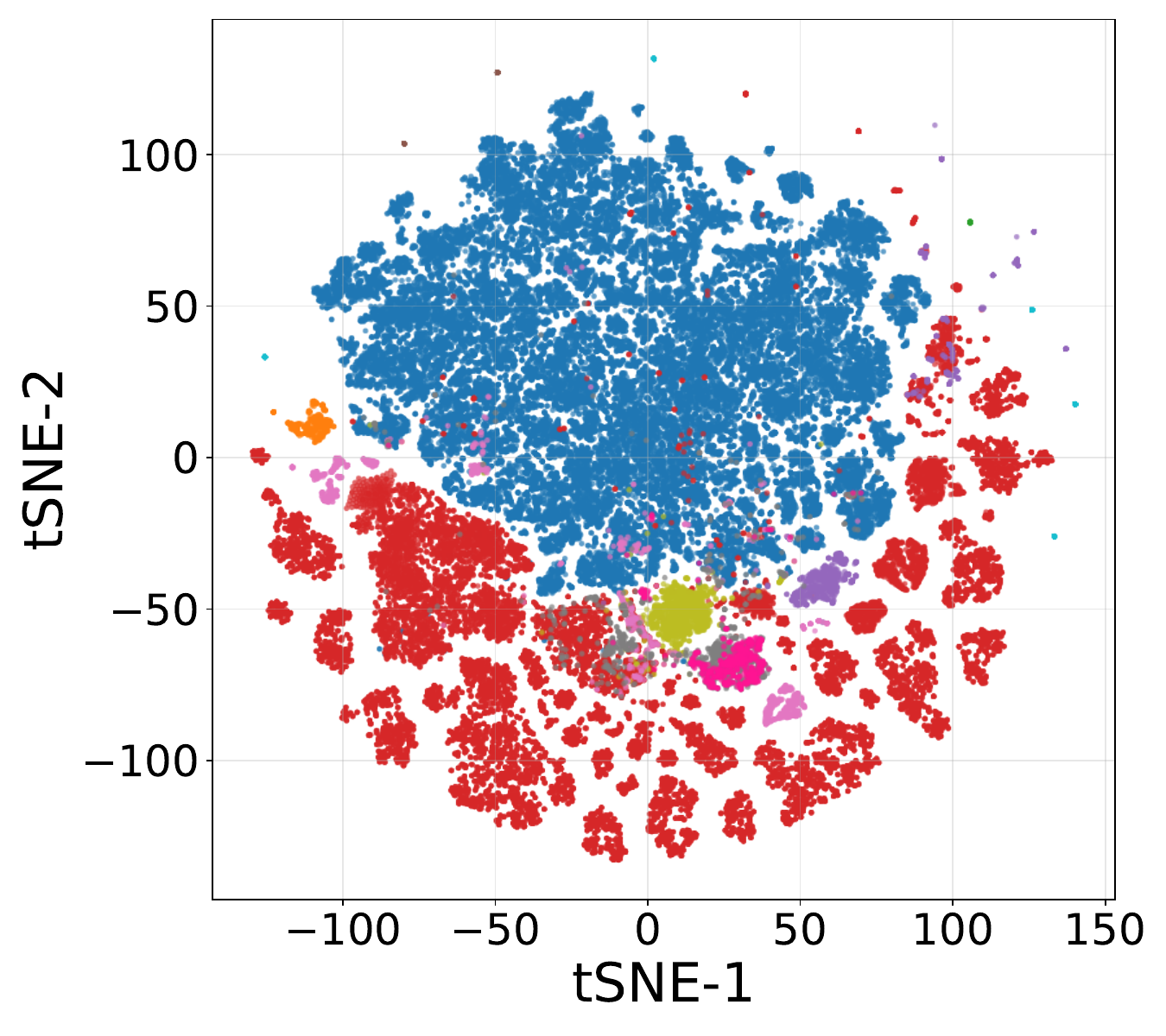}
\subcaption{Layer 31}
\end{subfigure}

\caption{t-SNE visualization across layers 0--31 of Vicuna-7B}
\label{fig:vicuna_tsne}
\end{figure*}

\begin{figure*}[t]
\centering

\includegraphics[width=0.9\textwidth, trim={0 0.8cm 0 0 cm}, clip]{latex/Figure/legend.pdf}

\begin{subfigure}[b]{0.18\linewidth}
\includegraphics[width=\linewidth]{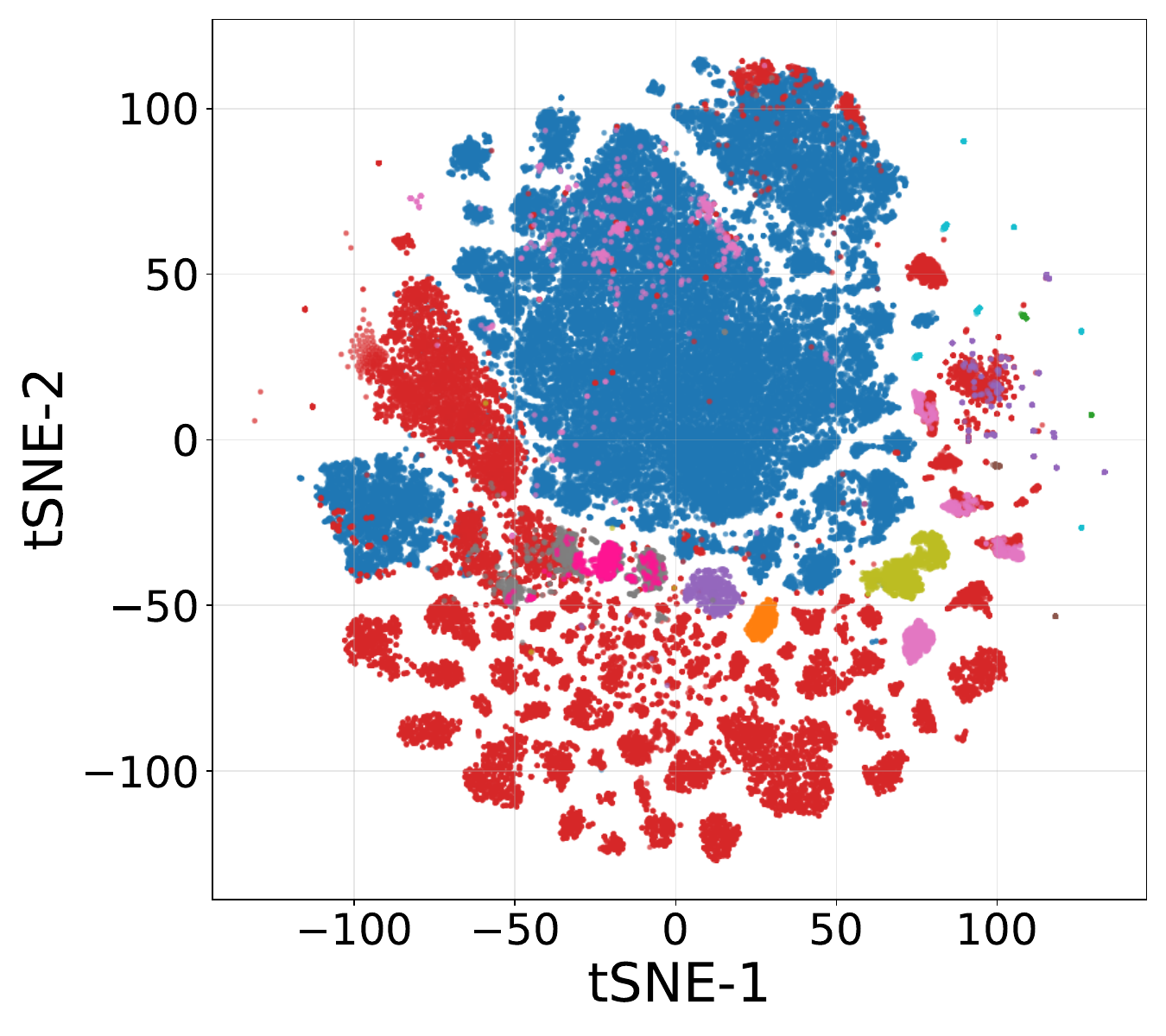}
\subcaption{Layer 0}
\end{subfigure}\hfill
\begin{subfigure}[b]{0.18\linewidth}
\includegraphics[width=\linewidth]{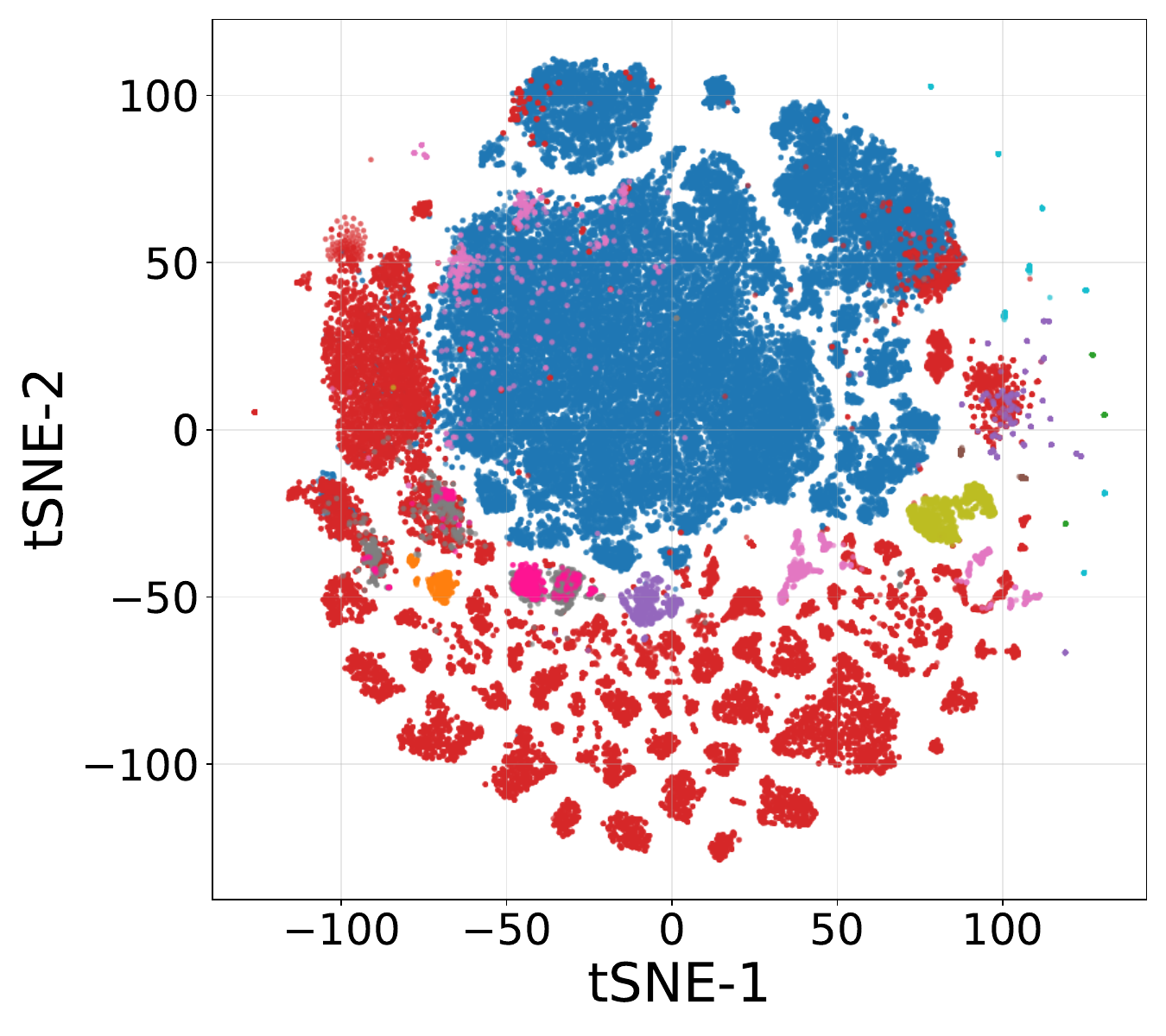}
\subcaption{Layer 1}
\end{subfigure}\hfill
\begin{subfigure}[b]{0.18\linewidth}
\includegraphics[width=\linewidth]{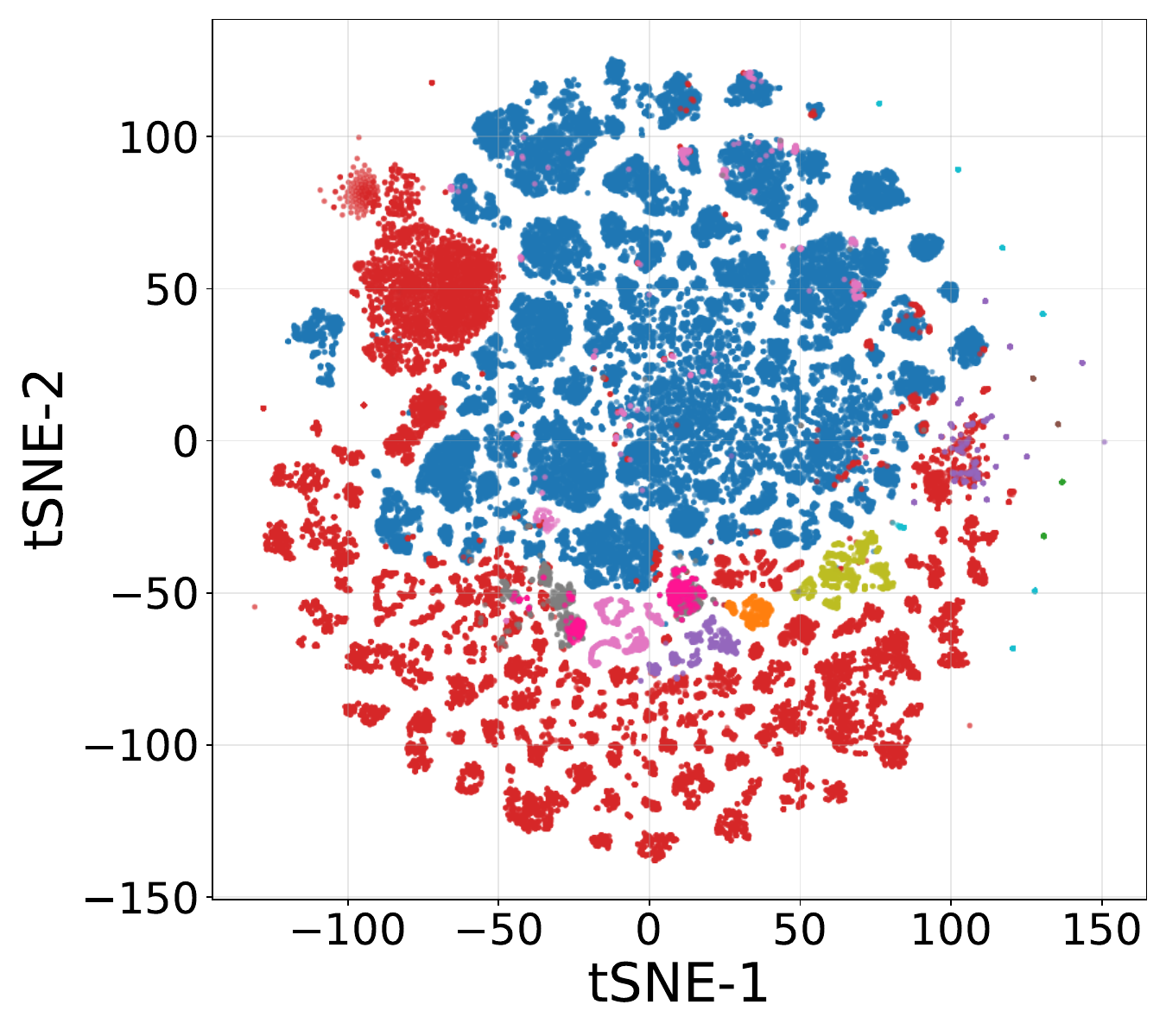}
\subcaption{Layer 2}
\end{subfigure}\hfill
\begin{subfigure}[b]{0.18\linewidth}
\includegraphics[width=\linewidth]{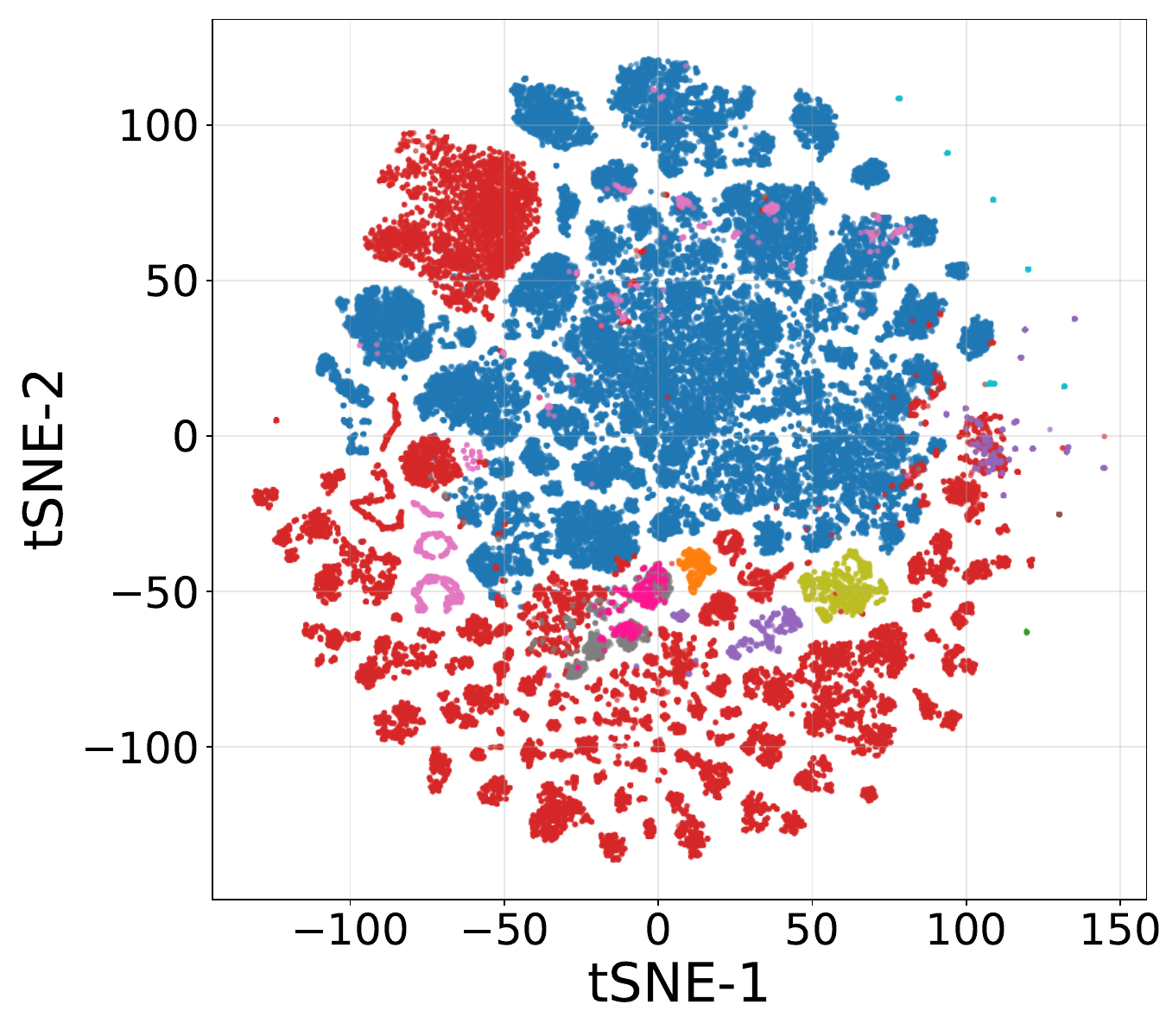}
\subcaption{Layer 3}
\end{subfigure}\hfill
\begin{subfigure}[b]{0.18\linewidth}
\includegraphics[width=\linewidth]{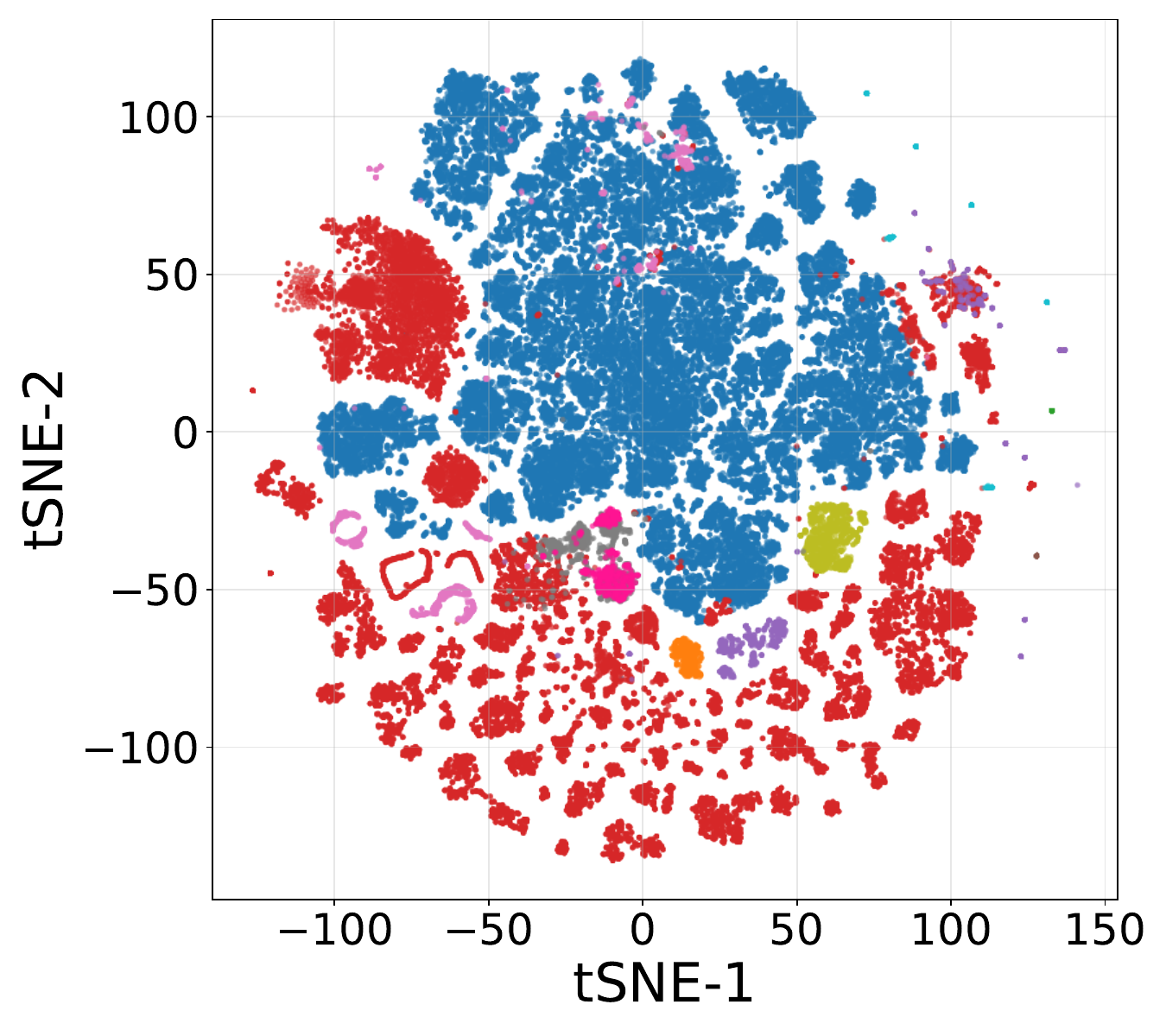}
\subcaption{Layer 4}
\end{subfigure}

\vspace{3pt}

\begin{subfigure}[b]{0.18\linewidth}
\includegraphics[width=\linewidth]{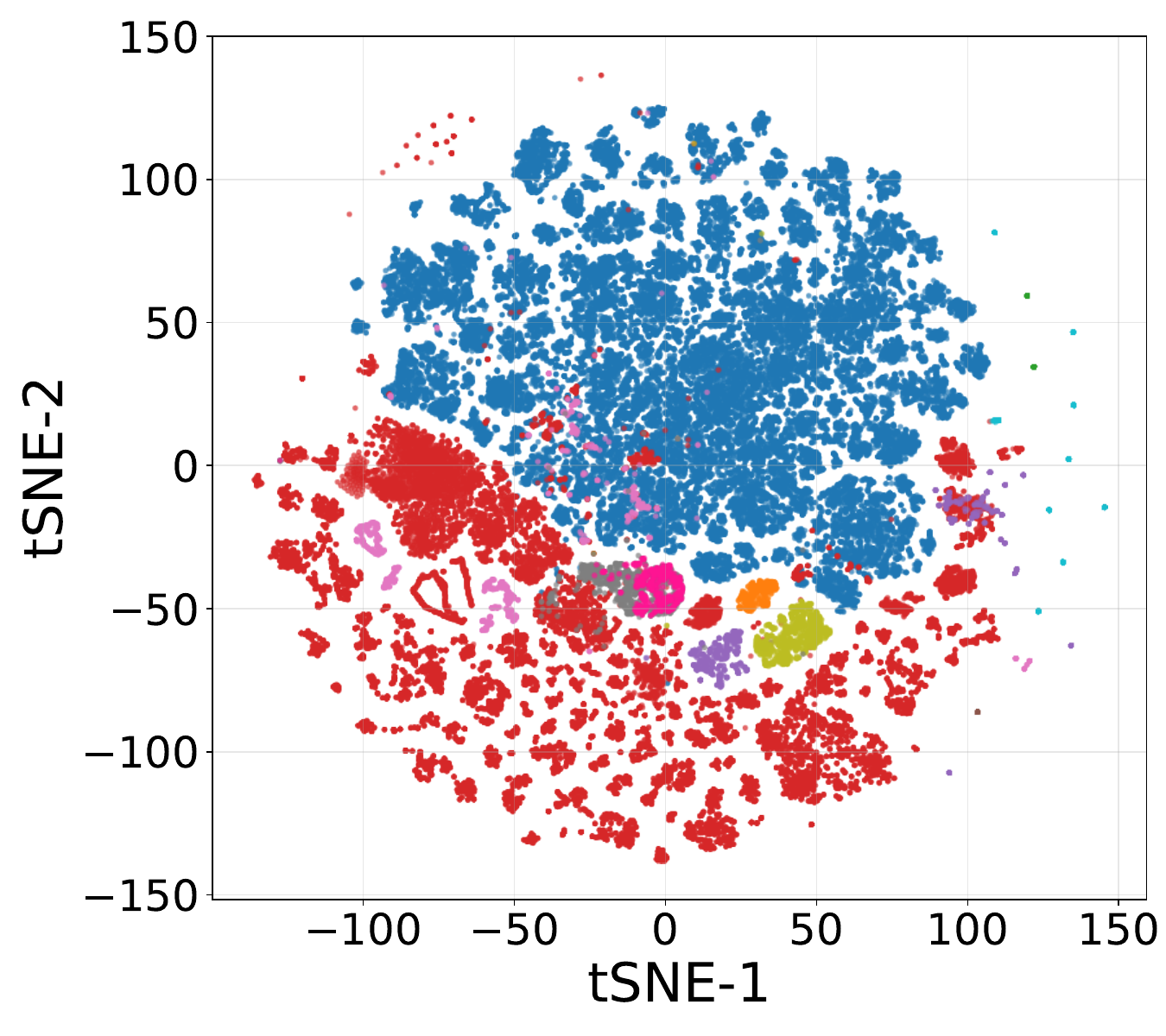}
\subcaption{Layer 5}
\end{subfigure}\hfill
\begin{subfigure}[b]{0.18\linewidth}
\includegraphics[width=\linewidth]{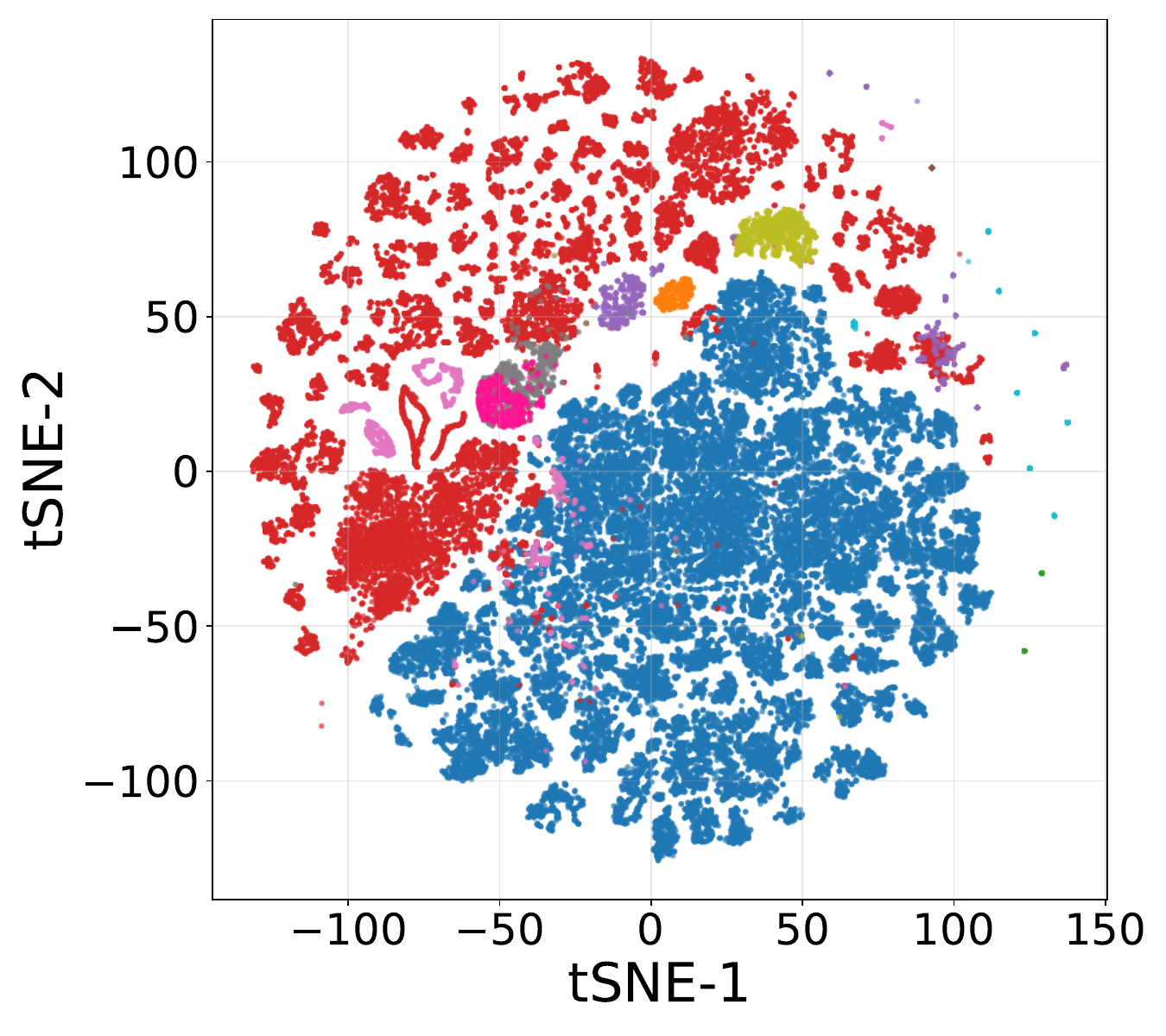}
\subcaption{Layer 6}
\end{subfigure}\hfill
\begin{subfigure}[b]{0.18\linewidth}
\includegraphics[width=\linewidth]{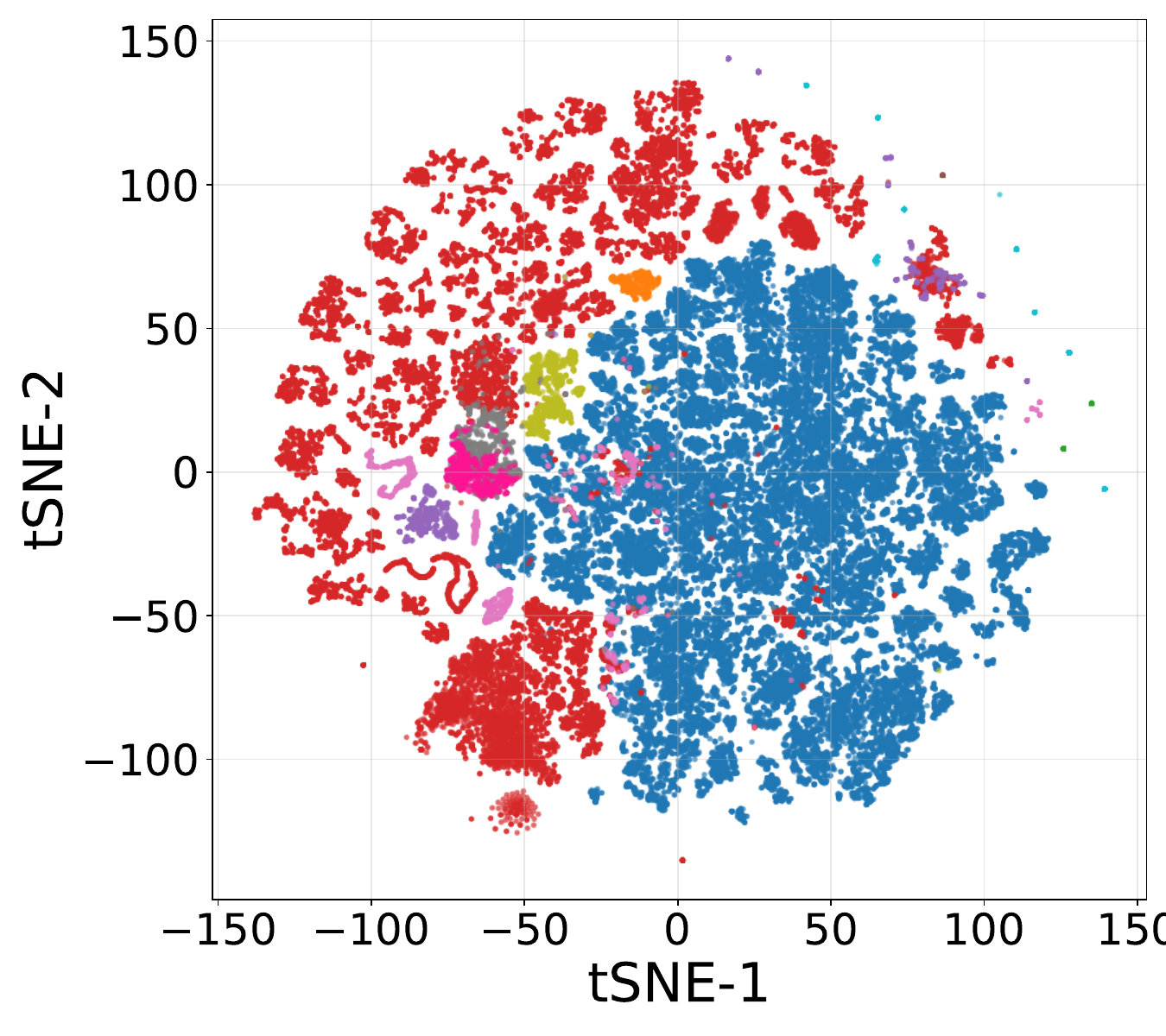}
\subcaption{Layer 7}
\end{subfigure}\hfill
\begin{subfigure}[b]{0.18\linewidth}
\includegraphics[width=\linewidth]{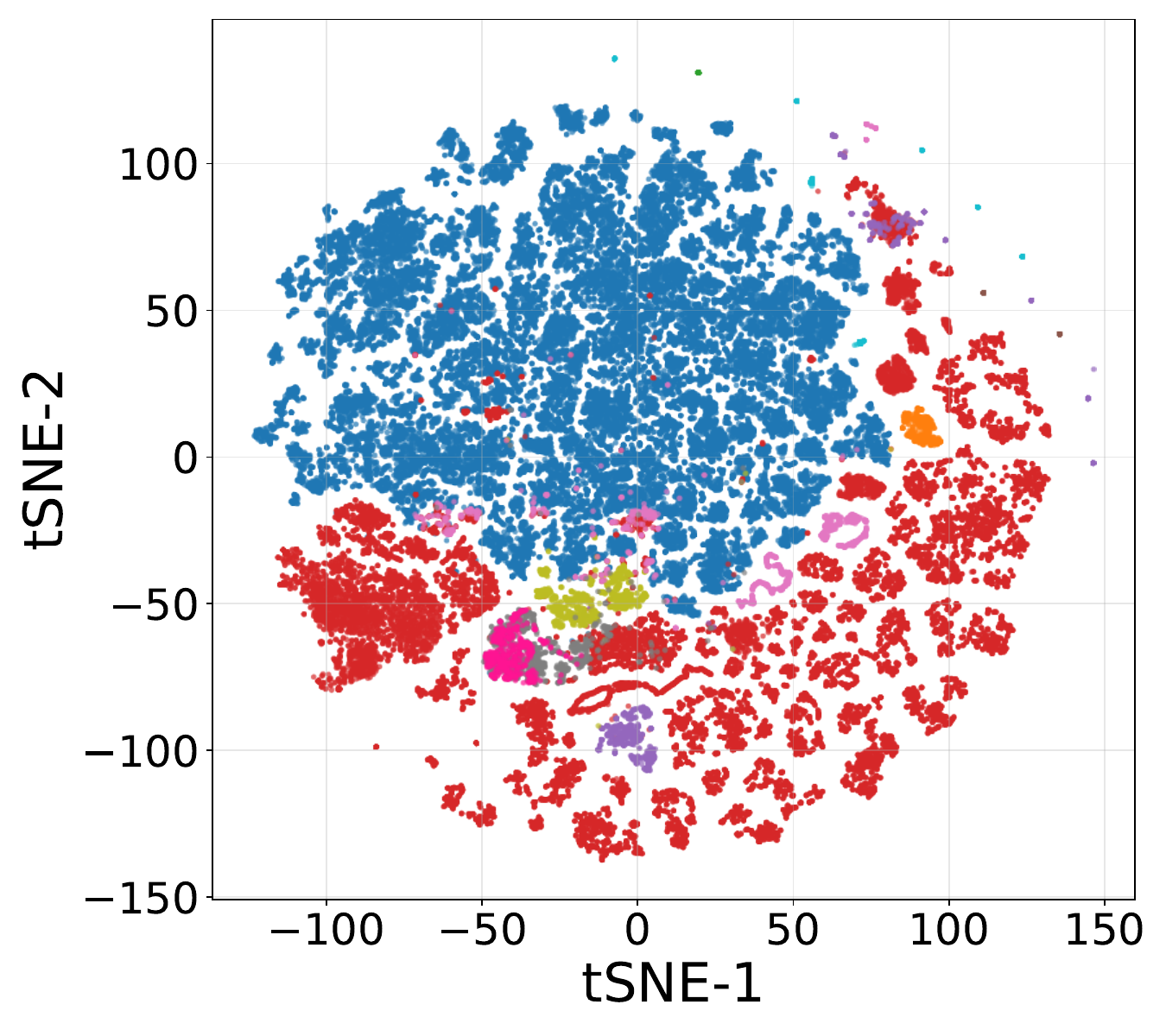}
\subcaption{Layer 8}
\end{subfigure}\hfill
\begin{subfigure}[b]{0.18\linewidth}
\includegraphics[width=\linewidth]{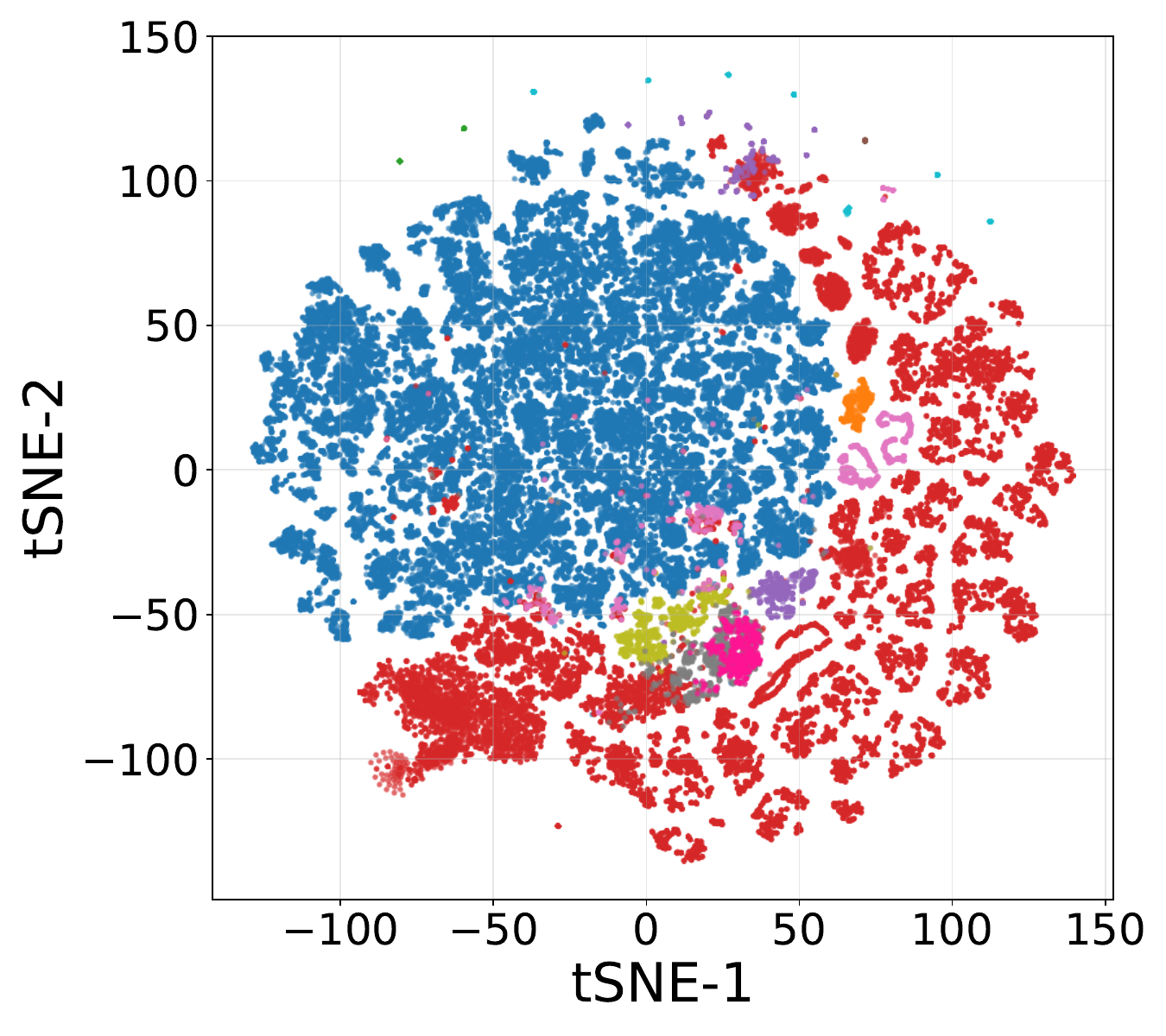}
\subcaption{Layer 9}
\end{subfigure}

\vspace{3pt}

\begin{subfigure}[b]{0.18\linewidth}
\includegraphics[width=\linewidth]{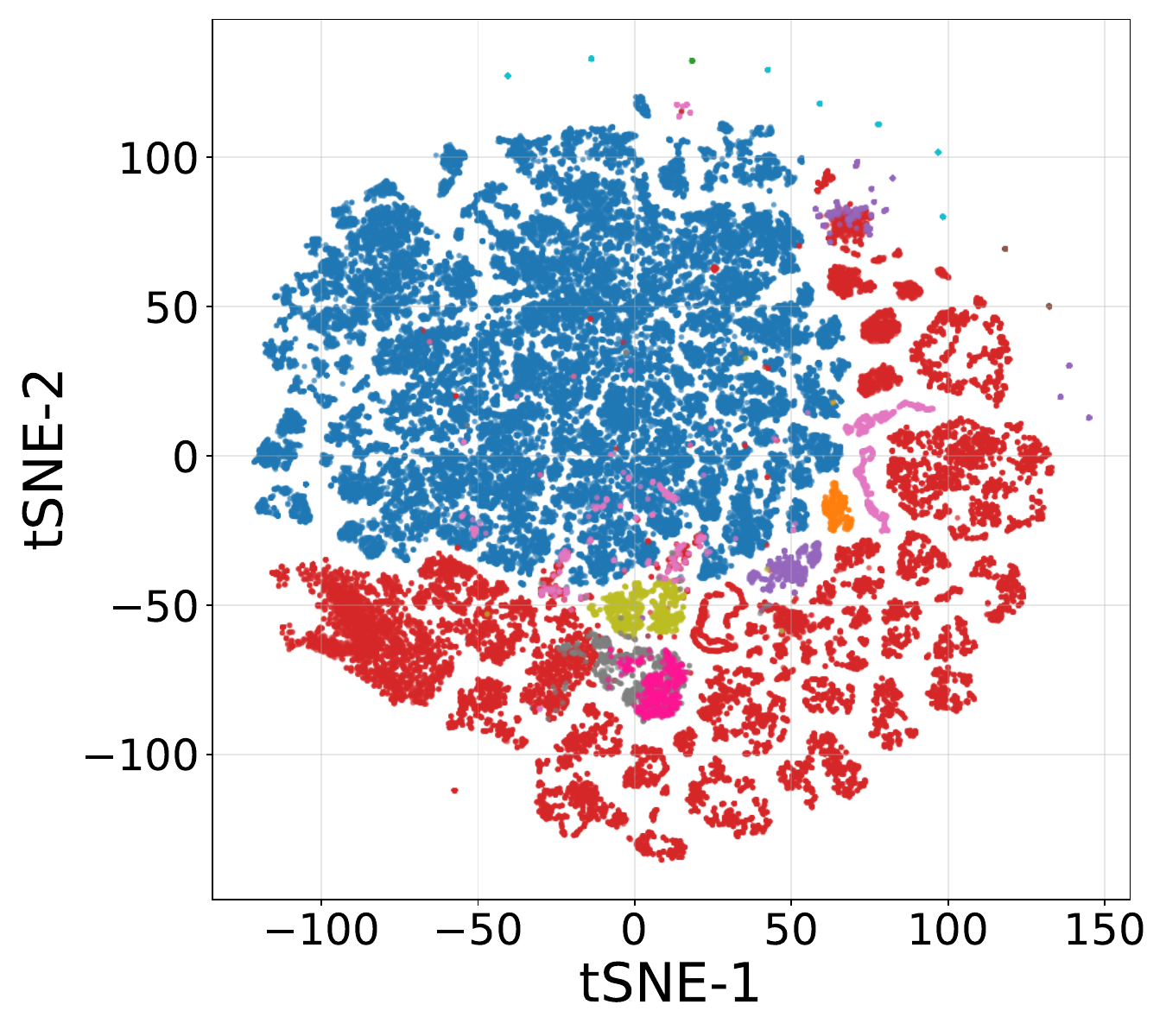}
\subcaption{Layer 10}
\end{subfigure}\hfill
\begin{subfigure}[b]{0.18\linewidth}
\includegraphics[width=\linewidth]{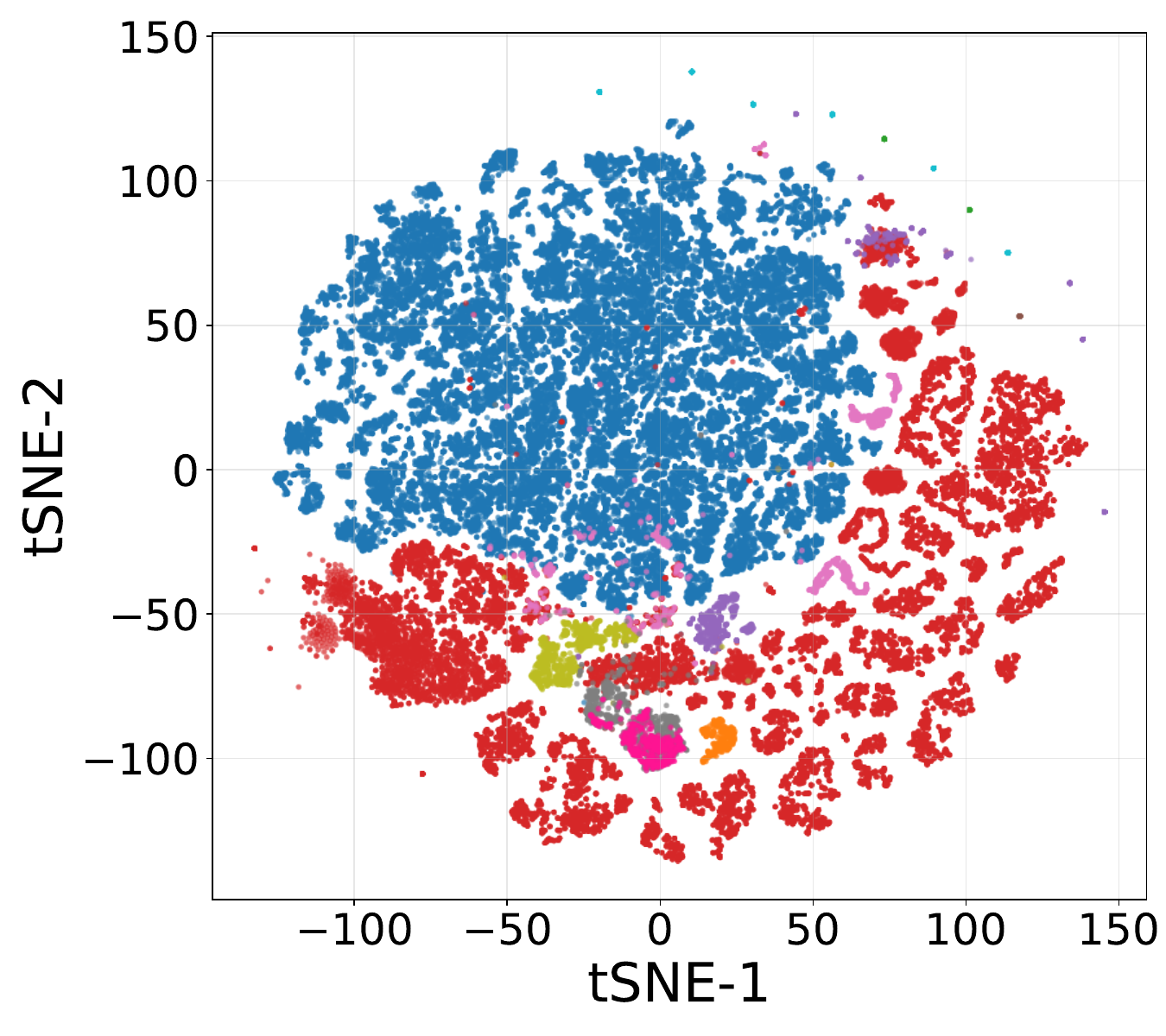}
\subcaption{Layer 11}
\end{subfigure}\hfill
\begin{subfigure}[b]{0.18\linewidth}
\includegraphics[width=\linewidth]{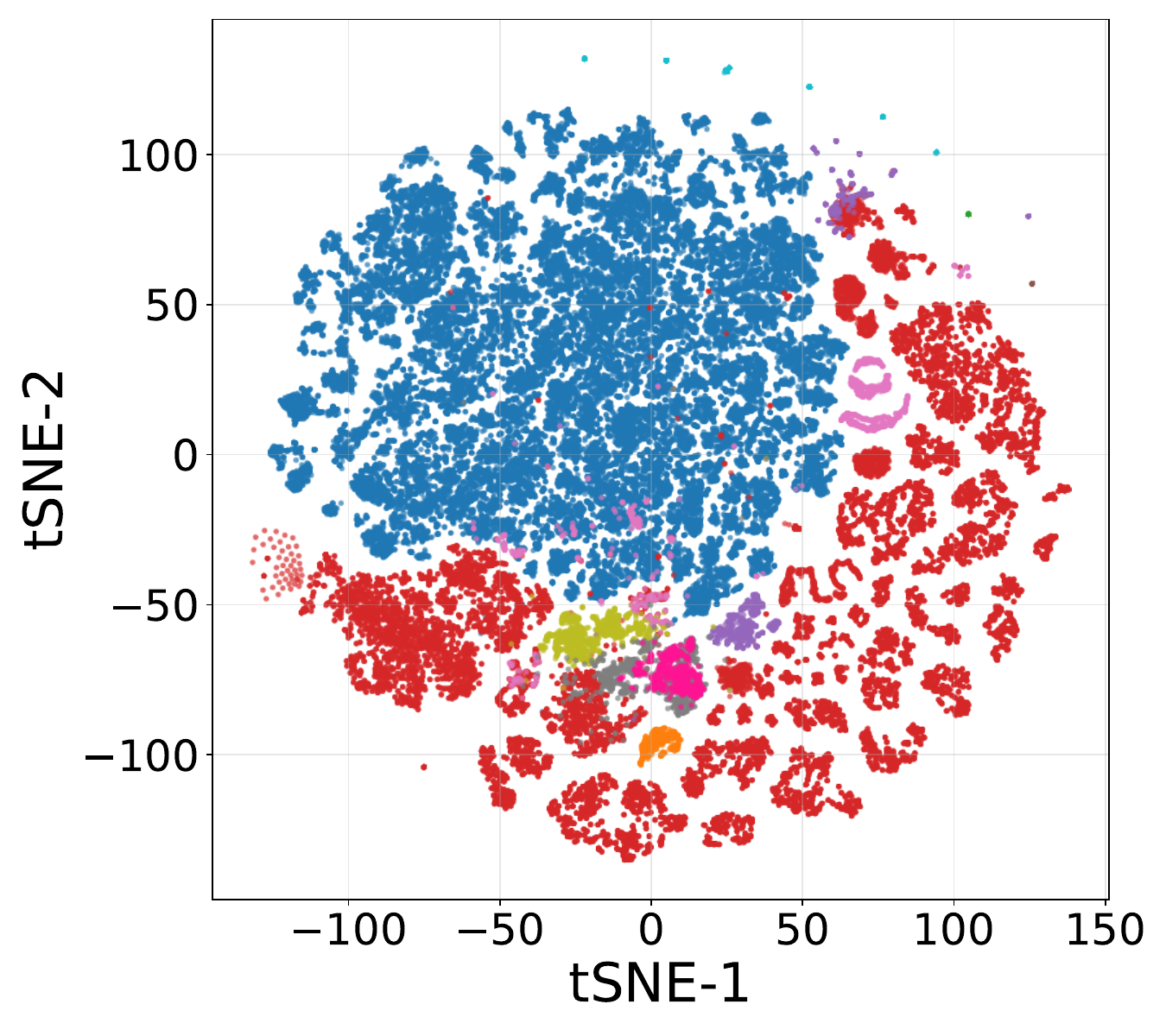}
\subcaption{Layer 12}
\end{subfigure}\hfill
\begin{subfigure}[b]{0.18\linewidth}
\includegraphics[width=\linewidth]{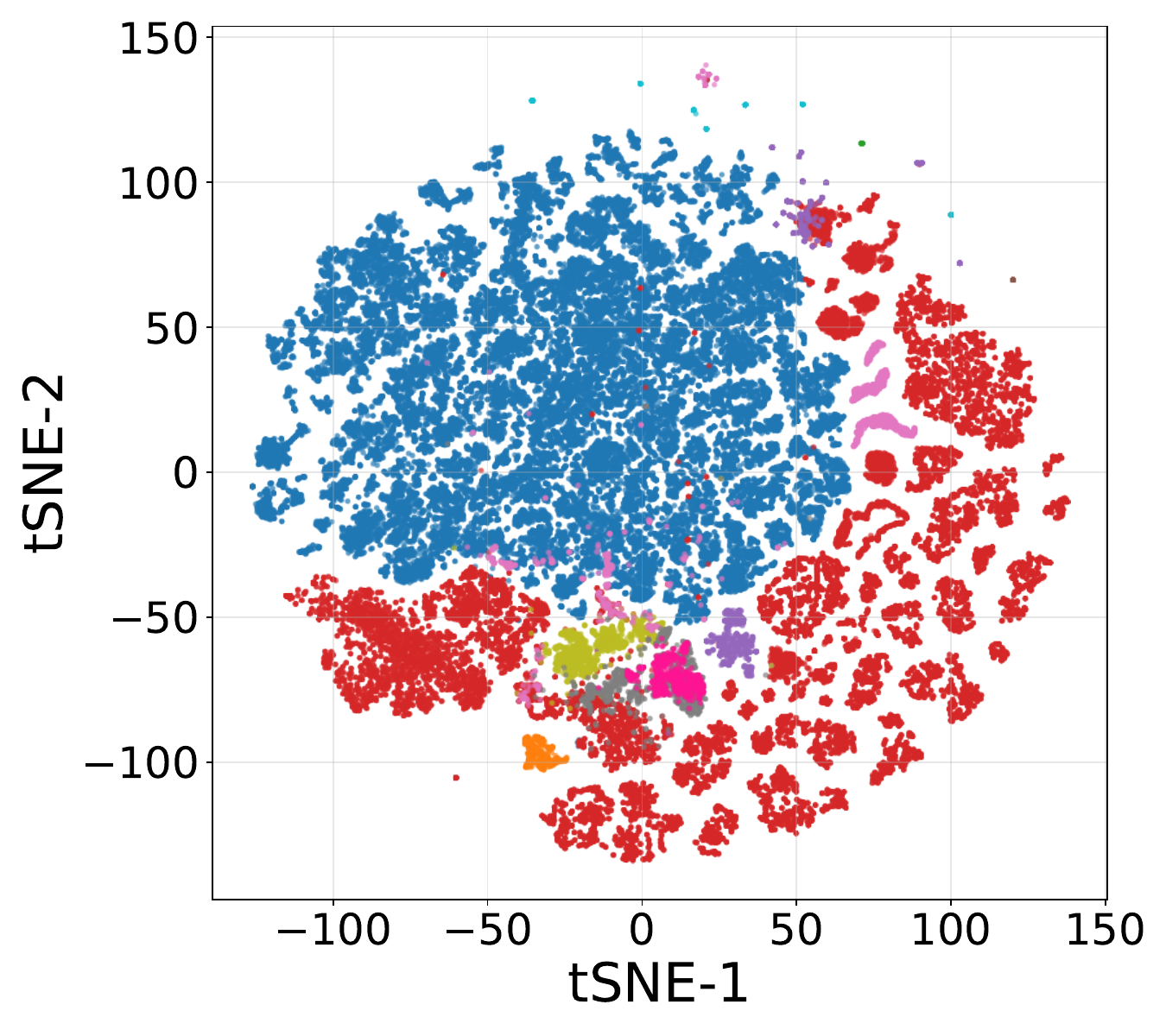}
\subcaption{Layer 13}
\end{subfigure}\hfill
\begin{subfigure}[b]{0.18\linewidth}
\includegraphics[width=\linewidth]{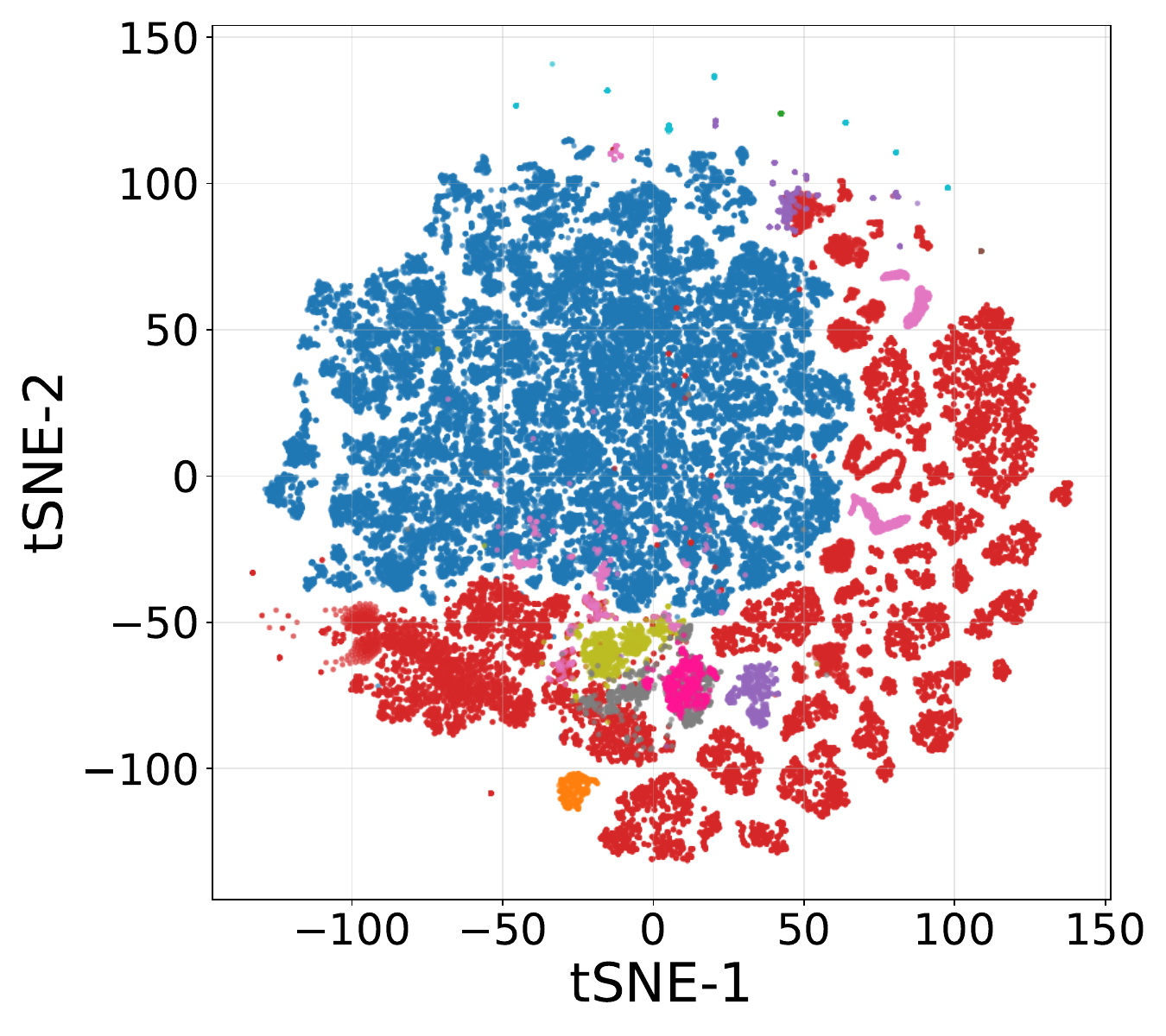}
\subcaption{Layer 14}
\end{subfigure}

\vspace{3pt}

\begin{subfigure}[b]{0.18\linewidth}
\includegraphics[width=\linewidth]{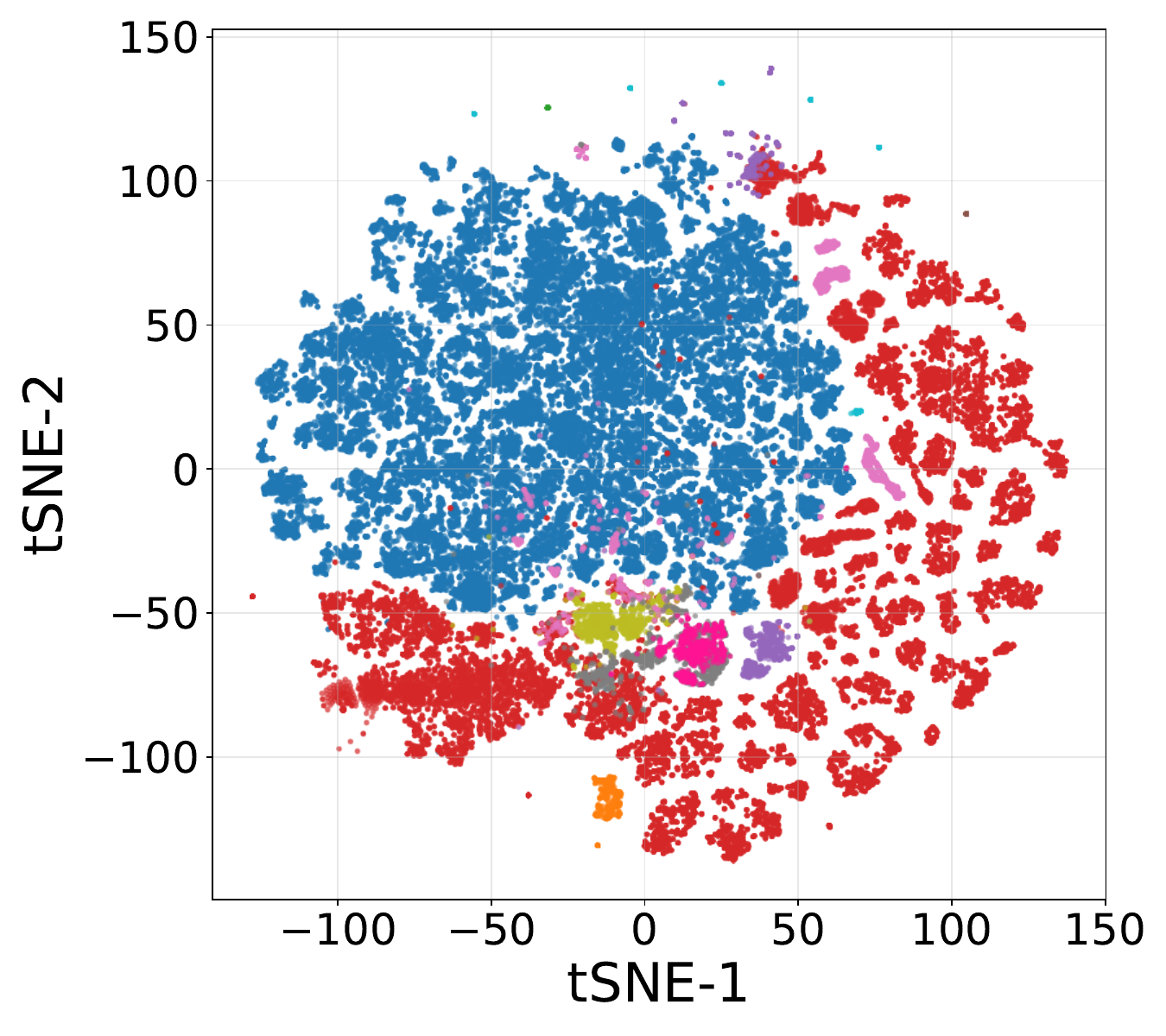}
\subcaption{Layer 15}
\end{subfigure}\hfill
\begin{subfigure}[b]{0.18\linewidth}
\includegraphics[width=\linewidth]{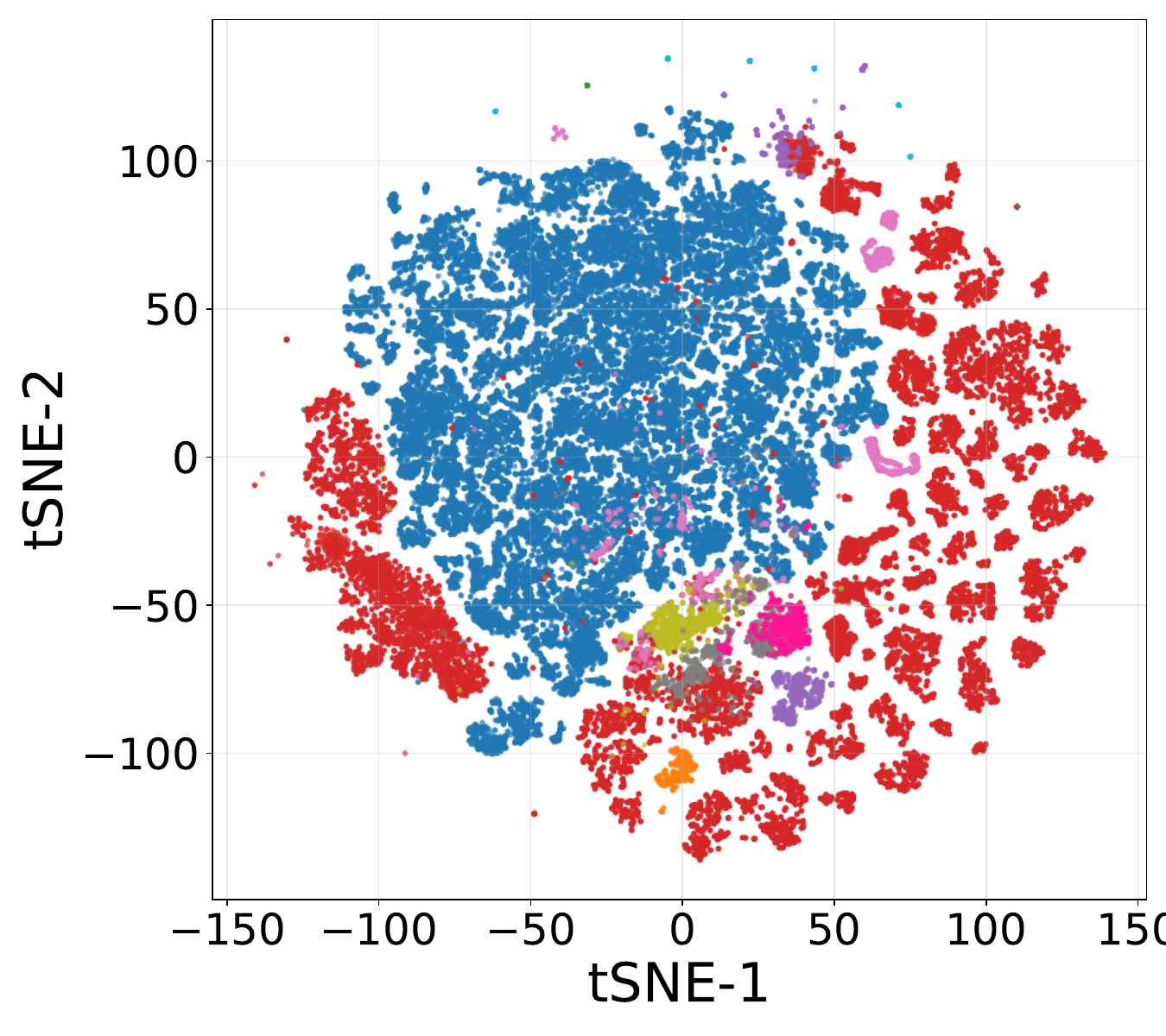}
\subcaption{Layer 16}
\end{subfigure}\hfill
\begin{subfigure}[b]{0.18\linewidth}
\includegraphics[width=\linewidth]{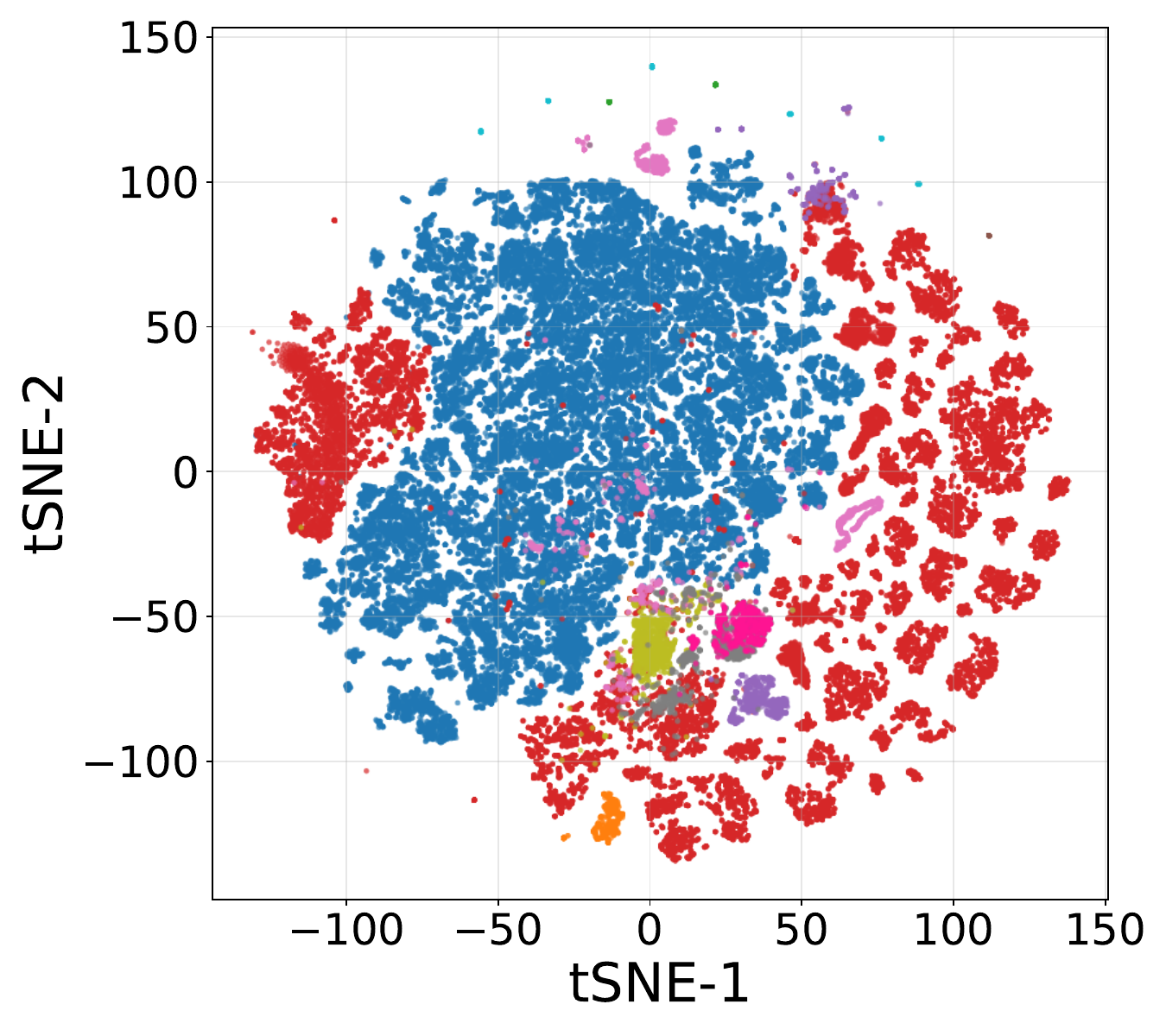}
\subcaption{Layer 17}
\end{subfigure}\hfill
\begin{subfigure}[b]{0.18\linewidth}
\includegraphics[width=\linewidth]{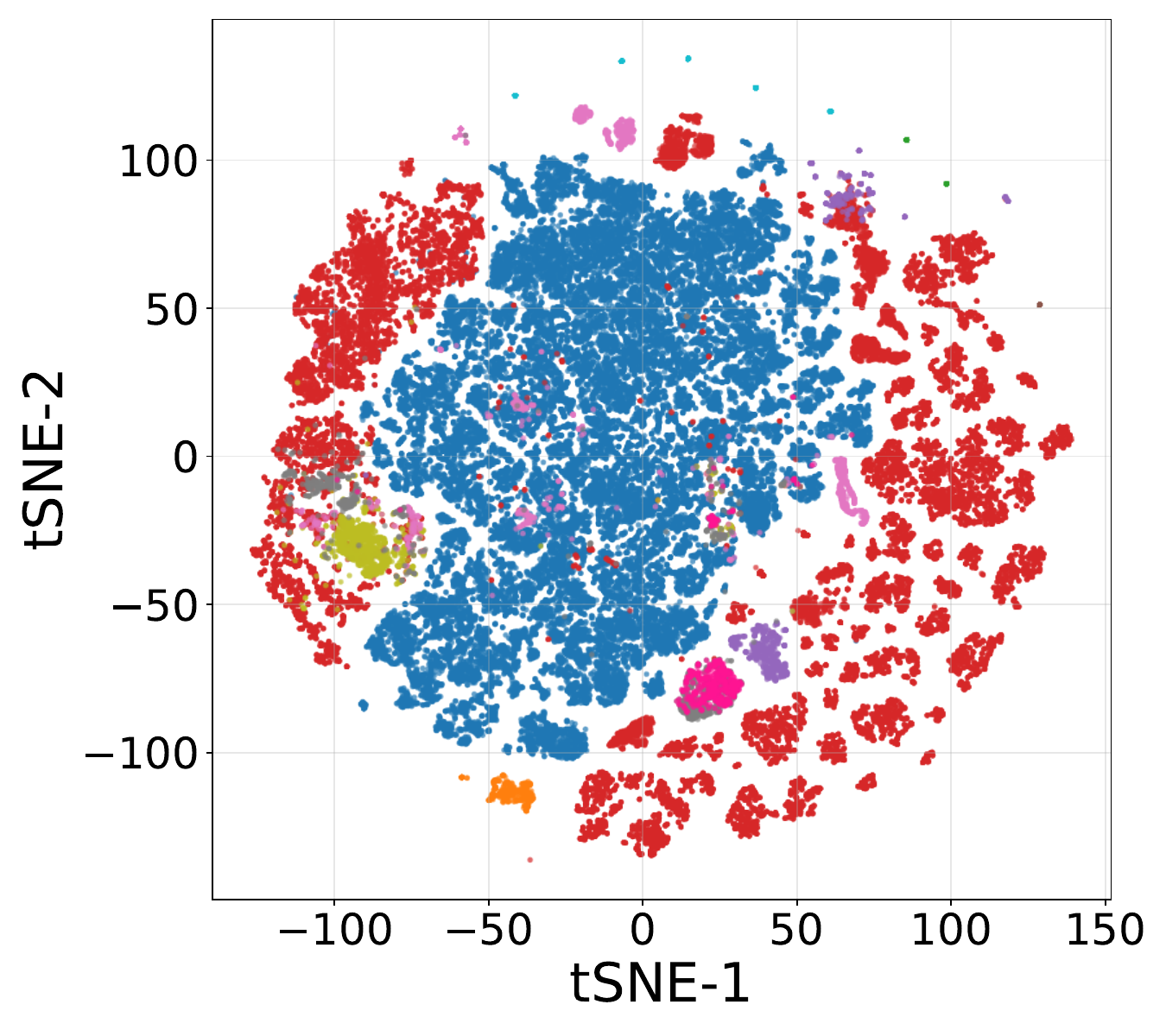}
\subcaption{Layer 18}
\end{subfigure}\hfill
\begin{subfigure}[b]{0.18\linewidth}
\includegraphics[width=\linewidth]{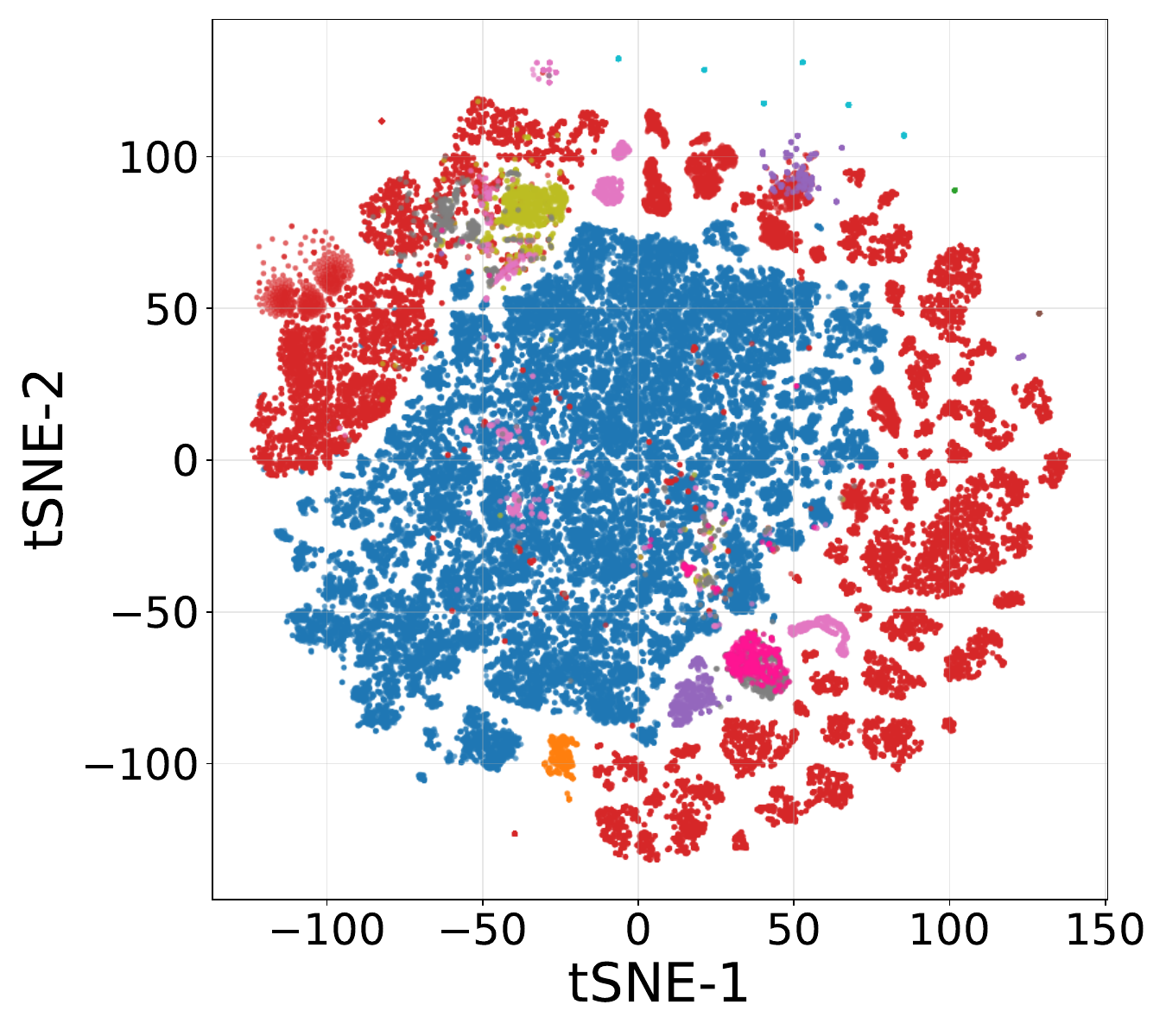}
\subcaption{Layer 19}
\end{subfigure}

\vspace{3pt}

\begin{subfigure}[b]{0.18\linewidth}
\includegraphics[width=\linewidth]{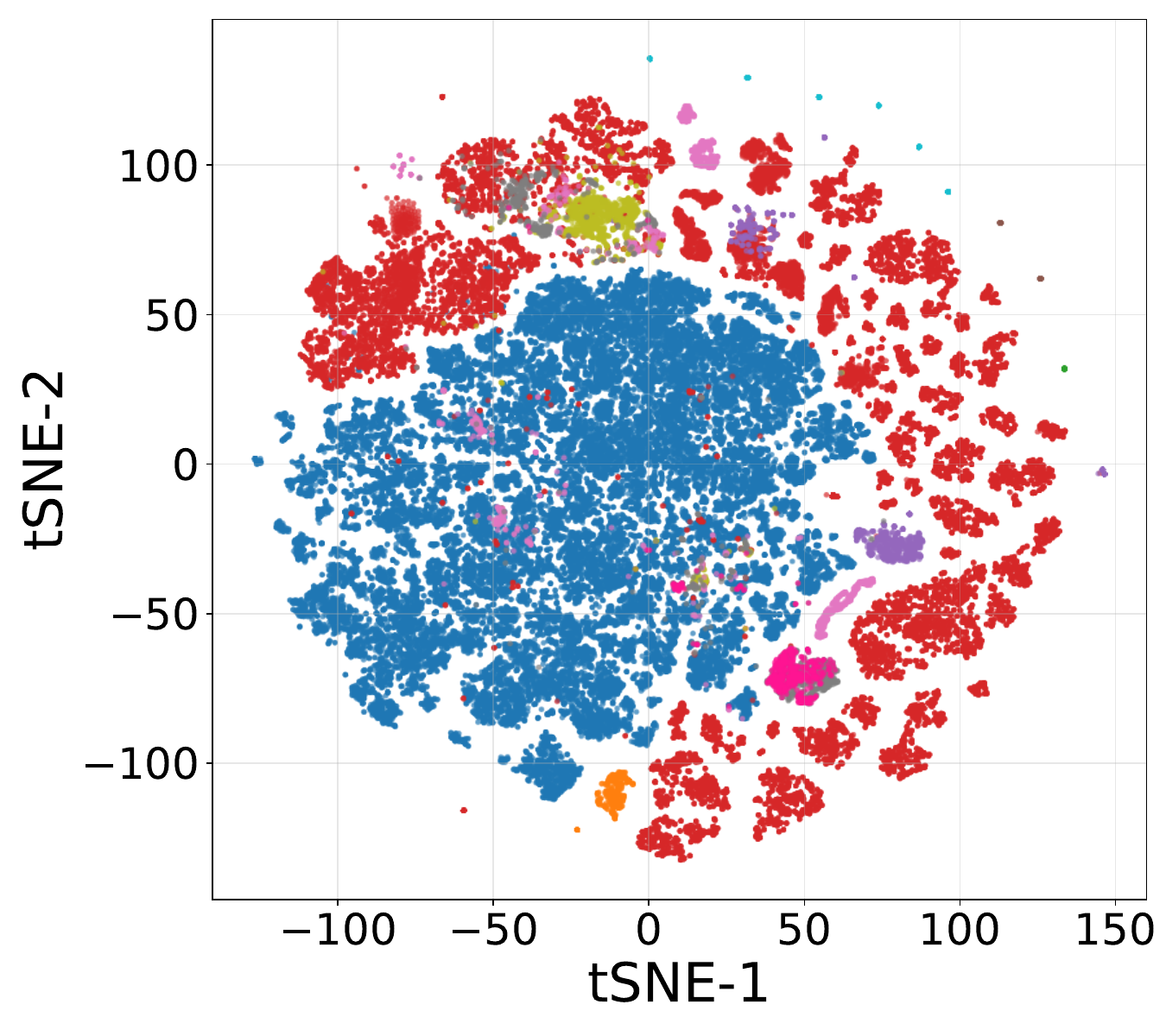}
\subcaption{Layer 20}
\end{subfigure}\hfill
\begin{subfigure}[b]{0.18\linewidth}
\includegraphics[width=\linewidth]{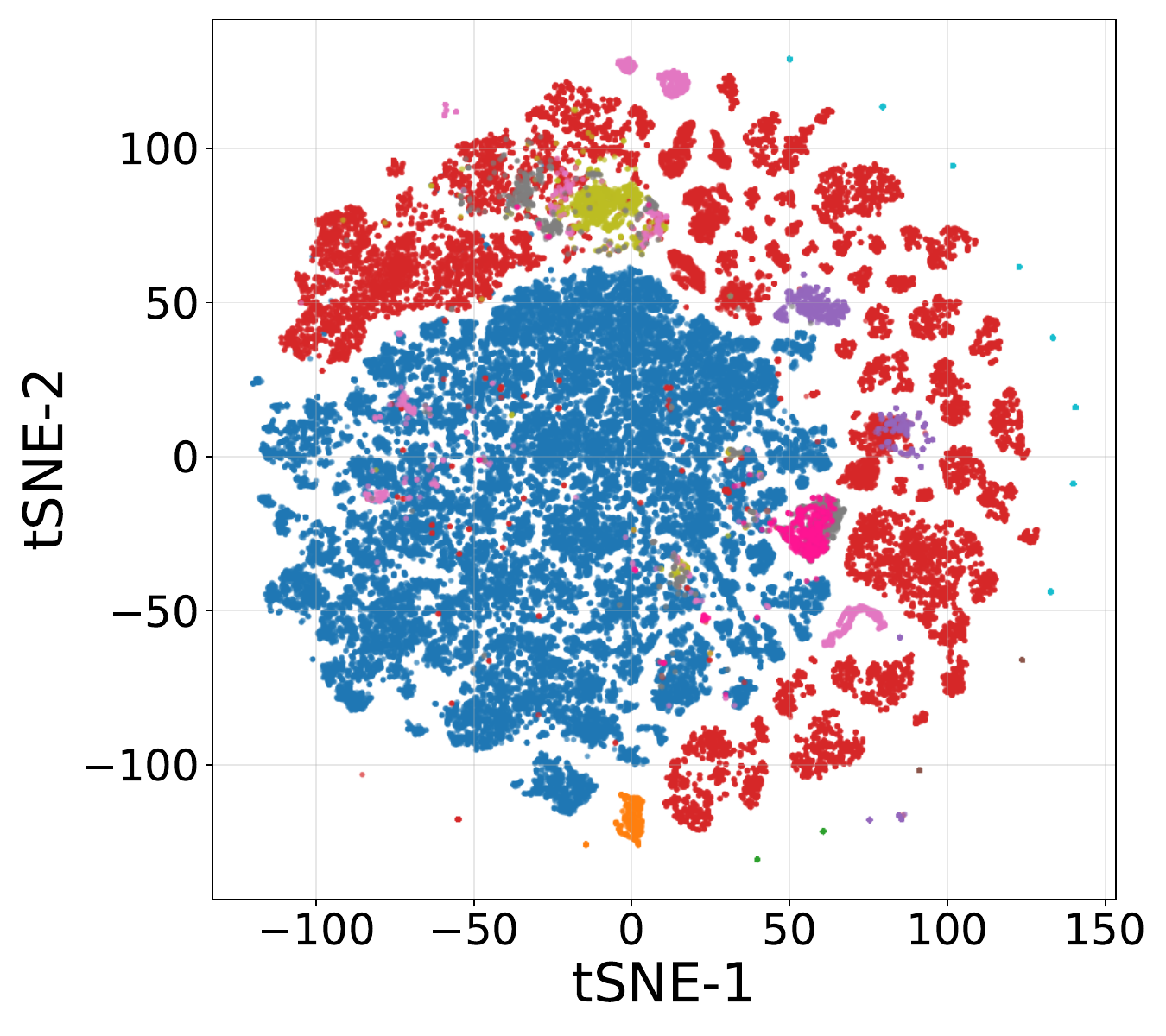}
\subcaption{Layer 21}
\end{subfigure}\hfill
\begin{subfigure}[b]{0.18\linewidth}
\includegraphics[width=\linewidth]{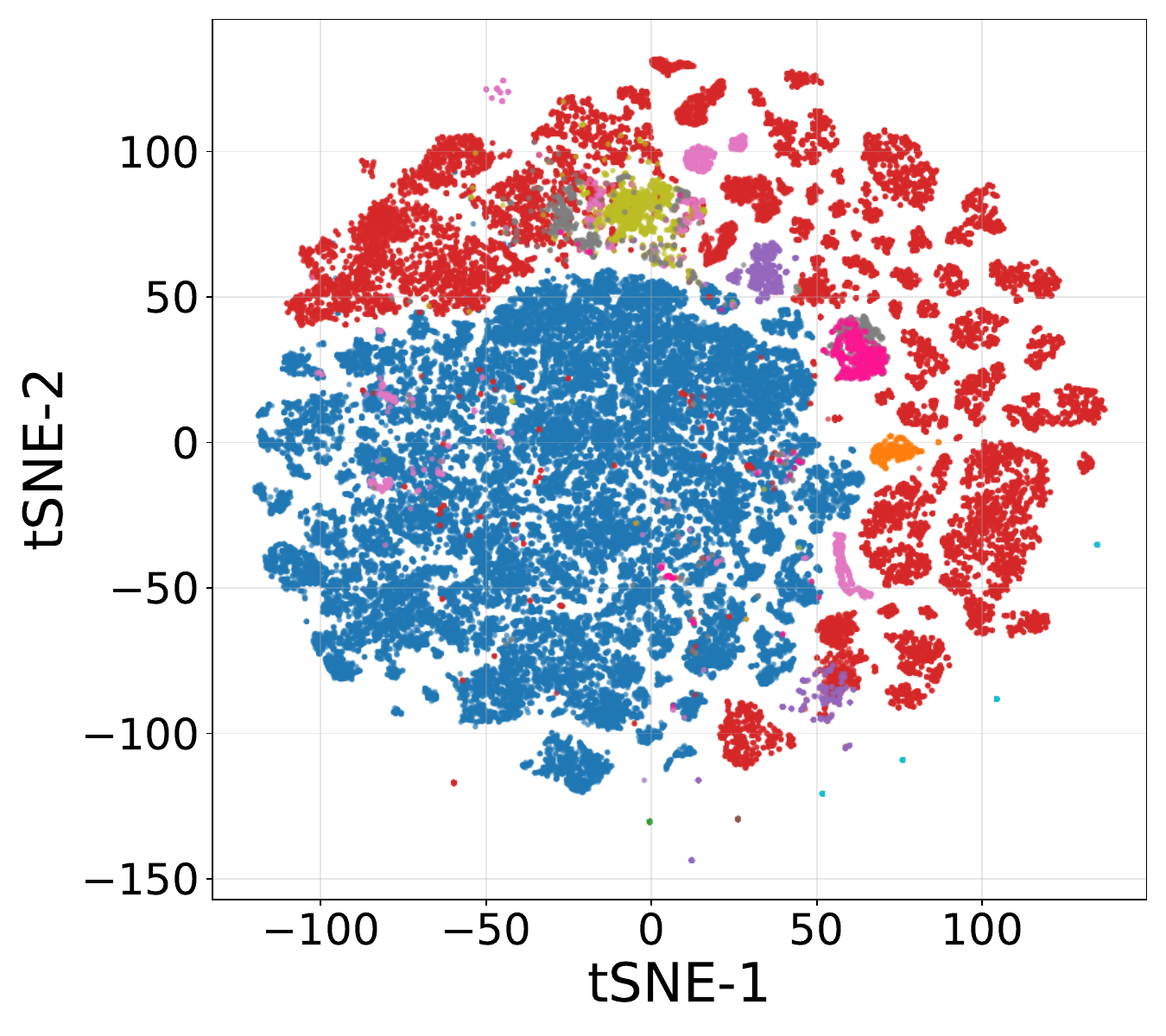}
\subcaption{Layer 22}
\end{subfigure}\hfill
\begin{subfigure}[b]{0.18\linewidth}
\includegraphics[width=\linewidth]{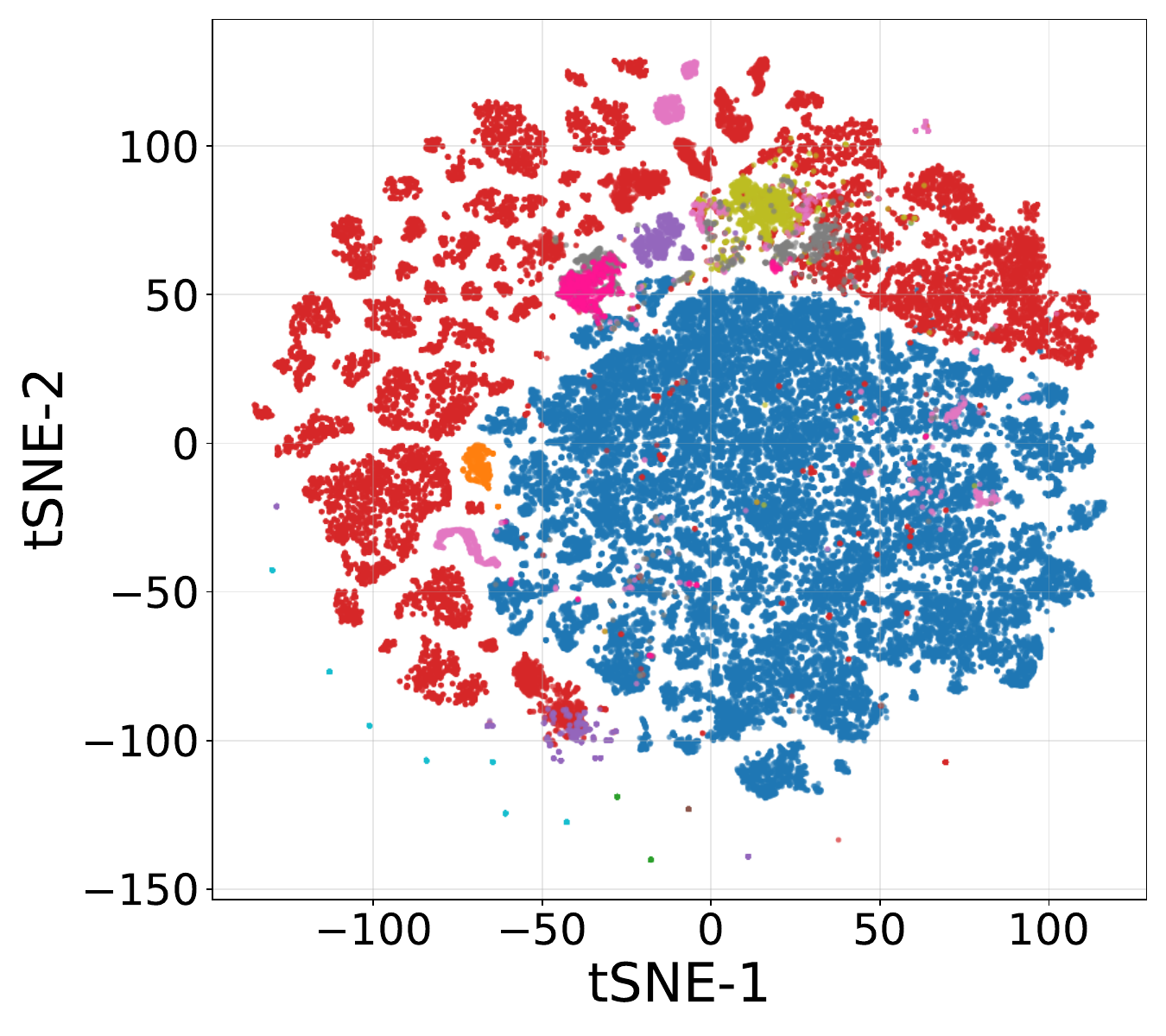}
\subcaption{Layer 23}
\end{subfigure}\hfill
\begin{subfigure}[b]{0.18\linewidth}
\includegraphics[width=\linewidth]{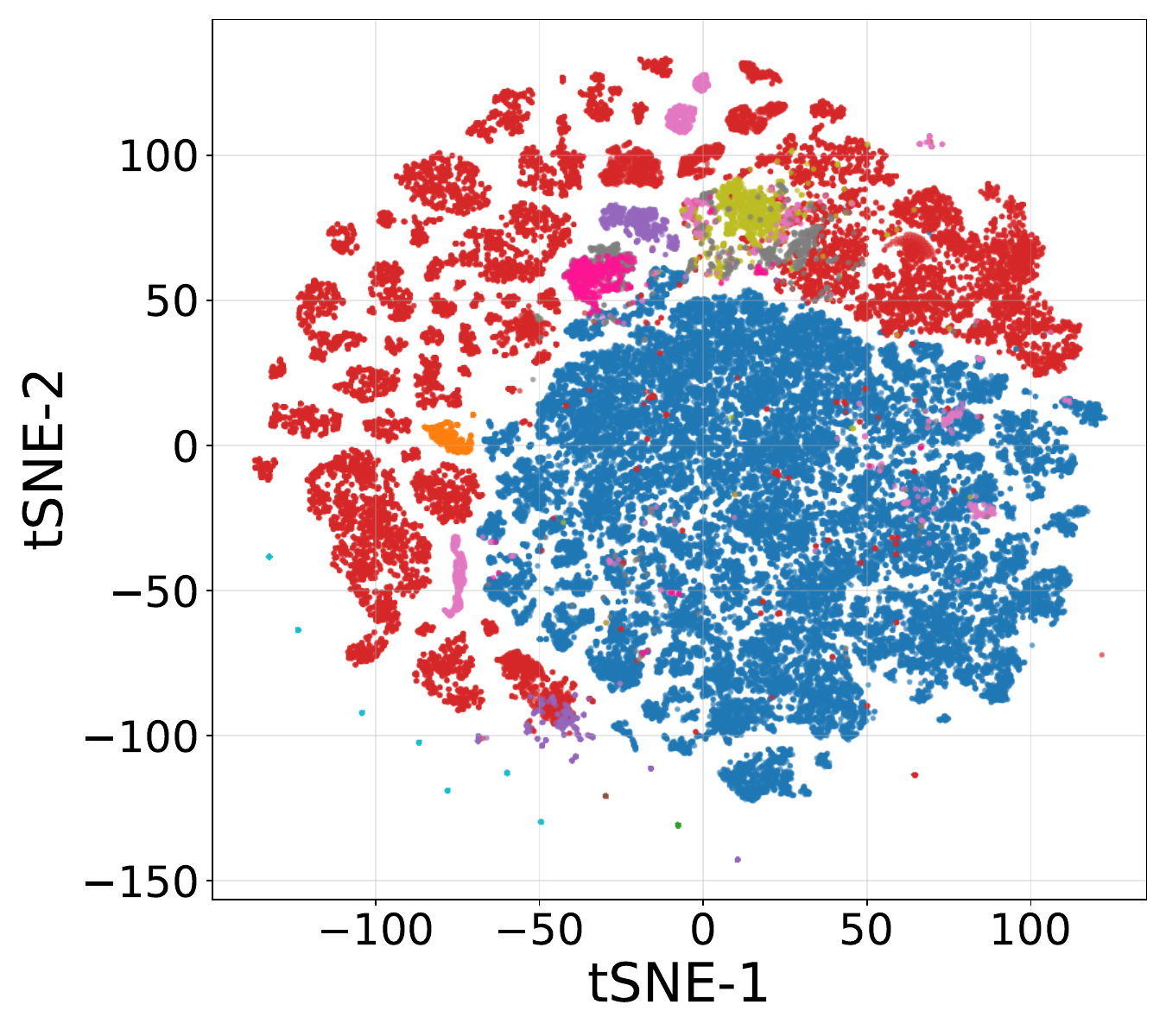}
\subcaption{Layer 24}
\end{subfigure}

\vspace{3pt}

\begin{subfigure}[b]{0.18\linewidth}
\includegraphics[width=\linewidth]{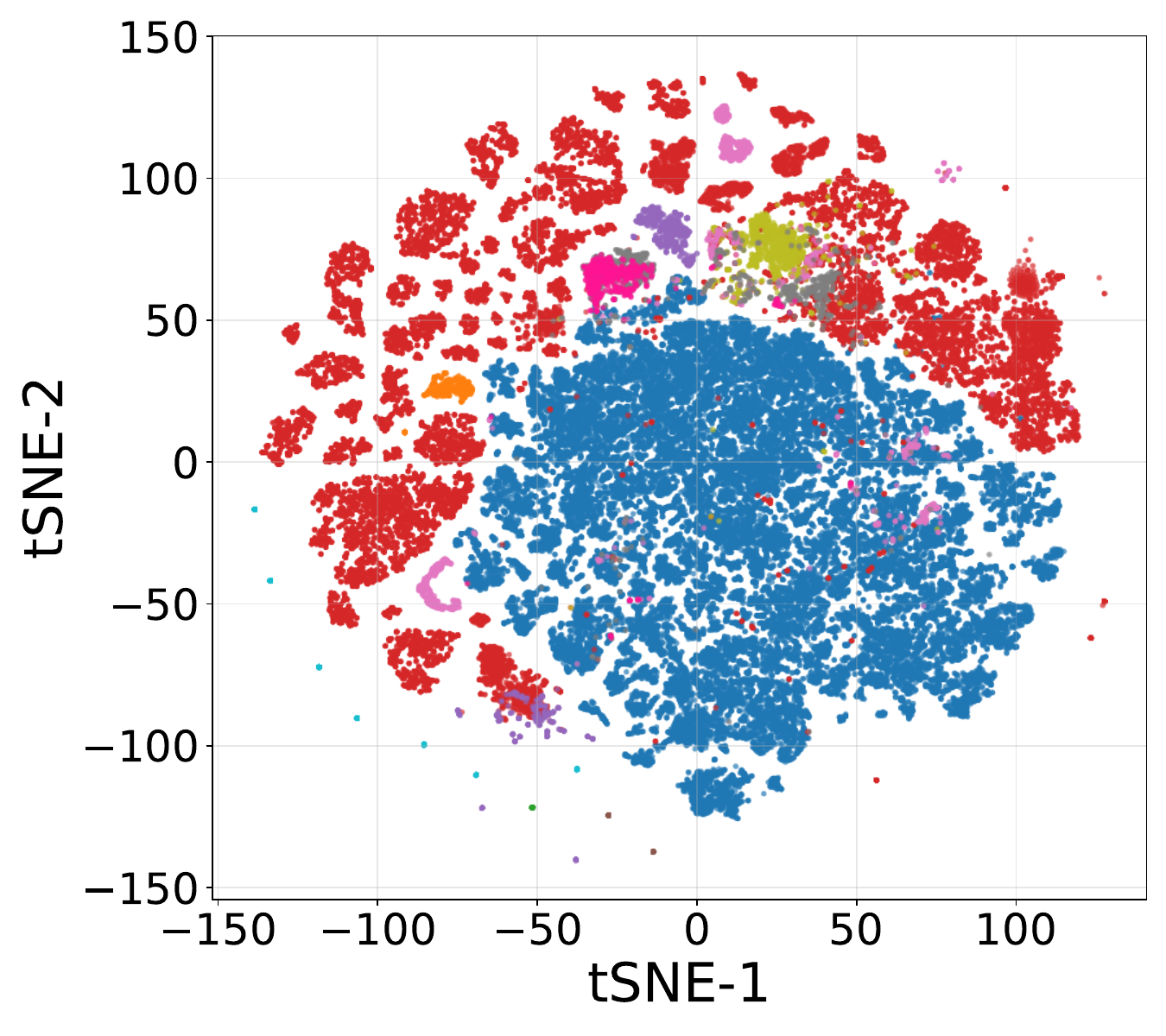}
\subcaption{Layer 25}
\end{subfigure}\hfill
\begin{subfigure}[b]{0.18\linewidth}
\includegraphics[width=\linewidth]{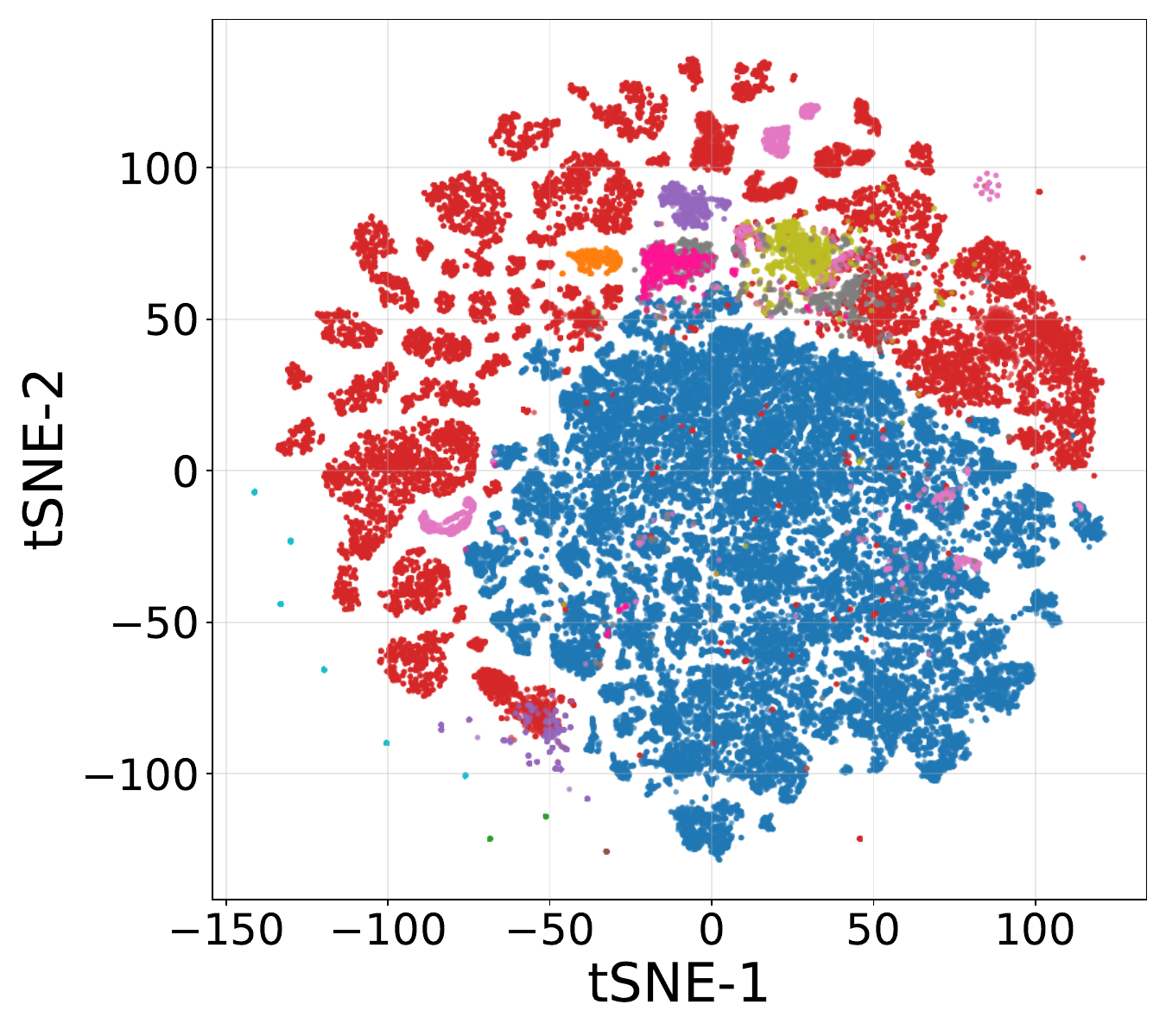}
\subcaption{Layer 26}
\end{subfigure}\hfill
\begin{subfigure}[b]{0.18\linewidth}
\includegraphics[width=\linewidth]{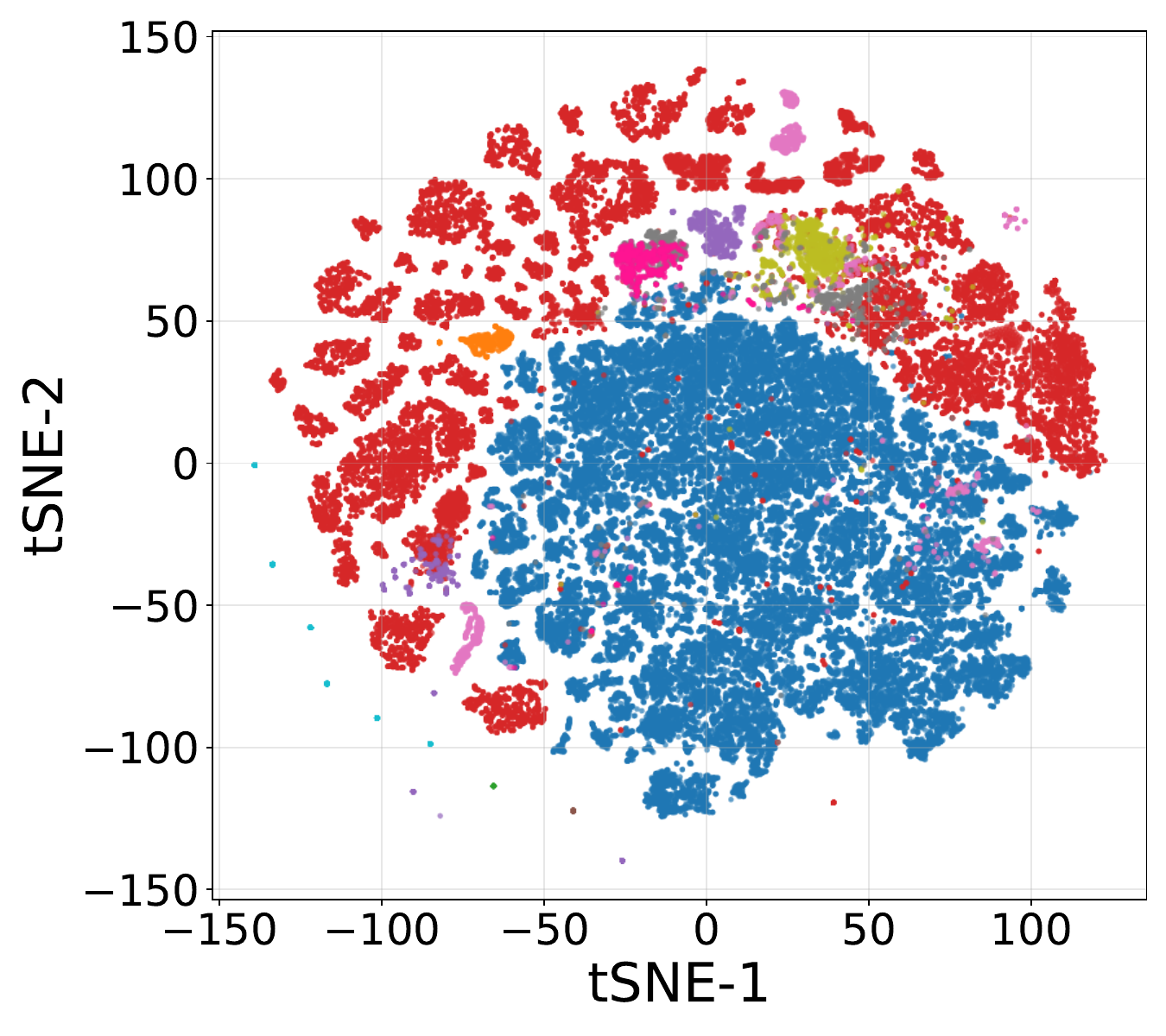}
\subcaption{Layer 27}
\end{subfigure}\hfill
\begin{subfigure}[b]{0.18\linewidth}
\includegraphics[width=\linewidth]{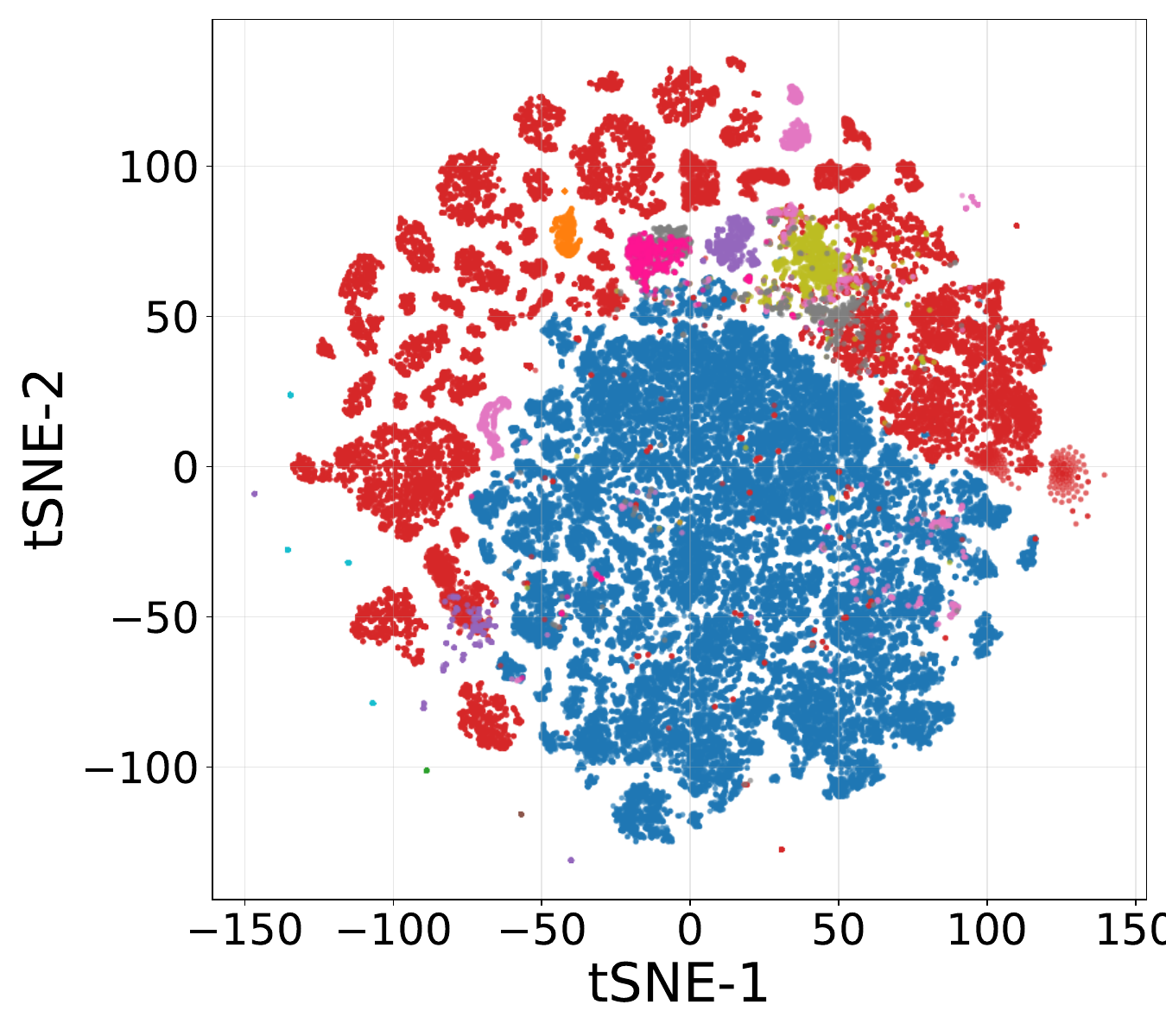}
\subcaption{Layer 28}
\end{subfigure}\hfill
\begin{subfigure}[b]{0.18\linewidth}
\includegraphics[width=\linewidth]{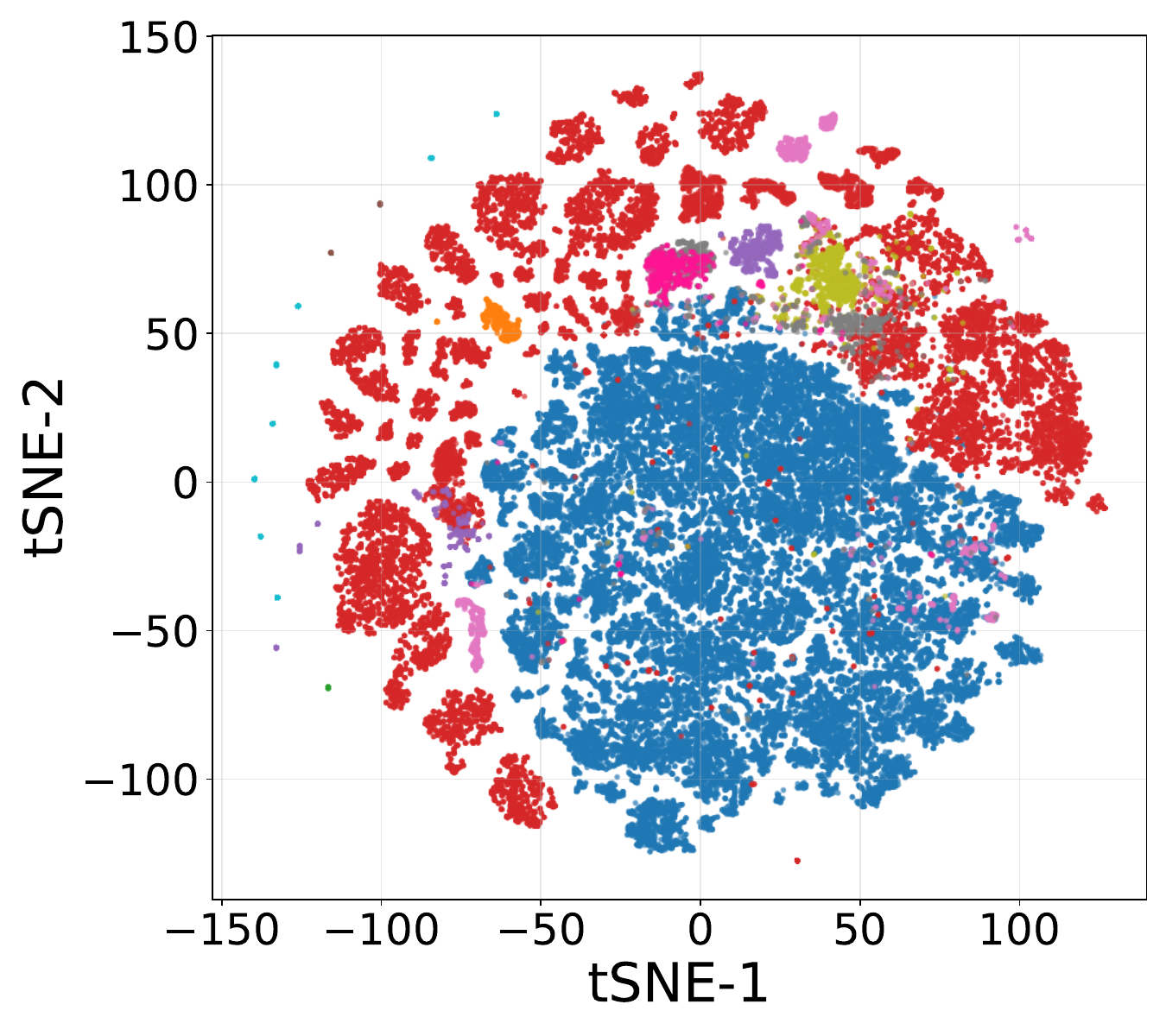}
\subcaption{Layer 29}
\end{subfigure}

\vspace{3pt}

\begin{subfigure}[b]{0.18\linewidth}
\includegraphics[width=\linewidth]{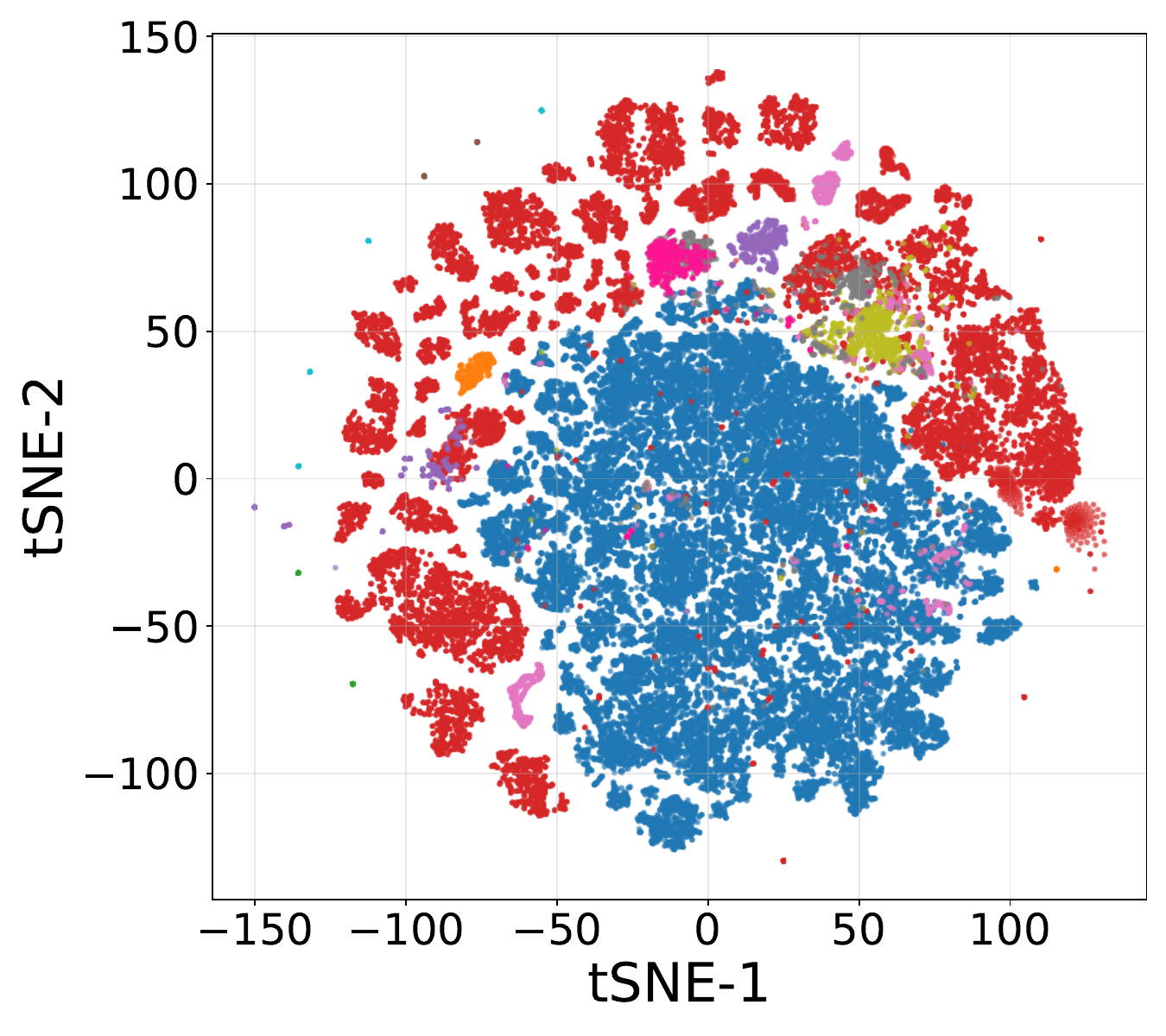}
\subcaption{Layer 30}
\end{subfigure}\hfill
\begin{subfigure}[b]{0.18\linewidth}
\includegraphics[width=\linewidth]{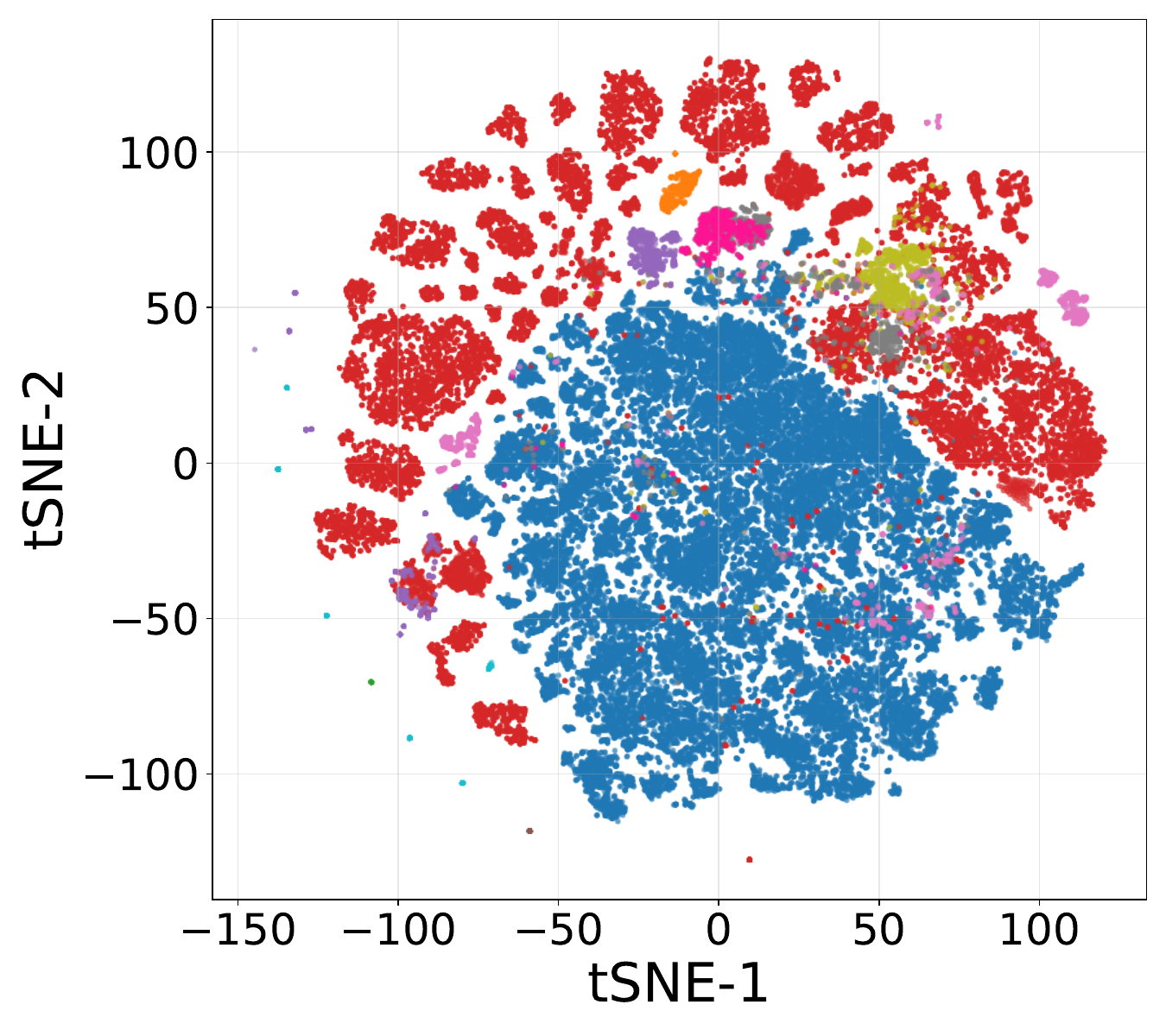}
\subcaption{Layer 31}
\end{subfigure}

\caption{t-SNE visualization across layers 0--31 of Mistral-7B}
\label{fig:mistral_tsne}
\end{figure*}

In this section, we provide additional analysis of the internal representation space of SLMs to understand how different types of inputs are encoded across network layers. We perform a layer-wise sensitivity analysis on three instruction-tuned SLMs: LLaMA-2-7B-Chat, Vicuna-7B-v1.5, and Mistral-7B-Instruct. For each model, all transformer layers from 0 to 31 are analyzed. At every layer, the hidden representation of the last token is extracted and projected into a two-dimensional space using t-SNE to visualize the distribution of different prompt categories. The complete layer-wise visualizations are shown in Figure~\ref{fig:llama_tsne}, Figure~\ref{fig:vicuna_tsne}, and Figure~\ref{fig:mistral_tsne}. The analysis considers three input types: benign prompts, direct malicious prompts, and nine optimized jailbreak attack categories. To better interpret the behavior across the network, layers are grouped into early (0--10), middle (11--21), and late (22--31) stages.

Across all three models and all layers, jailbreak-related features remain visible in the representation space, as shown in Figures~\ref{fig:llama_tsne}--\ref{fig:mistral_tsne}. Optimized jailbreak prompts begin to diverge from benign representations even in early layers, although partial overlap with direct malicious prompts may still exist. In the middle layers, the separation becomes more structured, where benign prompts form compact and stable clusters, direct malicious prompts move away from the benign region, and optimized jailbreak attacks produce wider and more scattered distributions. In the late layers, the distinction becomes the most pronounced, with optimized jailbreak attacks forming clearly identifiable clusters that remain separated from benign representations across all three models.

We further observe that benign prompts maintain consistent and compact clusters across layers, while direct malicious prompts deviate from the benign region with occasional overlap depending on the model and layer. In contrast, optimized jailbreak attacks introduce larger shifts in the representation space, resulting in distinct clusters that are easier to separate from benign inputs. The separation between benign and optimized jailbreak representations is generally larger than the separation between benign and direct malicious prompts. Similar distribution patterns are observed across LLaMA, Vicuna, and Mistral, indicating that the behavior is model-independent. Overall, the results confirm that distinguishable activation patterns appear throughout the network, showing that useful signals for jailbreak detection exist across multiple layers rather than being limited to a specific depth.

\begin{figure*}[t]
\centering
\includegraphics[width=\textwidth]{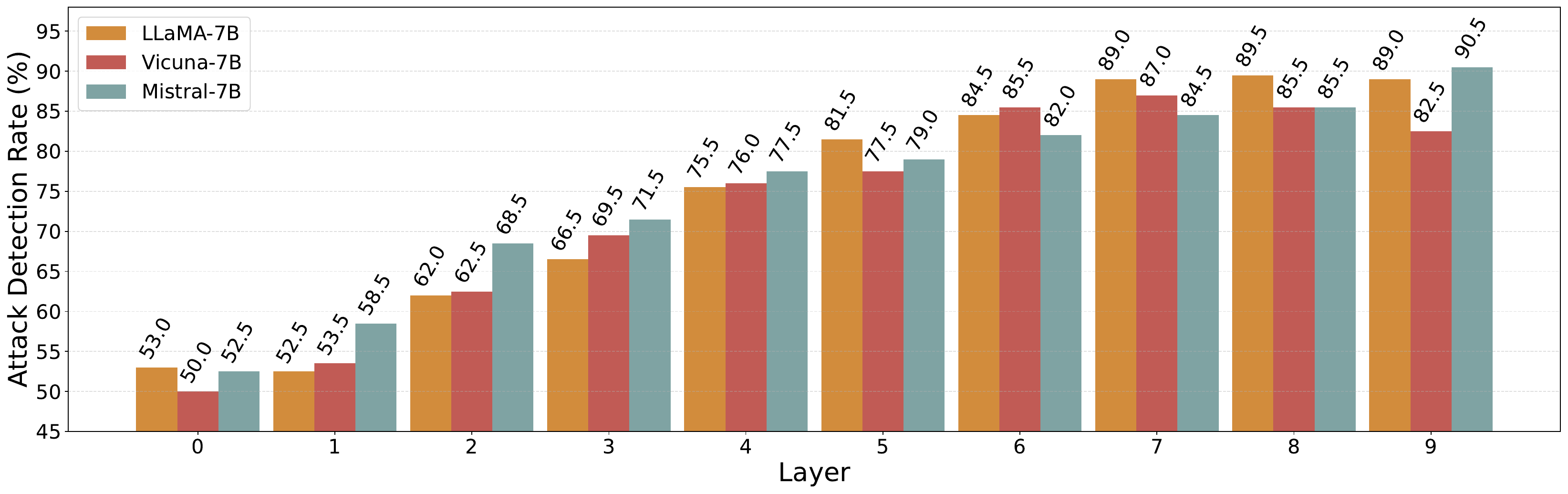}

\caption{Attack detection accuracy in the representation space using GUARD-SLM across early layers (0--10), Dataset: HarmBench , model: meta-
llama/Llama-2-7b-chat-hf }
\label{fig:guard_slm_harmbench_early}
\end{figure*}

\begin{figure*}[t]
\centering
\includegraphics[width=\textwidth]{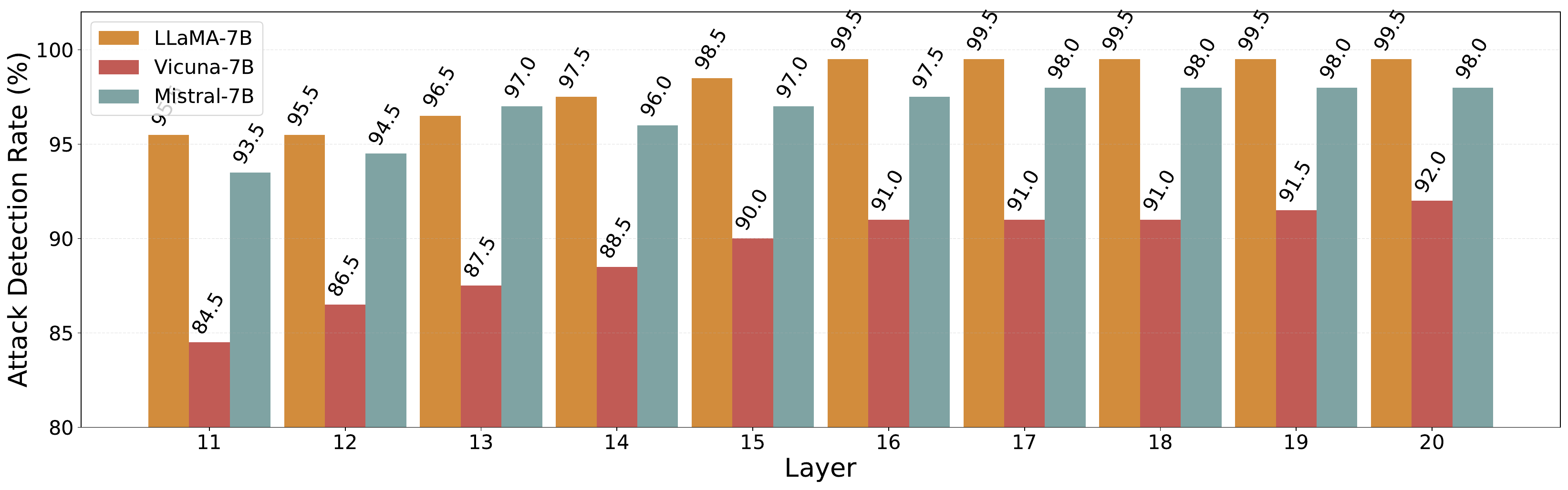}

\caption{Attack detection accuracy in the representation space using GUARD-SLM across middle layers (11--21), Dataset: HarmBench, model: meta-
llama/Llama-2-7b-chat-hf}
\label{fig:guard_slm_harmbench_middle}
\end{figure*}

\begin{figure*}[t]
\centering
\includegraphics[width=\textwidth]{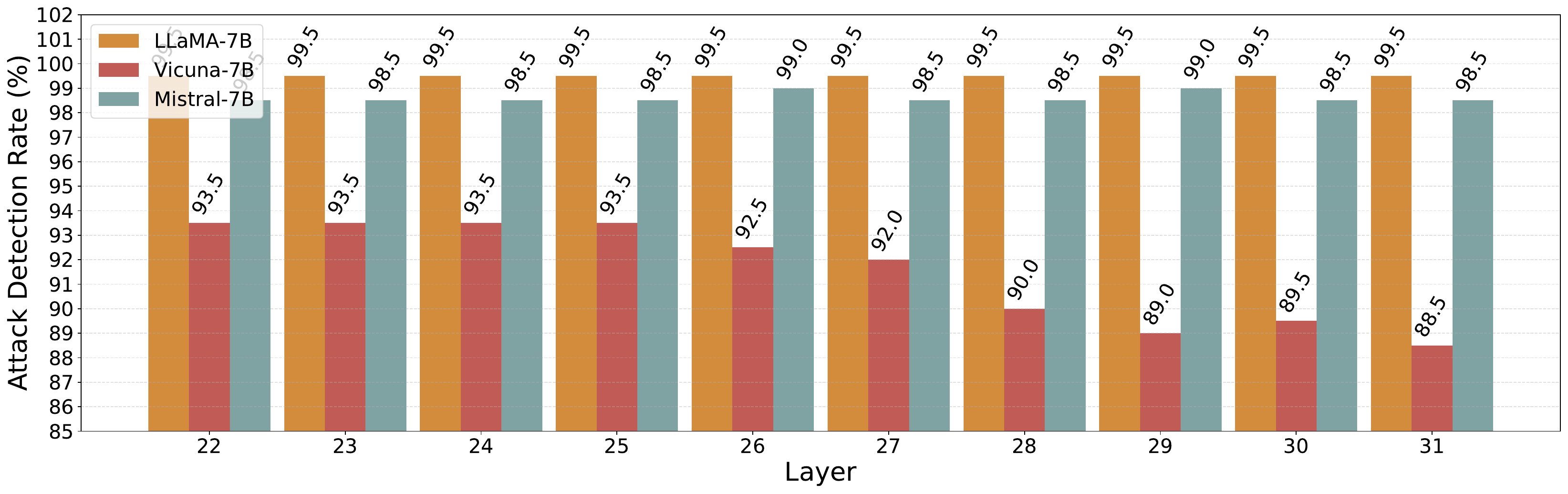}

\caption{Attack detection accuracy in the representation space using GUARD-SLM across late layers (22--31), Dataset: HarmBench, model: meta-
llama/Llama-2-7b-chat-hf}
\label{fig:guard_slm_harmbench_late}
\end{figure*}

\FloatBarrier

\balance
\subsection{GUARD-SLM in-depth result analysis}

For in-depth analysis of the proposed defense technique, we consider three instruction-tuned SLMs: LLaMA-2 (meta-llama/Llama-2-7b-chat-hf), Vicuna (lmsys/vicuna-7b-v1.5), and Mistral (mistralai/Mistral-7B-Instruct-v0.2). Tables~\ref{tab:layer_harmauto}--\ref{tab:ica_jailbroken} present the layer-wise jailbreak detection accuracy of GUARD-SLM across different attack categories for these models. As shown in Tables~\ref{tab:layer_harmauto} and \ref{tab:pair_tap}, the proposed method achieves consistently high detection accuracy across all transformer layers for both direct malicious prompts and optimized jailbreak attacks. For the HarmBench direct prompt setting, the detection accuracy gradually increases from early layers to late layers, exceeding 99\% in most middle and late layers for all three models.
For better understanding, we visualize the attack detection rate of our model in Figure~\ref{fig:guard_slm_harmbench_early}- \ref{fig:guard_slm_harmbench_late}. In contrast, for optimized jailbreak attacks such as AutoDAN, PAIR, and TAP, the detection accuracy remains almost perfect across all layers, reaching nearly 100\% for all evaluated models. This observation indicates that optimized jailbreak prompts produce highly distinguishable activation patterns in the representation space, allowing GUARD-SLM to reliably detect them even in early transformer layers.

A similar trend is observed for the remaining jailbreak categories shown in Tables~\ref{tab:gcg_cipher}, \ref{tab:deep_code}, and \ref{tab:ica_jailbroken}. For GCG, Cipher, DeepInception, CodeChameleon, ICA, and Jailbroken attacks, the detection accuracy stays close to 100\% across almost all layers for all three models. Compared to direct malicious prompts, which may partially overlap with benign representations in early layers, optimized jailbreak attacks form more consistent and separable regions throughout the network. These results confirm that jailbreak-related features are present across the entire representation space and can be detected without relying on a specific layer. Overall, the experimental results demonstrate that GUARD-SLM provides stable and robust detection performance across different models, attack strategies, and transformer layers, achieving near-perfect detection accuracy for most optimized jailbreak attacks.

\begin{table*}[t]
\centering
\caption{Layer-wise jailbreak detection accuracy using GUARD-SLM across LLaMA-7B, Vicuna-7B, and Mistral-7B models.}
\label{tab:layer_harmauto}
\small

\begin{tabular}{c|ccc|ccc}
\hline

 & \multicolumn{3}{c|}{\textbf{Acc(\%) on HarmBench}}
 & \multicolumn{3}{c}{\textbf{Acc(\%) on AutoDAN}} \\

\cline{2-7}

\textbf{Layers}
& \textbf{LLaMA-7B}
& \textbf{Vicuna-7B}
& \textbf{Mistral-7B}
& \textbf{LLaMA-7B}
& \textbf{Vicuna-7B}
& \textbf{Mistral-7B}
\\
\hline
\rowcolor{early}
0  & 53.00\% & 50.00\% & 52.52\% & 100.0\% & 100.0\% & 100.0\% \\
\rowcolor{early}
1  & 52.50\% & 53.50\% & 58.50\% & 100.0\% & 100.0\% & 100.0\% \\
\rowcolor{early}
2  & 62.00\% & 62.50\% & 68.50\% & 100.0\% & 100.0\% & 100.0\% \\
\rowcolor{early}
3  & 66.50\% & 69.50\% & 71.50\% & 100.0\% & 100.0\% & 100.0\% \\
\rowcolor{early}
4  & 75.50\% & 76.00\% & 77.50\% & 100.0\% & 100.0\% & 100.0\% \\
\rowcolor{early}
5  & 81.50\% & 77.50\% & 79.00\% & 100.0\% & 100.0\% & 100.0\% \\
\rowcolor{early}
6  & 84.50\% & 85.50\% & 82.00\% & 100.0\% & 100.0\% & 100.0\% \\
\rowcolor{early}
7  & 89.00\% & 87.00\% & 84.50\% & 100.0\% & 100.0\% & 100.0\% \\
\rowcolor{early}
8  & 89.50\% & 85.50\% & 85.50\% & 100.0\% & 100.0\% & 100.0\% \\
\rowcolor{early}
9  & 89.00\% & 82.50\% & 90.50\% & 100.0\% & 100.0\% & 100.0\% \\
\rowcolor{early}
10 & 92.50\% & 86.50\% & 90.50\% & 100.0\% & 100.0\% & 100.0\% \\
\rowcolor{middle}
11 & 95.50\% & 84.50\% & 93.50\% & 100.0\% & 100.0\% & 100.0\% \\
\rowcolor{middle}
12 & 95.50\% & 86.50\% & 94.50\% & 100.0\% & 100.0\% & 100.0\% \\
\rowcolor{middle}
13 & 96.50\% & 87.50\% & 97.00\% & 100.0\% & 100.0\% & 100.0\% \\
\rowcolor{middle}
14 & 97.50\% & 88.50\% & 96.00\% & 100.0\% & 100.0\% & 100.0\% \\
\rowcolor{middle}
15 & 98.50\% & 90.00\% & 97.00\% & 100.0\% & 100.0\% & 100.0\% \\
\rowcolor{middle}
16 & 99.50\% & 91.00\% & 97.50\% & 100.0\% & 100.0\% & 100.0\% \\
\rowcolor{middle}
17 & 99.50\% & 91.00\% & 98.00\% & 100.0\% & 100.0\% & 100.0\% \\
\rowcolor{middle}
18 & 99.50\% & 91.00\% & 98.00\% & 100.0\% & 100.0\% & 100.0\% \\
\rowcolor{middle}
19 & 99.50\% & 91.50\% & 98.00\% & 100.0\% & 100.0\% & 100.0\% \\
\rowcolor{middle}
20 & 99.50\% & 92.00\% & 98.00\% & 100.0\% & 100.0\% & 100.0\% \\
\rowcolor{middle}
21 & 99.50\% & 92.50\% & 98.50\% & 100.0\% & 100.0\% & 100.0\% \\
\rowcolor{late}
22 & 99.50\% & 93.50\% & 98.50\% & 100.0\% & 100.0\% & 100.0\% \\
\rowcolor{late}
23 & 99.50\% & 93.50\% & 98.50\% & 100.0\% & 100.0\% & 100.0\% \\
\rowcolor{late}
24 & 99.50\% & 93.50\% & 98.50\% & 100.0\% & 100.0\% & 100.0\% \\
\rowcolor{late}
25 & 99.50\% & 93.50\% & 98.50\% & 100.0\% & 100.0\% & 100.0\% \\
\rowcolor{late}
26 & 99.50\% & 92.50\% & 99.00\% & 100.0\% & 100.0\% & 100.0\% \\
\rowcolor{late}
27 & 99.50\% & 92.00\% & 98.50\% & 100.0\% & 100.0\% & 100.0\% \\
\rowcolor{late}
28 & 99.50\% & 90.00\% & 98.50\% & 100.0\% & 100.0\% & 100.0\% \\
\rowcolor{late}
29 & 99.50\% & 89.00\% & 99.00\% & 100.0\% & 100.0\% & 100.0\% \\
\rowcolor{late}
30 & 99.50\% & 89.50\% & 98.50\% & 100.0\% & 100.0\% & 100.0\% \\
\rowcolor{late}
31 & 99.50\% & 88.50\% & 98.50\% & 100.0\% & 100.0\% & 100.0\% \\

\hline
\end{tabular}

\end{table*}

\begin{table*}[t]
\centering
\caption{Layer-wise jailbreak detection accuracy using GUARD-SLM across LLaMA-7B, Vicuna-7B, and Mistral-7B models.}
\label{tab:pair_tap}
\small

\begin{tabular}{c|ccc|ccc}
\hline

 & \multicolumn{3}{c|}{\textbf{Acc(\%) on PAIR}} 
 & \multicolumn{3}{c}{\textbf{Acc(\%) on TAP}} \\

\cline{2-7}

\textbf{Layers}
& \textbf{LLaMA-7B}
& \textbf{Vicuna-7B}
& \textbf{Mistral-7B}
& \textbf{LLaMA-7B}
& \textbf{Vicuna-7B}
& \textbf{Mistral-7B}

\\
\hline
\rowcolor{early}
0  & 100.0\% & 100.0\% & 100.0\% & 100.0\% & 100.0\% & 100.0\% \\
\rowcolor{early}
1  & 100.0\% & 99.50\% & 100.0\% & 100.0\% & 100.0\% & 100.0\% \\
\rowcolor{early}
2  & 100.0\% & 100.0\% & 100.0\% & 100.0\% & 100.0\% & 100.0\% \\
\rowcolor{early}
3  & 100.0\% & 100.0\% & 100.0\% & 100.0\% & 100.0\% & 100.0\% \\
\rowcolor{early}
4  & 100.0\% & 100.0\% & 100.0\% & 100.0\% & 100.0\% & 100.0\% \\
\rowcolor{early}
5  & 100.0\% & 100.0\% & 100.0\% & 100.0\% & 100.0\% & 100.0\% \\
\rowcolor{early}
6  & 100.0\% & 100.0\% & 100.0\% & 100.0\% & 100.0\% & 100.0\% \\
\rowcolor{early}
7  & 100.0\% & 100.0\% & 100.0\% & 100.0\% & 100.0\% & 100.0\% \\
\rowcolor{early}
8  & 100.0\% & 100.0\% & 100.0\% & 100.0\% & 100.0\% & 100.0\% \\
\rowcolor{early}
9  & 100.0\% & 100.0\% & 100.0\% & 100.0\% & 100.0\% & 100.0\% \\
\rowcolor{early}
10 & 100.0\% 
& 100.0\% & 100.0\% & 100.0\% & 100.0\% & 100.0\% \\
\rowcolor{middle}
11 & 100.0\% & 100.0\% & 100.0\% & 100.0\% & 100.0\% & 100.0\% \\
\rowcolor{middle}
12 & 100.0\% & 100.0\% & 100.0\% & 100.0\% & 100.0\% & 100.0\% \\
\rowcolor{middle}
13 & 100.0\% & 100.0\% & 100.0\% & 100.0\% & 100.0\% & 100.0\% \\
\rowcolor{middle}
14 & 100.0\% & 100.0\% & 100.0\% & 100.0\% & 100.0\% & 100.0\% \\
\rowcolor{middle}
15 & 100.0\% & 100.0\% & 100.0\% & 100.0\% & 100.0\% & 100.0\% \\
\rowcolor{middle}
16 & 100.0\% & 100.0\% & 100.0\% & 100.0\% & 100.0\% & 100.0\% \\
\rowcolor{middle}
17 & 100.0\% & 100.0\% & 100.0\% & 100.0\% & 100.0\% & 100.0\% \\
\rowcolor{middle}
18 & 100.0\% & 99.50\% & 100.0\% & 100.0\% & 100.0\% & 100.0\% \\
\rowcolor{middle}
19 & 100.0\% & 99.50\% & 99.50\% & 100.0\% & 100.0\% & 100.0\% \\
\rowcolor{middle}
20 & 100.0\% & 99.50\% & 99.50\% & 100.0\% & 100.0\% & 100.0\% \\
\rowcolor{middle}
21 & 100.0\% & 99.50\% & 99.50\% & 100.0\% & 99.50\% & 100.0\% \\
\rowcolor{late}
22 & 100.0\% & 99.50\% & 99.50\% & 100.0\% & 99.50\% & 100.0\% \\
\rowcolor{late}
23 & 100.0\% & 99.50\% & 99.50\% & 100.0\% & 99.50\% & 100.0\% \\
\rowcolor{late}
24 & 100.0\% & 99.50\% & 99.50\% & 100.0\% & 100.0\% & 100.0\% \\
\rowcolor{late}
25 & 100.0\% & 99.50\% & 99.50\% & 100.0\% & 100.0\% & 100.0\% \\
\rowcolor{late}
26 & 100.0\% & 99.50\% & 99.50\% & 100.0\% & 100.0\% & 100.0\% \\
\rowcolor{late}
27 & 100.0\% & 99.50\% & 99.50\% & 100.0\% & 100.0\% & 100.0\% \\
\rowcolor{late}
28 & 100.0\% & 99.50\% & 99.50\% & 100.0\% & 100.0\% & 99.50\% \\
\rowcolor{late}
29 & 100.0\% & 99.50\% & 99.50\% & 100.0\% & 100.0\% & 100.0\% \\
\rowcolor{late}
30 & 100.0\% & 99.50\% & 99.50\% & 100.0\% & 100.0\% & 100.0\% \\
\rowcolor{late}
31 & 100.0\% & 99.50\% & 99.50\% & 100.0\% & 100.0\% & 99.50\% \\

\hline
\end{tabular}

\end{table*}

\begin{table*}[t]
\centering
\caption{Layer-wise jailbreak detection accuracy using GUARD-SLM across LLaMA-7B, Vicuna-7B, and Mistral-7B models.}
\label{tab:gcg_cipher}
\small

\begin{tabular}{c|ccc|ccc}
\hline

 & \multicolumn{3}{c|}{\textbf{Acc(\%) on GCG}} 
 & \multicolumn{3}{c}{\textbf{Acc(\%) on Cipher}} \\

\cline{2-7}

\textbf{Layers}
& \textbf{LLaMA-7B}
& \textbf{Vicuna-7B}
& \textbf{Mistral-7B}
& \textbf{LLaMA-7B}
& \textbf{Vicuna-7B}
& \textbf{Mistral-7B}

\\
\hline
\rowcolor{early}
0  &100.0\%&100.0\%&100.0\% &100.0\%&100.0\%&100.0\%\\
\rowcolor{early}
1  &100.0\%&100.0\%&99.50\% &100.0\%&100.0\%&100.0\%\\
\rowcolor{early}
2  &100.0\%&100.0\%&100.0\% &100.0\%&100.0\%&100.0\%\\
\rowcolor{early}
3  &100.0\%&100.0\%&100.0\% &100.0\%&100.0\%&100.0\%\\
\rowcolor{early}
4  &100.0\%&100.0\%&100.0\% &100.0\%&100.0\%&100.0\%\\
\rowcolor{early}
5  &100.0\%&100.0\%&100.0\% &100.0\%&100.0\%&100.0\%\\
\rowcolor{early}
6  &100.0\%&100.0\%&100.0\% &100.0\%&100.0\%&100.0\%\\
\rowcolor{early}
7  &100.0\%&100.0\%&100.0\% &100.0\%&100.0\%&100.0\%\\
\rowcolor{early}
8  &100.0\%&100.0\%&100.0\% &100.0\%&100.0\%&100.0\%\\
\rowcolor{early}
9  &100.0\%&100.0\%&100.0\% &100.0\%&100.0\%&100.0\%\\
\rowcolor{early}
10 &100.0\%&100.0\%&100.0\% &100.0\%&100.0\%&100.0\%\\
\rowcolor{early}
11 &100.0\%&100.0\%&100.0\% &100.0\%&100.0\%&100.0\%\\
\rowcolor{middle}
12 &100.0\%&100.0\%&100.0\% &100.0\%&100.0\%&100.0\%\\
\rowcolor{middle}
13 &100.0\%&100.0\%&100.0\% &100.0\%&100.0\%&100.0\%\\
\rowcolor{middle}
14 &100.0\%&100.0\%&100.0\% &100.0\%&100.0\%&100.0\%\\
\rowcolor{middle}
15 &100.0\%&100.0\%&100.0\% &100.0\%&100.0\%&100.0\%\\
\rowcolor{middle}
16 &100.0\%&100.0\%&100.0\% &100.0\%&100.0\%&100.0\%\\
\rowcolor{middle}
17 &100.0\%&100.0\%&100.0\% &100.0\%&100.0\%&100.0\%\\
\rowcolor{middle}
18 &100.0\%&100.0\%&100.0\% &100.0\%&100.0\%&100.0\%\\
\rowcolor{middle}
19 &100.0\%&100.0\%&100.0\% &100.0\%&100.0\%&100.0\%\\
\rowcolor{middle}
20 &100.0\%&100.0\%&100.0\% &100.0\%&100.0\%&100.0\%\\
\rowcolor{middle}
21 &100.0\%&100.0\%&100.0\% &100.0\%&100.0\%&100.0\%\\
\rowcolor{late}
22 &100.0\%&100.0\%&100.0\% &100.0\%&100.0\%&100.0\%\\
\rowcolor{late}
23 &100.0\%&100.0\%&100.0\% &100.0\%&100.0\%&100.0\%\\
\rowcolor{late}
24 &100.0\%&100.0\%&100.0\% &100.0\%&100.0\%&100.0\%\\
\rowcolor{late}
25 &100.0\%&100.0\%&100.0\% &100.0\%&100.0\%&100.0\%\\
\rowcolor{late}
26 &100.0\%&100.0\%&100.0\% &100.0\%&100.0\%&100.0\%\\
\rowcolor{late}
27 &100.0\%&100.0\%&100.0\% &100.0\%&100.0\%&100.0\%\\
\rowcolor{late}
28 &100.0\%&100.0\%&100.0\% &100.0\%&100.0\%&100.0\%\\
\rowcolor{late}
29 &100.0\%&100.0\%&100.0\% &100.0\%&100.0\%&100.0\%\\
\rowcolor{late}
30 &100.0\%&100.0\%&100.0\% &100.0\%&100.0\%&100.0\%\\
\rowcolor{late}
31 &100.0\%&100.0\%&100.0\% &100.0\%&100.0\%&100.0\%\\

\hline
\end{tabular}

\end{table*}

\begin{table*}[t]
\centering
\caption{Layer-wise jailbreak detection accuracy using GUARD-SLM on DeepInception and CodeChameleon attacks across LLaMA-7B, Vicuna-7B, and Mistral-7B models.}
\label{tab:deep_code}
\small

\begin{tabular}{c|ccc|ccc}
\hline

 & \multicolumn{3}{c|}{\textbf{Acc(\%) on DeepInception}} 
 & \multicolumn{3}{c}{\textbf{Acc(\%) on CodeChameleon}} \\

\cline{2-7}

\textbf{Layers}
& \textbf{LLaMA-7B}
& \textbf{Vicuna-7B}
& \textbf{Mistral-7B}
& \textbf{LLaMA-7B}
& \textbf{Vicuna-7B}
& \textbf{Mistral-7B}

\\
\hline
\rowcolor{early}
0  &100.0\%&100.0\%&100.0\% &100.0\%&100.0\%&100.0\%\\
\rowcolor{early}
1  &100.0\%&100.0\%&100.0\% &100.0\%&100.0\%&100.0\%\\
\rowcolor{early}
2  &100.0\%&100.0\%&100.0\% &100.0\%&100.0\%&100.0\%\\
\rowcolor{early}
3  &100.0\%&100.0\%&100.0\% &100.0\%&100.0\%&100.0\%\\
\rowcolor{early}
4  &100.0\%&100.0\%&100.0\% &100.0\%&100.0\%&100.0\%\\
\rowcolor{early}
5  &100.0\%&100.0\%&100.0\% &100.0\%&100.0\%&100.0\%\\
\rowcolor{early}
6  &100.0\%&100.0\%&100.0\% &100.0\%&100.0\%&100.0\%\\
\rowcolor{early}
7  &100.0\%&100.0\%&100.0\% &100.0\%&100.0\%&100.0\%\\
\rowcolor{early}
8  &100.0\%&100.0\%&100.0\% &100.0\%&100.0\%&100.0\%\\
\rowcolor{early}
9  &100.0\%&100.0\%&100.0\% &100.0\%&100.0\%&100.0\%\\
\rowcolor{early}
10 &100.0\%&100.0\%&100.0\% &100.0\%&100.0\%&100.0\%\\
\rowcolor{early}
11 &100.0\%&100.0\%&100.0\% &100.0\%&100.0\%&100.0\%\\
\rowcolor{middle}
12 &100.0\%&100.0\%&100.0\% &100.0\%&100.0\%&100.0\%\\
\rowcolor{middle}
13 &100.0\%&100.0\%&100.0\% &100.0\%&100.0\%&100.0\%\\
\rowcolor{middle}
14 &100.0\%&100.0\%&100.0\% &100.0\%&100.0\%&100.0\%\\
\rowcolor{middle}
15 &100.0\%&100.0\%&100.0\% &100.0\%&100.0\%&100.0\%\\
\rowcolor{middle}
16 &100.0\%&100.0\%&100.0\% &100.0\%&100.0\%&100.0\%\\
\rowcolor{middle}
17 &100.0\%&100.0\%&100.0\% &100.0\%&100.0\%&100.0\%\\
\rowcolor{middle}
18 &100.0\%&100.0\%&100.0\% &100.0\%&100.0\%&100.0\%\\
\rowcolor{middle}
19 &100.0\%&100.0\%&100.0\% &100.0\%&100.0\%&100.0\%\\
\rowcolor{middle}
20 &100.0\%&100.0\%&100.0\% &100.0\%&100.0\%&100.0\%\\
\rowcolor{middle}
21 &100.0\%&100.0\%&100.0\% &100.0\%&100.0\%&100.0\%\\
\rowcolor{late}
22 &100.0\%&100.0\%&100.0\% &100.0\%&100.0\%&100.0\%\\
\rowcolor{late}
23 &100.0\%&100.0\%&100.0\% &100.0\%&100.0\%&100.0\%\\
\rowcolor{late}
24 &100.0\%&100.0\%&100.0\% &100.0\%&100.0\%&100.0\%\\
\rowcolor{late}
25 &100.0\%&100.0\%&100.0\% &100.0\%&100.0\%&100.0\%\\
\rowcolor{late}
26 &100.0\%&100.0\%&100.0\% &100.0\%&100.0\%&100.0\%\\
\rowcolor{late}
27 &100.0\%&100.0\%&100.0\% &100.0\%&100.0\%&100.0\%\\
\rowcolor{late}
28 &100.0\%&100.0\%&100.0\% &100.0\%&100.0\%&100.0\%\\
\rowcolor{late}
29 &100.0\%&100.0\%&100.0\% &100.0\%&100.0\%&100.0\%\\
\rowcolor{late}
30 &100.0\%&100.0\%&100.0\% &100.0\%&100.0\%&100.0\%\\
\rowcolor{late}
31 &100.0\%&100.0\%&100.0\% &100.0\%&100.0\%&100.0\%\\

\hline
\end{tabular}

\end{table*}

\begin{table*}[t]
\centering
\caption{Layer-wise jailbreak detection accuracy using GUARD-SLM on ICA and Jailbroken attacks across LLaMA-7B, Vicuna-7B, and Mistral-7B models.}
\label{tab:ica_jailbroken}
\small

\begin{tabular}{c|ccc|ccc}
\hline

 & \multicolumn{3}{c|}{\textbf{Acc(\%) on ICA}} 
 & \multicolumn{3}{c}{\textbf{Acc(\%) on Jailbroken}} \\

\cline{2-7}

\textbf{Layers}
& \textbf{LLaMA-7B}
& \textbf{Vicuna-7B}
& \textbf{Mistral-7B}
& \textbf{LLaMA-7B}
& \textbf{Vicuna-7B}
& \textbf{Mistral-7B}

\\
\hline
\rowcolor{early}
0  &100.0\%&100.0\%&100.0\% &86.36\%&87.64\%&86.50\%\\
\rowcolor{early}
1  &100.0\%&100.0\%&100.0\% &88.57\%&85.33\%&88.02\%\\
\rowcolor{early}
2  &100.0\%&100.0\%&100.0\% &86.97\%&86.03\%&80.60\%\\
\rowcolor{early}
3  &100.0\%&100.0\%&100.0\% &84.78\%&85.50\%&86.66\%\\
\rowcolor{early}
4  &100.0\%&100.0\%&100.0\% &86.14\%&87.03\%&93.21\%\\
\rowcolor{early}
5  &100.0\%&100.0\%&100.0\% &89.78\%&92.24\%&93.93\%\\
\rowcolor{early}
6  &100.0\%&100.0\%&100.0\% &92.57\%&94.31\%&95.62\%\\
\rowcolor{early}
7  &100.0\%&100.0\%&100.0\% &93.26\%&93.98\%&93.86\%\\
\rowcolor{early}
8  &100.0\%&100.0\%&100.0\% &96.47\%&93.22\%&94.71\%\\
\rowcolor{early}
9  &100.0\%&100.0\%&100.0\% &97.97\%&92.19\%&96.24\%\\
\rowcolor{early}
10 &100.0\%&100.0\%&100.0\% &98.48\%&93.78\%&97.12\%\\
\rowcolor{early}
11 &100.0\%&100.0\%&100.0\% &98.83\%&95.14\%&96.88\%\\
\rowcolor{middle}
12 &100.0\%&100.0\%&100.0\% &99.00\%&95.66\%&98.33\%\\
\rowcolor{middle}
13 &100.0\%&100.0\%&100.0\% &99.26\%&97.17\%&98.72\%\\
\rowcolor{middle}
14 &100.0\%&100.0\%&100.0\% &99.24\%&94.95\%&98.74\%\\
\rowcolor{middle}
15 &100.0\%&100.0\%&100.0\% &99.12\%&94.69\%&98.67\%\\
\rowcolor{middle}
16 &100.0\%&100.0\%&100.0\% &99.19\%&94.67\%&97.86\%\\
\rowcolor{middle}
17 &100.0\%&100.0\%&100.0\% &99.10\%&94.78\%&98.05\%\\
\rowcolor{middle}
18 &100.0\%&100.0\%&100.0\% &99.05\%&93.93\%&96.79\%\\
\rowcolor{middle}
19 &100.0\%&100.0\%&100.0\% &98.97\%&93.48\%&96.64\%\\
\rowcolor{middle}
20 &100.0\%&100.0\%&100.0\% &99.02\%&92.93\%&96.43\%\\
\rowcolor{middle}
21 &100.0\%&100.0\%&100.0\% &98.79\%&92.59\%&96.74\%\\
\rowcolor{late}
22 &100.0\%&100.0\%&100.0\% &98.52\%&93.28\%&96.09\%\\
\rowcolor{late}
23 &100.0\%&100.0\%&100.0\% &98.10\%&92.31\%&95.72\%\\
\rowcolor{late}
24 &100.0\%&100.0\%&100.0\% &97.79\%&92.05\%&95.43\%\\
\rowcolor{late}
25 &100.0\%&100.0\%&100.0\% &97.81\%&92.09\%&95.53\%\\
\rowcolor{late}
26 &100.0\%&100.0\%&100.0\% &98.14\%&91.07\%&94.57\%\\
\rowcolor{late}
27 &100.0\%&100.0\%&100.0\% &98.21\%&90.64\%&94.62\%\\
\rowcolor{late}
28 &100.0\%&100.0\%&100.0\% &98.03\%&90.88\%&94.78\%\\
\rowcolor{late}
29 &100.0\%&100.0\%&100.0\% &98.22\%&91.17\%&94.33\%\\
\rowcolor{late}
30 &100.0\%&100.0\%&100.0\% &98.16\%&91.15\%&94.26\%\\
\rowcolor{late}
31 &100.0\%&100.0\%&100.0\% &98.47\%&90.76\%&93.38\%\\

\hline
\end{tabular}

\end{table*}

\end{document}